\documentclass[3p,times]{elsarticle}

\usepackage{ecrc}
\usepackage{amsmath}
\usepackage{color}
\usepackage{epstopdf}
\usepackage{comment}
\usepackage[section]{placeins} 
\def\beq{\begin{equation}}
\def\eeq{\end{equation}}

\newcommand{\gt}{>}

\renewcommand{\v}[1]{\textbf{#1}}
\renewcommand{\rm}[1]{\textrm{#1}}
\renewcommand{\d}{\mathrm{d}}

\def\esym{$E_{\rm{sym}}(\rho)$~}

\def\es0{$E_{\rm{sym}}(\rho_0)$~}

\def\us0{$U_{\rm{sym}}(\rho_0,k_{\rm{F}})$~}

\def\l0{$L(\rho_0)$~}

\def\emass{$m^*_{\rm{n-p}}(\rho_0,\delta)$~}

\def\beq{\begin{eqnarray}}
\def\eeq{\end{eqnarray}}
\def\bea{\begin{eqnarray}}
\def\eea{\end{eqnarray}}
\def\beqa{\begin{eqnarray}\begin{array}{l}}
\def\eeqa{\end{array}\end{eqnarray}}

\def\barr{\left(\begin{array}{c}}
\def\earr{\end{array}\right)}
\def\bmat{\left(\begin{array}{cc}}
\def\emat{\end{array}\right)}

\def\mathscr{\mathcal}

\def\3d{3-D}

\newcommand{\mev}{\,\textrm{MeV}}

\volume{00}

\firstpage{1}

\journalname{Progress in Particle and Nuclear Physics}

\runauth{Bao-An Li, Bao-Jun Cai, Lie-Wen Chen and Jun Xu}

\jid{}

\jnltitlelogo{\bf\small Progress in Particle and Nuclear Physics}

\CopyrightLine{2011}{Published by Elsevier Ltd.}

\usepackage{amssymb}
\usepackage{graphicx}

\usepackage[figuresright]{rotating}

\begin{document}
\renewcommand\thefigure{\arabic{figure}}

\begin{frontmatter}



\dochead{}

\title{Nucleon Effective Masses in Neutron-Rich Matter}

\author[TAMUc]{Bao-An~Li$^{*1,}$}\footnote{Email:Bao-An.Li@Tamuc.edu},
\author[SHU,SJTU]{Bao-Jun Cai$^{2,}$}\footnote{Email:bjcai87@gmail.com}
\author[SJTU]{Lie-Wen~Chen$^{3,}$}\footnote{Email:Lwchen@Sjtu.edu.cn}
\author[SINAP]{ Jun Xu$^{4,}$}\footnote{Email:xujun@sinap.ac.cn},

\address[TAMUc]{Department of Physics and Astronomy, Texas A\&M University-Commerce, Commerce,
Texas 75429-3011, USA}

\address[SHU]{Department of Physics, Shanghai University, Shanghai 200444, China}

\address[SJTU]{School of Physics and Astronomy and Shanghai Key Laboratory for Particle Physics and Cosmology,
Shanghai Jiao Tong University, Shanghai 200240, China}

\address[SINAP]{Shanghai Institute of Applied Physics, Chinese Academy of Sciences, Shanghai 201800, China}

\begin{abstract}
Various kinds of isovector nucleon effective masses are used in the literature to characterize the momentum/energy dependence of the nucleon symmetry potential or self-energy due to the space/time non-locality of the underlying isovector strong interaction in neutron-rich nucleonic matter. The multifaceted studies on nucleon isovector effective masses are multi-disciplinary in nature. Besides structures, masses and low-lying excited states of nuclei as well as nuclear reactions, studies of the isospin dependence of short-range correlations in nuclei from scatterings of high-energy electrons and protons on heavy nuclei also help understand nucleon effective masses especially the so-called E-mass in neutron-rich matter.  A thorough understanding of all kinds of nucleon effective masses has multiple impacts on many interesting issues in both nuclear physics and astrophysics. Indeed, essentially all microscopic many-body theories and phenomenological models with various nuclear forces available in the literature have been used to calculate single-nucleon potentials and the associated nucleon effective masses in neutron-rich matter. There are also fundamental principles connecting different aspects and impacts of isovector strong interactions. In particular, the Hugenholtz-Van Hove theorem connects analytically nuclear symmetry energy with both isoscalar and isovector nucleon effective masses as well as their own momentum dependences.  It also reveals how the isospin-quartic term in the equation of state of neutron-rich matter depends on the high-order momentum-derivatives of both isoscalar and isovector nucleon potentials. The Migdal--Luttinger theorem facilitates the extraction of nucleon E-mass and its isospin dependence from experimentally constrained single-nucleon momentum distributions. The momentum/energy dependence of the symmetry potential and the corresponding neutron-proton effective mass splitting also affect transport properties and the liquid-gas phase transition in neutron-rich matter. Moreover, they influence the dynamics and isospin-sensitive observables of heavy-ion collisions through both the Vlasov term and the collision integrals of the Boltzmann-Uehling-Uhlenbeck transport equation. We review here some of the significant progresses made in recent years by the nuclear physics community in resolving some of the hotly debated and longstanding issues regarding nucleon effective masses especially in dense neutron-rich matter. We also point out some of the remaining key issues requiring further investigations in the era of high precision experiments using advanced rare isotope beams.
\end{abstract}

\begin{keyword}
Nucleon effective mass \sep Isospin \sep Equation of state \sep Neutron-rich matter \sep Nuclear symmetry
energy \sep Heavy-ion reactions \sep Transport model \sep Viscosity\sep Thermal conductivity
\sep Optical potential \sep Short-range correlation \sep Neutron stars
\PACS 21.65.Cd \sep 21.65.Ef \sep 25.70.-z \sep 21.30.Fe \sep 21.10.Gv
\sep 26.60-c. \\
\end{keyword}
\end{frontmatter}
\tableofcontents

\begin{center}
\begin{table}[h!]
\centering\small
\centerline{\large Main Acronyms Used} \vspace*{.2cm}
\begin{tabular}{|l|l|l|l|}
\hline
$\delta$, $\beta$, $\alpha$&isospin asymmetry, $(\rho_{\rm{n}}-\rho_{\rm{p}})/(\rho_{\rm{n}}+\rho_{\rm{p}})$
&ANM&asymmetric nuclear matter\\\hline
BCPM&Barcelona-Catania-Paris-Madrid
&BCS&Bardeen-Cooper-Schrieffer\\\hline
BGBD&Bombaci-Gale-Bertsch-Das Gupta
&BHF&Brueckner-Hartree-Fock\\\hline
CBF&correlated-base function
&ChPT&chiral perturbative theory\\\hline
CMS&center of mass system
&CP&critical point\\\hline
DBHF&Dirac-Brueckner-Hartree-Fock
&EMC&European Muon Collaboration\\\hline
EOS&equation of state
&FFG&free Fermi gas\\\hline
GBD&Gale-Bertsch-Das Gupta
&GHF&Gogny-Hartree-Fock\\\hline
HMT&high momentum tail
&HVH&Hugenholtz-Van Hove\\\hline
IBUU&isospin Boltzmann-Uehling-Uhlenbeck
&ImMDI&improved momentum-dependent interaction\\\hline
ImQMD&improved quantum molecular dynamics
&IPM&independent particle model\\\hline
ISGQR&isoscalar giant quadrupole resonance
&IVGDR&isovector giant dipole resonance\\\hline
LIGO& Laser Interferometer Gravitational-Wave Observatory
&MA&maximum asymmetry\\\hline
MDI&momentum-dependent interaction
&MFP&mean free path\\\hline
MID&momentum-independent interaction
&NLO&next-to-leading order\\\hline
NN&nucleon-nucleon
&OPE&one-pion-exchange\\\hline
PNM&pure neutron matter
&QMC&quantum Monte Carlo\\\hline
QMD&quantum molecular dynamics
&RFG&relativistic Fermi gas\\\hline
RHF&relativistic Hartree-Fock
&RIA&relativistic impulse approximation\\\hline
RMF&relativistic mean field
&RPA&random phase approximation\\\hline
SCGF&self-consistent Green's function
&SEP&Schr$\ddot{\rm{o}}$dinger equivalent potential\\\hline
SHF&Skyrme-Hartree-Fock
&SNM&symmetric nuclear matter\\\hline
SRC & short range correlation
&TBF&three body force\\\hline
TOV&Tolman-Oppenheimer-Volkoff
&UNEDF&universal nuclear energy density functional\\\hline
VCS&variational chain summation
&VMC&variational Monte Carlo\\\hline
\end{tabular}
\end{table}
\end{center}

\centerline{\large Notations Used for Nucleon Effective Masses}
To avoid causing confusions while use as much as possible the original conventions of various authors to describe the different kinds of nucleon
effective masses, we use the capital $M^*$ with various subscripts or superscripts to denote the isospin and density-dependent effective masses. The $M^*_0=M^*_{\rm{s}}$ are reserved for the effective mass at arbitrary densities
in SNM. While their values at the saturation density $\rho_0$ are denoted with a lower-case $m^*$ with different subscripts. For example, $m^*_0=m^*_{\rm{s}}$ and $m^*_{\rm{v}}$ are respectively the nucleon isoscalar and isovector effective masses at $\rho_0$. $M$ and sometimes $m$ are used interchangeably ($M=m$) as the average nucleon mass in free-space. Since we do not change the labels of figures adopted from others, there are still some inconsistencies that we shall try to reduce in the figure captions. The $p$, $k$ and $q$ are used interchangeably as the momentum or wave number.

\section{Introduction}
The concept of nucleon effective mass $M^*$ was originally developed by Brueckner\,\cite{Bru} to describe equivalently the motion of nucleons in a momentum-dependent potential with the motion of a quasi-nucleon of mass
$M^*$ in a momentum-independent potential. It has been generalized later and widely used to characterize the momentum and/or energy dependence of the single-nucleon potential or the real part of the nucleon self-energy in nuclear medium\,\cite{Jeu76,Mah85,Jam89}. Basically, the nucleon effective mass reflects leading effects of the space-time non-locality of the underlying nuclear interactions and the Pauli exchange principle. It is a fundamental quantity characterizing the nucleon's propagation in nuclear medium\,\cite{Jeu76,Jam89,Sjo76,Neg81}. In particular, the so-called k-mass and E-mass describe, respectively, the momentum and energy dependence of the  single-nucleon potential while the total effective mass is the product of the two. Moreover, the E-mass is also directly related to the discontinuity of the single-nucleon momentum distribution at the Fermi surface of nuclear matter\,\cite{Jeu76,Mig57,Gal58,Lut60,Czy61,Bel61,Sar80,Bla81,Kro81,Jac82,Heb09}. Because of its fundamental nature, importance and broad impacts in many areas of nuclear physics and astrophysics, the study on nucleon effective masses has a long and rich history. Accompanying essentially every effort of understanding properties of complex nuclei, their collective motions and excitations as well as reactions using various theories and interactions, one often encounters the concept and issues of nucleon effective masses. Interested readers can find several historical reviews on nucleon effective masses in the literature, see, e.g.,\,\cite{Jeu76,Mah85,Jam89}.

Thanks to the advancement in astrophysical observations, especially properties of neutron stars, and new experiments at various rare isotope beam facilities, more efforts were devoted in recent years to understanding nucleon effective masses in neutron-rich matter. One particularly interesting quantity is the total neutron-proton effective mass splitting $m^*_{\rm{n-p}}$ and its dependence on the density and isospin asymmetry of the medium. As we shall discuss in detail, the most critical but still poorly known physics for understanding nucleon effective masses in neutron-rich matter is the momentum/energy dependence of nucleon isovector (symmetry) potential. The latter determines the  $m^*_{\rm{n-p}}$ which has many ramifications in both nuclear physics and astrophysics. In fact, whether the effective mass for neutrons is higher, equal to or lower than that for protons in neutron-rich matter has been a longstanding question with conflicting answers. Interestingly, several studies in recent years have found circumstantial evidences that the $m^*_{\rm{n-p}}$ is positive in neutron-rich matter and proportional to the isospin asymmetry of the medium at nuclear matter saturation density $\rho_0$. However, there are still many issues especially at supra-saturation densities for high-momentum nucleons. Further investigations into nucleon effective masses at various densities and momenta in neutron-rich matter will have important ramifications in addressing many unresolved issues\,\cite{mrs,Cha05,LCK08,LiChen15,NST16}. For example, the equilibrium neutron/proton ratio in primordial nucleosynthesisis depends on the $m^*_{\rm{n-p}}$\,\cite{Ste06}. It is also important for understanding neutrino opacities\,\cite{Bur06} and several properties of neutron stars\,\cite{Yak01,Pag06,Bal14}, thermal and transport properties as well as the liquid-gas phase transition in neutron-rich matter\,\cite{Jxu15a,Beh11,Jxu15b}, properties of mirror nuclei\,\cite{Nol69}, locations of the neutron and proton drip-lines\,\cite{Wod97}, energy level densities of reaction partners\,\cite{Cha05}, reaction dynamics and several isospin-sensitive observables in heavy-ion collisions\,\cite{LiBA04,LiChen05,Riz05,Gio10,Feng12,Zhang14,Xie14}.

Indeed, some impressive progresses have been made over the last decade while many challenging issues remain to be resolved to fully understand nucleon effectives masses in neutron-rich matter. Brief reviews by some of us on the major issues of this topic can be found in refs.\,\cite{LCK08,LiChen15,NST16}. The multifaceted studies on nucleon effective masses are multi-disciplinary, with many interesting results obtained by many people and they have multiple impacts in both nuclear physics and astrophysics. Given our limited knowledge in this field,  while trying to be inclusive we shall focus on the selected topics listed in the Table of Contents of this review.
We notice that earlier predictions on nucleon effective masses by various many-body theories were reviewed extensively by some of us about 10 years ago\,\cite{LCK08} and are thus mostly skipped here unless necessary for comparisons.

\section{Definitions of different kinds of nucleon effective masses}\label{definition}
In this section, after first recalling the basic definitions of several different kinds of nucleon effective masses used in the literature, we define the neutron-proton effective mass splitting and discuss the main physics behind it.

\subsection{Nucleon effective masses in non-relativistic models}
We focus on non-relativistic nucleon effective masses or the ones derived from the Schr\"odinger equivalent single-particle potential in relativistic models. The k-mass $M^{\ast,\rm{k}}_{J}$ and E-mass $M^{\ast,\rm{E}}_{J}$ of a nucleon $J=\rm{n/p}$ can be obtained from the momentum and energy dependence of the single-nucleon potential $U_{J} (\rho,\delta, k,E)$ in nucleonic matter of density $\rho$ and isospin asymmetry $\delta\equiv (\rho_{\rm{n}}-\rho_{\rm{p}})/\rho$ via\,\cite{Jeu76,Mah85,Jam89}
\begin{equation}
\frac{M^{\ast,\rm{k}}_{J}}{M}=\left[1+\frac{M}{k}\frac{\partial U_{J}}{\partial k}\right]^{-1}~\rm{and}~
\frac{M^{\ast,\rm{E}}_{J}}{M}=1-\frac{\partial U_{J}}{\partial E}
\end{equation}
where $M$ (and sometimes $m$ in some literature) is the average mass of nucleons in free-space and $k=|\v{k}|$ is the magnitude of nucleon momentum. They have been shown to reflect respectively the space and time non-locality of the underlying nuclear interactions\,\cite{Jam89}.
Once an on-shell energy-momentum dispersion relation $E(k)$ or $k(E)$ is obtained from solving the equation $\rm{E}_J=k^2/2M+U_{J} (\rho,\delta,k,\rm{E})$, an equivalent single-particle potential depending on either momentum or energy can be obtained. The total effective mass $M^{\ast}_{J}$
\begin{equation}\label{em1}
\frac{M^{*}_{J}}{M}=1-\frac{d U_{J}(\rho,\delta,k(E),E)}{dE}\Bigg|_{E(k_{\rm{F}}^{J})}
=\left[1+\frac{M}{k_{\rm{F}}^J}\frac{d U_{J}(\rho,\delta,k,E(k))}{d k}\Bigg|_{k_{\rm{F}}^J}\right]^{-1}
\end{equation}
then characterizes equivalently either the momentum or energy dependence of the single-nucleon potential. Consequently, the total effective mass $M^{\ast}_{J}$ is the only one extractable from either the first or the second part of the above equation. Nevertheless, as we shall discuss later, the so-called Migdal jump at the Fermi surface allows an independent extraction from experimentally constrained quasi-nucleon momentum distribution.
In the above equation, one has $k_{\rm{F}}^J=(1+\tau_3^J\delta)^{1/3} k_{\rm{F}}$ with
$k_{\rm{F}}=(3\pi^2\rho/2)^{1/3}$ being the nucleon Fermi momentum
in symmetric nuclear matter (SNM) at density $\rho$, $\tau_3^{J}=+1$ ($-1$) for neutrons (protons).
Noticing that\,\cite{Jeu76}
\begin{equation}
\frac{dE}{dk}\equiv \frac{k}{M_J^*}=\frac{k}{M}+\frac{\partial U}{\partial k}+\frac{\partial U}{\partial E}\cdot \frac{dE}{dk},
\end{equation}
the well-known relation $\frac{M^{\ast}_{J}}{M}=\frac{M^{\ast,\rm{E}}_{J}}{M}\cdot \frac{M^{\ast,\rm{k}}_{J}}{M}$
can then be readily proven.

For instance, the low-energy dispersion relation in the nucleon optical models\,\cite{ZHLI,LiX13,LiX15} reads $ E={k^2}/{2M}+U(E)$ where $U(E)=S_0+S_1E$ is the single nucleon potential
with $S_0$ and $S_1$ being two constants determined by experiments. If we rewrite $U(E)$ without changing its value by adding two terms, i.e., $-\gamma E+\gamma E$ with a constant
$\gamma$ ($\neq-1$), then the energy can be rewritten as $ E={k^2}/{2M}+S_0-\gamma E+(S_1+\gamma)E$. After substituting $E=(k^2/2M+S_0)/(1-S_1)$ from the original dispersion relation into the $(S_1+\gamma)E$ term,
an equivalent potential $\widehat{U}(k,E)$ with clearly separated momentum and energy dependences can then be defined through
\begin{equation}
E=\frac{k^2}{2M}+\widehat{U}(k,E) ,~~{\rm{with}}~~
\widehat{U}(k,E)=S_0-\gamma
E+\frac{S_1+\gamma}{1-S_1}\left(\frac{k^2}{2M}+S_0\right).
\end{equation}
Then a $\gamma$-dependent E-mass and k-mass is readily obtained as, respectively,
\begin{equation}
{M^{\ast,\rm{E}}}/M=1+\gamma,~~{M^{\ast,\rm{k}}}/M=\frac{1-S_1}{1+\gamma},\end{equation}
while the total effective mass remains as $M^{\ast}/M=({M^{\ast,\rm{E}}}/M)\cdot({M^{\ast,\rm{k}}}/M)=1-S_1$. It illustrates that although the
${M^{\ast,\rm{E}}}/M$ and ${M^{\ast,\rm{k}}}/M$ can take almost any value (by varying the parameter $\gamma$),
the total effective mass $M^{\ast}/M$ is fixed by the on-shell dispersion relation itself independent of the parameter $\gamma$ introduced.

We notice that most experiments and phenomenological models probe only the total
effective mass\,\cite{Jeu76,Mah85,Jam89,LiChen15,ZHLI,LiX13,LiX15,Zha16}.
It is also worth noting that in simulating heavy-ion collisions using transport models, people often directly use phenomenological potentials as an input. In the recent literature especially those exploring the density dependence of nuclear symmetry energy, various forms are used to model the density and momentum dependences of the isoscalar and isovector potentials. The above expressions facilitate the exploration of the nucleon effective masses underlying the phenomenological potentials employed in transport models.

The nucleon effective masses have been studied very extensively within various energy density functionals especially with many Skyrme interactions, see, e.g., refs.\,\cite{Zha16,Cha97,Far01,Ben03,Cao06,Les06,Sto07,Klu09,Ou11,Dut12}. In the standard Skyrme-Hartree-Fock (SHF) approaches, the total nucleon effective masses are often written in terms of the non-relativistic
nucleon isoscalar and isovector masses, respectively, given by
\begin{eqnarray}
&&\frac{M^*_{\rm{s}}}{M}=\left(1+\frac{M}{8\hbar^2}\rho\Theta_{\rm{s}}\right)^{-1}\equiv(1+\kappa_{\rm{s}})^{-1},\\
&&\frac{M^*_{\rm{v}}}{M}=\left(1+\frac{M}{4\hbar^2}\rho\Theta_{\rm{v}}\right)^{-1}\equiv(1+\kappa_{\rm{v}})^{-1},
\end{eqnarray}
where $\Theta_{\rm{s}}=[3t_1+(5+4x_2)t_2]$ and $\Theta_{\rm{v}}=t_1(x_1+2)+t_2(x_2+2)$ in terms of the standard Skyrme parameters\,\cite{Cha97}. By definition, the $\kappa_{\rm{s}}$ and $\kappa_{\rm{v}}$ measure respectively the reduction of $M^*_{\rm{s}}$ and $M^*_{\rm{v}}$ with respect to the nucleon free mass $M$. In the literature, the $\kappa_{\rm{v}}$ is also known as the enhancement factor of the Thomas-Reiche-Kuhn sum rule (e.g., due to the exchange and momentum dependent force), see, e.g., refs.\,\cite{Sto07,Klu09,Dut12,Peter,Rei15,Ois16}.
In asymmetric nuclear matter (ANM), the effective mass of nucleon $J$ can be expressed in terms of $M^*_{\rm{s}}$ and $M^*_{\rm{v}}$ as
\begin{equation}
\frac{1}{M^*_J} \equiv \frac{1}{M^*_{\rm{s}}}
		+ \tau_3^J\delta~\left(\frac{1}{M^*_{\rm{s}}} - \frac{1}{M^*_{\rm{v}}} \right).
\end{equation}
Using the standard Skyrme parameters, it can be rewritten as
\begin{equation}
\frac{M^*_J}{M}=\left[1+\frac{M}{8\hbar^2}\rho\Theta_{\rm{s}}-\frac{M}{8\hbar^2}\tau_3^J(2\Theta_{\rm{v}}-\Theta_{\rm{s}})\delta\rho\right]^{-1}.
\end{equation}
In the Landau Fermi-liquid theory, the isoscalar and isovector nucleon effective masses are related to two Landau parameters $F_1$ and $F^{\prime}_1$ via
\begin{eqnarray}
&&\frac{M^*_{\rm{s}}}{M}=1+\frac{F_1}{3},\\
&&\frac{M^*_{\rm{v}}}{M}=1+\frac{F^{\prime}_1}{3}.
\end{eqnarray}
The stability of nuclear matter requires $F_1\gt -3$ and $F^{\prime}_1\gt -3$, which is equivalent to requiring both isoscalar and isovector
nucleon effective masses stay positive at all densities \cite{zzcl16}.

\subsection{Non-relativistic nucleon effective masses from relativistic models}\label{Rmass}
As discussed extensively in the literature, see, e.g., refs.\,\cite{Jam89,mu04,Fuc05,zuo05,Ron06,Fuc06,Long06,Che07,Ang16},
while several different kinds of nucleon effective masses with very different physical meanings
have been widely used for various purposes in relativistic models, it is the so-called Lorentz mass\,\cite{Jam89}
\begin{equation}
M_{\mathrm{Lorentz},J}^{\ast }\equiv M(1-dU_{\rm{SEP},J}/dE_J),
\end{equation}
that should be compared with the non-relativistic effective mass $M^{*}_{J}$ of nucleon $J$ with a free mass of $M_J$.
The Schr\"{o}dinger-equivalent potential $U_{\mathrm{SEP},J}$ is determined by the nucleon scalar self-energy $\Sigma _{J}^{\rm{S}}$ and the
time-like component of the vector self-energy $\Sigma _{J}^{0}$ via\,\cite{Jam89}
\begin{equation}
U_{\mathrm{SEP},J } =\Sigma_{J }^{\rm{S}}+\frac{1}{2M}\left[\left(\Sigma _{J}^{\rm{S}}\right)^{2}-\left(\Sigma _{J}^{0}\right)^{2}\right]+\frac{\Sigma
_{J}^{0}} {M}E_{J }
=\Sigma _{J}^{\rm{S}}+\Sigma _{J }^{0}+\frac{1}{2M}\left[\left(\Sigma _{J }^{\rm{S}}\right)^{2}-\left(\Sigma _{J }^{0}\right)^{2}\right]
+\frac{\Sigma _{J }^{0}}{M}E_{\mathrm{kin}}
\label{Usep}
\end{equation}
where $E_{\mathrm{kin}}=E_{J}-M$ with $E_{J}$ being its
total single-particle energy. A non-relativistic mass $M_{\rm{NR},J}^{\ast }$
defined as
\begin{eqnarray}
\frac{M_{\rm{NR},J }^{\ast }}{M}=\left[ 1+\frac{M}{k}\frac{dU_{\mathrm{SEP},J}}{dk}\right] ^{-1}
\end{eqnarray}
was introduced in ref.\,\cite{Fuc05}. It reduces to the Lorentz mass $M_{\mathrm{Lorentz},J}^{\ast }$
when relativistic corrections to the kinetic energy in the
single-particle energy can be neglected\,\cite{LCK08,Che07}.  A similarly defined non-relativistic effective mass using the Schr\"{o}dinger-equivalent potential is also
called the Lorentz-vector effective mass\,\cite{Ron06}.

To avoid possible confusions, we also quote here the definitions of several other effective masses often used in relativistic models.
The Dirac mass $M_{\mathrm{Dirac}}^{\ast }$ (also called the Lorentz-scalar effective mass) is defined as
\begin{eqnarray}
M_{\mathrm{Dirac},J}^{\ast }=M+\Sigma _{J}^{\rm{S}}.
\end{eqnarray}
The Landau mass $M_{\mathrm{Landau}}^{\ast }$ is defined as
\begin{equation}
M_{\mathrm{Landau},J }^{\ast } =p\frac{dp}{dE_{J}}
\end{equation}
in terms of the single-particle density of state $dE_{J}/dp$\,\cite{Sjo76}. In non-relativistic models, it has been identified as the total effective mass $M^*_J$\,\cite{Sto07}.
In relativistic models, it is given by\,\cite{Typ05}
\begin{equation}
M_{\mathrm{Landau},J}^{\ast } =\left(E_{J}-\Sigma _{J
}^{0}\right)\left(1- \frac{d\Sigma_{J}^{0}}{dE_{J}}\right)-\left(M+\Sigma
_{J}^{\rm{S}}\right)\frac{d\Sigma _{J}^{\rm{S}}}{dE_{J}}.
\label{MLandau}
\end{equation}
More discussions on effective masses in relativistic models can be found in refs.\,\cite{Jam89,Che07}.

\subsection{Neutron-proton effective mass splitting in neutron-rich nucleonic matter}
With momentum and/or energy dependent single-particle potentials $U_{J} (\rho,\delta, k,E)$ from microscopic or phenomenological models, one can then investigate nucleon effective masses. As we mentioned earlier,
in neutron-rich matter an interesting quantity is the neutron-proton effective mass splitting. For various physics questions, one needs to study its dependences on the nucleon momentum as well as the density and isospin asymmetry of the medium. Conventionally, the nucleon effective masses are normally taken at their respective Fermi momenta and their splitting
\begin{equation}
m^*_{\rm{n-p}}(\rho,\delta)\equiv(M_{\rm n}^*-M_{\rm p}^*)/M
\end{equation}
can then be expressed in terms of the single-nucleon potentials as
\begin{align}\label{em2}
m^*_{\rm{n-p} }=\frac{\displaystyle M\left(\left.\frac{1}{k_{\rm{F}}^{\rm{p}}}\frac{dU_{\rm{p}}}{dk}\right|_{k_{\rm{F}}^{\rm{p}}}-\left.\frac{1}{k_{\rm{F}}^{\rm{n}}}\frac{dU_{\rm{n}}}{dk}\right|_{k_{\rm{F}}^{\rm{n}}}\right)}{\displaystyle\left[1+\left.\frac{M}{k_{\rm{F}}^{\rm{p}}}\frac{dU_{\rm{p}}}{dk}\right|_{k_{\rm{F}}^{\rm{p}}}\right]\left[1+\left.\frac{M}{k_{\rm{F}}^{\rm{n}}}\frac{dU_{\rm{n}}}{dk}\right|_{k_{\rm{F}}^{\rm{n}}}\right]}.
\end{align}
To reveal the interesting physics underlying the $m^*_{\rm{n-p}}$, it is instructive to expand the single-nucleon potential $U_{J}(k,\rho,\delta)$ in isospin-asymmetric matter in the well-known Lane form\,\cite{Lan62}
\begin{align}\label{sp}
U_{J}(k,\rho,\delta)&\approx U_0(k,\rho)+\tau^J_3 U_{\rm{sym},1}(k,\rho)\delta+U_{\rm{sym},2}(k,\rho)\delta^2+\tau^J_3U_{\rm{sym},3}(k,\rho) \delta^3+\mathcal{O}(\delta^4),
\end{align}
where $U_0(k,\rho)$, $U_{\rm{sym},1}(k,\rho)$ and $U_{\rm{sym},2}(k,\rho)$ are the isoscalar, isovector (first-order symmetry) and isoscalar (second-order symmetry) potentials, respectively.
The first two terms are normally used in optical model analyses and the $U_{\rm{sym},1}(k,\rho)
\equiv U_{\rm{sym}}(k,\rho)$ is often referred as the Lane potential. As we shall discuss in detail, the $U_{\rm{sym},1}(k,\rho)$ is normally positive but may become negative at high momentum/energies. The opposite contribution of the $U_{\rm{sym},1}(k,\rho)$ to the neutron and proton potential is responsible essentially for all isospin effects and its momentum dependence determines the isospin-dependence of nucleon effective masses.The isovector (third-order symmetry) potential $U_{\rm{sym},3}(k,\rho)$ is normally neglected. However, it has been found to contribute significantly to the isospin-quartic symmetry energy in some models\,\cite{XuC11}. Since the $\delta$-dependent terms are always much smaller than the isoscalar potential $U_0(\rho,k)$, the denominator in Eq.\,(\ref{em2}) can be well approximated by
\begin{equation}
\left(1+\frac{M}{k_{\rm F}}\frac{dU_{\rm{p}}}{dk}\right)
\left(1+\frac{M}{k_{\rm F}}\frac{dU_{\rm{n}}}{dk}\right)\approx\left(1+\frac{M}{k_{\rm F}}\frac{dU_0}{dk}\right)^2\equiv\left(\frac{M}{M^*_0}\right)^2
\end{equation}
where $M^*_0\equiv M^*_{\rm{s}}$ is the nucleon isoscalar effective mass corresponding to the momentum dependence of $U_0$\,\cite{LiBA13}. Up to the first-order in isospin-asymmetry $\delta $, the $m^*_{\rm{n-p}}$ can thus be simplified approximately to
\begin{equation}\label{npe1}
m^*_{\rm{n-p}}\approx 2\delta\frac{M}{k_{\rm{F}}}\left[-\frac{dU_{\rm{sym},1}}{dk}-\frac{k_{\rm{F}}}{3}\frac{d^2U_0}{dk^2}+\frac{1}{3}\frac{dU_0}{dk}\right]_{k_{\rm{F}}}\left(\frac{M^*_0}{M}\right)^2
\equiv s(\rho)\delta,
\end{equation}
which also defines the linear splitting function $s(\rho)$. We emphasize that the above expression is valid at arbitrary densities. It indicates that the $m^*_{\rm{n-p}}$ depends apparently on the momentum dependence of  both the isovector and isoscalar potentials. Interestingly, however, calculations using nucleon optical potentials at $\rho_0$ from nucleon-nucleus scattering experiments have shown that the
last two terms, i.e., $-k_{\rm{F}}/3\cdot d^2U_0/dk^2$ and $1/3\cdot dU_0/dk$, almost completely cancel each other, leaving the momentum dependence of the isovector potential $dU_{\rm{sym},1}/dk$ as the dominating factor\,\cite{LiX15}.
Eq.\,(\ref{npe1}) gives the linear isospin-splitting of the effective mass. In fact, one can even obtain the higher
order isospin-splitting functions and the next term is the cubic one defined via $m_{\rm{n-p}}^{\ast}\approx s(\rho)\delta+t(\rho)\delta^3$ with $t(\rho)$
given by\,\cite{CaiBJPhDThesis14}
\begin{align}
t(\rho)=&\frac{2M}{k_{\rm{F}}}\left(\frac{M_{\rm{0}}^{\ast}}{M}\right)^2\Bigg\{\frac{2M_{\rm{0}}^{\ast}}{k_{\rm{F}}}\cdot\left[\frac{\partial
U_{\rm{sym}}}{\partial
{k}}+\frac{k_{\rm{F}}}{3}\frac{\partial^2
U_0}{\partial{k}^2}-\frac{1}{3}\frac{\partial
U_0}{\partial{k}}\right]\notag\\
&\times\left[-\frac{k_{\rm{F}}^2}{9}\frac{\partial^3U_0}{\partial{k}^3}
+\frac{k_{\rm{F}}}{3}\frac{\partial^2U_{\rm{sym}}}{\partial{k}^2}+\frac{\partial
U_{\rm{sym,2}}}{\partial{k}}
-\frac{k_{\rm{F}}}{9}\frac{\partial^2U_0}{\partial{k}^2}-\frac{1}{3}\frac{\partial
U_{\rm{sym}}}{\partial{k}}+\frac{2}{9}\frac{\partial
U_0}{\partial{k}}\right]\notag\\
&\hspace*{0.5cm}-\frac{M_{\rm{0}}^{\ast2}}{k_{\rm{F}}^2}\cdot\left[\frac{\partial
U_{\rm{sym}}}{\partial
{k}}+\frac{k_{\rm{F}}}{3}\frac{\partial^2
U_0}{\partial{k}^2}-\frac{1}{3}\frac{\partial
U_0}{\partial{k}}\right]^3-\frac{5k_{\rm{F}}^3}{81}\frac{\partial^4
U_0}{\partial{k}^4}\notag\\
&\hspace*{0.5cm} +\frac{k_{\rm{F}}^2}{9}\frac{\partial^3
U_{\rm{sym}}}{\partial{k}^3}-\frac{k_{\rm{F}}}{3}\frac{\partial^2
U_{\rm{sym,2}}}{\partial{k}^2}-\frac{\partial
U_{\rm{sym,3}}}{\partial{k}}
-\frac{k_{\rm{F}}^2}{27}\frac{\partial^3U_0}{\partial{k}^3}
+\frac{k_{\rm{F}}}{9}\frac{\partial^2U_{\rm{sym}}}{\partial{k}^2}\notag\\
&\hspace*{0.5cm}+\frac{1}{3}\frac{\partial
U_{\rm{sym,2}}}{\partial{k}}
-\frac{2k_{\rm{F}}}{27}\frac{\partial^2U_0}{\partial{k}^2}-\frac{2}{9}\frac{\partial
U_{\rm{sym}}}{\partial{k}}+\frac{14}{81}\frac{\partial
U_{0}}{\partial{k}}
 \Bigg\}_{k_{\rm{F}}}.
\end{align}
Obviously, it is determined by the first, second and third order momentum derivatives of both the isoscalar and isovector parts of the nucleon potential in a complicated way.
Generally, the higher order isospin-splitting functions are expected to play small roles in finite nuclei,
however, their effects could be as important as the linear isospin-splitting function, e.g.,
in neutron stars, since the isospin asymmetry there is close to 1.

Sj\"oberg studied the Landau effective masses (i.e., the total effective mass in non-relativistic models) in the liquid phase of neutron stars\,\cite{Sjo76}. The neutron-proton effective Landau mass splitting can be written as
\begin{eqnarray}
m^*_{\rm{n-p}}= \frac{M_{\rm{n}}^* k^{\rm{n}}_{\rm{F}}}{3 \pi^2} \left[ f_1^{\rm{nn}} +
\left(\frac{k^{\rm{p}}_{\rm{F}}}{k^{\rm{n}}_{\rm{F}}}\right)^2 f_1^{\rm{np}} \right]-\frac{ M_{\rm{p}}^* k^{\rm{p}}_{\rm{F}}}{3\pi^2} \left[
f_1^{\rm{pp}} + \left(\frac{k^{\rm{n}}_{\rm{F}}}{k^{\rm{p}}_{\rm{F}}}\right)^2 f_1^{\rm{np}} \right], \label{sjo}
\end{eqnarray}
where $f_1^{\rm{nn}}$, $f_1^{\rm{pp}}$ and $f_1^{\rm{np}}$ are the
neutron-neutron, proton-proton and neutron-proton quasiparticle
interactions projected on the $\ell=1$ Legendre polynomial, as for
the effective mass in a one-component Fermi liquid. Sj\"oberg
has shown that all $f_1$'s are negative in SNM at normal density and
also in asymmetric matter at tree-level, and predicted that the proton effective mass is smaller than
that of neutrons ($m_{\rm{n}}^*>m_{\rm{p}}^*$) in neutron-rich matter\,\cite{Sjo76}.

Within the non-relativistic energy density functional approaches, the $m^*_{\rm{n-p}}$ can be expressed in terms of the $\kappa_{\rm{s}}$ and $\kappa_{\rm{v}}$\ as in ref. \cite{Les06}.
First, we notice that
\begin{equation}\label{sign0}
\frac{M^*_{\rm{n}}-M^*_{\rm{p}}}{M^*_{\rm{n}}M^*_{\rm{p}}}= 2\delta\frac{M^*_{\rm{s}}-M^*_{\rm{v}}}{M^*_{\rm{s}}M^*_{\rm{v}}}
\end{equation}
such that $m^*_{\rm{n-p}} > 0$  for $M^*_{\rm{s}} > M^*_{\rm{v}}$ or equivalently
\begin{equation}\label{mnp2}
m^*_{\rm{n-p}}= 2\delta\frac{\kappa_{\rm{v}} - \kappa_{\rm{s}}}{(1 + \kappa_{\rm{s}})^2 -\delta^2 (\kappa_{\rm{v}} - \kappa_{\rm{s}})^2}
\approx 2\delta\left(\frac{M^*_{\rm{s}}}{M}\right)^2\left[\frac{M}{M^*_{\rm{v}}}-\frac{M}{M^*_{\rm{s}}}\right]
\end{equation}
with the approximation that $(M^*_{\rm{n}}/M)(M^*_{\rm{p}}/M)\approx (M^*_{\rm{s}}/M)^2$ valid when $\delta$ is small.
For pure neutron matter (PNM), it reduces to the expression given in ref.\,\cite{Les06} by setting $\delta=1$.

\FloatBarrier
\section{Empirical values of nucleon isosclar and isovector effective masses as well as the neutron-proton effective mass splitting in neutron-rich matter at saturation density}\label{emass-values}
To extract information about nucleon effective masses from experiments has been a longstanding goal of nuclear physics. Some detailed discussions about the current values of the nucleon isoscalar and isovector effective masses from fitting saturation properties of nuclear matter, masses and level densities of ground state as well as low-energy collective models of nuclei can be found in refs.\,\cite{Dut12,Rei15}. Unfortunately, often the extracted values are model dependent and they are valid only at the saturation density. Here we make a few observations and comments especially about the implications of existing results on the neutron-proton effective mass splitting in neutron-rich matter. So far, the nucleon isoscalar effective mass $M^*_{\rm{s}}/M$ seems to be better determined than the isovector one $M^*_{\rm{v}}/M$, while both are still suffering from some model and interaction dependences. They also depend somewhat on the data used in the analysis. This is partially because they depend on different combinations of several parameters in the underling interactions, and rarely there are known experimental observables that are uniquely sensitive to each of the model parameters.

\subsection{Mean field model analyses of structure properties and low-energy excitations of nuclei}
There seems to be a consensus that analyses of isoscalar giant quadrupole resonances (ISGQR) call for $m^*_{\rm{s}}/m\approx 0.8\pm 0.1$\,\cite{Jeu76,Zha16,Cha97,Ben03,Sto07,Rei15,Boh79}, while the uncertainties involved still need to be better quantified. For example, Kl\"upfel \textit{et al.} deduced an ``optimum'' $m^*_{\rm{s}}/m\approx 0.9$ from a global fit to several observables using SHF energy functionals\,\cite{Klu09}. However, it was emphasized that the final ``optimum" for $m^*_{\rm{s}}/m$ depends very much on the choice of the relative weight of the different observables. It is interesting to note that various many-body calculations support the value of $m^*_{\rm{s}}/m=0.8\pm 0.1$ for infinite nuclear matter at $\rho_0$, see, e.g., refs.\,\cite{Fri81,Wir88,Zuo99}, irrespective of their level of sophistications. For completeness, it is necessary to mention here an interesting study\,\cite{Peter} on a perceived contradiction historically between the  $m^*_{\rm{s}}/m$ extracted for infinite nuclear matter and the finding from mean-field calculations on all but light nuclei that one must uses $m^*_{\rm{s}}/m\approx 1$ to reproduce the observed density of single-particle states near the Fermi surface\,\cite{Bro63,Bar81,Shl92,Mug98}. Moreover, to precisely reproduce essentially all the mass data within SHF using a conventional Skyrme force it was also necessary to use $m^*_{\rm{s}}/m\approx1$, since without a correct spacing of the single-particle states near the Fermi surface it was found impossible to fit the masses of open-shell nuclei\,\cite{Ton00,Gor01}. The seemingly inconsistent conclusions about the $m^*_{\rm{s}}/m$ were resolved by either considering explicitly the coupling between particle modes and surface-vibration modes\,\cite{Ber68,Bern80,Lit06} or introducing a radius-dependent effective mass peaking at the nuclear surface\,\cite{Ma83,Pra83}. Ma and Wambach used in their seminal work\,\cite{Ma83} a phenomenological radius-dependent $m^*_{\rm{s}}/m$.  It has a value of about 0.7 in the interior and peaks at the nuclear surface to $1.2$-$1.5$ depending on the nucleus before falling to its asymptotic value of 1.0. The surface-peaking of $m^*_{\rm{s}}/m$ is due to the coupling between single-particle motion and the surface-vibration. Their result was later confirmed by Farine \textit{et al.}\,\cite{Far01} within self-consistent SHF calculations by adding new terms to the Skyrme force itself and Zalewski \textit{et al.}\,\cite{Zal10} by adding directly new terms to the SHF energy density functionals. Thus, considering the coupling between the single-particle motion and surface-vibration modes, the value of $m^*_{\rm{s}}/m\approx1$ extracted from single-particle level densities of finite nuclei was understood as an appropriate mean value over the nucleus\,\cite{Far01,Peter,Ma83,Zal10}, while for infinite nuclear matter at $\rho_0$
the isoscalar nucleon effective mass is $m^*_{\rm{s}}/m=0.8\pm 0.1$.

The isovector nucleon effective mass $m^*_{\rm{v}}/m$ can be determined from the isovector giant dipole resonance (IVGDR). For example, Kl\"upfel \textit{et al.} recently deduced a value of $m^*_{\rm{v}}/m\approx 0.7$ from analyzing the GDR data of $^{208}$Pb\,\cite{Klu09}, consistent with the earlier finding by Krivine \textit{et al.}\,\cite{Kri80}.
By analyzing the newest GDR data of $^{208}$Pb\,\cite{exp2,Roc15}, Zhang and Chen recently extracted a value of $m^*_{\rm{v}}/m\approx 0.80 \pm 0.03$\,\cite{Zha16}.
More recently, a systematic analysis of the GDR data in heavy rare-earth elements with $Z$ from 64 to 74 was carried out using the SkM$^*$, SV-Kap20, SV-bas and SV-Kap60 Skyrme energy density functionals\,\cite{Ois16}. They found a value of $\kappa_{\rm{v}} =0.53, 0.2, 0.4$ and 0.6 leading to $m^*_{\rm{v}}/m$=0.653, 0.834, 0.715 and 0.625, respectively, for the four models used. These results illustrate the model dependence of the extracted values for $m^*_{\rm{v}}/m$. One should notice that the results also have some dependence on the data sets used.
For example, it has been found that the values for $M^*_{\rm{v}}/M$ based on the GDR in heavy nuclei are not quite consistent with the GDR in $^{16}$O\,\cite{Erl10}. Moreover, we notice that Pearson and Goriely carried out Skyrme-Hartree-Fock-BCS calculations of nuclear masses for 416 spherical, or near-spherical, nuclei with $A\ge 36$. By holding the isoscalar mass $m^*_{\rm{s}}/m$ fixed at 1.05 which was necessary for them to fit earlier a larger set of experimental nuclear masses, they have shown that the selected nuclear mass data constrain the value of the isovector effective mass to be in the range of $m^*_{\rm{v}}/m=0.9\pm 0.2$\,\cite{Pea01}. Moreover, consistent with the constraints mentioned above, several SHF calculations meeting essentially all known empirical constraints predict a $m^*_{\rm{v}}/m$ value in the range of $0.603\sim 0.930$\,\cite{Dut12}.

As shown in Eq.\,(\ref{sign0}), the sign of the neutron-proton effective mass splitting in neutron-rich matter is determined by the difference between the isoscarlar and isovector masses, i.e., $(M^*_{\rm{n}}-M^*_{\rm{p}})\propto (M^*_{\rm{s}}-M^*_{\rm{v}})$. Given the compatible values of the isoscalar and isovector nucleon effective masses within their uncertainties, it is hard to make an absolute conclusion about the sign of the neutron-proton effective mass splitting at $\rho_0$ based solely on  the empirical evidences mentioned above. We also notice that rarely the two kinds of effective masses are determined simultaneously from analyzing the same sets of data within the same model\,\cite{Zha16}. Of course, constraints on the $M^*_{\rm{s}}/M$ and $M^*_{\rm{v}}/M$ and thus the $(M^*_{\rm{n}}-M^*_{\rm{p}})$ are expected to be improved as more precise data of more neutron-rich nuclei are analyzed consistently with more advanced theories. In this regard, it is very interesting to cite the major conclusions of a rather comprehensive study by Lesinski \textit{et al.}\,\cite{Les06}. By adding a density-dependent term to the ``Saclay-Lyon" SLy series of the Skyrme energy density functional, they constructed three new sets of Skyrme parameterizations having the same $m^*_{\rm{s}}/m=0.7$ and other features but different neutron-proton effective mass splittings of $m^*_{\rm{n-p}}=-0.284, 0.001$ and $0.170$ corresponding to $\kappa_{\rm{v}}=0.15,0.43$ and 0.60, respectively. By construction, they all reproduce approximately the same EOSs of both PNM and SNM predicted by the Variational Chain Summation (VCS) method\,\cite{APR}. They found that effects of the $m^*_{\rm{n-p}}$ on single-particle energies (i.e., level densities), pairing gaps and binding energies, are noticeable and consistent, but limited. However, GDR analyses were found more fruitful for probing the $m^*_{\rm{n-p}}$, and a clear tendency favoring a positive $m^*_{\rm{n-p}}$ was obtained. As already suggested in ref.\,\cite{Bern80}, they also pointed out that the main reason for not seeing a dramatic modification of the studied observables when altering the $m^*_{\rm{v}}/m$ is the limited amount of strongly isospin-asymmetric nuclear matter at high enough density in the ground state of nuclei\,\cite{Les06}. This feature was used by the UNEDF (Universal Nuclear Energy Density Functional) collaboration in justifying their setting of the $m/m^*_{\rm{v}}$ to a constant value of 1.249 as in the original SLy4 force\,\cite{Sly4} in their covariant analyses of nuclear binding energies and pairing gaps\,\cite{UNDEF}. They found that the $m/m^*_{\rm{s}}$ has an optimal value of 0.95728 within the 95\% confidence interval of $[0.832,1.083]$, thus leading to a positive neutron-proton effective mass splitting of approximately $m^*_{\rm{n-p}}\approx 0.637\delta$ according to Eq. (\ref{mnp2}).

Overall, to summarize our observations about the $m^*_{\rm{n-p}}$ based on extensive investigations of various properties of nuclear structures and giant resonances by many people in the community, it is probably appropriate to repeat here the conclusions reached by Lesinski \textit{et al.}\,\cite{Les06}. Namely, clearly there are some empirical evidences favoring a positive $m^*_{\rm{n-p}}$  and there is no reason to omit this constraint in future studies.

\subsection{Dynamical model analyses of nuclear reactions and low-energy excitations of nuclei}
On the other hand, as we shall discuss in more detail later, nuclear reactions especially those involving
neutron-rich nuclei can provide invaluable information about the neutron-proton effective mass splitting not only at $\rho_0$ but also at higher or lower densities. The information at abnormal densities will be useful for understanding some properties of neutron stars. In fact, some observables in nuclear reactions directly probe the momentum dependence of the single-nucleon potential. Especially, some isospin-sensitive observables are known to be sensitive to the momentum-dependence of the isovector potential, thus the isovector nucleon effective mass\,\cite{LCK08}. For example, the momentum/energy dependence of the isovector potential at $\rho_0$ has been extracted recently from optical model analyses\,\cite{LiX13,LiX15} of all 2249 data sets of reaction and angular differential cross sections of neutron and proton scattering on 234 targets at beam energies from 0.05 to 200 MeV available in the EXFOR database at the Brookhaven National Laboratory\,\cite{Exfor}. The study found clearly a neutron-proton total effective mass splitting of $m^{*}_{\rm{n-p}}(\rho_0,\delta)\approx (0.41\pm0.15)\delta$. It is definitely positive. Moreover, as we shall also discuss in more detail, the  Hugenholtz--Van Hove (HVH) theorem\,\cite{Hug58} establishes a direct connection between the neutron-proton effective mass splitting and the density dependence of nuclear symmetry energy\,\cite{XuC10,XuC11,Rchen,Cai-EL,CXu14}. Analyses of 28 empirical constraints on the density dependence of nuclear symmetry energy from both nuclear physics and astrophysics, have led to an estimate of $m^*_{\rm{n-p}}(\rho_0,\delta)\approx (0.27\pm 0.25)\delta$\,\cite{LiBA13}. Thus, the sign of the neutron-proton effective mass splitting in neutron-rich matter at $\rho_0$ is rather solidly determined while its amplitude is still subject to some uncertainties.

Furthermore, experiments using electron-nucleus and/or proton-nucleus scatterings have recently put strong constraints on the strength and isospin dependence of nucleon-nucleon short-range correlations (SRC) as well as the associated quasi-nucleon momentum distribution in neutron-rich matter\,\cite{Ant88,Ant93,Arr12,Cio15,Ryc15,Hen16x}.
This enables an independent constraint on the nucleon E-mass\,\cite{CaiLi16a} and
its isospin dependence through the Migdal--Luttinger theorem\,\cite{Mig57,Lut60}.
Thus, nuclear reactions provide some interesting and useful information about nucleon effective masses
on top of what we have learned from studying nuclear structures, masses and giant resonances.
In turn, the extracted neutron-proton effective mass splitting may help constrain some parameters
in nuclear forces or energy density functionals.

\begin{figure}[h!]
\includegraphics[scale=0.32]{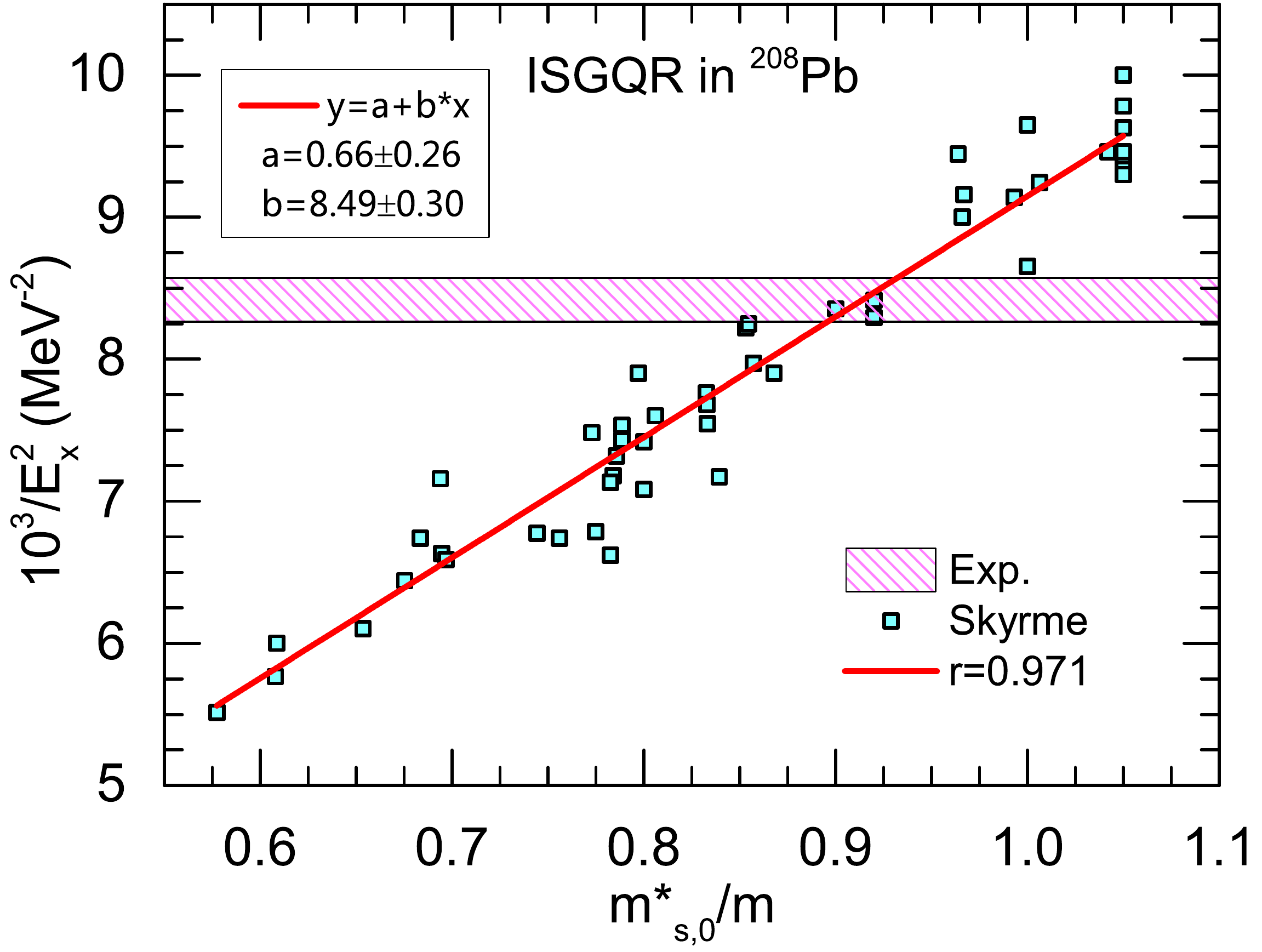}
\includegraphics[scale=0.3]{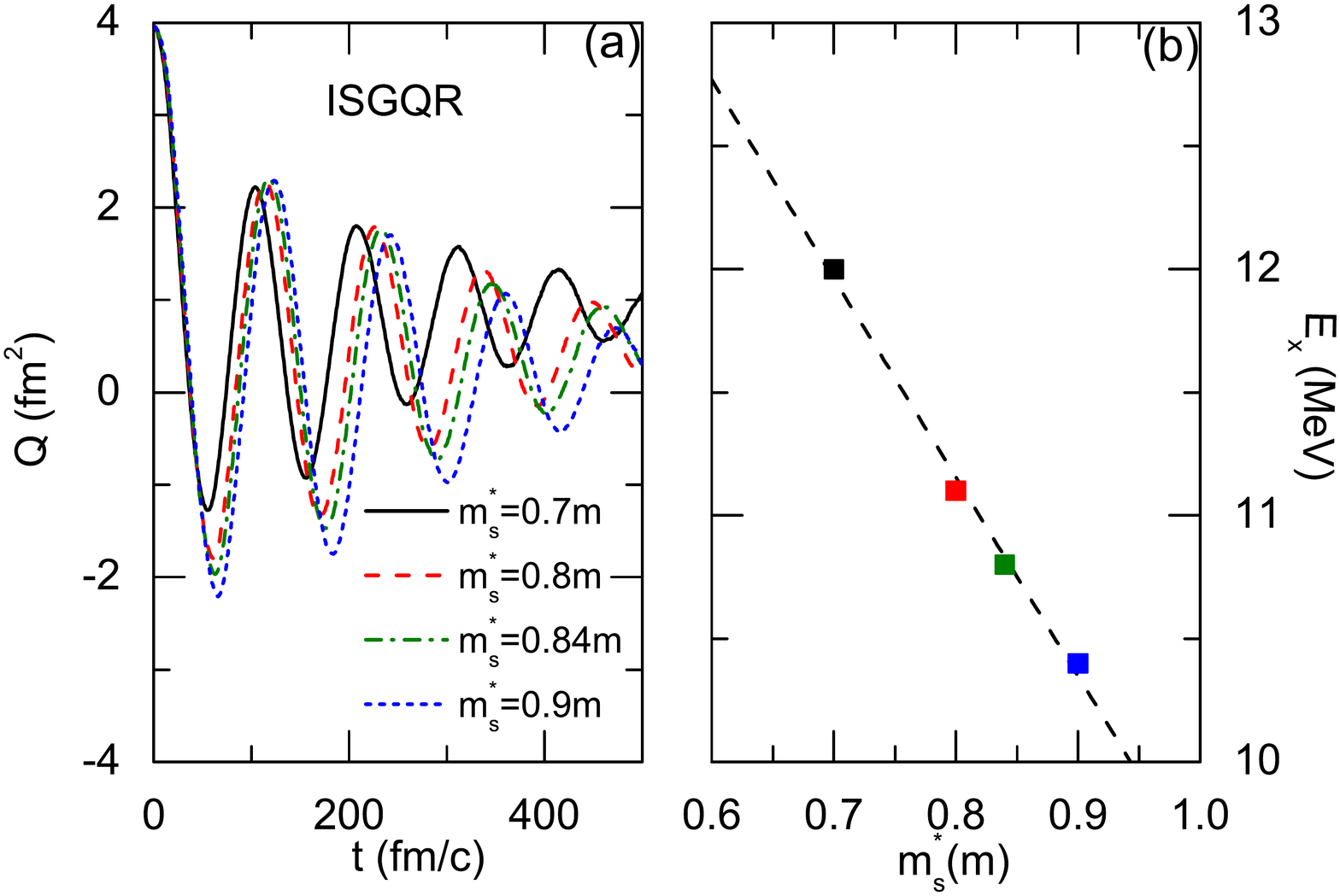}
\caption{Left: Relation between the excitation energy $E_{\rm{x}}$ of ISGQR in $^{208}$Pb
and the isoscalar effective mass $m_{\rm{s}}^{\ast}/m$ predicted by $50$ Skyrme forces from the SHF+RPA method, with the
the hatched band corresponding to the experimental
value $E_{\rm{x}}=10.9\pm0.1$\,MeV.
The linear fit as well as the Perason correlation coefficient $r$ is also displayed. Taken from
ref.\,\cite{Zha16}. Right: Time evolution of the ISGQR moment (a) and the relation between
$E_{\rm{x}}$ of ISGQR in $^{208}$Pb and $m_{\rm{s}}^*$ (b) from IBUU transport model simulations using the
ImMDI interaction. Taken from ref.\,\cite{Kong17prc}. }
\label{fig:ISGQR}
\end{figure}
\subsection{A comparison of nucleon effective masses extracted from two approaches: SHF+RPA versus transport model}
As illustrations of the various analyses mentioned above, here we compare nucleon effective masses extracted from the same sets of data using the SHF+RPA and a transport for nuclear reactions.
As noticed earlier, the isoscalar and isovector effective masses are rarely extracted simultaneously from the same data sets within the same model.
One exception is the recent analyses\,\cite{Zha16} of ISGQR\,\cite{exp3}, IVGDR\,\cite{exp1} and the electric dipole polarizability $\alpha_{\rm{D}}$\,\cite{exp2} in $^{208}$Pb
simultaneously within the SHF+RPA approach. The Random Phase Approximation (RPA) was used to calculate the resonance properties based on the SHF model which predicts different nucleon effective masses
with various Skyrme forces. As an example of dynamical approaches, we use the recent work  of ref.\,\cite{Kong17prc} where the same sets of data were analyzed simultaneously by using an improved isospin- and momentum-dependent interaction (ImMDI)\,\cite{Xu15prc} within an isospin-dependent Boltzmann--Uehling--Uhlenbeck (IBUU) transport model for nuclear reactions\,\cite{LiBA04}.

As shown in the left panel of Fig.\,\ref{fig:ISGQR}, the inverse excitation energy squared $1/E_{\rm{x}}^2$ of ISGQR
in $^{208}$Pb varies with the isoscalar effective mass $m_{\rm{s}}^*$ approximately linearly, i.e, $1/E_{\rm{x}}^2 \propto m_{\rm{s}}^*$,
based on the SHF+RPA calculations \cite{Zha16}. In the panel (a) of the right part of Fig.\,\ref{fig:ISGQR}, the time evolution of the
ISGQR moment from the IBUU simulations shows a characteristic ISGQR behavior. After a Fourier transformation, the excitation energy $E_{\rm{x}}$ of
the ISGQR in $^{208}$Pb shows an approximately inverse relation with $m_{\rm{s}}^*$\,\cite{Kong17prc}.
Although both approaches show a decreasing $E_{\rm{x}}$ with increasing $m_{\rm{s}}^*$, one goes down as $(m^*_s)^{-1/2}$ following the semi-empirical relation from the harmonic oscillator model \cite{Bohr69}
while the other one varies more like $1/m^*_{\rm{s}}$. Since numerically the two functions give very similar values in the region of $m_{\rm{s}}^*/m=[0.7,0.9]$, one may not attach much physics significance to the difference of the two fitting functions used here. Comparing the calculations with the experimental data shown in the left window, the SHF+RPA approach favors an isoscalar nucleon effective mass of
$m_{\rm{s}}^*/m = 0.91 \pm 0.05$, while on the right, the IBUU dynamical model favors a value of $m_{\rm{s}}^{\ast}/m\sim 0.84$. The difference in the extracted values from these two approaches
exemplifies the typical model dependence in determining the $m_{\rm{s}}^{\ast}/m$ at $\rho_0$.

\begin{figure}[h!]
\hspace*{0.5cm}\includegraphics[scale=0.32]{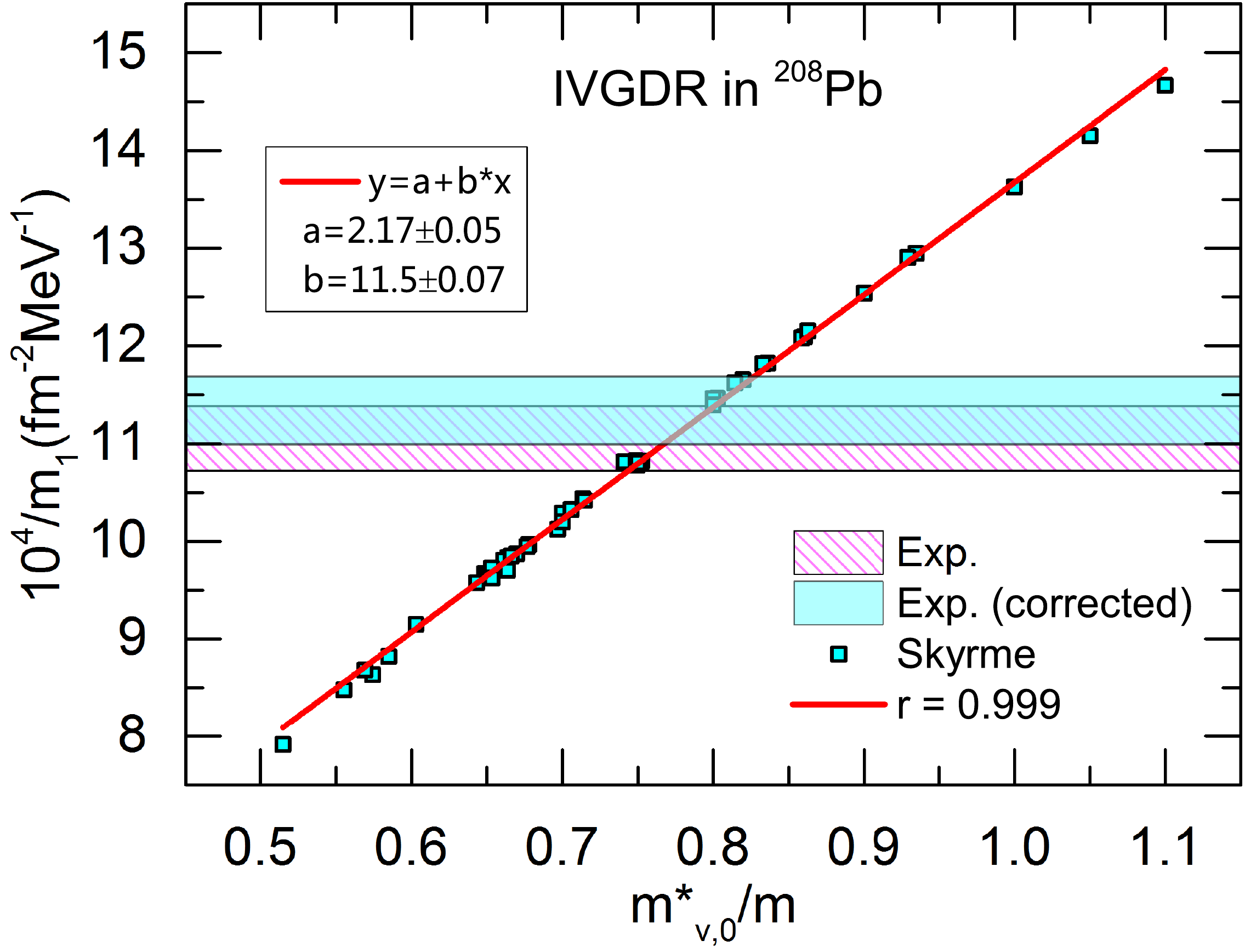}
\includegraphics[scale=0.3]{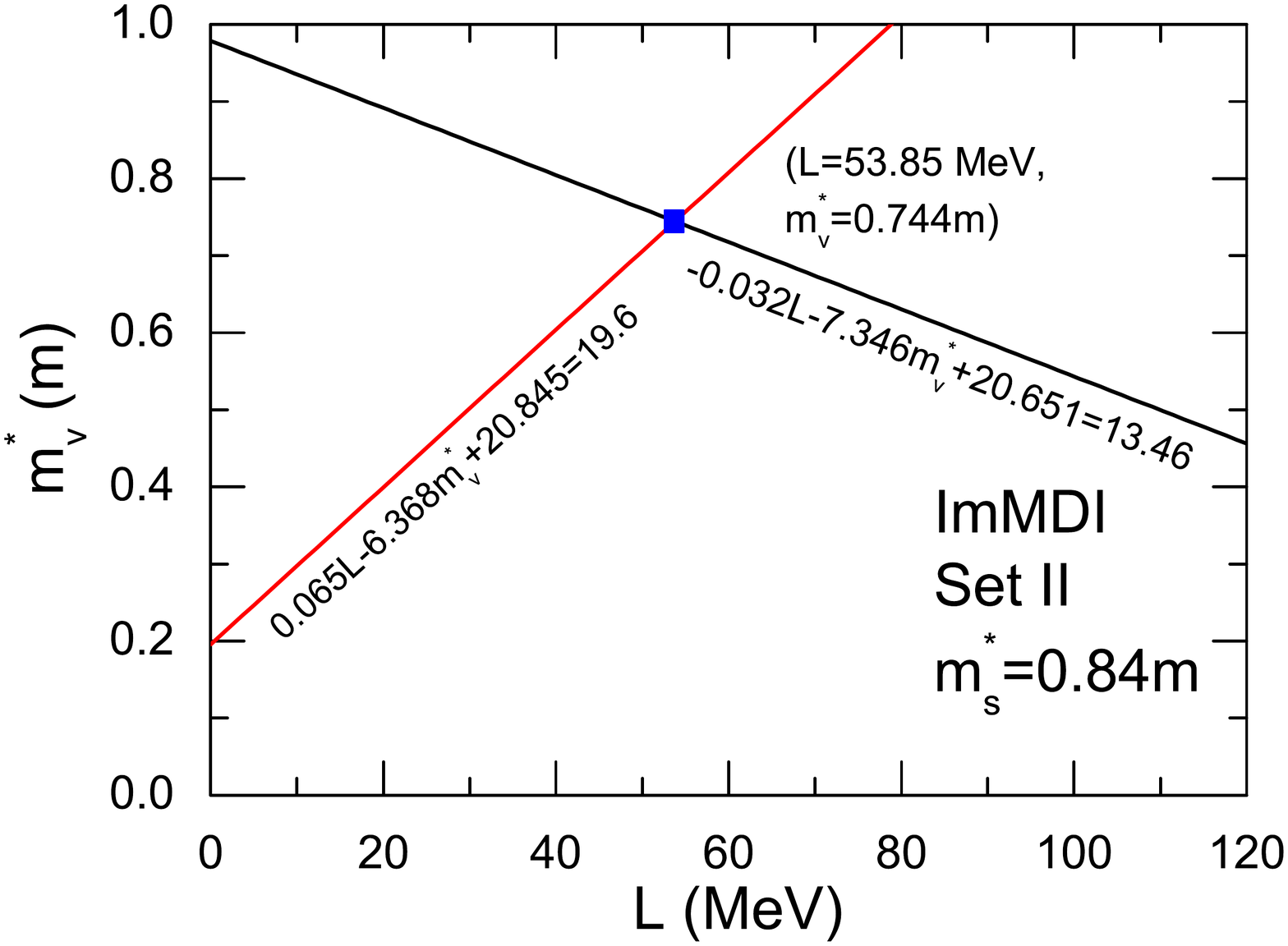}
\caption{Left: Relation between the EWSR $m_1$ of the IVGDR in $^{208}$Pb and the
isovector nucleon effective mass $m_{\rm{v}}^{\ast}/m$ predicted by 50 Skyrme interactions,
with the hatched band (cyan band) corresponding to the (corrected)
experimental value. The linear fit as well as the
Perason correlation coefficient $r$ is also displayed.
Taken from ref.\,\cite{Zha16}. Right: The values of $m_{\rm{v}}^*$
and density slope $L$ of the symmetry energy that satisfy respectively the experimental constraints on the centroid energy (red)
of IVGDR and the electric dipole polarizability (black) in $^{208}$Pb
from IBUU transport model simulations using the ImMDI interaction. Taken from ref.\,\cite{Kong17prc}. }
\label{fig:IVGDR}
\end{figure}

The energy weighted sum rule (EWSR)
\begin{eqnarray}
m_{k}=\int_0^\infty dE E^{k}S(E),
\end{eqnarray}
with $S(E)$ the strength function, is normally used to characterize properties of collective excitations of nuclei.
The $m_{-1}$ is related to the electric dipole polarizability, and the centroid energy of IVGDR can be
expressed as $E_{-1}=({m_{1}/m_{-1}})^{1/2}$.
The relation between $m_1$ of the IVGDR in $^{208}$Pb
and the isovector nucleon effective mass $m_{\rm{v}}^{\ast}/m$
from the SHF+RPA calculations is shown in the left panel of Fig.\,\ref{fig:IVGDR}.
It is seen that $1/m_1$ shows a linear relation with the $m_{\rm{v}}^*$. Comparing the calculations with the experimental data leads to the extraction of $m_{\rm{v}}^*/m = 0.80 \pm 0.03$.
This value of $m_{\rm{v}}^*/m$ together with the $m_{\rm{s}}^*/m = 0.91 \pm 0.05$ obtained from analyzing the ISGQR data shown in the left panel of Fig.\,\ref{fig:ISGQR} lead to a neutron-proton effective mass splitting of $m_{\rm{n-p}}^{\ast}=(0.27 \pm 0.15)\delta$, which is in nice agreement with that obtained from the optical model analysis of nucleon-nucleus scatterings\,\cite{LiX15}.
It is worth noting that the isospin-cubic term in the neutron-proton effective mass splitting was also evaluated in the same analysis and was found very small\,\cite{Zha16}.

On the other hand, shown in the right panel of Fig.\,\ref{fig:IVGDR} are the $m_{\rm{v}}^*$ as functions of the slope parameter $L$ of nuclear symmetry energy based on the IBUU transport model simulations
to reproduce respectively the experimental centroid energy of IVGDR and the electric dipole polarizability in $^{208}$Pb. Both functions are approximately linear, with a crossing point at $L=53.85\,\text{MeV}$ and $m_{\rm{v}}^*/m=0.744$ satisfying both experimental constraints. The isovector nucleon effective mass extracted this way together with the $m_{\rm{s}}^{\ast}/m\sim0.84$ from analyzing the ISGQR also using the IBUU
leads to a neutron-proton effective mass splitting of $m_{\rm{n-p}}^{\ast}=(0.216 \pm 0.114)\delta$ after considering statistical and fitting errors, consistent with that obtained from the SHF+RPA studies discussed above.

In summary of this section, many analyses indicate that the nucleon isoscalar and isovector effective masses in nuclear matter at $\rho_0$  are $m^*_{\rm{s}}/m\approx 0.8\pm 0.1$ and $m^*_{\rm{v}}/m=0.6\sim 0.93$, respectively. Their uncertainties, especially those of the $m^*_{\rm{v}}/m$ are still poorly determined. There are empirical evidence especially from nuclear reactions supporting a positive neutron-proton effective mass splitting $m^*_{\rm{n-p}}$ in neutron-rich matter at saturation density while its magnitude still has large uncertainties. Moreover, as we shall discuss in the following, many-body theory predictions for the positive sign of $m^*_{\rm{n-p}}$ in neutron-rich matter are rather solid. Certainly, additional work is needed in both theory and experiment
to better understand the nucleon isovector effective mass $m^*_{\rm{v}}/m$
and the underlying momentum dependence of the isovector interaction.
More reliable knowledge about them is required for the elucidation of the highly neutron-rich
systems especially at supra-saturation densities of astrophysical interest\,\cite{Far01,Dut12}
and nuclear reactions with radioactive beams\,\cite{LCK08,LiChen15,NST16,Cou14}.
As we shall discuss in detail, the momentum dependence and nucleon effective masses affect heavy-ion
reactions through both the mean-field potential and in-medium nucleon-nucleon scattering cross
section or viscosity. Their imprints on the isospin-dependent reaction dynamics and observables in heavy-ion collisions will be reviewed in Section \ref{HI}.

\FloatBarrier
\section{Theoretical predictions on the momentum dependence of the symmetry potential and neutron-proton effective mass splitting in neutron-rich matter}
As essentially all major issues related to the neutron-proton effective mass splitting and density dependence of nuclear symmetry energy in neutron-rich matter can be traced back to the poorly known momentum dependence of the single-nucleon symmetry (isovector) potential $U_{\rm{sym},1}(k,\rho)$ (or other symbols used in the literature), both microscopic nuclear many-body theories and phenomenological models using various interactions have been applied to calculate the $U_{\rm{sym},1}(k,\rho)$ as precisely as possible. As we shall discuss in more detail, the only relatively well-known constraint available is the symmetry potential at $\rho_0$ extracted from optical model analyses of large sets of nucleon-nucleus scattering data. This constraint (boundary condition at $\rho_0$) has been used in recent years to check the validity of model predictions in some studies. However, predictions using models obviously violating this boundary condition have also been put forward in various studies. In the following, we make some observations about the major features of the symmetry potential using a few examples from the literatures.
Some of these features may be explored in the near future with reactions induced by rare isotopes. We notice that the community has been using many different symbols for the single-nucleon, isoscalar $U_0$ and isovector
$U_{\rm{Lane}}, U_{\rm{sym},1}, U_{\rm{sym}}, U_{\rm{iso}}$ or $U_I$ potentials at isospin-asymmetry $\delta$, $\alpha$ or $\beta$. In this review, we may use different symbols for the same variable only if necessary and try to minimize the possible confusions.

\subsection{Symmetry (isovector or Lane) potentials predicted by nuclear many-body theories}\label{Umodel}
\vspace{-2.5cm}
\begin{figure}[htb]
\begin{center}
\includegraphics[width=7cm,height=6cm]{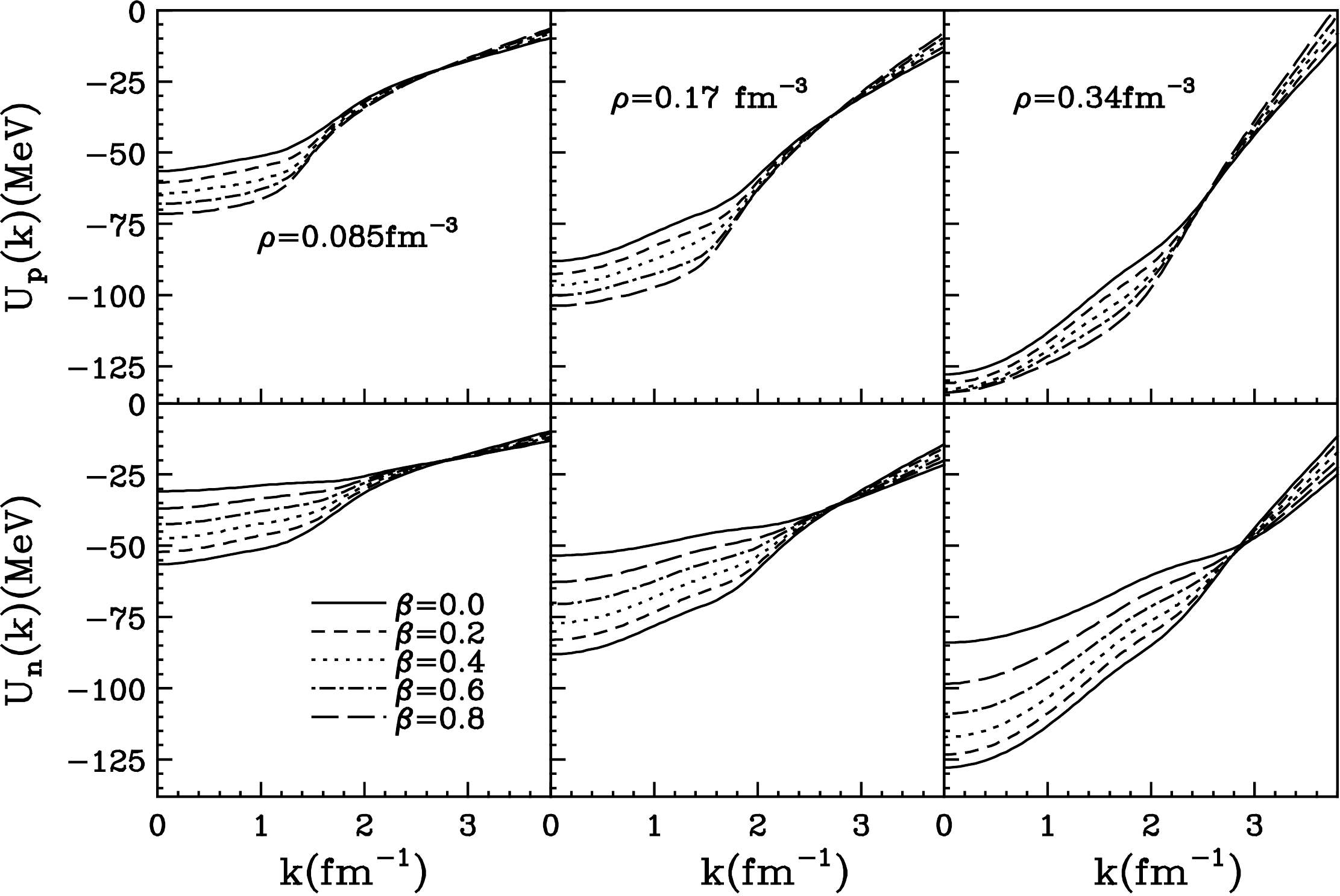}
\hspace{1.cm}
\includegraphics[width=8cm,height=9cm]{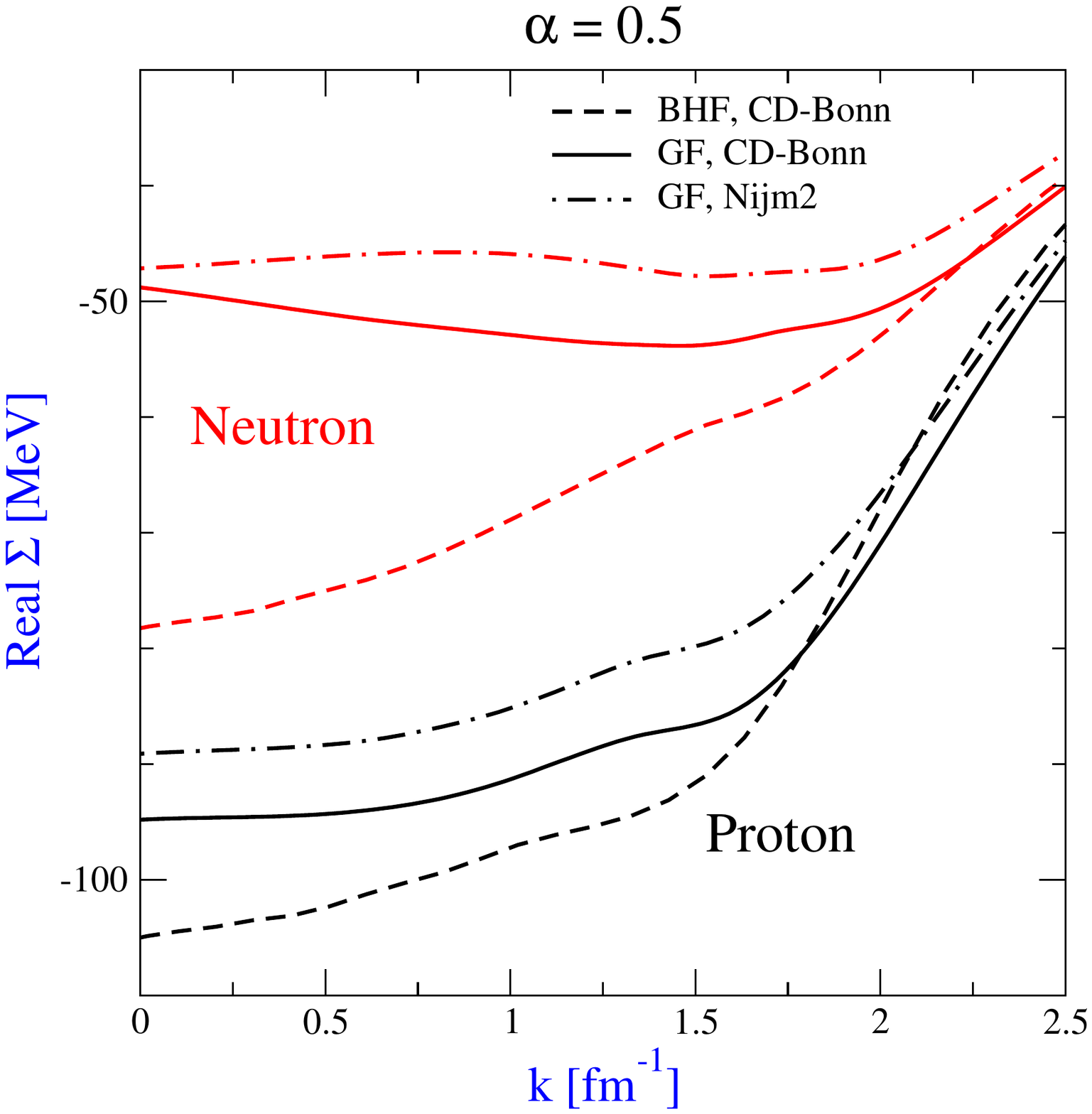}
\caption{\small{Left: Single-nucleon potentials as a function of momentum for different values of the isospin-asymmetry parameter $\delta=\beta=\alpha$ at three densities. Taken from ref.\,\cite{zuo05}.
Right: Single-particle potential determined
from the BHF and the Green's function (GF) approach using the CD-Bonn and Nijm2 interaction at a density of $\rho$ = 0.17 fm$^{-1}$ for an isospin-asymmetry
$\delta=\beta=\alpha$=0.5. Taken from ref.\,\cite{mu04}.}}
\label{Zuo-fig1}
\end{center}
\end{figure}

Let's first look at the isospin-dependence of the single-nucleon potentials. Shown in the left window of Fig.\,\ref{Zuo-fig1} is the single-proton (upper windows) and single-neutron (lower windows) potentials at $0.5\rho_0, \rho_0$ and $2\rho_0$ and several values of the isospin-asymmetry $\beta$, respectively, obtained by Zuo \textit{et al.} using the extended Brueckner--Hartree--Fock (BHF) Approach\,\cite{Zuo99}. Several major features shared by predictions using essentially all models are worth emphasizing. First of all, the mean field $U_{\rm{p}} (k)$ of protons (neutrons) becomes more attractive (repulsive),  going from isospin symmetric ($\beta=0$) to pure neutron ($\beta=1$) matter. The $U_{\rm{n}}$ and $U_{\rm{p}}$ varify almost linearly with the isospin asymmetry $\beta$ as the Lane potential of Eq.\,(\ref{sp}).
It is well known that the current predictions of the single-particle potential depend quantitatively on both the many-body theories and interaction used, while they all predict some common features. For example, shown in the right window of Fig.\,\ref{Zuo-fig1} are comparisons of the single-particle potentials predicted by the BHF and Green's function (GF) approaches using the CD-Bonn and Nijm2 interaction at a density of $\rho$ = 0.17 fm$^{-1}$ for an isospin asymmetry of $\alpha$=0.5. While they all predict that protons feel more attractive potentials than neutrons, the magnitude of these potentials are significantly model and interaction dependent. It was shown in ref.\,\cite{mu04} that the hole-hole ladder contribution leads to less attractive single-particle energies in the GF approach as compared to the BHF model. The difference tends to be larger for neutrons than protons and decreases with increasing momentum in neutron-rich matter.

 \begin{figure}[tbh]
\begin{center}
\includegraphics[width=10.cm,angle=-90]{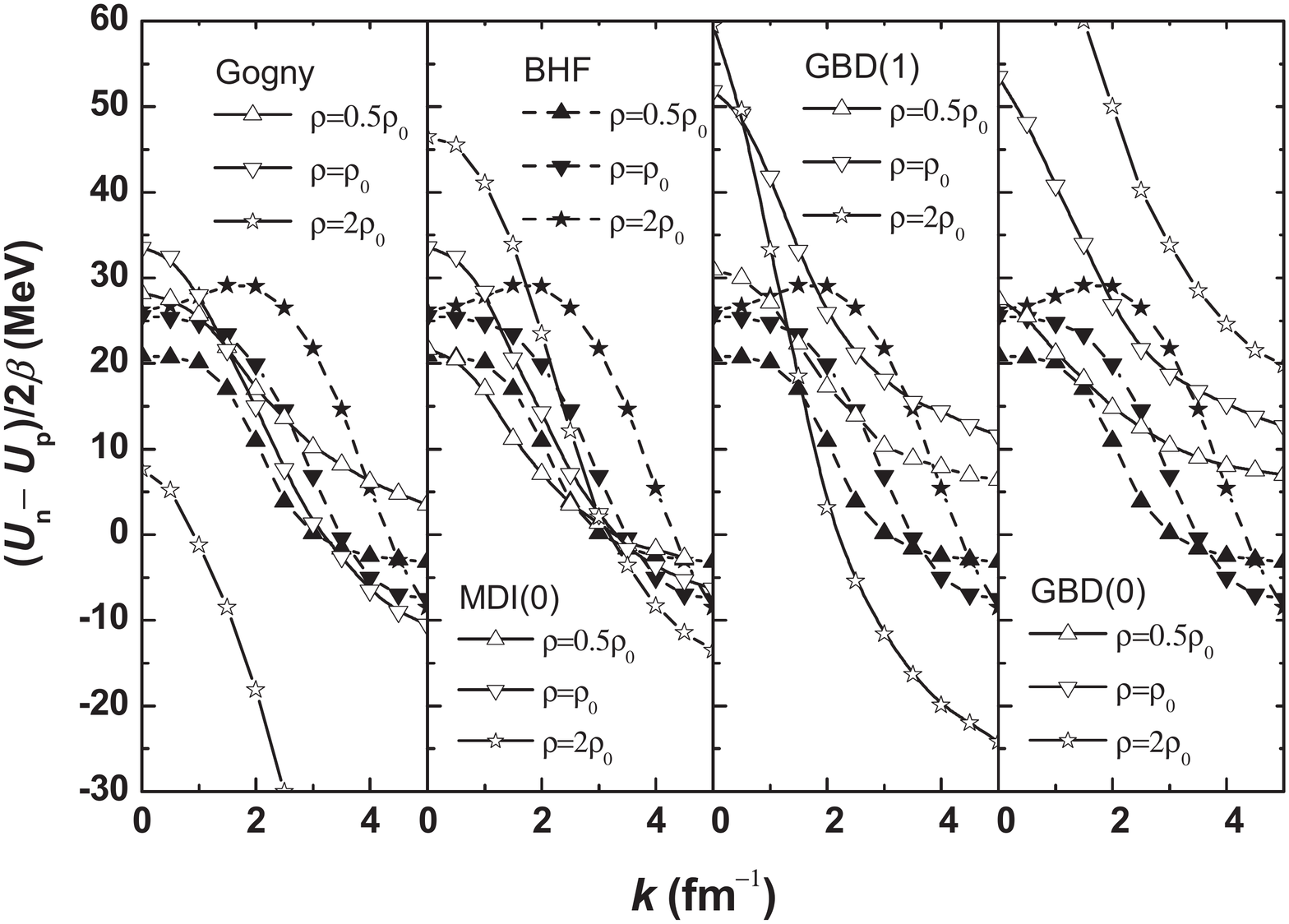}
\caption{Comparison of the BHF symmetry potential with parameterizations based on Hartree-Fock using several phenomenological interactions at three densities. Taken from ref.\,\cite{zuo05}.}
\label{Zuo-fig8}
\end{center}
\end{figure}
\begin{figure}[htb]
\begin{center}
\includegraphics[width=15cm,height=7.5cm]{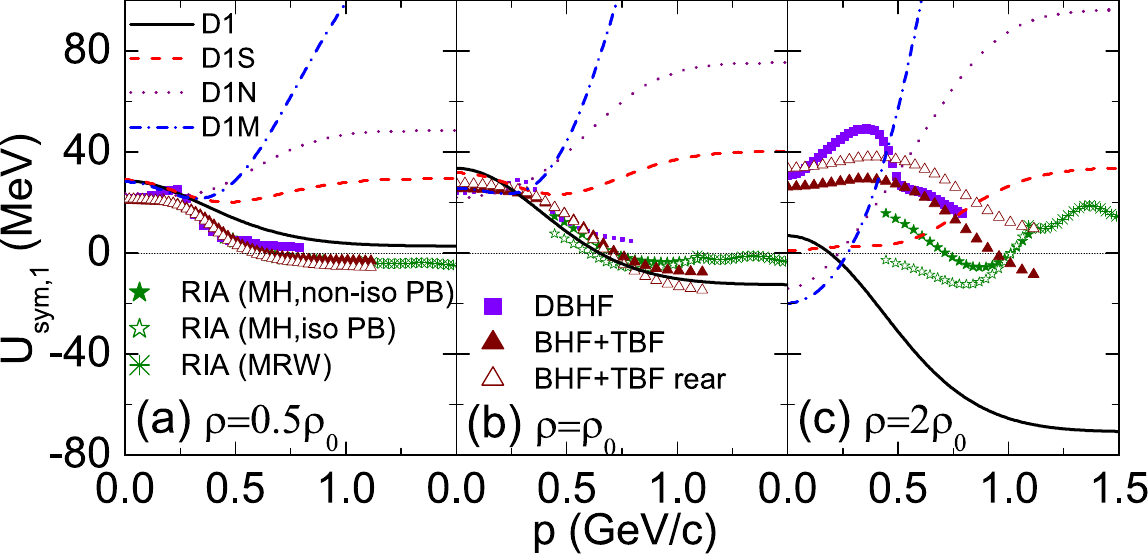}
\caption{Density and momentum dependence of the nucleon isovector potential predicted by the Gogny--Hartree--Fock (GHF)
calculations using the D1, D1S, D1M and D1N interactions, Dirac--Brueckner--Hartree--Fock (DBHF) and
Relativistic Impulse Approximation (RIA) with various two-body and three-body forces (TBF).
Taken from ref.\,\cite{Rchen}.}\label{Usym1}
\end{center}
\end{figure}

In isospin asymmetric matter, it is the difference of single-particle potentials for neutrons and protons that is responsible for various isospin-dependent phenomena. Thus, instead of studying the single-nucleon potentials, interesting features can be revealed more easily by examining individually the isoscalar potential
\begin{equation}
U_0\approx (U_{\rm{n}}+U_{\rm{p}})/2
\end{equation}
and the symmetry (isovector) potential
\begin{equation}
U_{\rm{Lane}}=U_{\rm{sym},1}=U_{\rm{sym}}=U_{\rm{iso}}=U_I\approx (U_{\rm{n}}-U_{\rm{p}})/2\delta.
\end{equation}
Interestingly, it is seen from the left window of Fig.\,\ref{Zuo-fig1} that the sign of the symmetry potential
reverses when the $U_{\rm{p}}$ and $U_{\rm{n}}$ cross at certain momenta. The crossing was explained earlier in terms of some phase-space arguments\,\cite{LOM} and has also been found in many other calculations. We emphasize that those phenomenological models predicting an increasing symmetry potential with increasing energy/momentum, corresponding to a negative neutron-proton effective mass splitting, do not predict such a crossing. Therefore, searching for signatures of the flipping sign of the symmetry potential at high nucleon energy/momentum in nuclear reactions will be useful. We note that while there are common qualitative features, the
quantitative values of the symmetry potentials predicted by different models are still quite diverse. For example, shown in Fig.\,\ref{Zuo-fig8} are the symmetry potentials vs momentum at three densities, $\rho = 0.085$, $0.17$, and $0.34$ fm$^{-3}$ obtained within the extended BHF approach using the Argonne $V_{18}$ two-body interaction\,\cite{wiringa:1995} and a microscopic three-body
 force (TBF)\,\cite{grange:1989} (curves with filled symbols)\,\cite{zuo05} in comparison with parameterizations based on Hartree-Fock calculation using several phenomenological interactions, e.g, the Gogny force, MDI (Momentum Dependent Interaction) and Gale--Bertsch--Das Gupta (GBD) interaction\,\cite{Das03,GBD,Bom01,Riz04,Li04}.

As discussed in detail by Zuo \textit{et al.}\,\cite{zuo05}, while most predictions show a decreasing symmetry potential with increasing nucleon energy, they quantitatively differ significantly especially at supra-saturation densities,
In particular, the phenomenological symmetry potentials drop much faster especially at $\rho = 0.34$ fm$^{-3}$ compared to the BHF prediction. The two sets of the GBD parameterizations, i.e., GBD (0) and GBD (1), differ strongly.
 It is encouraging to see that at the saturation density, the momentum dependence of the Gogny and MDI(0) parametrization is closer to the BHF prediction. However, the original Gogny force predicts
deeply negative symmetry potentials at supra-saturation densities. Moreover, the Gogny--Hartree--Fock (GHF)
predictions depend sensitively on the force parameters.
For example, shown in Fig.\,\ref{Usym1} are the momentum dependences of the symmetry potentials using the GHF,
Dirac--Brueckner--Hartree--Fock (DBHF) and Relativistic Impulse Approximation (RIA) with various two-body and three-body
interactions at sub-saturation, saturation and supra-saturation densities, respectively\,\cite{Rchen}. While they all basically agree with each other at saturation density, they disagree significantly especially at high densities.
The GHF predictions using the four widely used parameter sets D1, D1S, D1N and D1M are particularly interesting.
Unlike the result with the D1 parameter set, calculations with the  D1S, D1N and D1M predict increasing symmetry potentials
with increasing momentum even at saturation density. As we  shall discuss in more detail,
this is in contrast with the momentum dependence of the symmetry potential at $\rho_0$ extracted from optical
model analysis of nucleon-nucleus scatterings.

\begin{figure}[htb]
\vspace{-0.4cm}
\includegraphics[width=15cm,height=9cm]{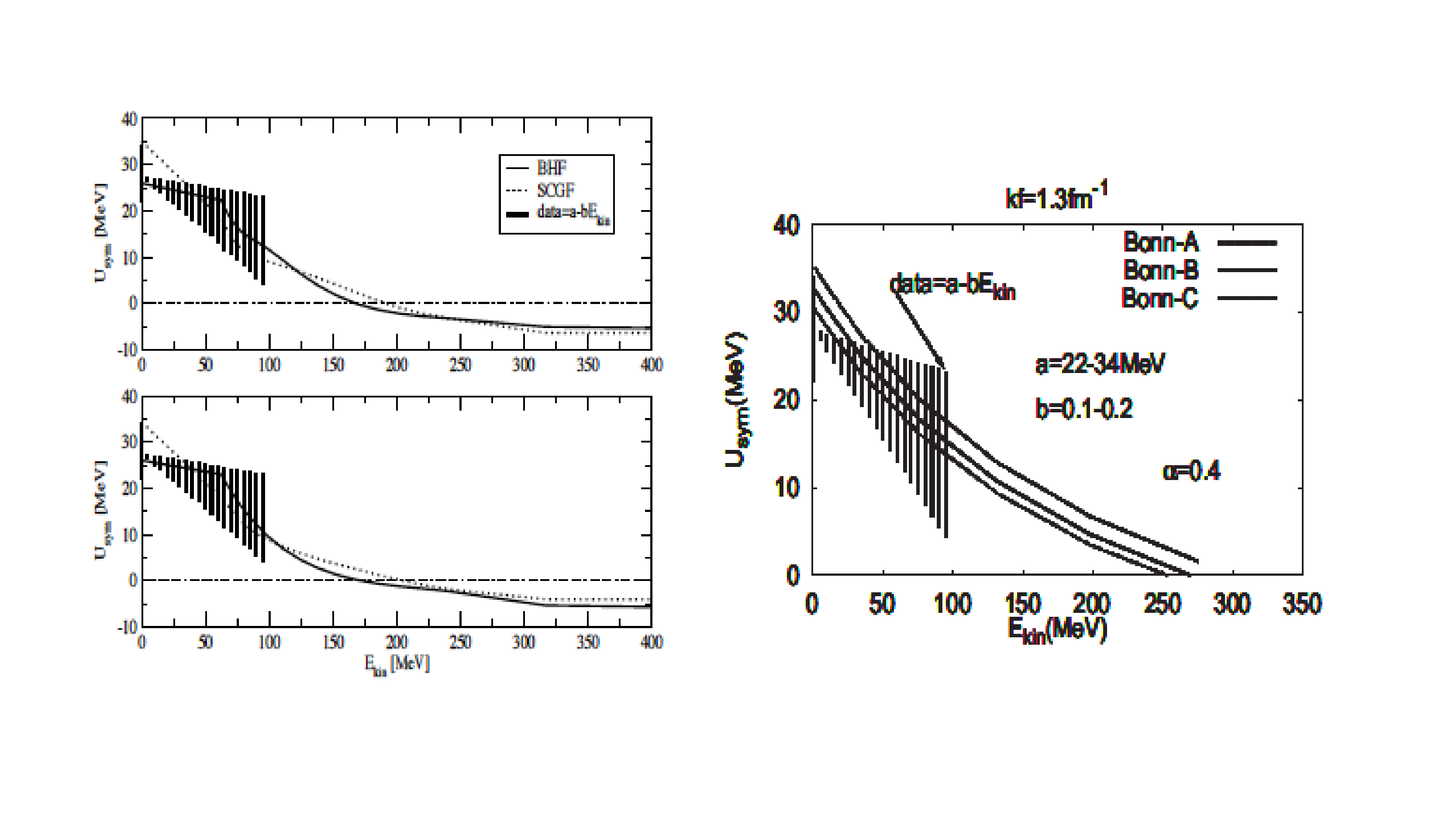}
\vspace{-1cm}
\caption{Left: The symmetry potential as a function of nucleon kinetic energy at nuclear matter saturation density and an isospin-asymmetry $\delta=0.2$ (upper panel) and at $\delta=0.4$ (lower panel) in comparison with the empirical symmetry potential (shaded area) from optical model analyses of old nucleon-nucleus scattering experiments. Taken from ref. \protect\,\cite{Egypt}. Right: The symmetry potential versus nucleon kinetic energy at saturation density within the DBHF approach using the Bonn A, B and C potentials. The shaded area represents the empirical symmetry potential at normal density from earlier analyses of nucleon-nucleus scatterings using optical models. Taken from ref.\,\cite{Sam05}.}\label{Uscgf}
\end{figure}

Predictions for the symmetry potential from the BHF using the CD-Bonn potential\,\cite{CDBonn} and the Self-Consistent Green's Function (SCGF) method\,\cite{Egypt} are compared in the left window of Fig.\,\ref{Uscgf} with the empirical symmetry potential from analyzing experimental data. The latter was extracted from optical model analyses of old neutron-nucleus scatterings, (p,n) charge exchange reactions as well as proton scatterings on isotope chains of several elements\,\cite{Hod94}. For nucleon kinetic energies  $E_{\rm{kin}}$ below about 100 MeV,  the empirical isovector optical potential, traditionally also known as the Lane potential, decreases approximately linearly with
increasing $E_{\rm{kin}}$ (in MeV) according to
\begin{equation}\label{ULane}
U_{\rm{Lane}} = a-bE_{\rm{kin}}
\end{equation}
where $a\approx 22-34$ MeV and $b\approx 0.1-0.2$\,\cite{Li04,Hod94}. We note here that this isovector optical potential should be carefully transformed into the symmetry potential of ANM at $\rho_0$ as we shall discuss in detail in section \ref{Exp-cs} before being used as a boundary condition for testing many-body theories. However, the uncertainties of this old empirical constraint at low-energies are larger than the difference caused by the transformation.
It thus still provides a useful reference at least qualitatively. This empirical symmetry potential was first used in ref.\,\cite{Li04} to constrain the phenomenological symmetry potentials and the corresponding neutron-proton effective mass splittings used in transport model simulations of heavy-ion collisions. As we shall discuss in detail in Section \ref{Exp-cs}, recent analyses of more complete and new data of nucleon-nucleus scatterings have not only extended the energy range but also constrained the symmetry potential more tightly. Interestingly, the old and new constrains on the symmetry potential are consistent qualitatively.

It is interesting to see that predictions of both the BHF and SCGF are consistent with the empirical symmetry potential. It is also interesting to see that the predicted nuclear symmetry potentials change from positive to negative values at $E_{\rm{kin}}$ higher than the crossing energy of about 200 MeV, implying that protons (neutrons) feel an attractive (repulsive) symmetry potential at lower energies but a repulsive (attractive) one at high energies\,\cite{Egypt}. As shown earlier, most of the calculations compared in Fig.\,\ref{Usym1} predict that the symmetry potential vanishes or changes sign at high momentum. Since the symmetry potential is generally very weak at high momenta, except in the GHF calculations, it is very challenging to find signatures of its sign inversion.
We comment here that the decreasing symmetry potential with increasing energy may make isospin effects in heavy-ion collisions weaker at high energies. On the other hand, the symmetry potential is also density dependent.   Indeed, as shown in Fig.\,\ref{Usym1}, some models predict an increasing symmetry potential at higher densities reachable with higher beam energies in heavy-ion collisions. Thus, it is currently unclear and remains an interesting question as to the optimal beam energy where effects of the symmetry potential maximize.

\begin{figure}[htb]
\begin{center}
\includegraphics[width=0.65\textwidth] {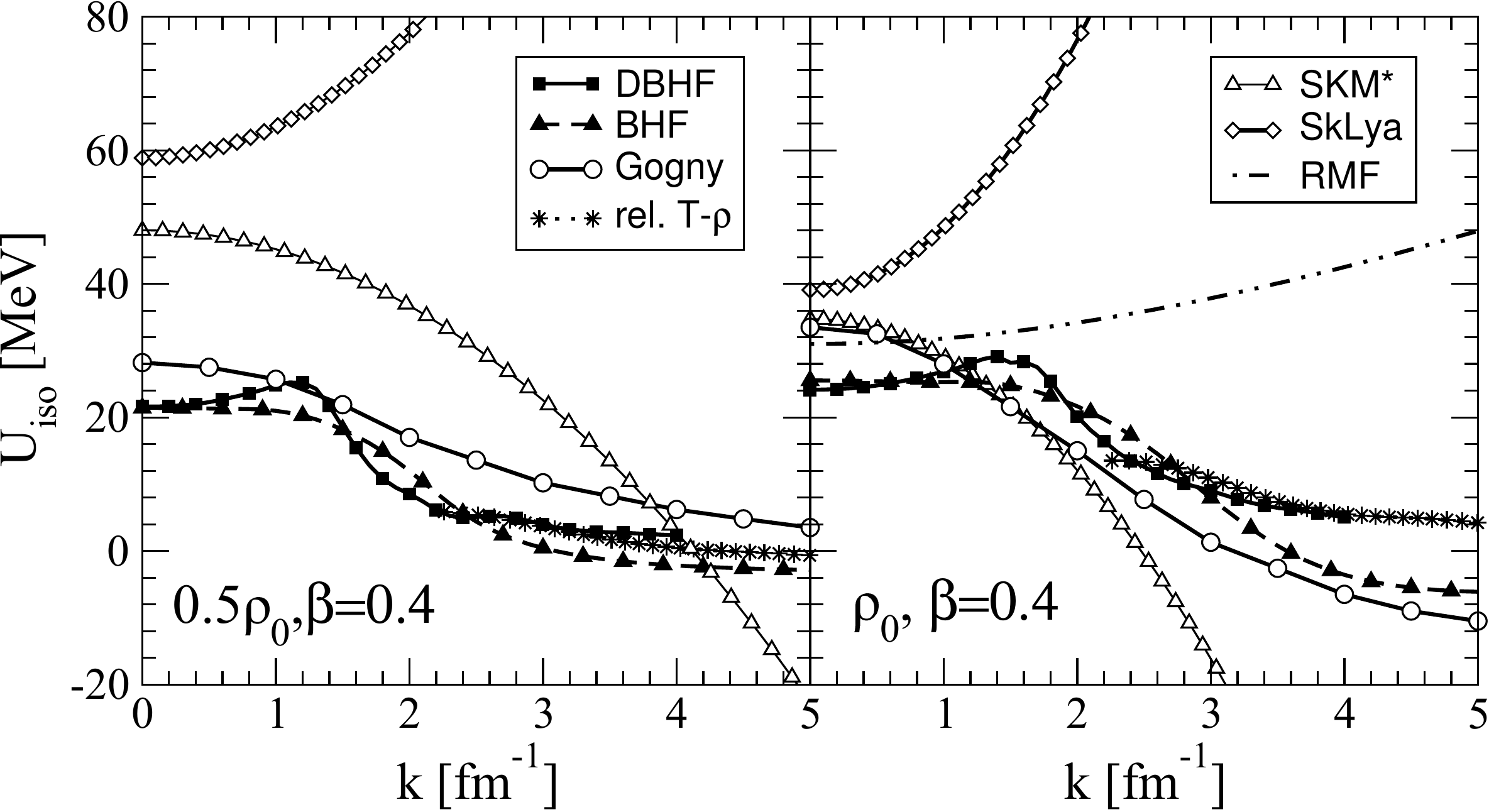}
\caption{The isovector optical potentials from the various models in asymmetric nuclear matter ($\beta=0.4$) as a function of momentum
$k$ at densities of $\rho=0.5 \rho_0$ (left) and $\rho_0$ (right) where $\rho_0$ is the saturation density of nuclear matter. The isovector optical potential from the DBHF approach using the Bonn-A interaction is compared with the ones from the non-relativistic BHF approach\,\cite{zuo05}, the phenomenological RMF\,\cite{gaitanos04}, Gogny-Hartree-Fock\,\cite{gogny02} and Skyrme-Hartree-Fock\,\cite{skyrme04} predictions as well as from a relativistic $T-\rho$ approximation\,\cite{chen05b}. Taken from ref.\,\cite{Fuc05}.
\label{Fuchs-fig1}}
\end{center}
\end{figure}

\begin{figure}[htb]
\begin{center}
\includegraphics[scale=0.5]{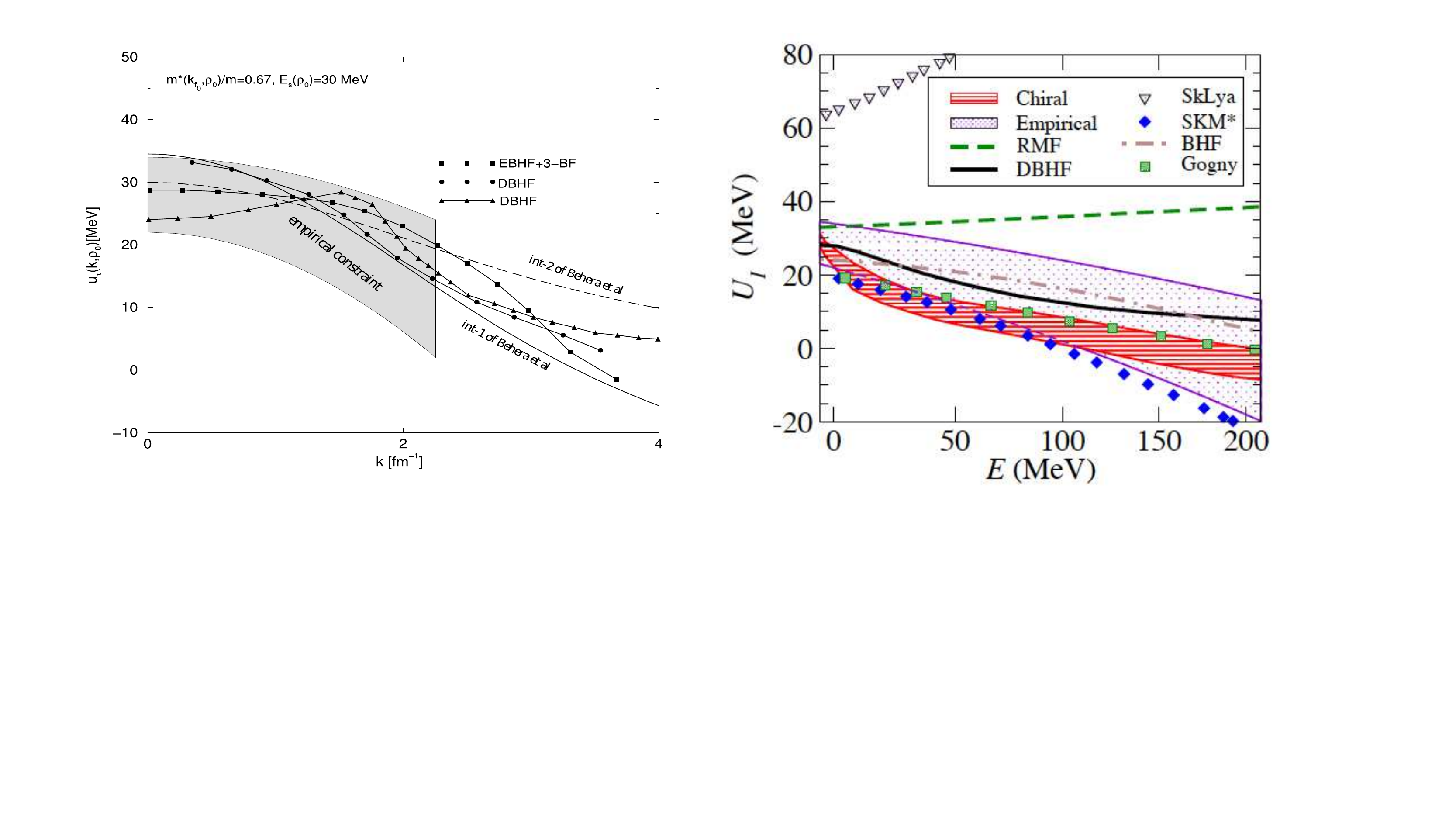}
\vspace{-4.5cm}
\caption{\protect\small \label{India}
Left: Comparisons of symmetry potentials at saturation density as a function of momentum $k$ predicted using (1) a phenomenological model with two different forms of the isospin-dependent finite-range interaction by Behera \textit{et al.}\,\cite{Beh11}, (2) the extended BHF calculations by Zuo \textit{et al.}\,\cite{zuo05}, (3) the DBHF calculations by van Dalen \textit{et al.}\,\cite{Fuc05} and (4) the DBHF calculations by Sammarruca \textit{et al.}\,\cite{Sam05}, Modified from a figure in ref.\,\cite{Beh11}. Right: Comparing the chiral effective field theory prediction on the kinetic energy dependence of the symmetry potential by Holt \textit{et al.} with results from other models and the empirical constraints. Taken from ref.\,\cite{Hol16}.}
\end{center}
\end{figure}

Several versions of the relativistic approaches using different approximations based on realistic nucleon-nucleon interactions, e.g., the DBHF approach\,\cite{Muther1,Sam03,DBHF}, have been used to study various properties of isospin-asymmetric nuclear matter with fruitful results. In particular, the nuclear symmetry potential and the corresponding nucleon effective masses in neutron-rich matter have been investigated within the DBHF using several interactions\,\cite{mu04,Fuc05,Ron06,Sam05}. The results are qualitatively consistent with each other, with the empirical constraint and some non-relativistic calculations. For example, shown in the right window of Fig.\ \ref{Uscgf} are symmetry potentials obtained by using the Bonn A, B, and C potentials\,\cite{Mac89}. The three interactions differ mainly in the strength of the tensor force involved\,\cite{Sam05}. The predicted symmetry potentials all decrease with increasing energy/momentum and approach negative values at high energy/momentum. At low energies, they are also consistent with the empirical constraint of Eq.\,(\ref{ULane}). Compared with the BHF predictions using the CD-Bonn interaction shown in the left panel of Fig.\,\ref{Uscgf}, the crossing energies where the symmetry potential becomes zero predicted in the DBHF using the Bonn A, B and C interactions are about 50 MeV higher, indicating an appreciable model and interaction dependence in the calculations. In Fig.\,\ref{Fuchs-fig1}, predictions using another version of the DBHF theory with the Bonn-A interaction are compared with the those from the non-relativistic BHF\,\cite{zuo05}, phenomenological Gogny\,\cite{gogny02} and Skyrme\,\cite{skyrme04} Hartree-Fock calculations as well as a relativistic $T-\rho$ approximation\,\cite{chen05b} based on empirical relativistic NN scattering amplitudes\,\cite{neil83}.
The DBHF results are in reasonably good agreement with those from the non-relativistic BHF \,\cite{zuo05}, the relativistic $T-\rho$ approximation\,\cite{chen05b} and the Hartree-Fock calculation using the default Gogny interaction\,\cite{gogny02}. It is well known that SHF predictions for the symmetry potential are extremely diverse depending strongly on the Skyrme parameterizations used. For example, some of the recent Skyrme-Lyon
parameterisations\,\cite{skyrme04}, e.g., the SkLya, predicts an increasing symmetry potential, while the SkM$^*$ predicts a decreasing one albeit with a much stronger slope than those predicted by the microscopic approaches.
It is also well-known that relativistic mean field models assuming normally momentum independent self-energy components predict that the symmetry potential $U_{\rm{iso}}$ increases linearly (quadratically) with energy
(momentum), see, e.g.,  refs.\,\cite{Che07,gaitanos04,baran05}. These predictions are similar to the Hartree-Fock calculations using the D1S, D1M and D1N Gogny as well as some Skyrme interactions. Unfortunately, the trend of
these predictions is opposite to that of the empirical experimental constraints discussed earlier.

Curiously, how do the relativistic microscopic many-body theory predictions quantitatively compare with each other? This is a meaningful question as it has been known that the DBHF predictions depend strongly
on approximation schemes and techniques used to determine the Lorentz and the isovector structure of the nucleon self-energy\,\cite{Fuc05}. To answer this question, shown in Fig.\,\ref{India} are comparisons of the symmetry potentials at saturation density as a function of momentum predicted using (1) a phenomenological model with two different forms of the isospin-dependent finite-range interaction by Behera \textit{et al.}\,\cite{Beh11}, (2) the extended BHF calculations by Zuo \textit{et al.}\,\cite{zuo05}, (3) the DBHF calculations by van Dalen \textit{et al.}\,\cite{Fuc05} and (4) the DBHF calculations by Sammarruca \textit{et al.}\,\cite{Sam05}. It is interesting to see that indeed they all fall into the still quite large uncertain range of the empirical constraint. However, they do have appreciable numerical differences in the whole momentum range considered. Obviously, the empirical constraint needs to be
refined, and indeed it has been as we shall see in section \ref{Exp-cs}. It is currently unclear if the quantitative differences in the predictions have any appreciable effect on any experimental observables.  Most of the predictions for the symmetry potential mentioned above are further compared with the prediction of the chiral effective field theory in the right window of Fig.\,\ref{India}. It is seen that the chiral effective field theory prediction (between the two red lines) is also consistent with the empirical experimental consistent of Eq.\,(\ref{ULane}) as well as the BHF and DBHF results. Moreover, the crossing energy is predicted to lie in the range of $155 \pm 45$ MeV, comparable with the BHF and SCGF predictions but somewhat lower than the DBHF predictions using the Bonn forces.

In summary of this subsection, essentially all nuclear many-body theories, relativistic and non-relativistic, microscopic and phenomenological in nature using basically all available two-body and three-body forces or model Lagrangians have been used to calculate the single-nucleon potential in neutron-rich nucleonic mater. Despite of the great successes of these models in describing saturation properties of nuclear matter and various structures of nuclei, mainly because of our poorly knowledge of isovector properties of nuclear interactions, predictions of these models on the symmetry potential/energy are still rather diverse especially at high densities/momentum.
In particular, most of the calculations predict that the nuclear symmetry potential decreases with increasing nucleon energy/momentum at saturation density qualitatively in agreement with the empirical constraint from analyses of nucleon-nucleus scattering, (p,n) and ($^3$He,t) charge exchange reactions. However, the quantitative values of the symmetry potential are still model and interaction dependent. For example, the crossing energy where the symmetry potential changes its sign is rather model and interaction dependent. Moreover, the predictions diverge quite widely at abnormal densities, especially at supra-saturation densities.  As the momentum/energy dependence of the symmetry potential is the key quantity affecting most of the isospin-dependent features of both nuclear structures and reactions, and it is the
fundamental origin of the neutron-proton effective mass splitting, its more thorough understanding and experimental constraints deserves continuous and special efforts of the nuclear physics community.

\subsection{Nucleon effective masses in neutron-rich matter predicted by nuclear many-body theories}
Given the single-nucleon potentials, various kinds of nucleon effective masses can be calculated straightforwardly using the definitions given in Section \ref{definition}. Because of the diverse predictions of nuclear many-body theories and the fact that the effective masses are determined by the slopes of the single-particle potentials with respect to momentum or energy, one can imagine that the corresponding predictions for the nucleon effective masses are even more diverse especially for the neutron-proton effective mass splitting in neutron-rich matter. In this section, we make a few observations about the major features of the predictions available in the literature by expanding a brief recent review about the nucleon effective masses in neutron-rich matter\,\cite{LiChen15}. The focus of many recent studies has been on the neutron-proton effective masse splittings and their dependences on the isospin asymmetry and density of the neutron-rich medium encountered in heavy-ion collisions\,\cite{NST16} and in some astrophysical situations\,\cite{mrs,Pag06}, such as in neutron stars and neutrino spheres of supernova explosions. As to the empirical constraints as we discussed in Section \ref{emass-values}, we have the isoscalar effective mass $m^*_{\rm{s}}/m\approx 0.8\pm 0.1$, the isovector effective mass $m^*_{\rm{v}}/m=0.6\sim 0.93$ and the neutron-proton effective mass splitting $m^*_{\rm{n-p}}\geq 0$ in neutron-rich matter at $\rho_0$, respectively. The latter and its more quantitative empirical constraints existing in the literature, i.e., $m^{*}_{\rm{n-p}}(\rho_0,\delta)\approx (0.41\pm0.15)\delta$ from ref.\,\cite{LiX15} and/or $m^*_{\rm{n-p}}(\rho_0,\delta)\approx (0.27\pm 0.25)\delta$ from ref.\,\cite{LiBA13} have not been widely used in constraining many-body theory predictions yet.
On the other hand, it is interesting to note here that chiral effective field theory of refs. \cite{Hol16,Hol13} predicted that $m^{*}_{\rm{s}}/m = 0.82\pm0.08$ and $m^{*}_{\rm{v}}/m = 0.69\pm0.02$, thus a
positive $m^*_{\rm{n-p}}\approx (0.309\pm 0.227)\delta$ in very good agreement with the empirical constraints. These results from the chiral effective field theory have been used recently to calibrate mean-field models using Skyrme interactions \cite{Kim,ZZhang}.

\begin{figure}[htb]
\centering
\includegraphics[width=7.5cm,height=8cm]{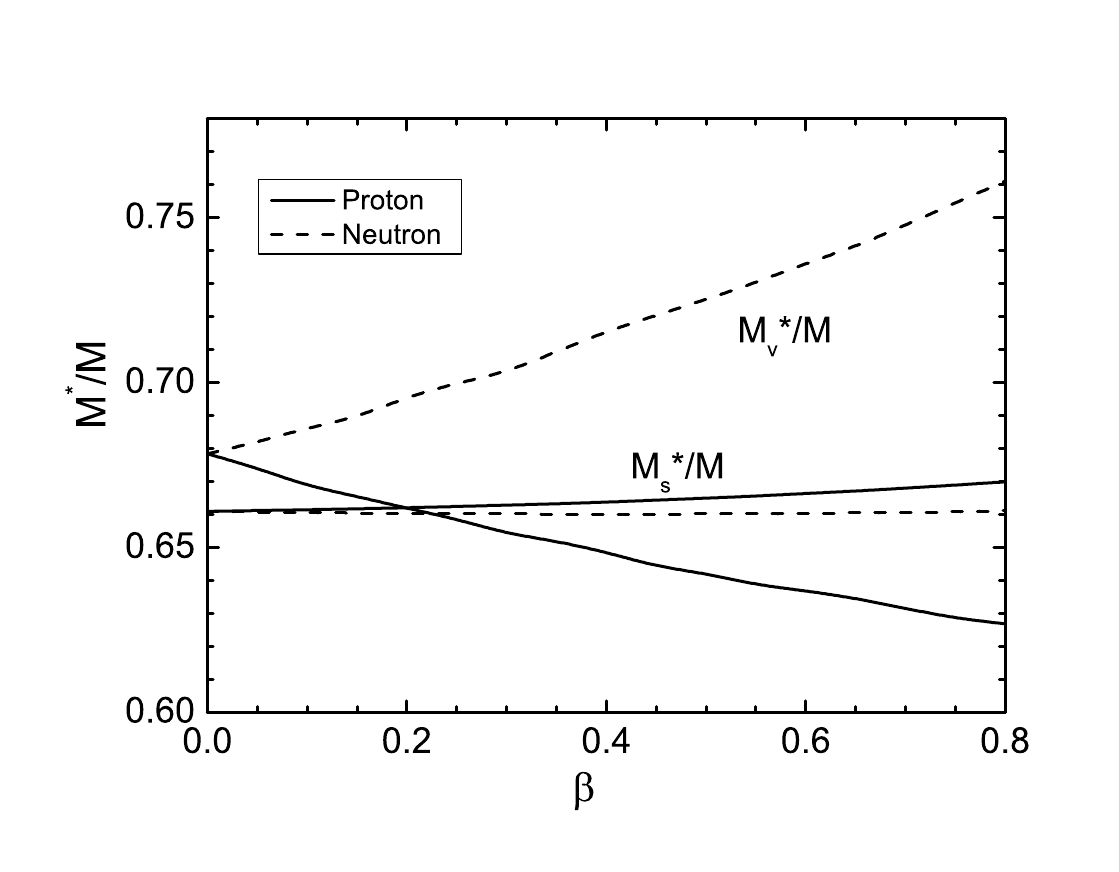}
\hspace{0.5cm}
\includegraphics[width=6.5cm,height=7.0cm]{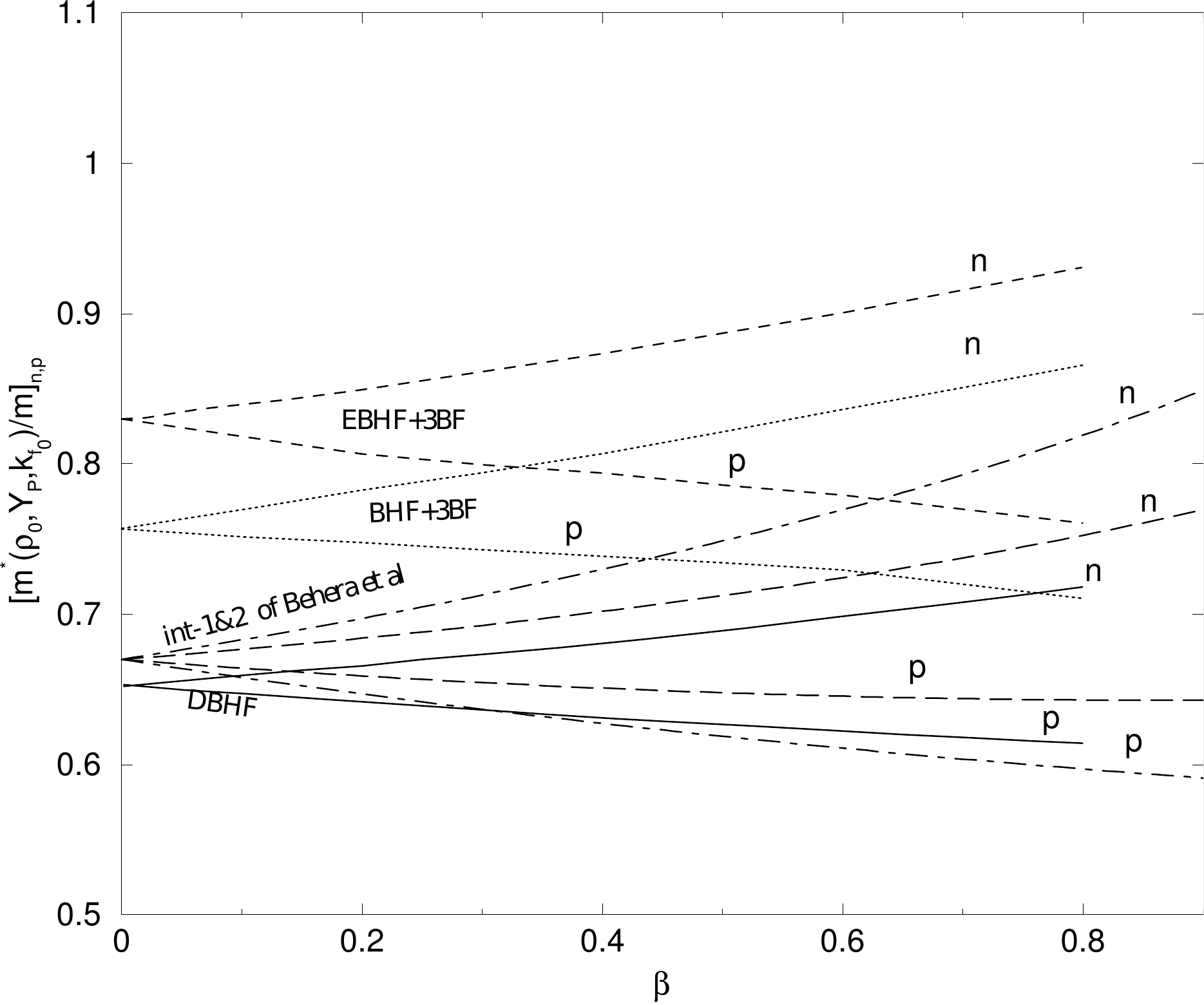}
\caption{Left: Nucleon Lorentz-scalar (Dirac) effective mass $M^*_{\rm{s}}$ and Lorentz-vector effective masses $M^*_{\rm{v}}$ at a nucleon energy of $E=50$ MeV in neutron-rich matter of isospin-asymmetry $\beta$ with $k_{\rm{F}}=1.36\,\rm{fm}^{-1}$. Taken from ref.\,\cite{Ron06}. Right: Comparisons of nucleon effective masses as a function of isospin asymmetry $\beta$ at saturation density predicted by (1) the DBHF theory\,\cite{Sam05}, (2) the BHF+3-body force, (3) the extended BHF+3-body force\,\cite{zuo05} as well as the energy density functional using two forms (int-1$\&$2) of isospin-dependent interactions\,\cite{Beh11}. Modified from a figure in ref.\,\cite{Beh11}.}
\label{MaFig}
\end{figure}
It is necessary to note that in standard RMF models, the scalar and vector nucleon self-energies are independent of momentum or energy. When only a vector isovector $\rho$-meson is included in RMF models, the Dirac mass has no isospin splitting, while the inclusion of the scalar-isovector $\delta$-meson leads to a negative neutron-proton Dirac mass splitting\,\cite{Liu02}. Because of the neglect of non-locality in RMF models, their predictions for various kinds of non-relativistic neutron-proton effective mass splittings using Schr\"odinger equivalent potentials are generally negative (see more detailed discussion in subsection \ref{rel-D}). A comprehensive review of different kinds of effective masses within 23 RMF models using different coupling schemes can be found in ref.\,\cite{Che07} or Chapter 4 of ref.\,\cite{LCK08}. We skip discussing most of the RMF results, except for comparisons in several places, as the conclusions in ref.\,\cite{LCK08} are still valid currently.

\begin{figure}[htb]
\begin{center}
\vspace{-3cm}
\includegraphics[scale=0.45]{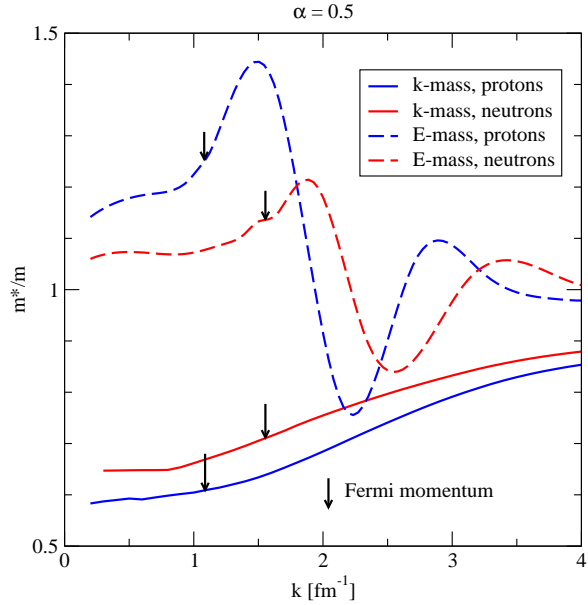}
\caption{Effective k-masses (solid lines) and E-masses (dashed lines) of neutrons (red) and protons (blue) derived from the BHF self-energies using the CD-Bonn interactions for nucleonic matter with an isospin asymmetry of 0.5 at saturation density. Taken from ref.\,\cite{mu04}.}
\label{Fig1-Muther}
\end{center}
\end{figure}

As mentioned earlier and emphasized in several papers, see, e.g., refs.\,\cite{Jam89,mu04,Fuc05,zuo05,Ron06,Fuc06,Che07}, effective masses in relativistic and non-relativistic models have different meanings and should not be directly compared with each other. For example, we notice here that while there is still some model dependence, as it was discussed in detail in ref.\,\cite{Fuc05}, the neutron-proton Dirac mass splitting is generally negative, i.e, $M_{\mathrm{Dirac}}^{\ast }$ for neutrons is less than that for protons in neutron-rich matter, while it is the opposite for the non-relativistic $m^*_{\rm{n-p}}$ obtained from the Schr\"odinger equivalent potential. For example, shown in the left window of Fig.\ \ref{MaFig} are the nucleon Lorentz-scalar effective masses (Dirac masses) (lower curves) and Lorentz-vector effective masses (upper curves) for neutrons (solid) and protons (dashed), respectively, from the DBHF calculations by Ma \textit{et al.}\,\cite{Ron06} using the DBHF approach originally developed by Schiller and M\"uther\,\cite{Muther1} and the Bonn B interaction\,\cite{Mac89}. Please note that these effective masses should not be confused with the isoscalar and isovector masses defined in Section \ref{definition} although they may be denoted by the same symbols in some papers. The neutron Lorentz-scalar effective mass was found smaller than that of protons due to the stronger scalar potential of neutrons in neutron-rich matter. However, the difference is very small. On the other hand, as it has been emphasized earlier, it is the Lorentz-vector effective masses that should be compared with the non-relativistic total nucleon effective masses. Indeed, it is interesting to see that the neutron Lorentz-vector mass is higher than that of protons due to the stronger energy dependence of the neutron vector potential in neutron-rich nuclear matter. This result is consistent with predictions of the Landau-Fermi liquid theory\,\cite{Sjo76} and the BHF approach\,\cite{LOM} as well as the empirical constraints discussed earlier.

Shown in the right window of Fig.\ \ref{MaFig} are comparisons of total nucleon effective masses as a function of isospin asymmetry $\beta$ at saturation density predicted by (1) the DBHF theory by Sammarruca \textit{et al.}\,\cite{Sam05}, (2) the BHF+3-body force, (3) the extended BHF+3-body force including the renormalization contributions\,\cite{zuo05} as well as the phenomenological energy density functional approach using two forms of isospin-dependent interactions\,\cite{Beh11}. Some of these results correspond to the momentum dependence of the symmetry potentials shown in Fig.\,\ref{India}. First of all, the two DBHF calculations shown on the left and right windows are consistent. It is also seen that the BHF and DBHF predict significantly different effective masses because their predictions for the slopes of the symmetry potential at Fermi momenta are quite different although their magnitudes are all consistent with the empirical constraints as already shown in Fig.\,\ref{India}. The results from the regular BHF and the extended BHF are also quite different basically at all isospin asymmetries. However, it is unclear to us what specific physics ingredients are responsible for the different results from some of the models. The model and interaction dependencies of the predictions for nucleon effective masses are expected to grow at higher densities. It has been discussed earlier that besides the different assumptions and approximations adopted in different many-body approaches, the poorly known spin-isospin dependence of the three-body force, short-range behavior of the tensor force and the isospin-dependence of nucleon-nucleon short-range correlations are among the most important origins of the uncertain high density/momentum behavior of the symmetry potential, and thus the neutron-proton effective mass splitting in dense neutron-rich matter\,\cite{Xu-tensor,LiNews}. The individual roles of these ingredients have been investigated within some models, but more comprehensive and in-depth comparisons will be very useful.
\begin{figure}[h!]
\centering
\includegraphics[width=5.5cm,height=6.5cm]{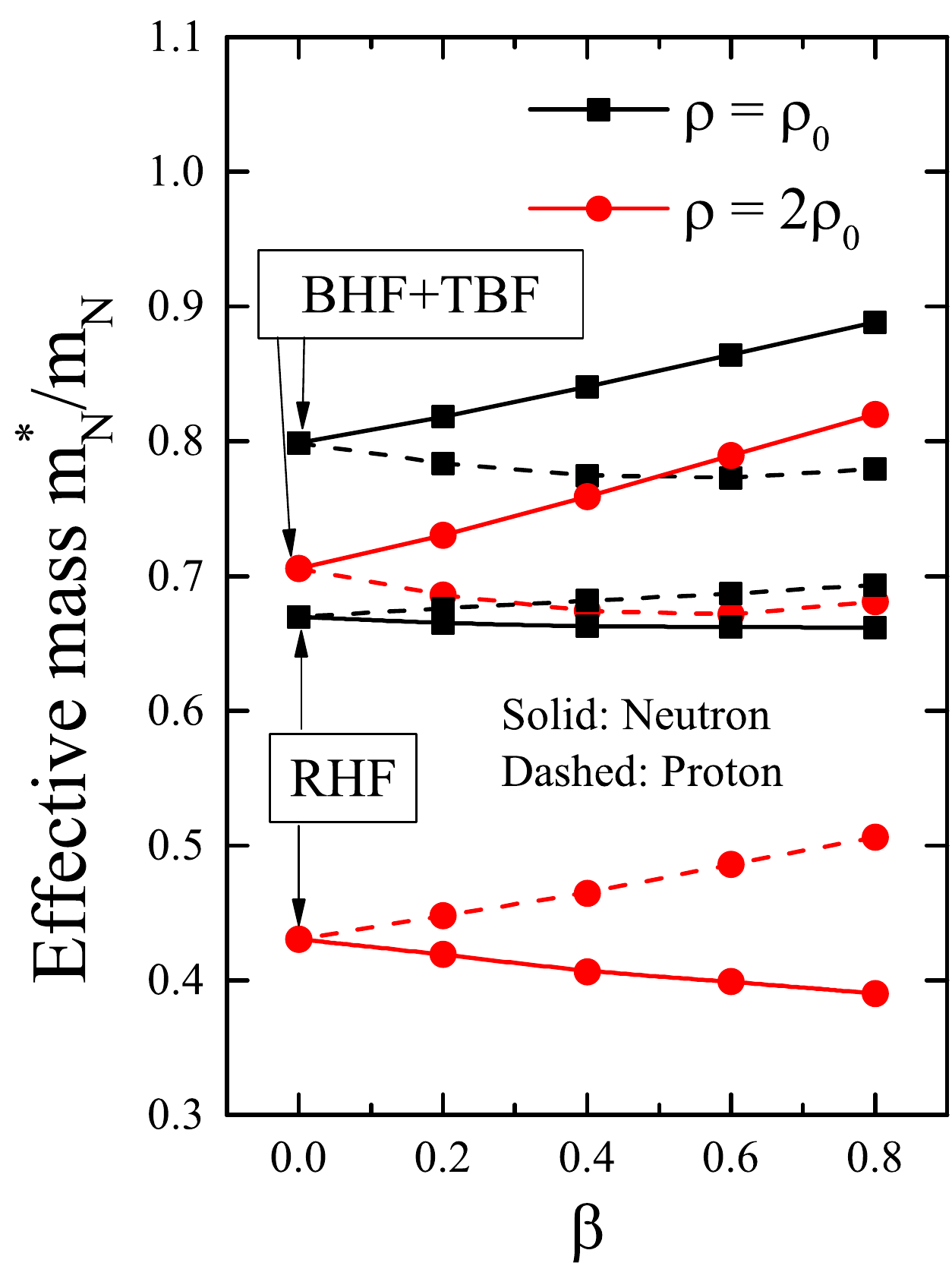}~~~~
\includegraphics[width=10.cm,height=6.5cm]{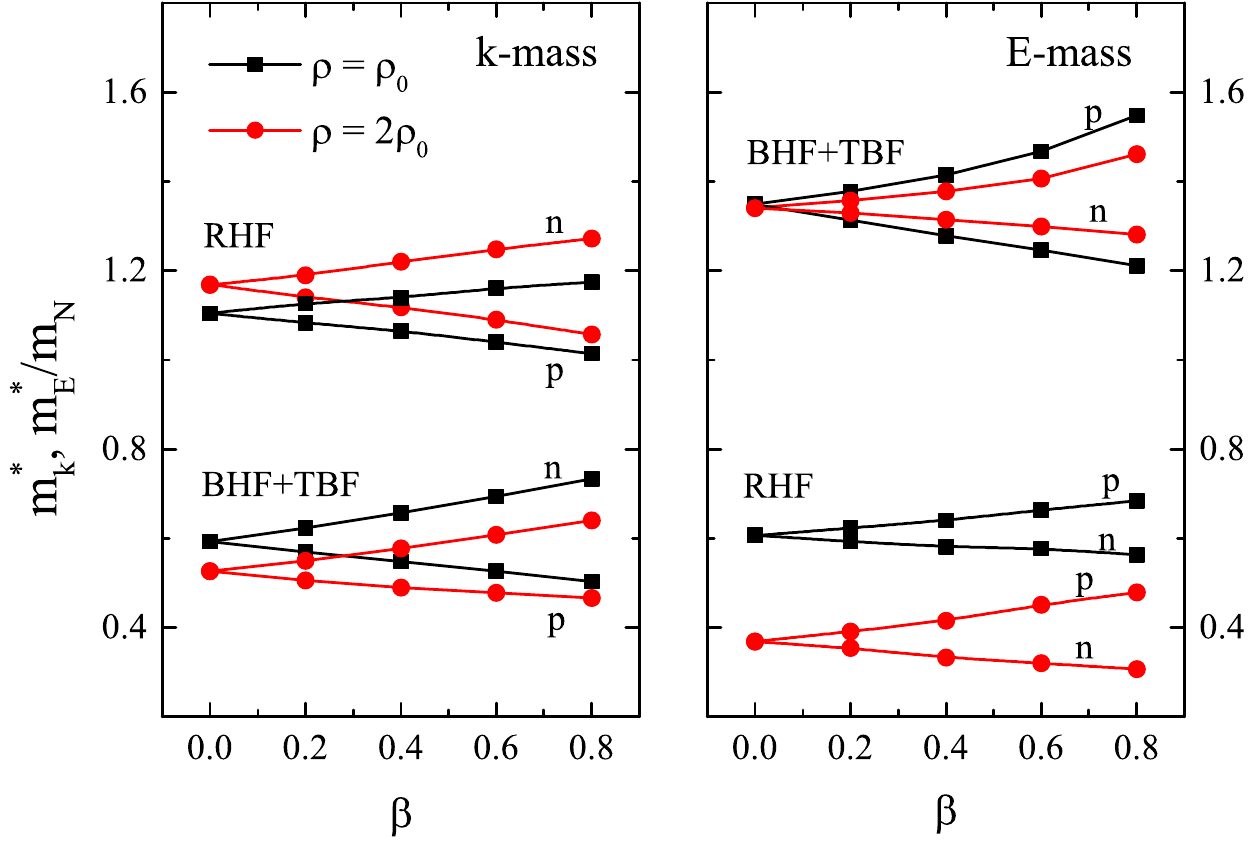}
\caption{The total nucleon effective mass (left),  k-mass (middle) and E-mass (right) as functions of isospin-asymmetry $\beta$ within the BHF + TBF model and the RHF model at two densities of $\rho = \rho_0$ (squares) and $\rho = 2\rho_0$ (dots). Taken from ref.\,\cite{Ang16}.}\label{AngFig}
\end{figure}

Many predictions have been made about nucleon effective masses using various non-relativistic many-body theories. As an example, shown in Fig.\,\ref{Fig1-Muther} are the E-mass and k-mass of neutrons and protons as functions of nucleon momentum in neutron-rich matter of isospin asymmetry $\alpha=0.5$ predicted by the BHF using the CD-Bonn interaction\,\cite{mu04}. It is seen that the k-mass of neutrons is higher than that of protons at all momentum considered, while the E-mass peaks near the Fermi momenta and is higher for protons than neutrons. The E-masses (k-masses) of both neutrons and protons at their respective Fermi momenta are higher (lower) than their free masses. These general features are shared by most BHF and DBHF calculations using various interactions and approximations. However, the quantitative results are still model and interaction dependent as indicated earlier from looking at the momentum/energy dependence of the symmetry potential in the previous section. As an example, it is useful to look at detailed comparisons between results from the extended BHF+TBF and the RHF (Relativistic Hartree-Fock of Long \textit{et al.}\,\cite{Long06}) in a recent study by Li \textit{et al.}\,\cite{Ang16}. Shown in Fig.\ \ref{AngFig}
are the total effective masses (left), k-mass (middle) and E-mass (right) at two densities of $\rho = \rho_0$ and $\rho = 2\rho_0$. In contrast to the BHF prediction, the RHF predicts that $m^*_{\rm{p}} > m^*_{\rm{n}}$ and the splitting is larger at high densities. However, it is interesting to see in the middle and right windows that the two models predict the same neutron-proton splitting trend for both the k-mass and E-mass, namely, $m^{\ast,\rm{E}}_{\rm{p}} > m^{\ast,\rm{E}}_{\rm{n}}$ and $m^{\ast,\rm{k}}_{\rm{n}} > m^{\ast,\rm{k}}_{\rm{p}}$. Since the total effective mass is the product of the k-mass and E-mass, the final neutron-proton effective mass splitting is determined by the smaller component. It was found that the RHF calculations exhibit a strong density dependence for the neutron-proton total effective mass splitting\,\cite{Long06,Ang16}. In fact, the splitting at low densities ($< 0.8\rho_0$) is different from the ones shown at $\rho = \rho_0$ or $2\rho_0$. Nevertheless, the RHF result at $\rho_0$ is inconsistent with the empirical constraint on the total neutron-proton effective mass splitting. This was attributed to the missing short-range correlations in the RHF model\,\cite{Ang16}.
\begin{figure}[h!]
\centering
\vspace{-3cm}
\includegraphics[scale=0.55]{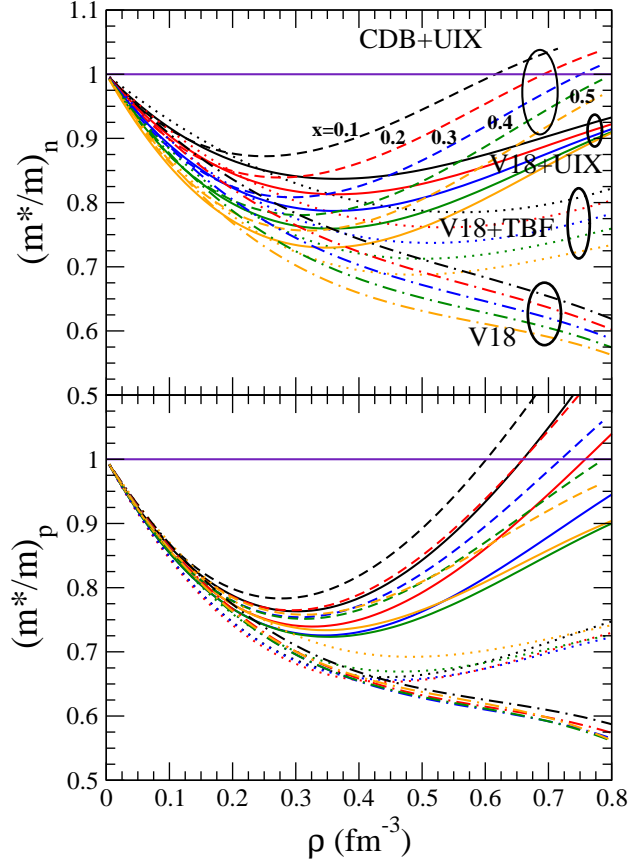}
\caption{Density dependence of neutron (top) and proton (bottom) effective
mass from BHF calculations at a proton fraction of: $x = 0.1, 0.2, 0.3, 0.4$, and $0.5$ using several  different two- and three-body forces. Taken from ref.\,\cite{Bal14}}
\label{EffMass3BFBaldo14}
\end{figure}
\begin{figure}
\centering
\includegraphics[width=18.cm, height=10.cm]{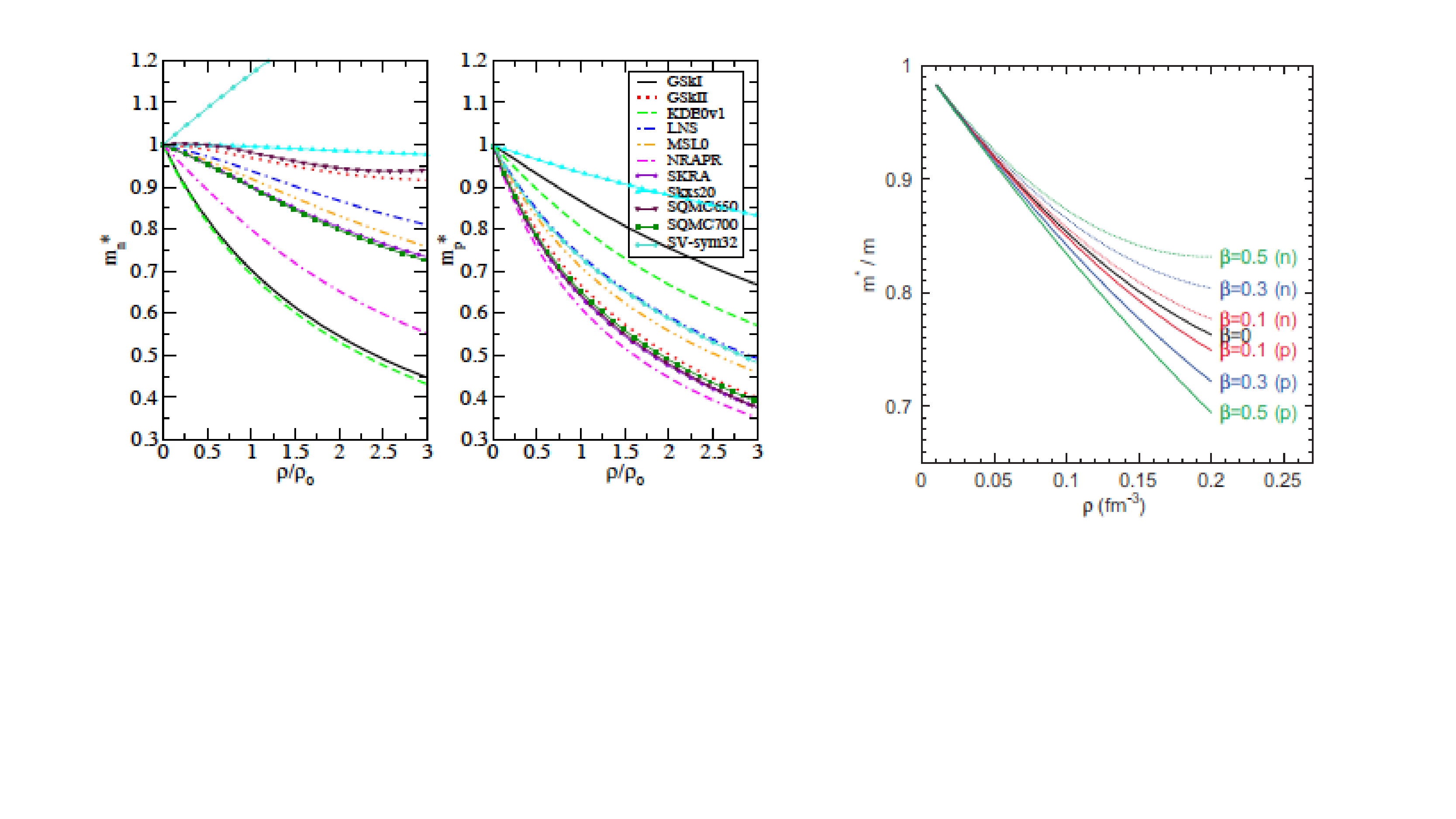}
\vspace{-4.cm}
\caption{\protect\small Left: Density dependence of neutron (left panel) and proton (right panel) effective mass in $\beta$-equilibrium matter as calculated using Skyrme interactions consistent with all macroscopic constraints considered.Taken from ref.\,\cite{Dut12}. Right: Proton (p, full line) and neutron (n, dotted line) effective masses as a function of the density for different values of the isospin asymmetry parameter $\beta$ from
the Barcelona-Catania-Paris-Madrid (BCPM) energy functional. Taken from ref.\,\cite{Baldo16a}.}
\label{Dutra-m}
\end{figure}

To investigate effects of the three-body force on high-density nucleon effective masses, Baldo \textit{et al.} have recently carried out a study using the BHF with different two- and three-body forces (TBFs).
Shown in Fig.\,\ref{EffMass3BFBaldo14} are their results using the Argonne V18 potential without TBFs (V18; dash-dotted lines), with microscopic TBFs (V18+TBF; dotted lines), with phenomenological Urbana TBFs (V18+UIX; solid lines) and for the CD-Bonn potential plus the Urbana TBFs (CDB+UIX; dashed lines). It is interesting to see that without the TBFs the nucleon effective masses all decrease with increasing density, whereas the TBFs leads to a rapid enhancement of the nucleon effective masses at densities above $0.3$-$0.4$ fm$^{-3}$ for both protons and neutrons and all interactions considered. These results clearly demonstrate that the three-body forces significantly influences the nucleon effective masses at high densities. Thus, the current theoretical uncertainties regarding nuclear TBFs contribute significantly to the uncertain nucleon effective masses and their isospin dependences especially at high densities.

It is well known that energy density functionals based on SHF or GHF using various interactions predict rather diverse behaviors of the nucleon effective masses as functions of the isospin asymmetry and/or density of the system. Again, this is partially because of our poorly knowledge about isovector interactions and their space-time no-localities as well as the spin-isospin dependence of the three-body force especially at high densities, see, e.g., reviews in refs.\,\cite{LCK08,Dut12}.  As mentioned earlier, much efforts have been devoted to systematic optimization using advanced computational and statistical techniques. Hopefully, the results will help improve our understanding and constraining of the model parameters involved by using the latest constraints from both terrestrial laboratories and observations of neutron stars properties. As to the studies on the nucleon effective masses in neutron-rich matter within these approaches, unfortunately, the empirical constraints we discussed earlier have not been taken into account in the optimization process in most studies except for some very recent work (see, e.g., ref.\,\cite{ZhangZ16b}). We point out here that even the positive sign of the neutron-proton effective mass splitting in neutron-rich matter at saturation density supported by many experimental evidences and theoretical calculations can help further limit the model parameters used in the energy density functionals. For example, in a very comprehensive study of 240 Skyrme energy density functionals, only 16 were found to meet all macroscopic constraints available\,\cite{Dut12}. Shown in the left of Fig.\,\ref{Dutra-m} are the nucleon effective masses as functions of density in $\beta$-equilibrium matter using 11 of the Skyrme interactions consistent with all macroscopic constraints considered in ref.\,\cite{Dut12}. It is seen that even at saturation density, some of these forces predict large negative values for the neutron-proton effective mass splitting. Some of these forces should be ruled out if one accepts even the positive sign of the neutron-proton effective mass splitting at saturation density as indicated by many analyses of data and theories . In this regard,
it is interesting to see in the right window of Fig.\,\ref{Dutra-m} results of a very recent studying within the Barcelona-Catania-Paris-Madrid (BCPM) functional. Meeting essentially all known constraints considered by the authors, this functional gives a neutron-proton effective mass splitting of $m^{*}_{\rm{n-p}}(\rho_0,\delta)\approx 0.2\delta$ consistent with the empirical constraint discussed earlier\,\cite{Baldo16a}.

In another very recent theoretical study by Mondal \textit{et al.}\,\cite{India17}, the density-dependent nucleon effective masses in isospin-asymmetric matter are parameterized as
\begin{equation}
\frac{M}{M_{J}^*(\rho,\delta)}=1+\frac{k_+}{2}\rho+\frac{k_-}{2}\rho \tau^J_3 \delta
\end{equation}
by generalizing the form $M/M^*=1+k\rho $ usually used in non-relativistic prescriptions of SNM\,\cite{Bohr69}. The neutron-proton effective
mass splitting at saturation density is then
\begin{equation}
m^{*}_{\rm{n-p}}(\rho_0,\delta)\approx -k_-\rho_0\left(\frac{m^*_{\rm{s}}}{m}\right)^2\delta
\end{equation}
where the approximation $ (M_{\rm{n}}^*\cdot M_{\rm{p}}^*)\approx(m^*_{\rm{s}})^2$ is made. By using an approximately universal correlation between the energy curvature and a combination of the magnitude and slope of nuclear symmetry energy from analyzing about 500 nuclear energy density functionals available in the literature,  all known empirical constraints on the saturation properties of nuclear matter as well as some information about the symmetry energy at the so-called ``cross point" of $\rho_{\rm{c}}\approx 0.1$ fm$^{-3}$, they concluded that the neutron-proton effective mass splitting is $m^*_{\rm{n-p}}\approx (0.17\pm0.24)\delta $ with an isoscalar nucleon effective mass of $m_{\rm{s}}^*/m=0.70 \pm 0.05$ at the saturation density, consistent with some earlier conclusions.

In summary of this subsection, most of the models predict that neutrons (protons) have a higher k-mass (E-mass) in neutron-rich nucleonic matter. However, the magnitudes of the predicted k-mass and E-mass are still rather model and interaction dependent. As a result, there is no community consensus regarding the isospin and density dependence of the total nucleon effective mass which is the product of the k-mass and E-mass. While most models predict that the neutron-proton total effective mass splitting is positive in neutron-rich matter consistent with the empirical constraint available, there are some exceptions. There are clear indications that besides the different assumptions and approximations used in handing many-body problems, the spin-isospin dependence of the three-body force, short-range behavior of the tensor force and the isospin-dependence of nucleon-nucleon short-range correlations play significant roles in determining the high density/momentum behavior of nucleon effective masses in neutron-rich matter. Obviously, some interesting issues regrading the model and interaction dependence of the many-body theory predictions remain to be  resolved.

In addition, as a practical matter important for the community, we comment here that nuclear reactions especially those induced by energetic heavy rare isotopes at various radioactive beam facilities have the potential to test predictions by many-body theories and help distinguish model ingredients responsible for the different predictions. To facilitate efforts in this direction, it is not enough for many-body theorists to only publish their results (especially the isospin, density and momentum-dependent single-particle potential in neutron-rich matter) at a few selected values of some variables. Some forms of tables and/or parameterizations of predictions for each set of model parameters/assumptions will be practically very useful for the nuclear reaction community. It is encouraging such efforts are currently being made \cite{Jeremy-Arno}.

\FloatBarrier
\section{Experimental constraints on the symmetry potential and neutron-proton effective mass splitting at saturation density from nucleon-nucleus scatterings}\label{Exp-cs}
Given the model and interaction dependences of the predicted nucleon symmetry potentials and effective masses in neutron-rich matter, ultimately one has to use experimental observables to constrain the model predictions. Interestingly, nucleon-nucleus scattering data accumulated over several decades have already been used to constrain the energy/momentum dependence of both the isoscalar and isovector potentials, thus the corresponding
nucleon effective masses at saturation density. These provide important boundaries for the nucleon effective masses and may be used to constrain nuclear many-body theories and the interactions used.
This area of research has a long and fruitful history, see, e.g., ref.\,\cite{Hod94} for a historical overview and refs.\,\cite{Khoa14,Pawel17} for recent reviews. In this section, we make some observations and comments about some of the latest progresses and remaining issues.

Corresponding to the $\pmb{\tau_1}\cdot\pmb{\tau_2}$ term in the Heisenberg force between two nucleons of isospin $\pmb {\tau_1}$ and $\pmb {\tau_2}$, the Lane optical potential for a projectile of mass $a_{\rm{p}}$ scattering on a target of mass $A_T$ has the form of\,\cite{Lan62}
\beq
\label{eq:Lane}
\mathcal{U}({\pmb r}) = \mathcal{U}_0({\pmb r}) + 4\mathcal{U}_{\rm{Lane}}({\pmb r})\cdot\frac{{\pmb \tau_{\rm{p}}}\,\cdot{\pmb T}}{a_{\rm{p}}\cdot A_T}
\eeq
where $\mathcal{U}_0$ and $\mathcal{U}_{\rm{Lane}}$ are isoscalar and isovector (symmetry) optical potentials, respectively, and ${\pmb \tau_{\rm{p}}}$ and ${\pmb T}$ are isospin operators for the projectile and target nucleus, respectively. After averaging over all target nucleons, the last term in the above expression leads to the $\pm (N_T-Z_T)/A_T$ term of the optical potential felt by a projectile neutron/proton.
It has been used extensively in extracting information about the energy/momentum dependence of the Lane potential in many earlier\,\cite{Kon03,Jeu91,Rap79,Pat76,Dab64} and recent\,\cite{LiX13,LiX15,XuC10,Khoa14,Pawel17,Ngo16,Khoa15,Khoa07} studies of nucleon-nucleon scattering, (p,n) and/or ($^3$He,t) charge exchange reactions. As new data are accumulated, depending on the data sets, number of model parameters as well as the form factors and/or density profiles used for the target, the quantitative results from model analyses sometimes have some discrepancies. However, to our best knowledge, most of the analyses indicate a decreasing Lane potential with increasing nucleon kinetic energy described approximately by Eq.\,(\ref{ULane}).  In addition, folding model calculations of optical potentials for nucleus-nucleus scattering, see, e.g. ref.\,\cite{Khoa14}, coupled with many-body calculations of the EOS and single-particle potentials in neutron-rich matter using the same interactions as used for calculating the optical potentials have also put some constraints on the density dependence of nuclear symmetry energy as well as the
neutron-proton effective mass splitting\,\cite{Khoa14,Ngo16,Khoa15,Khoa07}. In the following we summarize the results from studies originally reported in refs.\,\cite{LiX13,LiX15}. In these analyses, the extended form of the Lane potential of Eq.\,(\ref{sp}) was used in an effort to get information not only about the symmetry potential $U_{\rm{sym},1}(\rho_0,k)$ but also the second-order symmetry potential $U_{\rm{sym},2}(\rho_0,k)$.

\subsection{From nucleon optical potential to its potential in nuclear matter at saturation density}

\begin{figure}[h!]
\centering
\includegraphics[width=8cm,height=14cm,angle=90]{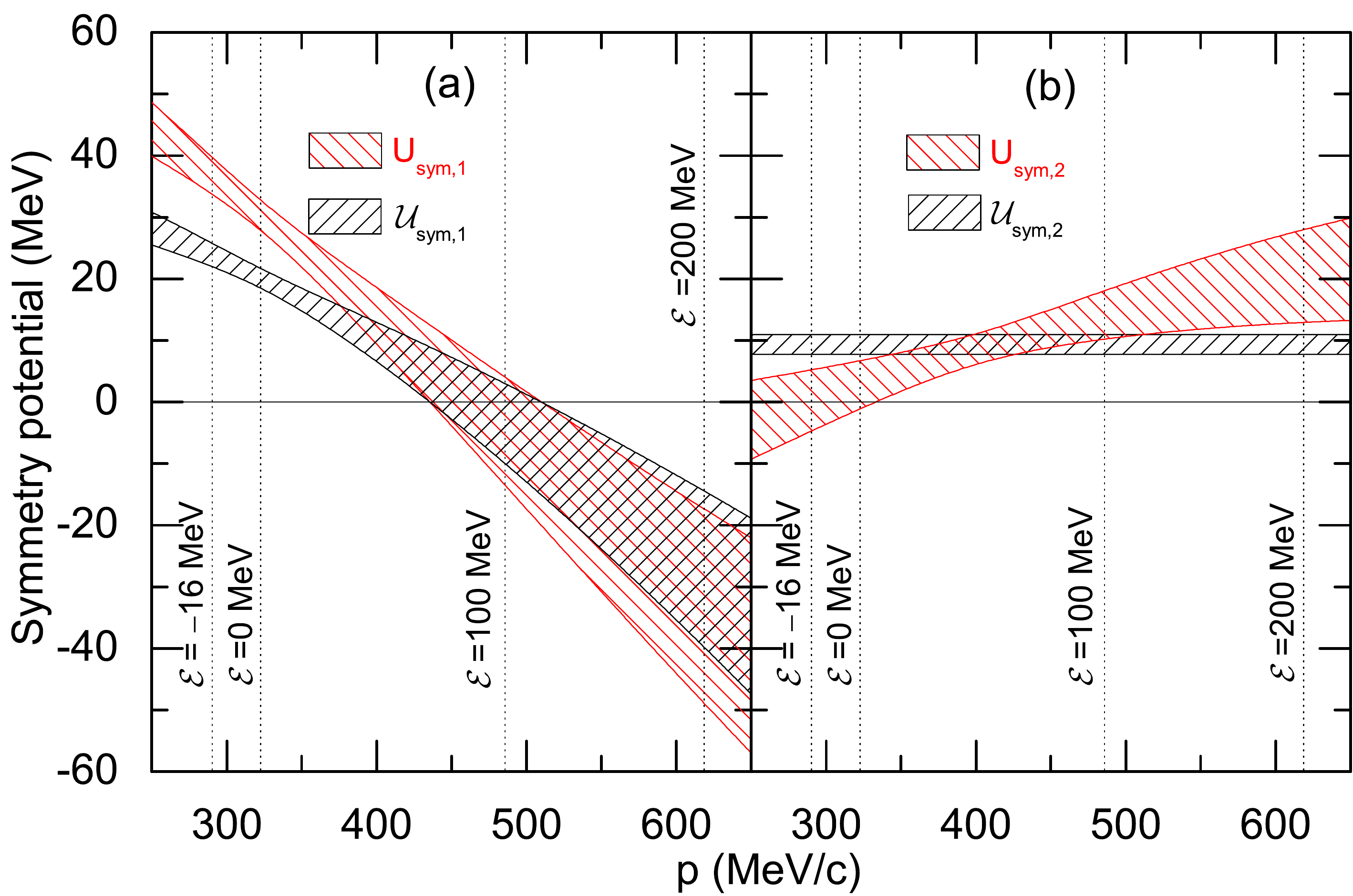}
\caption{Momentum dependence of $\mathcal{U}_{\mathrm{sym,1}}$ ($U_\mathrm{sym,1}$) (a)
and $\mathcal{U}_{\mathrm{sym,2}}$
($U_\mathrm{sym,2}$) (b). The corresponding momenta at
${E}=-16$, $0$, $100$ and $200$ MeV are indicated with dotted lines. Taken from ref.\,\cite{LiX13}.}
\label{Usym13}
\end{figure}

First of all, both the optical potential and single-nucleon potential in nuclear matter have the Lane form of Eq.\,({\ref{sp}). However, they are not necessarily identical. It is necessary to first briefly recall how one can get the $U_0(\rho_0,k)$, $U_{\rm{sym},1}(\rho_0,k)$ and $U_{\rm{sym},2}(\rho_0,k)$ in nuclear matter at saturation density from the optical model potentials $\mathcal{U}_0({E})$, $\mathcal{U}_{\rm{sym},1}({E})$ and $\mathcal{U}_{\rm{sym},2}({E})$ at a beam energy ${E}$. The question was discussed in refs.\,\cite{LiX13,Dab64} and some useful relationship for conveniently transforming the potentials were given. Using the kinetic energy $T_{J}$ and $\delta$ as two independent variables necessary in expressing the three parts of the nucleon potential given in Eq.\,(\ref{sp}), one has
\begin{equation}\label{Uu}
\mathcal{U}_{J}({E},\delta)=U_{J}(T_{J}({E}),\delta).
\end{equation}
In isospin asymmetric matter, neutrons and protons have different dispersion relationship $T_{J}({E})$ because of the momentum dependence of the isovector potential. The dispersion relationship $T({E})$ in SNM can be readily obtained from manipulating the single-nucleon energy
\begin{align}\label{Dispersion2}
{E}=T+U_0(T)
\end{align}
once the momentum dependence of the isoscalar potential $U_0(T)$ is known.
For the same nucleon energy ${E}$, the  $U_{J}(T_{J})$ can be expanded to the first-order in $\delta$.  Then the kinetic energy $T_{J}({E})$ for protons and neutrons in asymmetric matter can be written in terms of the $T({E})$ as
\begin{align}\label{Disp3}
T_{J}({E})=T({E})-\tau_3^JU_{\rm{sym},1}(T)\mu(T)\delta
\end{align}
where $\mu=(1+dU_0/dT)^{-1}$.
Inserting the above relationship into Eq.\,(\ref{sp}) and expanding all terms up to $\delta^2$, the Eq.\,(\ref{Uu}) gives the following transformation relations\,\cite{LiX13,Dab64}
\begin{align}\label{Tran}
U_0=&\mathcal{U}_0,\\
U_{\rm{sym},1}=&\frac{\mathcal{U}_{\rm{sym},1}}{\mu},\\
U_{\rm{sym},2}=&\frac{\mathcal{U}_{\rm{sym},2}}{\mu}+\frac{\vartheta\mathcal{U}_{\rm{sym},1}}{\mu^2}+\frac{\zeta\mathcal{U}^2_{\rm{sym},1}}{2\mu^3},
\end{align}
where
\begin{equation}
\mu=1-\frac{\partial \mathcal{U}_0}{\partial {E}},~~\vartheta=\frac{\partial \mathcal{U}_{\rm{sym},1}}{\partial {E}},~~\zeta=\frac{\partial^2 \mathcal{U}_0}{\partial {E}^2}.
\end{equation}

Shown in Fig.\,\ref{Usym13} are comparisons of the $\mathcal{U}_{\mathrm{sym,1}}$ and $U_\mathrm{sym,1}$ (window a)
as well as $\mathcal{U}_{\mathrm{sym,2}}$ and $U_\mathrm{sym,2}$ (window b) obtained from analyszing a set of data from the EXFOR database\,\cite{Exfor}.
It is seen that the $U_\mathrm{sym,1}$ is larger than the $\mathcal{U}_{\mathrm{sym,1}}$
at lower momenta (energies) while the $U_\mathrm{sym,1}$ becomes smaller than
the $\mathcal{U}_{\mathrm{sym,1}}$ at higher momenta (energies). Namely, they have different slopes in energy/momentum, just would give different neutron-proton effective mass splittings. In addition, both the $\mathcal{U}_{\mathrm{sym,1}}$ and $U_\mathrm{sym,1}$ decrease with nucleon
momentum $p$ and become negative when the nucleon momentum is larger than about $p=470$ MeV/c
(i.e., ${E}=90$ MeV). These are in qualitative agreement with predictions of most many-body theories as we discussed in the previous section.
 Furthermore, while the $\mathcal{U}_{\mathrm{sym,2}}$ is almost a constant, $U_{\rm{sym,2}}$
increases with nucleon momentum $p$ and becomes comparable to the $U_\mathrm{sym,1}$, especially at higher energies.

We notice that in some analyses the above transformations between $\mathcal{U}_{\mathrm{sym,1}}$ ($\mathcal{U}_{\mathrm{sym,2}}$ ) and $U_\mathrm{sym,1}$ ($U_{\rm{sym,2}}$) were not considered. Indeed, it is unnecessary for the isoscalar potential. However, the above discussions clearly indicate the importance of doing the transformation for the symmetry potential. Based on the discussions in the previous section, it is clear that the isoscalar effective mass $m_0^*/m$ can be extracted directly from the nucleon isoscalar optical potential. However, to get the neutron-proton effective mass splitting, it is important to consider the factor $\mu$. Since the Coulomb potential is normally explicitly considered in the optical model analyses of proton-nucleus scattering and one normally does the calculations in the theoretically uncharged isospin asymmetric nucleonic matter, the above transformations are thus valid for both neutrons and protons. However, it was known that to transform to  the interior of nuclei in $\beta$ equilibrium an extra relationship between the Coulomb potential and the symmetry potential is required\,\cite{Dab64}.
\begin{figure}[h!]
\centerline{\includegraphics[width=14.cm,height=8cm]{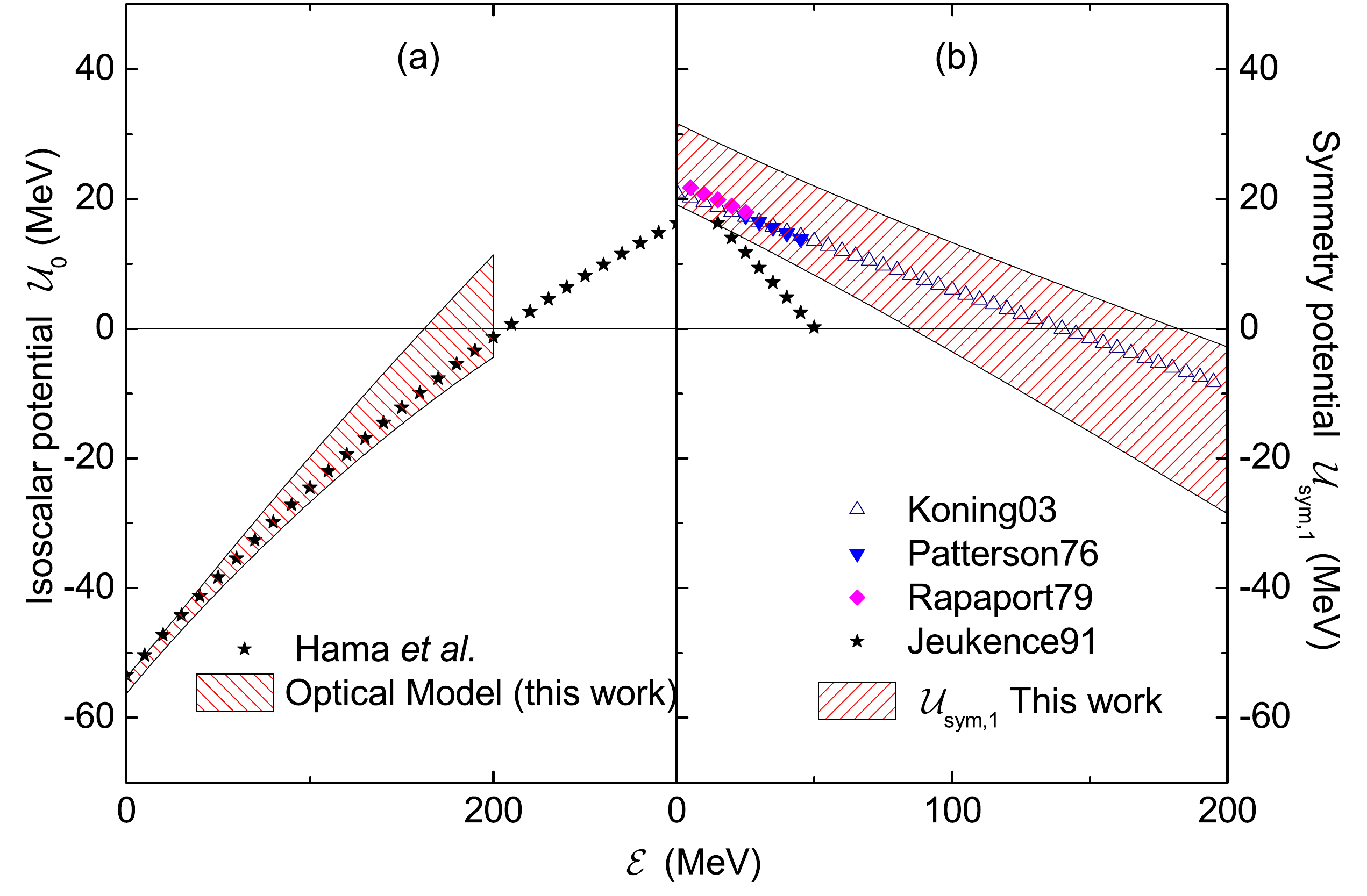}}
\caption{Energy dependent isoscalar $\mathcal{U}_0$(left) and isovector $\mathcal{U}_{\rm{sym}}$ (right) nucleon potentials from analyzing nucleon-nucleus scattering data. Taken from ref.\,\cite{LiX15}.}\label{Uoptical}
\end{figure}
\subsection{From the energy dependence of Lane potential to neutron-proton effective mass splitting in neutron-rich matter at saturation density}
An an example, shown in Fig.\,\ref{Uoptical} are the optical potentials from a global optical model analysis\,\cite{LiX15} of all 2249 data sets of reaction and angular differential cross sections of neutron and proton scattering on 234 targets at beam energies from 0.05 to 200 MeV available in the EXFOR database\,\cite{Exfor}.
Compared in the left window of Fig.\,\ref{Uoptical} are the nucleon isoscalar potentials from the analysis (hatched bands) in ref.\,\cite{LiX15} and the Schr$\ddot{\mathrm{o}}$dinger equivalent isoscalar potential obtained earlier by Hama \textit{et al.}\,\cite{Ham90}. They are consistent with each other and give an isoscalar effective mass of $m^{*}_0/m=0.65\pm 0.06$. Shown on the right is a comparison of the nucleon isovector potential $\mathcal{U}_{\rm{sym}}$ with results from several earlier studies\,\cite{Kon03,Jeu91,Rap79,Pat76}.  Albeit at different slopes and some of them have different energy ranges, the isovector potentials all clearly decrease with increasing energy.
Up to a kinetic energy of $\lesssim 200\,\rm{MeV}$, the average values of the nucleon optical potentials shown in Fig.\,\ref{Uoptical} from ref. \cite{LiX13,LiX15} can be parameterized as
\begin{align}
\mathcal{U}_0=S_0+S_1E+S_2E^2,~~
\mathcal{U}_{\rm{sym,1}}=a+bE\label{Usym-xh},~~
\mathcal{U}_{\rm{sym,2}}=d+fE
\end{align}
with $S_0\approx-55.06\pm 1.24\,\rm{MeV},S_1\approx 0.343\pm 0.030,
S_2\approx-(2.524\pm 1.224)\times10^{-4}\,\rm{MeV}^{-1},a\approx 25.4\pm 6.27\,\rm{MeV},b\approx-0.205\pm 0.056,d\approx 8.896\pm 4.864\,\rm{MeV}
$ and $f\approx-(3.844\pm 10.721)\times10^{-4}$\,\cite{LiX15}. We notice that the parameterization for the $\mathcal{U}_{\rm{sym,1}}$ in the above is consistent with that given in Eq.\,(\ref{ULane}).
Using the transformations discussed in the previous subsection, the above nucleon optical potentials can be translated to single-nucleon potential in
nuclear matter at saturation density as
\begin{align}
U_0(E)\approx&(-55.06\pm1.24)+(0.34\pm0.030)E+(-2.52\pm1.22)\times10^{-4}E^2,\\
U_{\rm{sym}}(E)\approx&(38.67\pm9.71)+(-0.34\pm0.089)E+(2.63\pm1.56)\times10^{-4}E^2\label{Usym-XH2},\\
U_{\rm{sym,2}}(E)\approx&(0.89\pm8.77)+(0.12\pm0.064)E+(-2.24\pm1.75)\times10^{-4}E^2.
\end{align}
These potentials provide useful boundary conditions for nuclear many-body theories.
In particular, we emphasize that the parameterization in Eq.\,(\ref{Usym-XH2}) instead of the
ones in Eq.\,(\ref{Usym-xh}) or Eq.\,(\ref{ULane}) (which are the nucleon optical potentials themselves) should be used as the nucleon symmetry potential
at saturation density in isospin asymmetric nuclear matter. It then leads to a neutron-proton total effective mass splitting of $m^{*}_{\rm{n-p}}(\rho_0)=(0.41\pm0.15)\delta$ at saturation density\,\cite{LiX15}.

We note that some earlier analyses of nucleon-nucleus scattering data, mostly from proton-nucleus scattering together with limited amount of neutron-nucleus scattering data, were carried out within a Dirac phenomenological optical model under the assumption that the isovector potential is either energy/momentum-independent or has the same energy/momentum dependence as the isoscalar potential\,\cite{Koz89}. A globally good reproduction of the experimental data in such analyses was cited by some people as an indication that the symmetry potential may increase with energy. This interpretation is obviously in contrast with the conclusions we discussed above. It is well known that the nucleon isoscalar optical potential increases with increasing nucleon energy. Because the isospin asymmetries of the target nuclei are normally very small, if the much weaker isovector potential, as shown in Eq.\,(\ref{Usym-xh}),  is pre-assumed to have the same energy dependence as the isoscalar potential, a good fit to the data does not necessarily mean that the isovector potential is actually increasing with energy.
In fact, to our best knowledge, so far there is no clear indication for the symmetry potential to increase with momentum/energy corresponding to a negative neutron-proton effective mass splitting at $\rho_0$ from any solid analysis of
available experimental data.

In summary of this section, optical model analyses of nucleon-nucleus and change exchange reactions have been very useful in constraining the momentum/energy dependence of the Lane optical potential at at $\rho_0$.
After a proper transformation of the optical potential into the single-nucleon potential in nuclear matter, the extracted information about the optical potentials constrains the nucleon isoscalar and isovector potential in nuclear matter at at $\rho_0$, and thus also the corresponding nucleon effective masses. A comprehensive analyses of all available nucleon-nucleus scattering data below 200 MeV indicates that the neutron-proton total effective mass splitting is $m^{*}_{\rm{n-p}}(\rho_0)=(0.41\pm0.15)\delta$ at $\rho_0$.

\FloatBarrier
\section{Isospin dependences of nucleon effective masses and their relationships with nuclear symmetry energy}
The equation of state (EOS) of neutron-rich matter and the isospin dependence of nucleon effective masses are determined by the same underlying interaction. In particular, the density dependence of nuclear symmetry energy
\begin{equation}
E_{\rm{sym}}(\rho )\equiv E_{\rm{sym},2}(\rho )=\left.\frac{1}{2}\frac{\partial ^{2}E(\rho,\delta )}{\partial \delta ^{2}}\right|_{\delta=0}
\end{equation}
appearing in the EOS
\begin{equation}\label{eos1}
E(\rho ,\delta )=E_0(\rho)+E_{\rm{sym},2}(\rho )\delta ^{2} +E_{\rm{sym},4}(\rho ) \delta ^{4} +\mathcal{O}(\delta^6)
\end{equation}
of isospin asymmetric nuclear matter is still uncertain especially at supra-saturation densities.
The symmetry energy $E_{\rm{sym}}(\rho)$ can
be expanded around $\rho_0$ to third order in density as
\begin{align}\label{Su}
  E_{\rm{sym}}(\rho)=E_{\rm{sym}}(\rho_0)+\frac{L}{3}\left(\frac{\rho}{\rho_0}-1\right)+\frac{K_{\rm{sym}}}{18}\left(\frac{\rho}{\rho_0}-1\right)^2 +\frac{J_{\rm{sym}}}{162}\left(\frac{\rho}{\rho_0}-1\right)^3+\mathcal{O}\left[\left(\frac{\rho}{\rho_0}-1\right)^4\right]
\end{align}
 in terms of its magnitude $E_{\rm{sym}}(\rho_0)$, slope $L=[3 \rho{\partial E_{\rm{sym},2}(\rho)}/{\partial\rho}]_{\rho_0}$,
 curvature $K_{\rm{sym}}=[9\rho^2(\partial^2E_{\rm{sym},2}(\rho)/\partial\rho^2)]_{\rho_0}$ and skewness $J_{\rm{sym},2}=[27\rho^3\partial^3E_{\rm{sym}}(\rho)/\partial\rho^3]_{\rho_0}$ at saturation density.
Similarly, the EOS of symmetric nuclear matter can be expanded
\begin{equation}\label{E0para}
  E_{0}(\rho)=E_0(\rho_0)+\frac{K_0}{18}\left(\frac{\rho}{\rho_0}-1\right)^2+\frac{J_0}{162}\left(\frac{\rho}{\rho_0}-1\right)^3+\mathcal{O}\left[
  \left(\frac{\rho}{\rho_0}-1\right)^4\right]
\end{equation}
in terms of its incompressibility $K_{\rm{0}}=[9\rho^2(\partial^2E_{\rm{0}}(\rho)/\partial\rho^2)]_{\rho_0}$ and skewness $J_{\rm{0}}=[27\rho^3\partial^3E_{\rm{0}}(\rho)/\partial\rho^3]_{\rho_0}$ at saturation density. The isospin-quartic (fourth-order) symmetry energy
\begin{equation}
E_{\mathrm{sym,4}}(\rho )\equiv \left.\frac{1}{24}\frac{\partial ^{4}E(\rho,\delta )}{\partial \delta ^{4}}\right|_{\delta=0} \label{Esyme4}
\end{equation}
may play significant roles in some properties of neutron stars, such as the
core-crust transition density and pressure\,\cite{SteA06,XuJ09}. However, it is theoretically very uncertain and there is no experimental constraint available so far. Of course, the uncertainties of the symmetry energies and the nucleon effective masses in neutron-rich matter are strongly correlated and they are normally predicted simultaneously by the same nuclear many-body theories. Nevertheless, it is interesting to examine analytically their direct relations. In particular, it would be instructive for getting insights about the poorly known isovector nuclear interaction if the relationships are model independent. In this section, we review some recent efforts in this direction. The focus is to identify the key underlying ingredients of the $E_{\mathrm{sym}}(\rho )$, $E_{\mathrm{sym,4}}(\rho )$, $L$ and the nucleon effective masses, especially the relations among them and their dependences on the momentum dependence of both the isoscalar and isovector single-nucleon potential.

\subsection{Hugenholtz--Van Hove theorem connecting nucleon potential, effective mass and symmetry energy in isospin asymmetric nuclear matter}
To our best knowledge, effects of the isoscalar nucleon effective mass $M^*_0$ on nuclear symmetry energy were first studied in some details as early as in 1958 by Brueckner and Gammel in ref.\,\cite{Bru58}. Later, in the sixties and seventies a number of works using the Bruckner theory\,\cite{bru64,Dab73} have studied the relationship among the symmetry energy, nucleon isoscalar effective mass and symmetry potential, leading to the expression
\begin{equation}
E_{\rm{sym}}(\rho) =\frac{1}{3} \frac{k_{\rm{F}}^2}{2 M^*_0(\rho,k_{\rm{F}})} +\frac{1}{2} U_{\rm{sym},1}(\rho,k_{\rm{F}}). \label{Esymexp2}
\end{equation}
The general HVH theorem\,\cite{Hug58,Sat99} or the Gibbs-Duhem relation required for all Fermonic many-body theories can be written as
\begin{eqnarray}\label{HVH}
E_{\rm{F}}=\frac{d\xi}{d \rho}=\frac{d (\rho E)}{d \rho} =  E+ \rho
\frac{d E}{d \rho} = E+ P/\rho
\end{eqnarray}
where $\xi=\rho E$ is the energy density and $P$ is the pressure at zero temperature. Applying the HVH theorem to isospin-asymmetric nuclear matter, Fritsch, Kaiser and Weise derived the following expression for nuclear
symmetry energy \cite{FKW}
\begin{equation}
E_{\rm{sym}}(\rho) =\frac{1}{3} \frac{k_{\rm{F}}^2}{2 M} +\frac{1}{2} U_{\rm{sym},1}(\rho,k_{\rm{F}})+\frac{k_{\rm{F}}}{6}\left(\frac{\partial U_0}{\partial k}\right)_{k_{\rm{F}}}-\frac{1}{6}\frac{k^4_{\rm{F}}}{2M^3}
\label{FKW}
\end{equation}
where the last term is a relativistic correction of less than 1 MeV at $\rho_0$ to the kinetic energy \cite{Kaiser02}. Neglecting the small relativistic correction and using the definition of isoscalar nucleon effective mass, the above expression reduces to the Eq. (\ref{Esymexp2}). To our best knowledge, it was the first time in 2005 that the HVH theorem was used to connection analytically the symmetry energy, nucleon isoscalar effective mass and symmetry potential in ref. \cite{FKW}. Later, by expanding both sides of the HVH theorem in isospin asymmetry, not only the above expression for $E_{\rm{sym}}(\rho)$ but also higher-order (e.g., the isospin-quartic) symmetry energy were systematically derived in terms of the nucleon isoscalar and isovector effective masses as well as their own momentum dependences by by Xu \textit{et al.}\,\cite{XuC11,XuC10,Rchen,CXu14}. Moreover, the density slope $L(\rho)$
at an arbitrary density $\rho$ was  expressed as \cite{XuC11,XuC10,Rchen,CXu14}
\begin{equation}
L(\rho)=
\frac{2}{3} \frac{ k_{\rm{F}}^2}{2 M_0^*(\rho,k_{\rm{F}})} + \frac{3}{2} U_{\rm{sym},1}(\rho,k_{\rm{F}})
- \frac{1}{6}\left.\left(\frac{k^3}{{M_0^*}^2}\frac{\partial M_0^*}{\partial k} \right)\right|_{k_{\rm{F}}}
+\left.\frac{dU_{\rm{sym},1}}{dk}\right|_{k_{\rm{F}}}
+ 3U_{\rm{sym},2}(\rho,k_{\rm{F}}).\label{Lexp2}
\end{equation}
Very recently, using a similar approach and assuming the nucleon potentials depend quadratically on its momentum, expressions for the $E_{\rm{sym}}(\rho), L(\rho)$ and their relations are also expressed in terms of parameters describing the density and isospin dependences of nucleon effective masses in ref.\,\cite{India17}.
It is seen from Eq.\,(\ref{Esymexp2}) that the \esym consists of the kinetic contribution to the symmetry energy (referred as kinetic symmetry energy in the following) equivalent to $1/3$ the Fermi energy of quasi-nucleons with an isoscalar effective mass $M^*_0(\rho,k_{\rm{F}})$
and the potential symmetry energy of $1/2$ the isovector potential $U_{\rm{sym},1}(\rho,k_{\rm{F}})$ at the Fermi momentum $k_{\rm{F}}$. The density slope $L(\rho)$ has five components depending on
\begin{itemize}
\item{the magnitude of the nucleon isoscalar effective mass $M^*_0$, $L_1(\rho)\equiv{ k_{\rm F}^2}/{3 M_0^{\ast}(\rho,k_{\rm F})}$},
\item{the momentum dependence of the nucleon isoscarlar effective mass, $L_2(\rho)\equiv-[{ k_{\rm F}^3}/{6 M_0^{\ast 2}(\rho,k_{\rm F})}]
[{\partial M_0^{\ast}(\rho, k)}/{\partial k}]_{k = k_{\rm F}}$},
\item{the magnitude of the nucleon isovector potential $U_{\rm{sym},1}(\rho,k_{\rm{F}})$, $L_3(\rho)\equiv3U_{\rm{sym}, 1}(\rho, k_{\rm F})/2$},
\item{the momentum dependence of the nucleon isovector potential, $L_4(\rho) \equiv k_{\rm{F}}[{\partial U_{\rm{sym}, 1}(\rho, k)}/{\partial k}]_{k = k_{\rm F}}$},
\item{the magnitude of the second-order symmetry potential $U_{\rm{sym},2}(\rho,k_{\rm{F}})$, $L_5(\rho) \equiv 3 U_{\rm{sym}, 2}(\rho, k_{\rm F})$}.
\end{itemize}
In the case that the single nucleon potential also depends on energy, one then uses the total differential to
replace the partial differential, i.e., $d/d k=\partial/\partial k+\partial k/\partial E\cdot\partial/\partial k\to\partial/\partial k$.
The above decomposition of $E_{\rm{sym}}(\rho)$ and $L(\rho)$ was applied in ref.\,\cite{XuC10} to the nucleon optical potential at $\rho_0$ extracted from analyzing some old data of nucleon-nucleus scatterings.
Later, it has been used with nucleon potentials from more extensive analyses of nucleon-nucleus scatterings and/or
predictions from HF calculations using Skyrme and/or modified Gogny forces\,\cite{XuC11,Rchen,CXu14,LiX13}. For example,
using the $U_0,U_\mathrm{sym,1}$ and $U_\mathrm{sym,2}$ at $\rho_0$ from transforming the parameterized nucleon optical potentials
in Eq.\,(\ref{Usym-xh}),
one can obtain the expression for the symmetry energy
\begin{equation}
E_{\rm{sym}}(\rho)=\frac{k_{\rm{F}}^2}{6M\mu}+\frac{a+b{E}}{2\mu}
=\frac{1}{\mu}\left(\frac{k_{\rm{F}}^2}{6M}+\frac{a+b{E}}{2}\right),
\end{equation}
and the decomposition of the slope parameter
\begin{align}
L_1(\rho)=&\frac{k_{\rm{F}}^2}{3M}+\frac{k_{\rm{F}}}{6}\left.\frac{\partial
U_0}{\partial
{k}}\right|_{k_{\rm{F}}}+\frac{k_{\rm{F}}}{6}\left.\frac{\partial{E}}{\partial
{k}}\frac{\partial
U_0}{\partial{E}}
\right|_{k_{\rm{F}}}
=\frac{k_{\rm{F}}^2}{3M}+\frac{k^2_{\rm{F}}}{6M}\frac{1-\mu}{\mu},\\
L_2(\rho)=&\frac{k^2_{\rm{F}}}{6}\left.\frac{\partial^2
U_0}{\partial
{k}^2}\right|_{k_{\rm{F}}}
+\frac{k_{\rm{F}}^2}{6}\cdot\left[
\frac{\partial^2{E}}{\partial {k}^2}\frac{\partial
U_0}{\partial{E}}+\left(\frac{\partial{E}}{\partial
{k}}\right)^2\frac{\partial^2U_0}{\partial{E}^2}\right]_{k_{\rm{F}}}
=\frac{k_{\rm{F}}^2}{3M}\frac{1-\mu}{\mu}
+\frac{S_2k_{\rm{F}}^4}{3M^2\mu^3},\\
L_3(\rho)=&\frac{3}{2}U_{\rm{sym},1}=\frac{3}{2}\frac{a+b{E}}{\mu},\\
L_4(\rho)=&k_{\rm{F}}\left.\frac{\partial
U_{\rm{sym}}}{\partial{k}}
\right|_{k_{\rm{F}}}+k_{\rm{F}}\left.\frac{\partial{E}}{\partial
{k}}\frac{\partial
U_{\rm{sym}}}{\partial{E}}
\right|_{k_{\rm{F}}}
=\frac{k_{\rm{F}}^2}{M\mu^2}\left[b+\frac{2S_2(a+b{E})}{\mu}\right],\\
L_5(\rho)=&3U_{\rm{sym,2}}=3\left[\frac{d+f{E}}{\mu}+\frac{b(a+b{E})}{\mu^2}
+\frac{S_2(a+b{E})^2}{\mu^3}\right],
\end{align}
where $\mu=1-S_1-2S_2E$. For example, using the parameters extracted from an earlier analysis of the available nucleon-nucleus scattering data\,\cite{LiX13}, the five components of $L(\rho_0)$ are, respectively,
$L_1(\rho_0)=37.76\pm 3.23$ MeV, $L_2(\rho_0)=-2.57\pm 1.29$ MeV,
$L_3(\rho_0)=55.08\pm 4.48$ MeV, $L_4(\rho_0)=-46.11\pm 11.87$ MeV, and
$L_5(\rho_0)=0.81\pm 14.48$ MeV.
It is interesting to note that the $L_2(\rho_0)$ is very small, indicating that the contribution of the momentum dependence of the isoscalar
nucleon effective mass is unimportant. Whereas the $L_4(\rho_0)$ contributes significantly negatively, demonstrating
the importance of the momentum dependence of the symmetry potential
$U_\mathrm{sym,1}$ which is also the one determining the neutron-proton effective mass splitting as we discussed earlier. This term, however, has a large error bar.
The resulting total $L(\rho_0)=44.98\pm22.31\,\textrm{MeV}$ is well within its boundary from analyzing many observables shown in Fig.\,\ref{Li-data}. The large uncertainty is mainly due to that of the isovector potential
$U_\mathrm{sym,1}$ and the second-order symmetry potential $U_\mathrm{sym,2}$.

\begin{figure}[tbh]
\centering
\includegraphics[width=10.5cm]{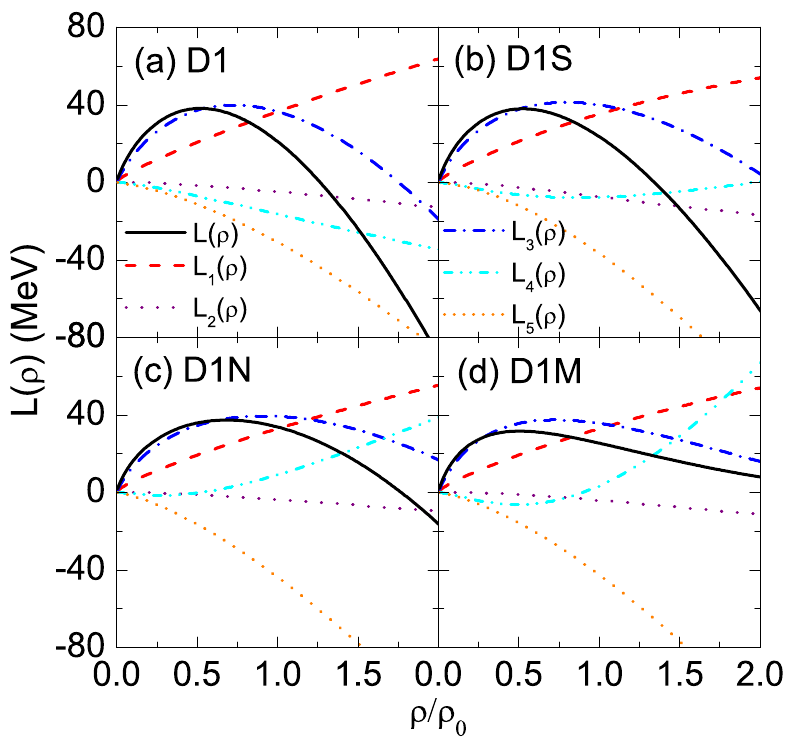}
\caption{Components of the symmetry energy slope parameter $L(\rho)$ from Gogny Hartree-Fock calculations with the D1 (a), D1S (b), D1N (c), and D1M (d) parameter sets. Taken from ref.\,\cite{Rchen}.}
\label{LGogny}
\end{figure}
Another interesting example shown in Fig.\,\ref{LGogny} is the decomposition of $L(\rho)$ within the HF approach using the same four Gogny parameter sets used in calculating the symmetry potentials shown in the left window of Fig.\,\ref{Usym1}. Again, the $L_{2}(\rho )$ contribution due to the momentum dependence of the isoscalar nucleon effective mass is negligible. The $L_{3}(\rho )$ or $U_{\rm{sym},1}(\rho ,k_{\rm{F}})$ exhibits different density dependences for different interactions used. The $L_{4}(\rho )$ reflecting explicitly the neutron-proton effective mass splitting is very different depending on the interaction used as one would expect from examining the symmetry potentials in Fig.\,\ref{Usym1}. It was also noted that the $L_5(\rho)$ due to the second-order symmetry potential $U_{\rm{sym}, 2}(\rho, k_{\rm F})$ in all four Gogny HF calculations are significant. Overall, it is seen that the momentum dependence of the symmetry potential plays a significant role in determining the $L(\rho)$.

\subsection{Isospin-quartic symmetry energy and its dependence on nucleon isoscalar and isovector effective masses}
The size and density dependence of the isospin-quartic term $E_{\rm{sym},4}(\rho)$ in the EOS of isospin asymmetric nuclear matter have been uncertain with conflicting predictions using various models, see, e.g., ref.\,\cite{Wang17} for a recent review. To understand the physics underlying the $E_{\rm{sym},4}(\rho)$, it is useful to decompose it in ways as model independent as possible. In this regard, we note that in deriving the Eqs.\,(\ref{Esymexp2}) and (\ref{Lexp2}) using the HVH theorem, the $E_{\rm{sym},4}(\rho)$ was simultaneously expressed by Xu \textit{et al.} as\,\cite{XuC11}
\begin{align}\label{Esym4}
E_{\rm{sym},4}(\rho) = &\frac{\hbar^2}{162M}
\left(\frac{3\pi^2}{2}\right)^{2/3} \rho^{2/3}+\Bigg[\frac{5}{324}\frac{\partial U_0(\rho,k)}{\partial k}
k - \frac{1}{108} \frac{\partial^2 U_0(\rho,k)}{\partial
k^2} k^2 +\frac{1}{648} \frac{\partial^3
U_0(\rho,k)}{\partial k^3}k^3\notag\\
&-
\frac{1}{36} \frac{ \partial U_{\rm{sym},1}(\rho,k)}{\partial k}
k + \frac{1}{72} \frac{ \partial^2 U_{\rm{sym},1}(\rho,k)}{\partial
k^2} k^2 + \frac{1}{12} \frac{\partial
U_{\rm{sym},2}(\rho,k)}{\partial k}k+ \frac{1}{4}
U_{\rm{sym},3}(\rho,k)\Bigg]_{k_{\rm{F}}}.
\end{align}
The first term is the well-known kinetic contribution to $E_{\rm{sym},4}(\rho)$, while the rest is determined by several terms describing the very fine details of the momentum dependences of each part of the single-nucleon potential:
the first-order, second-order and third-order derivatives of $U_0(\rho,k)$ with respect to momentum,  both the first-order and second-order derivatives of $U_{\rm{sym},1}(\rho,k)$, the first-order derivative of the $U_{\rm{sym},2}(\rho,k)$ as well as the coefficient $U_{\rm{sym},3}(\rho,k_{\rm{F}})$ of the $\delta^3$ term in expanding the single-particle potential according to Eq.\,(\ref{sp}). As we discussed earlier, our current knowledge of even the first-order derivative of single-nucleon potential is still very poor not to mention the high-order derivatives with respect to momentum. This explains why it is so difficult to accurately determine the isospin-quartic term and what unknown physics are at work. Since the HVH theorem is of general validity in the framework of Fermonic many-body theories, we emphasize that the decompositions of the $E_{\rm{sym},2}(\rho)$, $L(\rho)$ and $E_{\rm{sym},4}(\rho)$ are all general and they are independent of the specific models and interactions used.
\begin{figure}[h!]
\centering
\includegraphics[scale=0.45,clip]{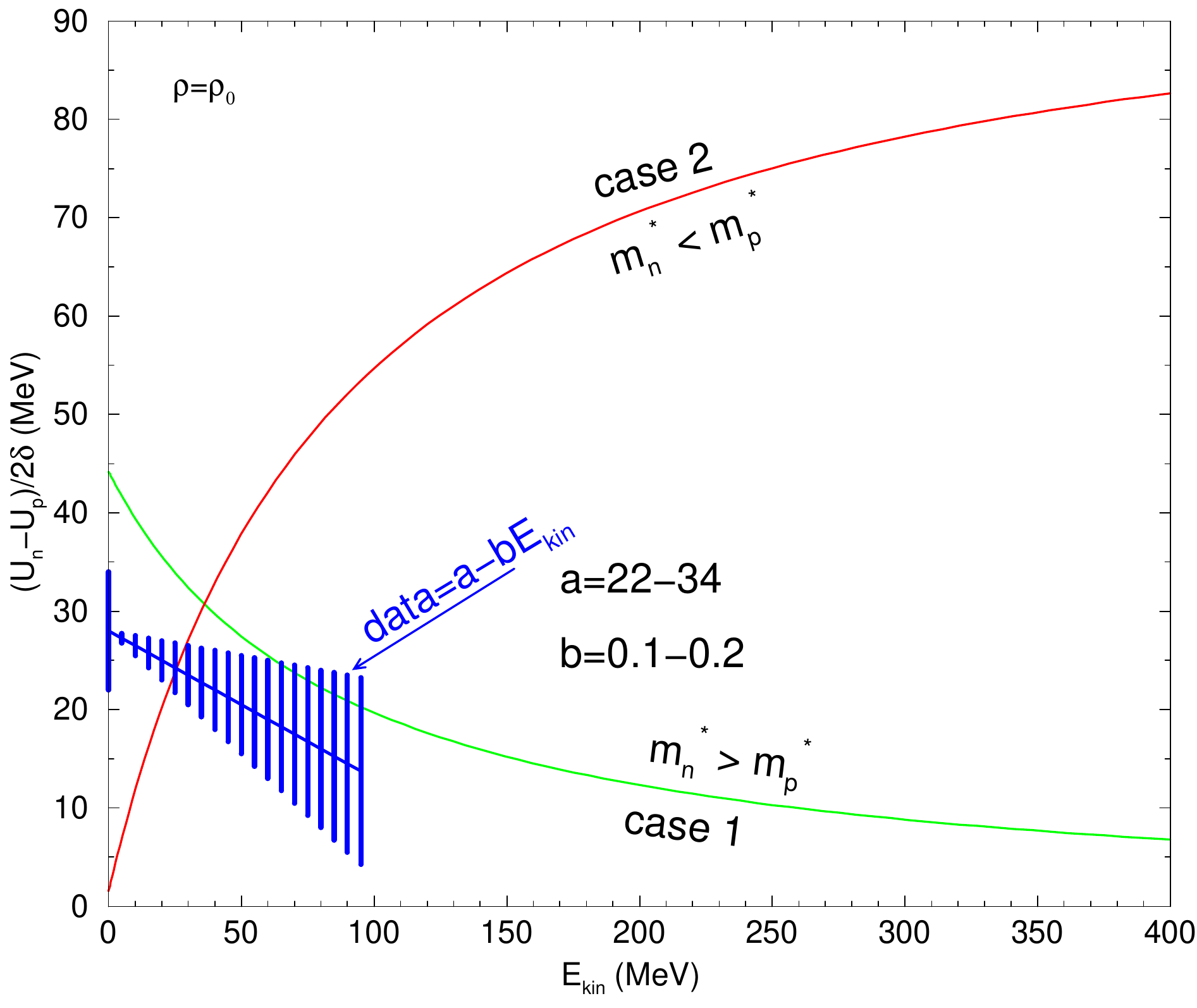}\\
\includegraphics[width=9.cm]{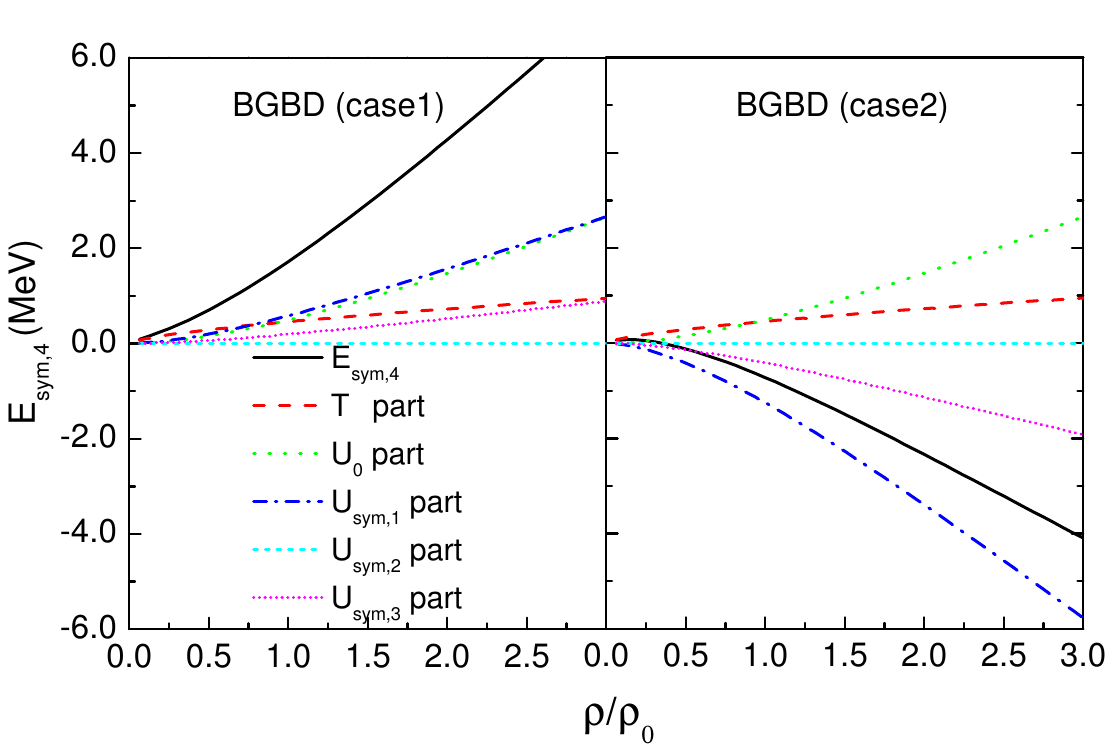}
\caption{Upper: Nuclear symmetry potential as a function of nucleon kinetic energy in nuclear matter at the saturation density. Taken from ref.\,\cite{Li04}.
Lower: The kinetic and various potential contributions to the fourth-order symmetry energy $E_{\rm{sym},4}$ with the BGBD potential for $m_{\rm{n}}^*>m_{\rm{p}}^*$ (left, case1) and for $m_{\rm{n}}^*< m_{\rm{p}}^*$ (right, case 2). Taken from ref.\,\cite{XuC11}.}
\label{BGBD}
\end{figure}

For illustrations, the BGBD phenomenological potential of Bombaci--Gale--Bertsch--Das Gupta\,\cite{Bom01} was used in ref.\,\cite{XuC11} to calculate the $E_{\rm{sym},4}(\rho)$ and examine its dependence on the neutron-proton effective mass splitting. The BGBD potential at a reduced density $u=\rho/\rho_0$ can be written as\,\cite{Bom01}
\begin{align}\label{Bomba}
U_J(u,\delta,k)=&Au+Bu^\sigma
-\frac{2}{3}(\sigma-1)\frac{B}{\sigma+1} \left(\frac{1}{2}+x_{3}\right)
u^{\sigma}\delta^2
\nonumber\\
&+\tau_3^J \left[{-\frac{2}{3}A\left(\frac{1}{2}+x_{0}\right) u -
\frac{4}{3}\frac{B}{\sigma+1}\left(\frac{1}{2}+x_{3}\right)u^{\sigma}\,}\right]\delta\nonumber\\
& +\frac{4}{5\rho_0}\left[\frac{1}{2} (3C-4z_1) \mathcal{I}_J
+ (C+2z_1)\mathcal{I}_{J'}\right]+ \left({C +\tau_3^J
\frac{C-8z_1}{5}\delta}\right)u g(k)
\end{align}
where $\mathcal{I}_J=[2/(2\pi)^3]\int d^3k f_{J}(k)g(k)$ with
$g(k)= 1/[{1+({{k}/{\Lambda}})^2 }]$ being a momentum
regulator of scale $\Lambda$ and $f_{J}(k)$ being the phase space distribution
function of nucleon $J=$ n or p, and $J^{\prime}=$p or n. This potential with different sets of parameters have been used in simulating heavy-ion collisions, see, e.g., ref.\,\cite{Riz04}.
In particular, the three parameters $x_0, x_3$ and $z_1$ can be adjusted to give different
symmetry energy $E_{\rm{sym},2}(\rho)$ and the neutron-proton effective
mass splitting $m^*_{\rm{n}}-m^*_{\rm{p}}$\,\cite{Bom01,Riz04,Li04}. For example,
shown in the upper window of Fig.\,\ref{BGBD} are the BGBD symmetry potentials versus nucleon kinetic energy for the two cases of $m_{\rm{n}}^*>m_{\rm{p}}^*$ (case 1) and $m_{\rm{n}}^*<m_{\rm{p}}^*$ (case 2), respectively.
As we have discussed earlier and shown in this figure, the tendency of case 2 is inconsistent with that of the empirical symmetry potential indicated by the blue bars.
As shown in the lower two windows of Fig.\,\ref{BGBD}, these two cases have completely opposite $E_{\rm{sym},4}(\rho)$ as a function of density.
Various contributions to the $E_{\rm{sym},4}(\rho)$ in the two cases can also be seen very clearly. While the kinetic contribution $T$ and $U_0$ contributions are the same in both cases, the $U_{\rm{sym},1}$ contribution
dominates the high-density behavior of $E_{\rm{sym},4}(\rho)$. Moreover, the $U_{\rm{sym},3}$ contributes appreciably and oppositely in the two cases.
Since the neutron-proton effective mass splitting $m_{\rm{n}}^*-m_{\rm{p}}^*$ is determined by the momentum dependence of the $U_{\rm{sym},1}$, it is then easy to understand the obvious correlation
between the $m_{\rm{n}}^*-m_{\rm{p}}^*$ and the $E_{\rm{sym},4}(\rho)$. Moreover, to our best knowledge, the $U_{\rm{sym},3}$ is completely unconstrained by neither theories nor experiments. Given the small $\delta$ one can reach even with radioactive beams, it will be a big challenge to experimentally probe this term in terrestrial laboratories. This also helps explain why the $E_{\rm{sym},4}(\rho)$ is still so uncertain.

Another interesting example illustrating the role of nucleon effective masses in determining the $E_{\rm{sym},4}(\rho)$, independent of the HVH decomposition,
is the results of the conventional and extended SHF calculations. It was shown very recently that the $E_{\text{sym},4}(\rho)$ is completely determined
by the isoscalar and isovector nucleon effective masses according to\,\cite{PuJ17}
\begin{equation}\label{Esym4ofmsmv}
E_{\text{sym},4}(\rho)=\frac{\hbar^2}{162M}\left(\frac{3\pi^2\rho}{2}\right)^{{2}/{3}}\left[\frac{3M}{M_{\rm{v}}^*(\rho)}-\frac{2M}{M_{\rm{s}}^*(\rho)}\right].
\end{equation}
It is seen that a larger $E_{\text{sym},4}(\rho)$ would require a very small $M^*_{\rm{v}}$ but larger $M^*_{\rm{s}}$\,\cite{PuJ17}.
Indeed, it was shown numerically that the $E_{\text{sym},4}(\rho_0)$ correlates strongly positively with $M_{\rm{s}}^*$ but negatively with $M_{\rm{v}}^*$.
Given the large uncertainties of the $M_{\rm{s}}^*$ and $M_{\rm{v}}^*$ even at the saturation density as we discussed earlier, no wonder why the values of $E_{\text{sym},4}(\rho)$ from various models are so different.
Moreover, as shown in Eq.\,(\ref{sign0}), $(M^*_{\rm{n}}-M^*_{\rm{p}})\propto (M^*_{\rm{s}}-M^*_{\rm{v}})$, the $E_{\rm{sym},4}(\rho)$ is thus closely related to the neutron-proton effective mass splitting in neutron-rich matter.
For instance,  using $m^{*}_{\rm{s}}/m = 0.82\pm0.08$ and $m^{*}_{\rm{v}}/m = 0.69\pm0.02$ predicted by the chiral effective field theory of refs. \cite{Hol16,Hol13},
then the isospin-quartic symmetry energy is about $E_{\text{sym},4}(\rho_0)=0.868\pm 0.142$ MeV at $\rho_0=0.16 \pm 0.02$ fm$^{-3}$.

\begin{figure}[htb]
\begin{center}
\includegraphics[scale=0.5]{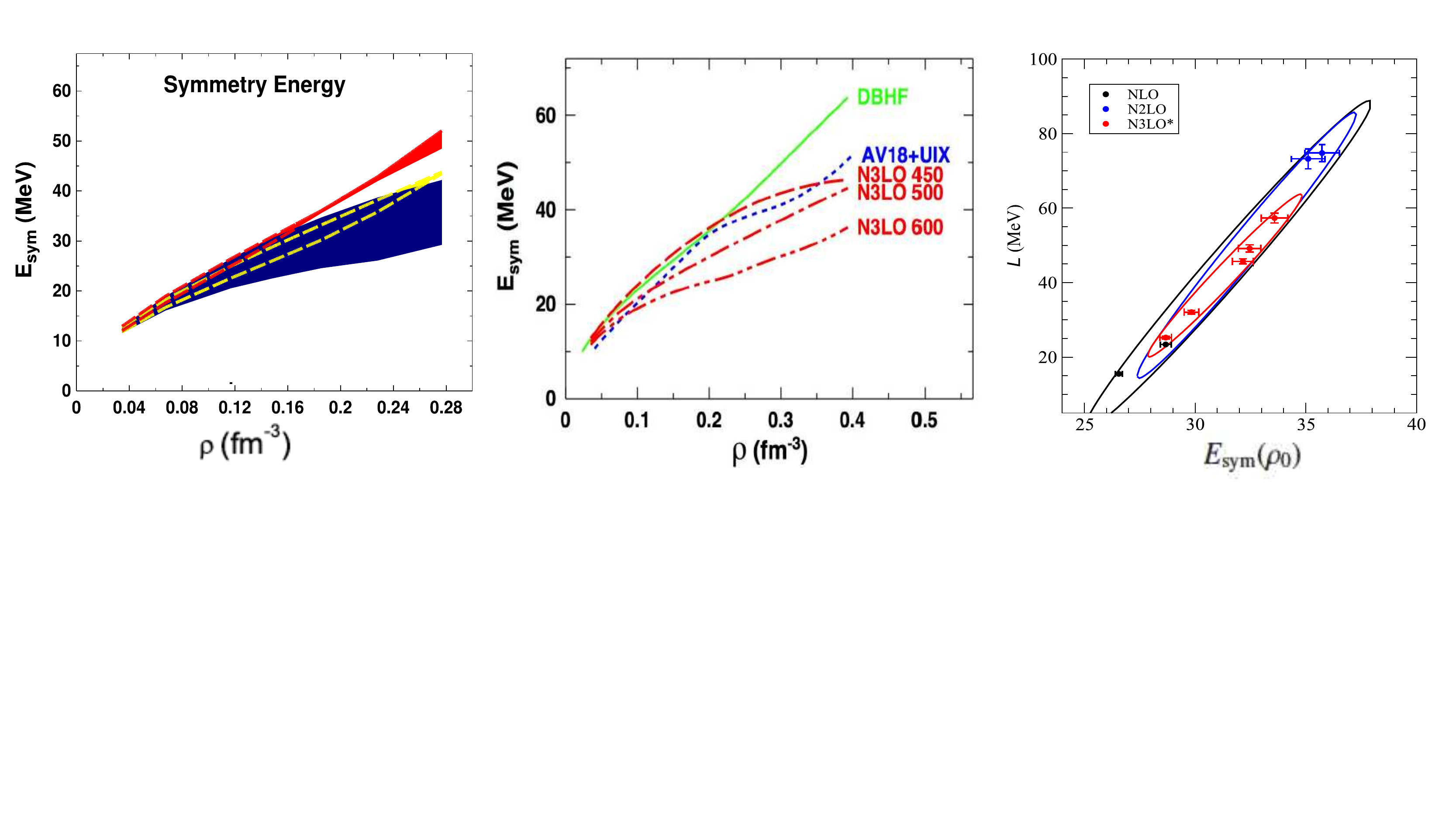}
\vspace{-4.6cm}
\caption{Left: The symmetry energy $E_{\rm{sym}}(\rho)$ from chiral effective field theory of ref. \cite{FS-c1}. Middle: comparing chiral effective field theory predictions using different high-momentum cutoff parameters with the DBHF result up to supra-saturation densities \cite{FS-c2}. Right: the correlation between the slope (L) and magnitude $E_{\rm{sym}}(\rho_0)$ of symmetry energy at saturation density with chiral effective forces within the many-body perturbation theory of ref. \cite{JH-Kaiser}.}\label{Chiral-F}
\end{center}
\end{figure}
\vspace{-0.5cm}
\subsection{A brief summary of the current status of constraining the density dependence of nuclear symmetry energy}
For completeness and ease of our discussions to relate the symmetry energy with the neutron-proton effective mass splitting, in this subsection we very briefly summarize the current situation of constraining the density dependence of nuclear symmetry energy. More comprehensive reviews on this topic can be found in refs. \cite{LCK08,baran05,ireview98,ibook01,Steiner05,Lynch09,Trau12,Tsang12,Lat13,Lat13b,Chuck14,Baldo16b,Blaschke16,EPJA,Oer17}.
Essentially all available many-body theories and interactions have been used to predict the density dependence of nuclear symmetry energy. The predictions more or less agree around the saturation density either as a prediction or by design. However, at both sub-saturation and supra-saturation densities the predictions often disagree especially at high densities where there is no experimental constraints available. Among the possible origins for the uncertain high-density behavior of the symmetry energy, the spin-isospin dependence of the three-body force in energy density functional approaches, the meson-baryon coupling schemes in relativistic mean-field models, the isospin dependent short-range correlations and the tensor force at short-distance in microscopic many-body theories are known to be very important.

Naturally, all nuclear many-body theories and interactions have their own advantages and disadvantages. While we are not in a position to make any judgement about them, we note that according to some experts chiral effective field theory approaches with a better controlled hierarchy of interactions are promising for eventually pinning down the EOS especially the density dependence of nuclear symmetry energy of neutron-rich matter. We thus use here predictions of some chiral effective field theories as an example. Reviews of recent progresses in calculating the EOS using chiral effective 2-body and 3-body forces within various many-body approaches can be found in refs. \cite{Heb14,Heb15,Mac16,Mut17}. As examples, shown in the left window of Fig. \ref{Chiral-F} are chiral effective field theory predictions for the symmetry energy $E_{\rm{sym}}(\rho)$ up to about $1.5\rho_0$ from ref. \cite{FS-c1}. The yellow and red bands represent the uncertainties in the predictions due to the high-momentum cutoff variations as obtained in complete calculations at NLO and N2LO, respectively. The blue band is the result of a calculation employing N3LO NN potentials together with the N2LO 3-body force.
The middle window compares chiral effective field theory predictions using different high-momentum cutoff parameters with that of the DBHF theory \cite{FS-c2} up to about $3\rho_0$. It is clearly shown that the predictions using different high-momentum cutoff parameters do not converge at supra-saturation densities. By design according to some experts, the chiral effective field theory is a low-energy theory that is more suitable at low densities/momenta. Indeed, as illustrated in the right window, the predicted correlation between the slope (L) and magnitude $E_{\rm{sym}}(\rho_0)$ at saturation density \cite{JH-Kaiser} are generally in good agreement with known experimental constraints that we shall discuss in detail in the following. It is interesting to note that recently the equation of state and nucleon effective masses in asymmetric nuclear matter from chiral two- and three-body forces as well as the binding energies of finite nuclei are used to construct/calibrate new Skyrme interactions within the SHF approach. The resulting interactions were used successfully in studying properties of neutron-skins and the dipole polarizability of nuclei as well as properties of neutron stars \cite{Kim,ZZhang}.
\begin{figure}[htb]
\begin{center}
\includegraphics[width=5cm]{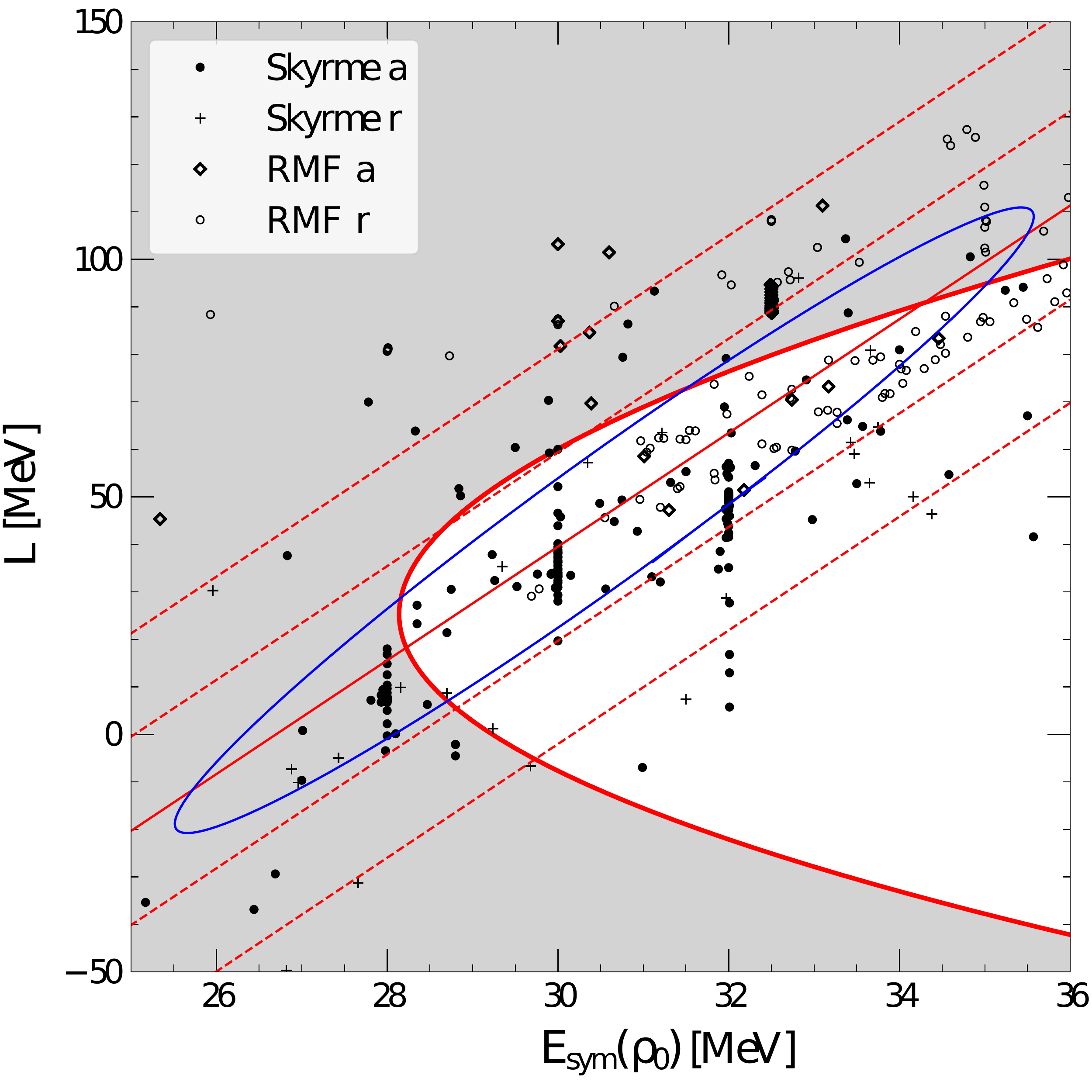}
\hspace{0.5cm}
\includegraphics[width=5cm]{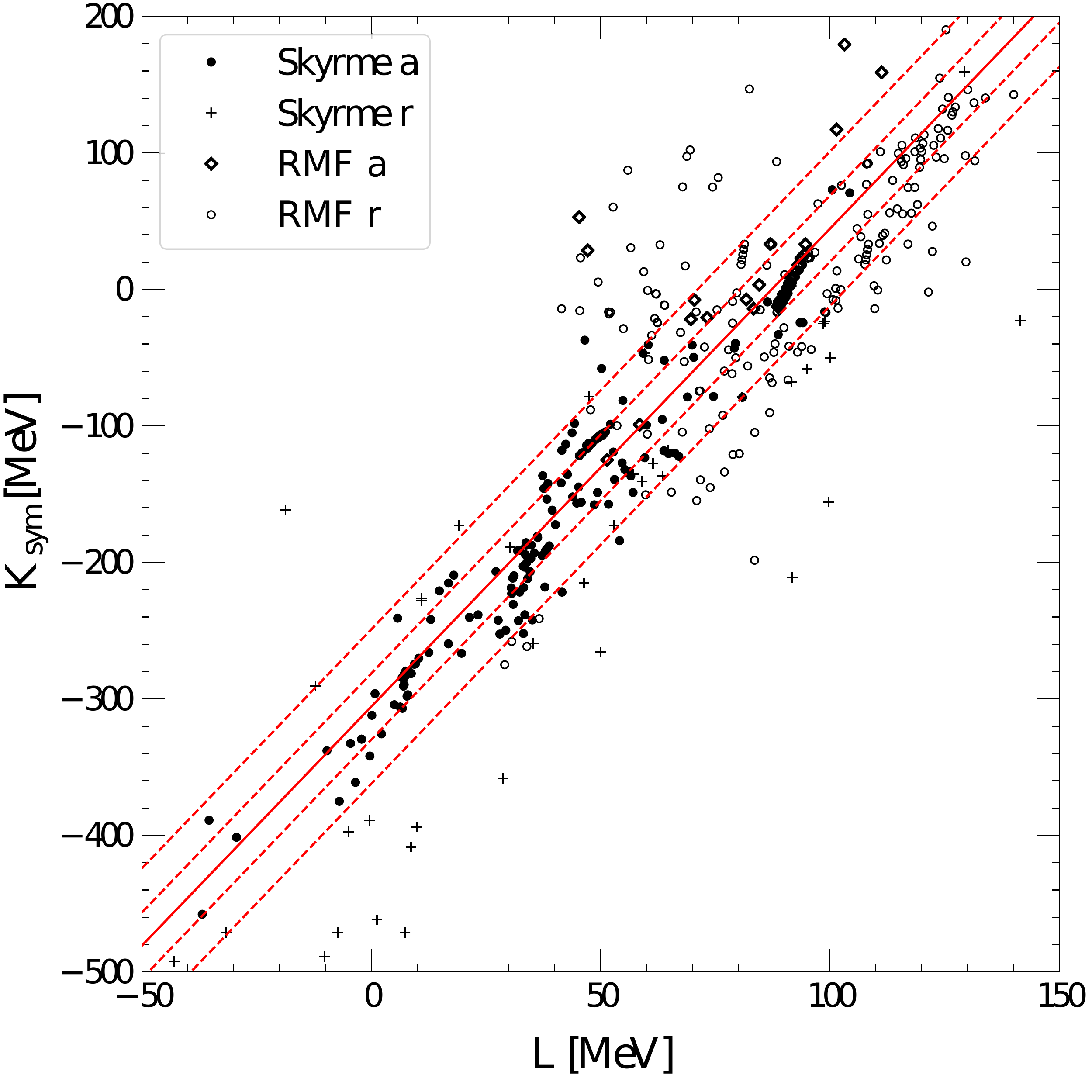}
\hspace{0.5cm}
\includegraphics[width=5cm]{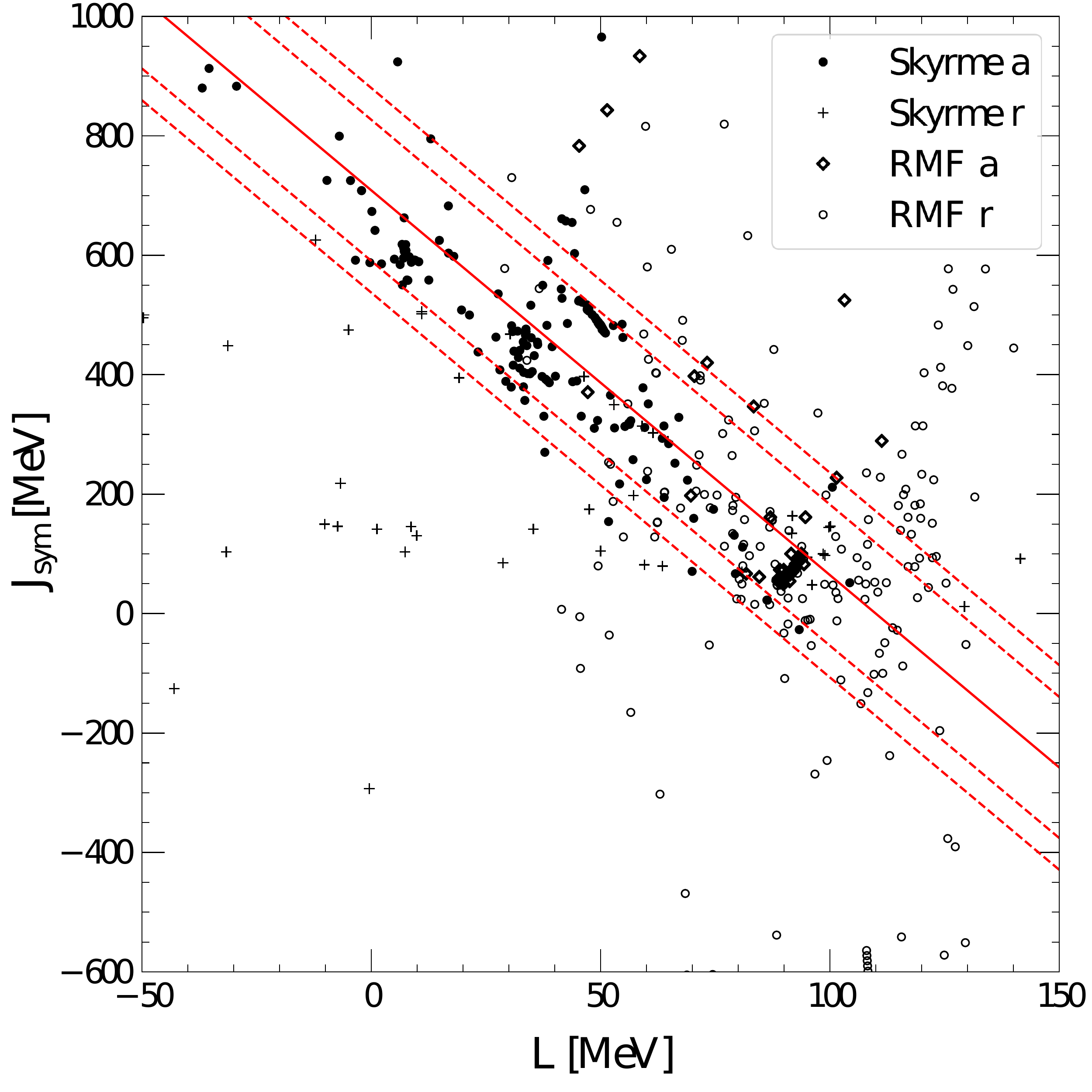}
\caption{
Left: The confidence ellipse (blue) for the $L-E_{\rm{sym}}(\rho_0)$ correlation as
  determined by the UNEDF Collaboration \cite{UNDEF} for an assumed fiducial binding
  energy error of 2~MeV.  The solid thick-red curve shows the conservative
  fiducial exclusion boundary (the shaded grey region is excluded) determined by Tews {\it et al.} from using the Unitary Fermi Gas model with parameters constrained by experimental data and/or theoretical predictions \cite{Tew17}.
  The symmetry energy parameters from the compilations of \cite{Dut12,Dutra14} are indicated
  with either the plus and open circle symbols (denoted `r') for interactions that were rejected, and the filled circles and open diamonds (denoted `a') showing interactions that were accepted.
  The correlation between these parameters for the `accepted' interactions of Equation~\eqref{eq:S0-l} is shown as the solid line, together with the dashed lines showing 68.3\% and 95.4\% enclosures.
The middle (right) panel shows the symmetry incompressibility $K_{\rm{sym}}$ (skewness $J_{\rm{sym}}$) versus the symmetry parameter $L$ for 240 Skyrme interactions compiled in \cite{Dut12} and 263 RMF forces compiled in \cite{Dutra14}. The solid lines show the correlations obtained from the `accepted' (a) realistic interactions after excluding the `rejected' (r) interactions; the dashed lines enclose, respectively, 68.3\% and 95.4\% of the accepted interactions. Taken from ref. \cite{Tew17}.}\label{Lat123}
\end{center}
\end{figure}
\begin{figure}[htb]
\begin{center}
\includegraphics[scale=0.6]{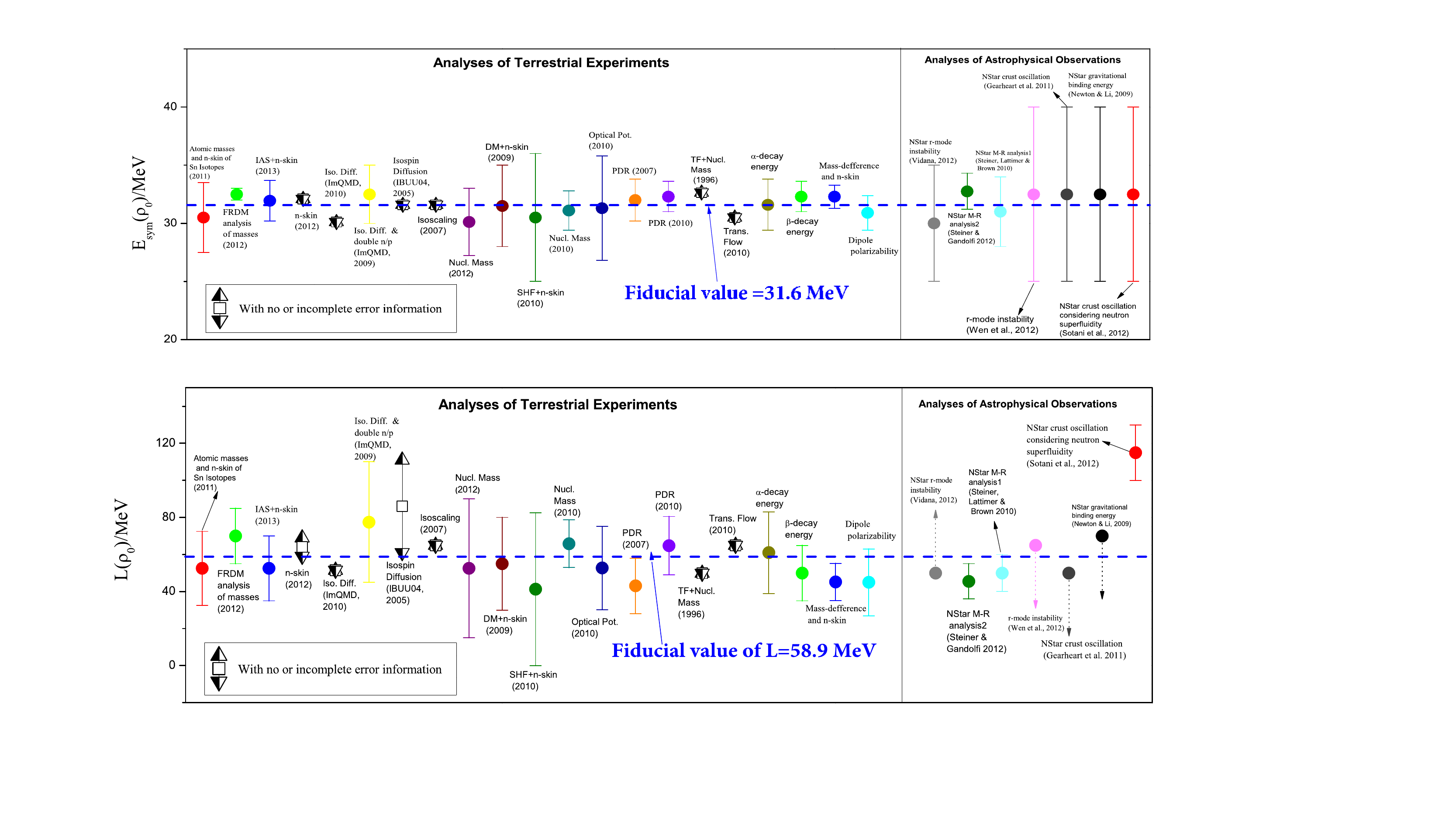}
\vspace{-2cm}
\caption{The reported magnitude $E_{\rm{sym}}(\rho_0)$ (upper window) and slope parameter $L$ (lower window) of nuclear symmetry energy at saturation density of nuclear matter from 28 analyses of terrestrial nuclear laboratory experiments and astrophysical observations.
Taken from ref.\,\cite{LiBA13}.}\label{Li-data}
\end{center}
\end{figure}
As shown in Eqs. (\ref{eos1}, \ref{Su} and \ref{E0para}), the EOS of ANM as well as its isospin-symmetric and asymmetric parts are normally expanded around $\delta=0$ and $\rho_0$, respectively, in terms of several characteristic functions and/or parameters. Naturally and mathematically intrinsic to the Taylor expansions used, the resulting values of the expansion coefficients are generally correlated. The poorly known high-density behavior of the symmetry energy or stiffness of the EOS are manifested in both the strengthes and ranges of the correlated coefficients. Often, it is impossible to accurately determine one coefficient without knowing the next-order coefficient and its uncertainty. These correlations have been studied extensively under various constraints using different approaches. For instance, in an effort of determining both the $E_{\rm{sym,2}}(\rho)$ and $E_{\rm{sym,4}}(\rho)$ from SHF model analyses of nuclear masses, Farine {\it et al.} already found in 1978 an approximately linear correlation between the slope parameter $L$ and the magnitude $E_{\rm{sym,2}}(\rho_0)$ of symmetry energy at saturation density \cite{Far78,Pearson14}. Another earlier and interesting example is the findings of Col\'o {\it et al.} that not only the $L$ and $E_{\rm{sym,2}}(\rho_0)$ but also the incompressibility $K_0$ and characteristics of $E_{\rm{sym}}(\rho)$ are correlated \cite{GC04}. Namely, while the incompressibility $K_0$ has been constrained to be around $K_0=240\pm 20$ MeV \cite{Shlomo06,Piekarewicz10,Khan12}, because of its correlations with $E_{\rm{sym}}(\rho)$ in finite nuclei it is impossible to narrow down the uncertainty of $K_0$ further without a better knowledge about the $E_{\rm{sym}}(\rho)$ near the saturation density. In turn, accurate knowledge about some of the coefficients in expanding the $E_0(\rho)$ and $E_{\rm{sym}}(\rho)$ from either experiments and/or theories, their correlations with other coefficients enable us to learn more about the latter.  Indeed, in recent years such correlations based on various models have been used extensively and fruitfully, see, e.g., refs. \cite{India17,Maz13,Col14,Tew17}. For example, as shown in Fig. \ref{Lat123} Tews {\it et al.} examined recently correlations among the EOS parameters from mean field model calculations using 240 Skyrme forces \cite{Dut12} and 263 RMF forces \cite{Dutra14} compiled by Dutra {\it et al.}. Excluding some of the unrealistic forces using the criteria: 0.149~fm$^{-3}<\rho_0<0.17$ fm$^{-3}$, $-17~\mev<E_0(\rho_0)<-15~\mev$, $25~\mev<E_{\rm{sym}}(\rho_0)<36~\mev$, and $180~\mev<K_0<275~\mev$ (accepted interactions), they found the following correlations with the respective correlation coefficients $r$ using 68.3\% and 95.4\% of the accepted interactions, respectively:
\begin{align}
L &=11.969 E_{\rm{sym}}(\rho_0) - (319.55 \pm 19.83 [41.56])~\mev, \qquad &  r= 0.74\,,\label{eq:S0-l}\\
K_{\rm{sym}}&=3.501~L-(305.67\pm24.26~[56.59])~\mev,\qquad & r=0.96\,,\\
J_{\rm{sym}}&=-6.443~L+ (708.74\pm118.14~[171.34])~\mev,\qquad &r=-0.86\
\end{align}
as depicted in Fig. \ref{Lat123}. These correlations were further used in a Unitary Fermi Gas model \cite{Tew17} to constrain the symmetry energy, leading to the fiducial exclusion boundary (solid thick-red curve) in the $L-E_{\rm{sym}}(\rho_0)$ plane shown in the left window of Fig. \ref{Lat123}. It is interesting to see that the obtained exclusion boundary tightens the confidence ellipse (blue) for the $L-E_{\rm{sym}}(\rho_0)$ correlation determined by the UNEDF Collaboration \cite{UNDEF} for an assumed fiducial binding energy error of 2~MeV. It is worth noting again that currently there is essentially no experimental constraint on the ranges of $K_{\rm{sym}}$ and $J_{\rm{sym}}$ characterizing the high-density behavior of nuclear symmetry energy.
Very recently by analyzing comprehensively the relative elliptical flows of neutrons and protons measured by the ASY-EOS and the FOPI-LAND collaborations at GSI using a Quantum Molecular Dynamics (QMD)
model \cite{Cozma}, a value of $K_{\rm{sym}} = 96 \pm 315(\rm{exp}) \pm 170(\rm{th}) \pm 166(\rm{sys})$ MeV was extracted. Unfortunately, such value does not help constrain model predictions.
We also notice that there have been continued efforts to determine the $K_{\rm{sym}}$ by studying directly the isospin dependence of nuclear incompressibility 
\begin{equation}
K(\delta)\approx K_0+K_{\tau}\delta^2+\mathcal{O}(\delta^4)
\end{equation}
where the $K_{\tau}$
\begin{equation}
K_{\tau}=K_{\rm{sym}}-6L-J_0L/K_0
\end{equation} 
can be extracted from studying giant resonances of neutron-rich nuclei\,\cite{TLi,Jorge-TLi}. Unfortunately, the current estimate of $K_{\tau}\approx -550\pm 100$\,MeV\,\cite{GC04} is still too rough to constrain the individual values of $J_0$ and/or $K_{\rm{sym}} $. New experiments with more neutron-rich beams have the potential of improving significantly the accuracy of measuring $K_{\tau}$. On the other hand, as we shall discuss in the following, the $L$ parameter has already been constrained to regions much narrower than that shown in the plots of Fig. \ref{Lat123} based on many analyses of different kinds of experimental data.

Over the last few years, significant efforts have been made by many people in both nuclear physics and astrophysics to constrain the density dependence of nuclear symmetry energy based on model analyses of various kinds of
experimental/observational data, including atomic masses, sizes of neutron-skins of heavy nuclei, various kinds of giant resonances, several observables and phenomena in heavy-ion collisions, systematics of isobaric analog states,
radii and cooling curve of neutron stars, etc. Unfortunately, the correlations among the symmetry energy coefficients ($E_{\rm{sym}}(\rho_0), L, K_{\rm{sym}}$ and $J_{\rm{sym}}$) are not always fully explored. Nevertheless, often the individual ranges/uncertainties of the $E_{\rm{sym}}(\rho_0)$ and $L$ are given. It is thus still useful to compare the individual values of $E_{\rm{sym}}(\rho_0)$ and $L$.
For example, shown in Fig.\,\ref{Li-data} are the 28 values of $E_{\rm{sym}}(\rho_0)$ and $L(\rho_0)$ extracted from analyzing terrestrial nuclear laboratory experiments and astrophysical observations up to 2013. Naturally, all analyses are based on some models and often different approaches are used in analyzing the same data or observations. However, it is interesting to see that there is a high level of consistency around the fiducial values of $E_{\rm{sym}}(\rho_0)=31.6\pm 2.66$ MeV and $L=59\pm 16$ MeV\,\cite{LiBA13}.

\begin{figure}
\begin{center}
\vspace{-4cm}
\includegraphics[width=0.45\columnwidth]{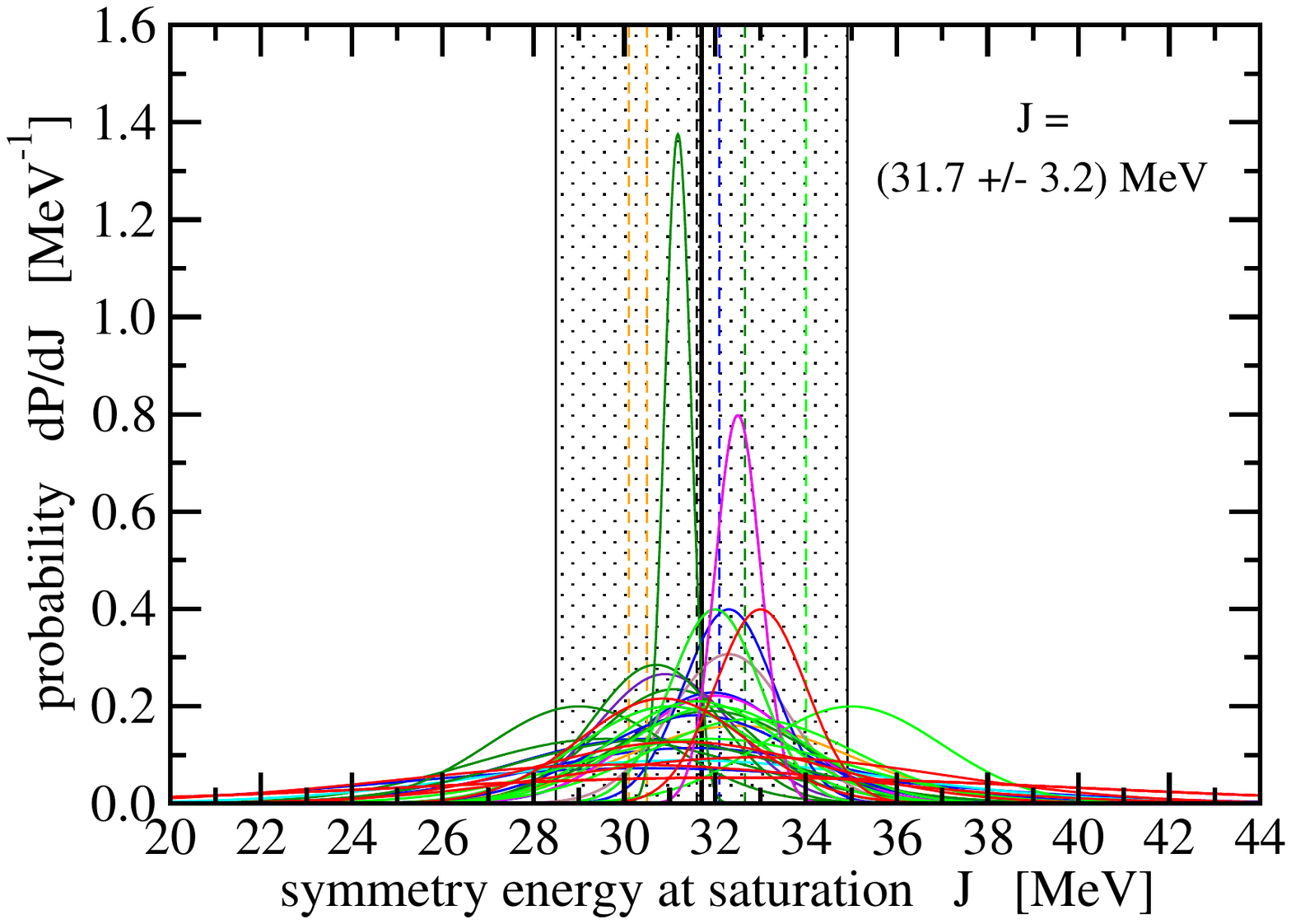}
\hspace{0.3cm}
\includegraphics[width=0.45\columnwidth]{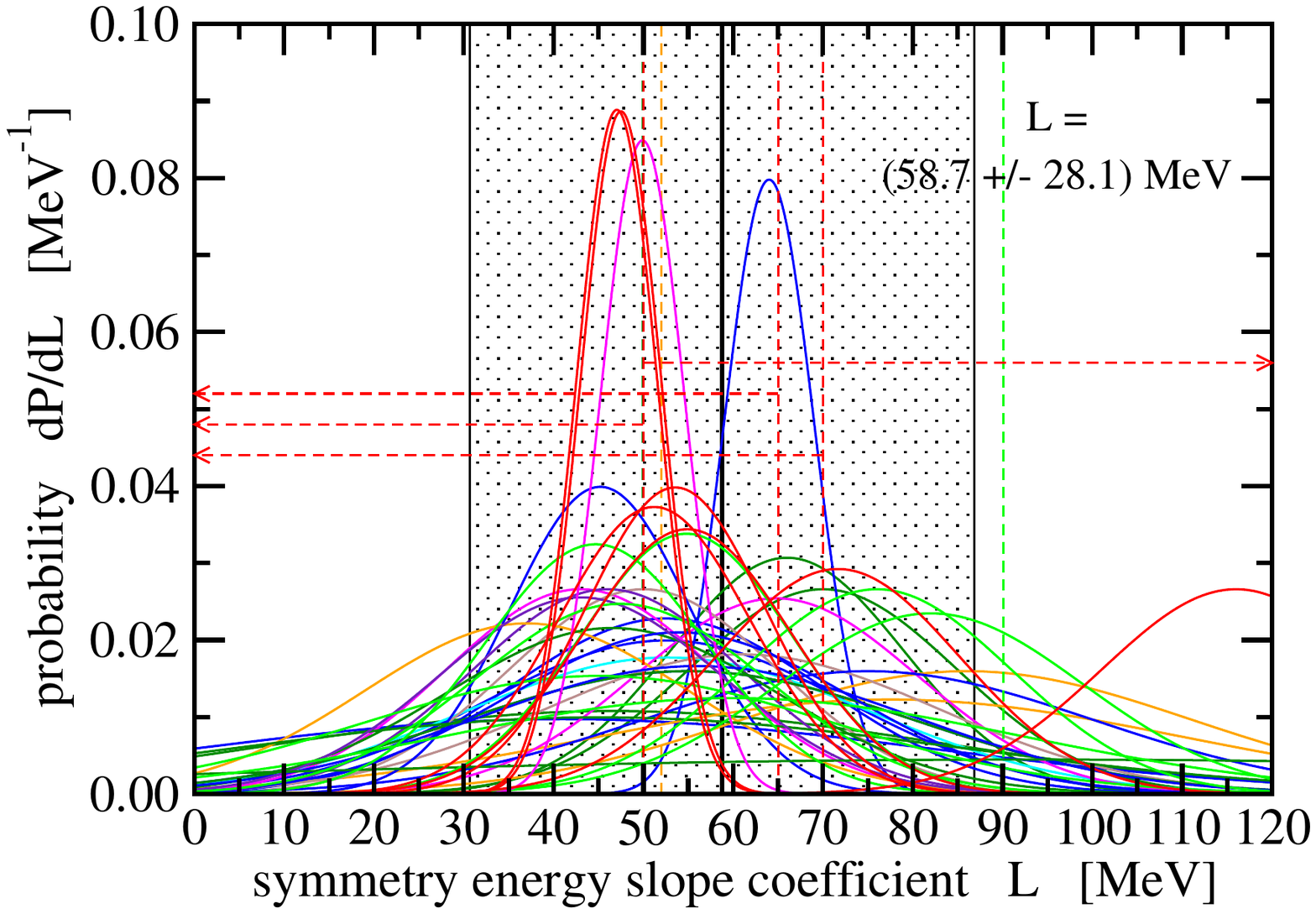}
\caption{Probability distribution of the symmetry energies at saturation $J\equiv E_{\rm{sym}}(\rho_0)$ (left panel)\protect\label{Typ}
and of the symmetry energy slope parameter $L$ (right panel) from 53 studies in the literature up to Oct. 2016. Taken from ref. \cite{Oer17}.}
\end{center}
\end{figure}
More recently,  Oertel {\it et al.} surveyed 53 analyses made up to 2016  \cite{Oer17}.
Shown in Fig. \ref{Typ} are the probability distributions of $J \equiv E_{\rm{sym}}(\rho_0)$ and $L$ values they obtained by assuming Gaussian forms with widths given by the errors/uncertainties of the individual studies.
The dashed vertical lines are used if no uncertainty information is available, while the allowed ranges with upper or
lower bounds are indicated by arrows.  Averaging over their selected results (excluding upper and lower
bounds), they found the fiducial values of $J = E_{\rm{sym}}(\rho_0)=(31.7 \pm 3.2)$~MeV and $L = (58.7 \pm 28.1)$~MeV. These results are consistent with those shown in Fig. \ref{Li-data} albeit with a larger ``error bar" for $L$ as a larger group of more diverse analyses is considered.  We also note that the fiducial values from both Fig. \ref{Li-data} and Fig. \ref{Typ} are consistent with the $E_{\rm{sym}}(\rho_0)$ versus $L(\rho_0)$ correlation plot using only 6 analyses selected by Lattimer and Steiner in ref. \cite{ste14}. Moreover, it is interesting to note that the recent LIGO measurement \cite{LIGO-PRL} of the tidal polarizability of spiraling neutron stars in GW170817 only limits the L value to $L<80$ MeV at the 90\% confidence level \cite{fatt17}. It is consistent but less restrictive than the existing constraints on L we discussed above.

\subsection{From symmetry energy to neutron-proton effective mass splitting in neutron-rich matter}
The above discussions focus on obtaining the symmetry energy from the density and momentum/energy dependence of single-nucleon potential or self-energy. In turn, if the density dependence of nuclear symmetry energy is well determined, one can also infer subsequently information about the neutron-proton effective mass splitting. For example, as an application of the non-relativistic expressions of $E_{\rm{sym}}(\rho)$ and $L(\rho)$ in Eqs.\,(\ref{Esymexp2}) and (\ref{Lexp2}), if one neglects the small contributions of the momentum dependence of the nucleon isoscalar effective mass itself and the second-order symmetry potential $U_{\rm{sym},2}$, the neutron-proton effective mass splitting at $\rho_0$ can then be readily expressed in terms of the  $E_{\rm{sym}}(\rho_0)$ and $L(\rho_0)$ as\,\cite{LiBA13}
\begin{equation}
m^*_{\rm{n-p}}(\rho_0,\delta)\approx\delta\cdot\frac{\displaystyle3E_{\rm{sym}}(\rho_0)-L(\rho_0)-3^{-1}({m}/{m^*_0})E_{\rm{F}}(\rho_0)}{\displaystyle
E_{\rm{F}}(\rho_0)\left({m}/{m_0^*}\right)^2}. \label{npemass2}
\end{equation}
A similar expression was derived very recently in ref.\,\cite{India17}.
It is interesting to see that the $M_{\rm n}^*$ is equal to, larger or smaller than the $M_{\rm p}^*$ depends on the symmetry energy and its slope. For example, using empirical values of \es0=31 MeV, $m_0^*/m=0.7$ and $E_{\rm{F}}(\rho_0)=36$ MeV, a value of \l0$\leq 76$ MeV is required to get a positive $m^*_{\rm{n-p}}(\rho_0,\delta).$ Interestingly, most of the extracted values of $E_{\rm{sym}}(\rho_0)$ and $L(\rho_0)$ from both terrestrial experiments and astrophysical observations satisfy this condition\,\cite{LiBA13,EPJA,Oer17}.
Moreover, using the decompositions of the symmetry energy and its slope parameter, the neutron-proton effective mass splitting $m_{\rm{n-p}}^{\ast}(\rho,\delta$ at an arbitrary density $\rho$ can be rewritten via
\begin{align}
m_{\rm{n-p}}^{\ast}(\rho,\delta)\approx
&-\delta\cdot \frac{2M_0^{\ast,2}}{Mk_{\rm{F}}^2}[2L_2(\rho)+L_4(\rho)]
=\delta\cdot\frac{2M_0^{\ast,2}}{Mk_{\rm{F}}}\left[\left.\frac{dU_{\rm{sym,1}}}{dk}\right|_{k_{\rm{F}}}
-\frac{k_{\rm{F}}}{3M_0^{\ast}}+\frac{6U_{\rm{sym,2}}(\rho)-2(L(\rho)-3E_{\rm{sym}}(\rho))}{k_{\rm{F}}}\right].
\end{align}
Thus, the empirical constraint that $M_{\rm{n}}^{\ast}>M_{\rm{p}}^{\ast}$ at $\rho_0$, then requires the inequality
\begin{equation}
\mathcal{D}\equiv U_{\rm{sym,2}}(\rho_0)+\left.\frac{k_{\rm{F}}}{6}\frac{dU_{\rm{sym}}}{dk}\right|_{k_{\rm{F}},\rho_0}
\geq \frac{L(\rho_0)}{3}-E_{\rm{sym}}(\rho_0)+\frac{k_{\rm{F}}^2}{18M_0^{\ast}}.
\end{equation}
The quantities on the right hand side are currently better constrained than those on the left hand side,
thus providing a useful constraint on the combination $\mathcal{D}$. For instance, $L(\rho_0)/3-E_{\rm{sym}}(\rho_0)+k_{\rm{F}}^2/18M_0^{\ast}\approx-5.5\pm5.8\,\rm{MeV}
$ with $L(\rho_0)=59\pm16\,\rm{MeV}$, $E_{\rm{sym}}(\rho_0)=31\pm2$\,MeV and $M_0^{\ast}/M=0.7\pm0.1$,
leading to $\mathcal{D}\gtrsim-5.5\,\rm{MeV}$.

From the 28 extracted values for \es0 and $L(\rho_0)$ shown in Fig.\,\ref{Li-data}, the mean values of $U_{\rm{sym},1}(\rho_0,k_{\rm{F}}) = 2[E_{\rm{sym}}(\rho_0)-3^{-1}({m}/{m^*_0})E_{\rm{F}}(\rho_0)]$
and $m^*_{\rm{n-p}}(\rho_0,\delta)$ from different studies were found\,\cite{LiBA13} to scatter very closely around $U_{\rm{sym},1}(\rho_0,k_F)=29$ MeV and \emass$=0.27\delta$, respectively. These results are remarkably consistent with those extracted from the optical model analyses we discussed in section \ref{Exp-cs}.
However, it is necessary to point out that currently it is impossible to estimate a scientifically sound error bar for the extracted  $m^*_{\rm{n-p}}(\rho_0,\delta)$ in this approach. This is because some reported constraints on the $E_{\rm{sym}}(\rho_0)$ and $L(\rho_0)$ are given in certain ranges where all values are equally probable, some others are given in terms of the means and the standard deviations, while the rest are given with only the mean values without any information about the associated uncertainties. Nevertheless, as an indication of uncertainties, a ``standard deviation" of $ 0.25\delta$ was estimated for
the \emass$=0.27\delta$ using the 28 mean values and the average sizes of the uncertainty ranges when available. Thus, based on this analysis alone, while the mean values indicate strongly that $m^*_{\rm{n}}$ is higher than $m^*_{\rm{p}}$ at $\rho_0$, its uncertainty remains to be quantified more scientifically.

\subsection{Decompositions of nuclear symmetry energy and neutron-proton effective mass splittings in relativistic approaches}\label{rel-D}
A relativistic version of the decomposition of the $E_{\rm{sym}}(\rho)$ and $L(\rho)$ in terms of Lorentz covariant nucleon self-energies was given in ref.\,\cite{Cai-EL}. Here the main conclusions as well as an example of applying it to the RMF model will be discussed briefly.  Very generally, the Lorentz covariant nucleon self-energy in ANM can be written as\,\cite{Jam81,Hor83,Ser86,Uec90}
\begin{equation}
\Sigma ^{J}(\rho ,\delta ,{k})=\Sigma _{\text{S}}^{J}(\rho ,\delta ,{k})+\gamma ^{0}\Sigma _{\text{V}}^{J}(\rho ,\delta ,{k})+\vec{\gamma}\cdot \mathbf{k}^{0}\Sigma _{\text{K}}^{J}(\rho ,\delta,k)  \label{DefSE2}
\end{equation}
where $\Sigma_{\text{S}}^{J}(\rho ,\delta ,{k})$ is the scalar self-energy, $\Sigma _{\text{V}}^{J}(\rho ,\delta,{k})\equiv -\Sigma ^{0,J}(\rho ,\delta ,{k})$ and
$\Sigma _{\text{K}}^{J}(\rho,\delta ,{k})$ are, respectively, the time and space component of the vector self-energy $\Sigma ^{\mu ,J}(\rho ,\delta ,{k})$,
and $\mathbf{k}^{0}=\mathbf{k}/{k}$ is the momentum unit vector. The symmetry energy has seven parts
\begin{align}
E_{\mathrm{sym}}(\rho )=E_{\mathrm{sym}}^{\mathrm{kin}}(\rho )
+E_{\mathrm{sym}}^{\mathrm{0,mom,K}}(\rho)+ E_{\mathrm{sym}}^{\mathrm{0,mom,S}
}(\rho)+E_{\mathrm{sym}}^{\mathrm{0,mom,V}}(\rho)
+E_{\mathrm{sym}}^{\mathrm{1st,K}}(\rho) +E_{\mathrm{sym}}^{\mathrm{1st,S}
}(\rho) +E_{\mathrm{sym}}^{\mathrm{1st,V}}(\rho)  \label{ExprEsym}
\end{align}
where $E_{\mathrm{sym}}^{\mathrm{kin}}(\rho)$, $E_{\mathrm{sym}}^{\mathrm{0,mom},%
\mathcal{O} }(\rho )$ and
$E_{\mathrm{sym}}^{\mathrm{1st},\mathcal{O}}(\rho ) $ (here
$\mathcal{O}$ denotes K, S or V) represent, respectively, the
contributions from the kinetic part, the momentum dependence of the
nucleon self-energies in SNM and the
first-order symmetry self-energies. They can be further expressed as
\begin{align}
E_{\text{sym}}^{\text{kin}}(\rho)=&\frac{k_{\text{F}}k_{\text{F}}^{\ast}(\rho)}{6
{E}_{\text{F}}^{\ast}(\rho)},  \label{DefEsymkin} \\
E_{\mathrm{sym}}^{\mathrm{0,mom,K}}(\rho)=&
\frac{k_{\text{F}}k_{\text{F}}^{\ast}(\rho)}{6{E}_{\text{F}}^{\ast}(\rho)}
\left.\frac{\partial \Sigma_{\text{K}}^0(\rho
,{k})}{\partial{k}}\right|_{k_{\rm{F}}},
\label{DefEsymmomK}
\end{align}
and
\begin{align}
E_{\mathrm{sym}}^{\mathrm{0,mom,S}}(\rho)=&
\frac{k_{\text{F}}M_0^{\ast}(\rho)}{6{E}_{\text{F}}^{\ast}(\rho)}
\left.\frac{\partial \Sigma_{\text{S}}^0(\rho
,{k})}{\partial{k}}\right|_{k_{\rm{F}}},
\label{DefEsymmomS} \\
E_{\mathrm{sym}}^{\mathrm{0,mom,V}}(\rho)=& \frac{k_{\text{F}}}{6}
\left.\frac{\partial \Sigma_{\text{V}}^0(\rho ,{k})}
{\partial{k}}\right|_{k_{\rm{F}}},  \label{DefEsymmomV} \\
E_{\mathrm{sym}}^{\mathrm{1st,K}}(\rho)=&\frac{1}{2}\frac{k_{\text{F}%
}^{\ast}(\rho) \Sigma_{\text{K}}^{\mathrm{sym,1}}(\rho
,k_{\rm{F}})}{{E}_{\text{F}}^{\ast}(\rho)},
\label{DefEsym1stK} \\
E_{\mathrm{sym}}^{\mathrm{1st,S}}(\rho)=&\frac{1}{2}\frac{M_0^{\ast}(\rho) \Sigma_{%
\text{S}}^{\mathrm{sym,1}}(\rho
,k_{\rm{F}})}{{E}_{\text{F}}^{\ast}(\rho)},
\label{DefEsym1stS} \\
E_{\mathrm{sym}}^{\mathrm{1st,V}}(\rho)=&\frac{1}{2}\Sigma_{\text{V}}^{%
\mathrm{sym,1}}(\rho ,k_{\rm{F}}) ,
\label{DefEsym1stV}
\end{align}%
where $k_{\text{F}}^{\ast }(\rho)=k_{\text{F}}+\Sigma
_{\text{K}}^{0}(\rho ,k_{\text{F}})$ and $M_{0}^{\ast }(\rho)$ is the Dirac mass in symmetric matter, i.e., $M_{0}^{\ast }(\rho)=M+\Sigma
_{\text{S}}^{0}(\rho ,k_{\text{F}})$, ${E}_{\text{F}}^{\ast
}(\rho)=({k_{\text{F}}^{\ast 2}+M_{0}^{\ast 2}})^{1/2}$, $\Sigma
_{\text{K}}^{0}(\rho ,{k})=\Sigma _{\text{K}}^{J}(\rho
,\delta =0,{k})$, $\Sigma _{\text{S}}^{0}(\rho
,{k})=\Sigma _{\text{S}}^{J}(\rho ,\delta
=0,{k})$, $\Sigma _{\text{V}}^{0}(\rho
,{k})=\Sigma _{\text{V}}^{J}(\rho ,\delta
=0,{k})$, and the $i$-th order symmetry self-energy is
defined as (here $\mathcal{O}=\text{K,S,V}$)
\begin{equation}
\Sigma _{\mathcal{O}}^{\mathrm{sym},i}(\rho ,{k})= \frac{1}{i!%
}\frac{\partial ^{i}}{\partial \delta ^{i}}\left[ \sum_{J=\text{p,n}}\frac{%
\tau _{3}^{Ji}\Sigma _{\mathcal{O}}^{J}(\rho ,\delta ,{k})}{2}%
\right]_{\delta =0}.  \label{SymmSE}
\end{equation}
Noticing that the nucleon Landau mass in SNM is
$M^{\ast}_{0,\textrm{Lan}}(\rho
,{k})={k}[{d{k}}/{d{E}^{0}(\rho
,{k})}]^{-1}$  with
${E}^{0}(\rho ,{k})= {E}^{J}(\rho ,\delta
=0,{k})$ the single nucleon energy in SNM, its expression at the Fermi momentum can be written as
\begin{equation}
M_{0,\rm{Lan}}^{\ast}(\rho)=k_{\rm{F}}\left\{\frac{k_{\rm{F}}^{\ast}}{{E}_{\rm{F}}^{\ast}}\left[1+\frac{\partial\Sigma_{\rm{K}}^0(\rho,{k})}{\partial
{k}}\right]
+\frac{M_0^{\ast}}{{E}_{\rm{F}}^{\ast}}\frac{\partial\Sigma_{\rm{S}}^0(\rho,{k})}{\partial
{k}}+\frac{\partial\Sigma_{\rm{V}}^0(\rho,{k})}{\partial
{k}}\right\}_{k_{\rm{F}}}^{-1}.\label{ss_129}
\end{equation}
Then considering the first four terms of the decomposition of
the symmetry energy, it can be directly shown that
${k_{\text{F}}^2}/{6M^{\ast}_{0,\textrm{Lan}}(\rho
,k_{\text{F}})}=E_{\textrm{sym}}^{\textrm{kin}}(\rho)+
E_{\mathrm{sym}}^{\mathrm{0,mom,K}}(\rho)+
E_{\mathrm{sym}}^{\mathrm{0,mom,S}}(\rho)+
E_{\mathrm{sym}}^{\mathrm{0,mom,V}}(\rho)$.
Finally, the symmetry energy in Eq.\,(\ref{ExprEsym}) can then be rewritten as
\begin{align}
E_{\mathrm{sym}}(\rho)&=\left.\frac{{k}^2}{6M^{\ast}_{0,\textrm{Lan}}(\rho
,{k})}
\right|_{k_{\rm{F}}}
+E_{\mathrm{sym}}^{\mathrm{1st,K}}(\rho) +E_{\mathrm{sym}}^{\mathrm{1st,S}%
}(\rho) +E_{\mathrm{sym}}^{\mathrm{1st,V}}(\rho)
\label{EsymDecomp}
\end{align}
Thus, at the mean-field level the symmetry energy can be
decomposed analytically in terms of the Lorentz covariant nucleon self-energies in asymmetric nuclear matter. This expression is general and reveals clearly the underlying physics determining the
density dependence of nuclear symmetry within relativistic approaches.

Because of the momentum dependence of the Landau mass itself besides the density dependence of all terms in Eq.\,(\ref{EsymDecomp}), the general and analytical expression for the slope parameter $L(\rho)$
has many more terms but all have clear physical meanings. Specifically,  $L(\rho)$ can be decomposed as\,\cite{Cai-EL}
\begin{align}  \label{ExprL}
L(\rho)=L^{\mathrm{kin}}(\rho)+L^{\mathrm{mom}}(\rho)+L^{\mathrm{1st}}(\rho) + L^{\mathrm{cross}}(\rho) +L^{\mathrm{2nd}}(\rho)
\end{align}
where
\begin{align}
L^{\mathrm{kin}}(\rho)&=\frac{k_{\text{F}}k_{\text{F}}^{\ast}}{6{E}_{\text{F}}^{\ast}}
+\frac{k_{\text{F}}^2M_0^{\ast,2}}{6{E}_{\text{F}}^{\ast,3}},\\
L^{\mathrm{mom}}(\rho)&=\frac{k_{\text{F}}^2M_0^{\ast,2}}
{3{E}_{\text{F}}^{\ast,2}}\left.\frac{\partial\Sigma_{\text{K}}^0}{\partial
{k}}\right|_{k_{\rm{F}}}
+\frac{k_{\text{F}}^2}{6}\left[\frac{k_{\text{F}}^{\ast}}{{E}_{%
\text{F}}^{\ast}} \frac{\partial^2\Sigma_{\text{K}}^0}{\partial k^2}+\frac{M_0^{\ast}}{{E}_{\text{F}}^{\ast}} \frac{%
\partial^2\Sigma_{\text{S}}^0}{\partial {k}^2}+\frac{%
\partial^2\Sigma_{\text{V}}^0}{\partial {k}^2}\right]_{k_{\text{F}}}
+\frac{k_{\text{F}}}{6}\left[\frac{k_{\text{F}}^{\ast}}{{E}_{\text{F%
}}^{\ast}} \frac{\partial\Sigma_{\text{K}}^0}{ \partial{k}} +\frac{%
M_0^{\ast}}{{E}_{\text{F}}^{\ast}}\frac{\partial\Sigma_{\text{S}}^0}{%
\partial{k}} +\frac{\partial\Sigma_{\text{V}}^0}{ \partial k}\right]_{k_{\rm{F}}}  \notag \\
&+\frac{k_{\text{F}}^2}{6{E}_{\text{F}}^{\ast,3}}\left[%
M_0^{\ast,2}\left( \frac{\partial\Sigma_{\text{K}}^0}{\partial {k}}%
\right)^2+k_{\text{F}}^{\ast,2}\left(\frac{\partial\Sigma_{\text{S}}^0}
{\partial {k}}\right)^2\right]_{k_{\rm{F}}}
-\frac{k_{\text{F}}^2k_{\text{F}}^{\ast}M_0^{\ast}}
{3{E}_{\text{F}}^{\ast,3}}\left[
\frac{\partial\Sigma_{\text{S}}^0}{\partial
{k}}\left(1+\frac{\partial\Sigma_{\text{K}}^0}{\partial
{k}}\right)\right]_{k_{\rm{F}}} ,
\label{DefLMDen}
\end{align}
and
\begin{align}
L^{\mathrm{1st}}(\rho)&=\frac{3}{2{E}_{\text{F}}^{\ast,3}}\left[%
M_0^{\ast}\Sigma_{\text{K}}^{\mathrm{{sym},1}}- k_{\text{F}}^{\ast}\Sigma_{%
\text{S}}^{\mathrm{{sym},1}}\right]^2
+\frac{3}{2}\left[\frac{k_{\text{F}}^{\ast}}{{E}_{\text{F}}^{\ast}}%
\Sigma_{\text{K}}^{\mathrm{{sym},1}}+ \frac{M_0^{\ast}}{{E}_{\text{F}%
}^{\ast}}\Sigma_{\text{S}}^{\mathrm{{sym},1}} +\Sigma_{\text{V}}^{\mathrm{{%
sym},1}}\right]
+\frac{k_{\text{F}}M_0^{\ast,2}\Sigma_{\text{K}}^{\text{sym,1}}}{ {E}%
_{\text{F}}^{\ast,3}} -\frac{k_{\text{F}}k_{\text{F}}^{\ast}M_0^{\ast}\Sigma_{%
\text{S}}^{\text{sym,1}}}{ {E}_{\text{F}}^{\ast,3}},
\label{DefLS1} \\
L^{\text{cross}}(\rho)&=
k_{\text{F}}\left[\frac{k_{\text{F}}^{\ast}}{{E}_{\text{F}}^{\ast}}%
\frac{\partial \Sigma_{\text{K}}^{\mathrm{{sym},1}}}{\partial {k}}+%
\frac{M_0^{\ast}}{{E}_{\text{F}}^{\ast}}\frac{\partial \Sigma_{\text{%
S}}^{\mathrm{{sym},1}}}{\partial{k}}+\frac{\partial \Sigma_{\text{V}%
}^{\mathrm{{sym},1}}}{\partial {k}}\right]_{{k}=k_{\text{F}%
}}
-\frac{k_{\text{F}}\Sigma_{\text{K}}^{\mathrm{{sym},1}}}{{E}_{\text{%
F}}^{\ast}} \left[\frac{k_{\text{F}}^{\ast,2}}{{E}_{\text{F}}^{\ast,2}}%
\left(\frac{\partial\Sigma_{\text{K}}^0}{ \partial{k}}+\frac{%
M_0^{\ast}}{k_{\text{F}}^{\ast}}\frac{\partial\Sigma_{\text{S}}^0}{ \partial k}\right) -\frac{\partial\Sigma_{\text{K}}^0}{ \partial{k}%
}\right]_{k_{\rm{F}}}  \notag \\
&-\frac{k_{\text{F}}\Sigma_{\text{S}}^{\mathrm{{sym},1}}}{{E}_{\text{%
F}}^{\ast}} \left[ \frac{M_0^{\ast,2}}{{E}_{\text{F}}^{\ast,2}}\left(%
\frac{k_{\text{F}}^{\ast}}{M_0^{\ast}} \frac{\partial\Sigma_{\text{K}}^0}{
\partial{k}}+\frac{\partial\Sigma_{\text{S}}^0}{ \partial k}\right)-\frac{ \partial\Sigma_{\text{S}}^0}{ \partial{k}}\right]%
_{k_{\rm{F}}} ,  \label{DefLCross} \\
L^{\mathrm{2nd}}(\rho)&=3\left[\frac{k_{\text{F}}^{\ast}}{{E}_{\text{%
F}}^{\ast}}\Sigma_{\text{K}}^{\mathrm{{sym},2}} +\frac{M_0^{\ast}}{{E%
}_{\text{F}}^{\ast}}\Sigma_{\text{S}}^{\mathrm{{sym},2}}+\Sigma_{\text{V}}^{%
\mathrm{{sym},2}}\right].  \label{DefLS2}
\end{align}
In the above, the density and momentum dependence of the three-types of self-energies have been denoted by $\Sigma
_{\mathcal{O}}^{\mathrm{sym},i}=\Sigma
_{\mathcal{O}}^{\mathrm{sym},i}(\rho ,k_{\rm{F}})$
($\mathcal{O}=\text{K,S,V}$). While lengthy, it is clearly seen that the $L(\rho)$ involves the first-order and second-order derivatives of self-energies with respect to momentum $k$,
as well as their first-order and second-order derivatives with respect to isospin asymmetry $\delta$ through the $\Sigma _{\mathcal{O}}^{\mathrm{sym},i}(\rho ,{k})$ terms.
Again, the above decompositions of the $E_{\mathrm{sym}}(\rho)$ and $L(\rho)$ are general. Different models may have intrinsically different components for the two quantities.
Thus, comparisons of these components may help better understand the known model and interaction dependences of predictions.

\begin{figure}[h!]
\centering
\includegraphics[scale=1.]{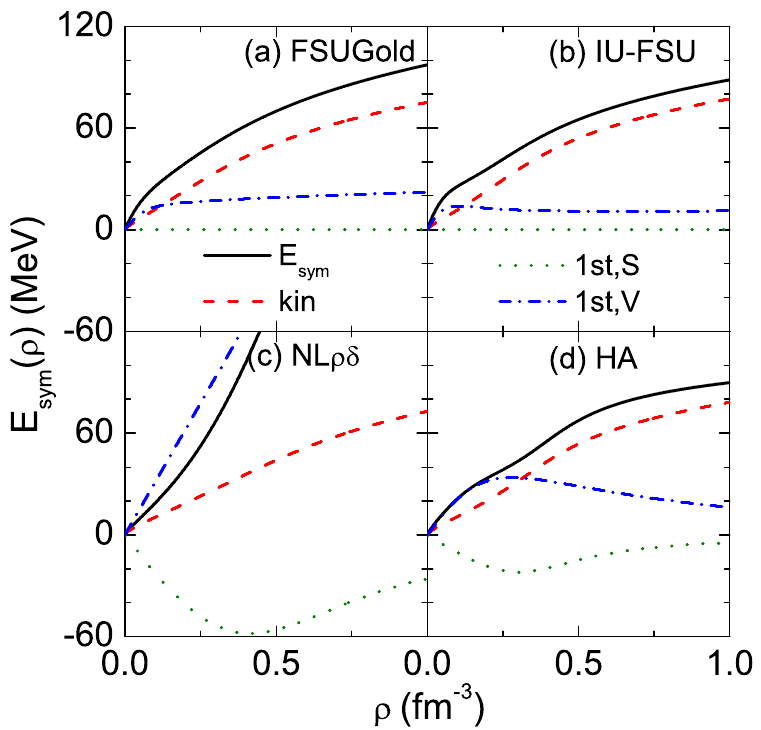}
\includegraphics[scale=1.]{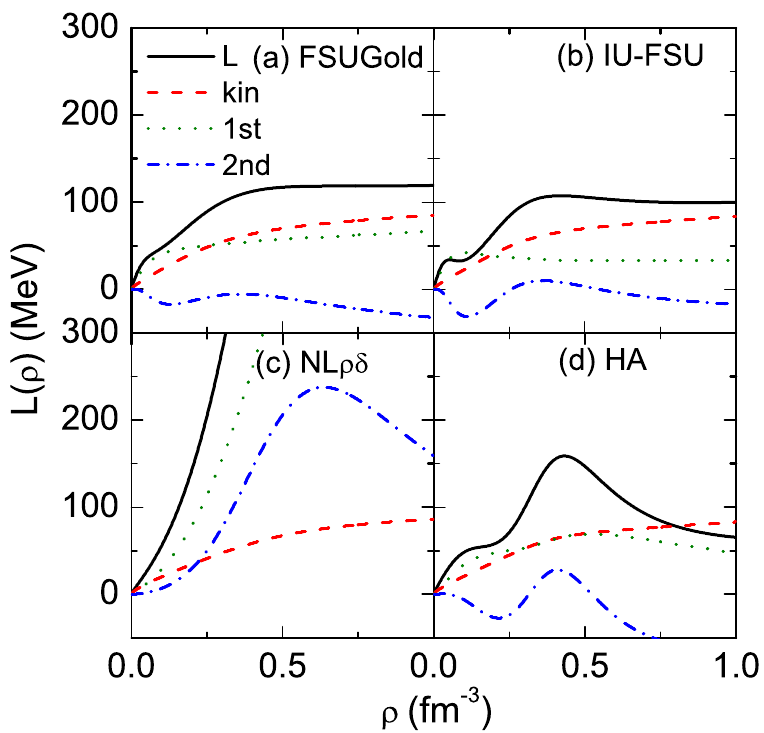}
\caption{(Color Online). Decompositions of the symmetry energy $E_{\mathrm{sym}}(\rho
)$ and its density-slope $L(\rho)$ in the nonlinear RMF model with different interactions. Taken from ref.\,\cite{Cai-EL}.} \label{FIGEsymDecom}
\end{figure}

In the RMF models, the general expression of the symmetry energy in Eq. (\ref{EsymDecomp}) is reduced to
\begin{equation}
E_{\rm{sym}}(\rho)=\frac{k_{\rm{F}}^2}{6M_{\rm{0,Lan}}^{\ast}(\rho,k_{\rm{F}})}+E_{\rm{sym}}^{\rm{1st,S}}(\rho)
+E_{\rm{sym}}^{\rm{1st,V}}(\rho)
\end{equation}
with $M_{\rm{0,Lan}}^{\ast}=E_{\rm{F}}^{\ast}=(k_{\rm{F}}^2+M_0^{\ast,2})^{1/2}$ and
\begin{equation}
E_{\rm{sym}}^{\rm{1st,S}}(\rho)=-\frac{g_{\delta}^2M_0^{\ast,2}\rho}{2{E}^{\ast,2}_{\rm{F}}Q_{\delta}},
~~E_{\rm{sym}}^{\rm{1st,V}}(\rho)=\frac{g_{\rho}^2\rho}{2Q_{\rho}}.
\end{equation}
While the slope parameter is reduced to
$L(\rho)=L^{\rm{kin}}(\rho)+L^{\rm{1st}}(\rho)+L^{\rm{2nd}}(\rho)
$ with
\begin{align}
L^{\rm{kin}}(\rho)=&\frac{k_{\rm{F}}^2({E}_{\rm{F}}^{\ast,2}+M_0^{\ast,2})}{6{E}_{\rm{F}}^{\ast,3}},\label{RMF_Lkin}\\
L^{\rm{1st}}(\rho)=&\frac{3g_{\rho}^2\rho}{2Q_{\rho}}-\frac{3g_{\delta}^2M_0^{\ast,2}\rho}{2{E}_{\rm{F}}^{\ast,2}Q_{\delta}}
\left(1-\frac{2k_{\rm{F}}^2}{3{E}_{\rm{F}}^{\ast,2}}-\frac{g_{\delta}^2k_{\rm{F}}^2\rho}{Q_{\delta}{E}_{\rm{F}}^{\ast,3}}\right),\\
L^{\rm{2nd}}(\rho)=&\frac{g_{\sigma}^2M_0^{\ast,2}k_{\rm{F}}^2\rho}{2Q_{\sigma}{E}_{\rm{F}}^{\ast,4}}
-\frac{3g_{\sigma}^3g_{\rho}^4\Lambda_{\rm{S}}\overline{\sigma}M_0^{\ast}\rho^2}{{E}_{\rm{F}}^{\ast}Q_{\sigma}Q_{\rho}^2}
-\frac{3g_{\omega}^3g_{\rho}^4\Lambda_{\rm{V}}\overline{\omega}_0\rho^2}{Q_{\omega}Q_{\rho}^2}
-\frac{3g_{\sigma}g_{\delta}^2M_0^{\ast,2}\rho^2}{2{E}_{\rm{F}}^{\ast,3}Q_{\sigma}Q_{\delta}}
\left(\frac{M_0^{\ast}\Gamma}{Q_{\delta}}-\frac{2g_{\sigma}k_{\rm{F}}^{2}}{{E}_{\rm{F}}^{\ast,2}}
\right)
\end{align}
where $
Q_{\sigma}=m_{\sigma}^2+3g_{\sigma}^2(\rho_{\rm{S}}/M_0^{\ast}
-\rho/{E}_{\rm{F}}^{\ast})+2b_{\sigma}Mg_{\sigma}^3\overline{\sigma}+3c_{\sigma}g_{\sigma}^4\overline{\sigma}^2$, $
Q_{\omega}=m_{\omega}^2+3c_{\omega}g_{\omega}^4\overline{\omega}_0^2$, $Q_{\rho}=m_{\rho}^2+\Lambda_{\rm{S}}g_{\sigma}^2g_{\rho}^2\overline{\sigma}^2+\Lambda_{\rm{V}}g_{\omega}^2g_{\rho}^2\overline{\omega}_0^2$,
$Q_{\delta}=m_{\delta}^2+3g_{\delta}^2({\rho_{\rm{S}}}/{M_0^{\ast}}
-{\rho}/{{E}_{\rm{F}}^{\ast}})$, and
\begin{equation}
\Gamma=3g_{\sigma}g_{\delta}^2\left(\frac{2\rho_{\rm{S}}}{M_0^{\ast,2}}
-\frac{3\rho}{M_0^{\ast}{E}^{\ast}_{\rm{F}}}+\frac{M_0^{\ast}\rho}{{E}_{\rm{F}}^{\ast,3}}\right).
\end{equation}
Among the relativistic approaches, the nonlinear $\sigma$-$\omega$-$\rho$-$\delta$ relativistic mean field
model has been widely used in the literature. However, many of the predictions depend on the structure and parameters of the RMF model Lagrangians, e.g.,
the FSUGold\,\cite{Tod05}, IU-FSU\,\cite{Fat10},
NL$\rho\delta$\,\cite{Liu02} and HA\,\cite{Bun03}. The FSUGold and IU-FSU
are two accurately calibrated interactions based on the ground state
properties of closed-shell nuclei, their linear response, and the
structure of neutron stars. Since they do not consider
the isovector-scalar $\delta$ meson field, the $\Sigma
_{\text{S}}^{\mathrm{{sym},1}}(\rho )$ is zero. On the other hand, both the
NL$\rho\delta$ and HA include the $\delta$ meson field.
The parameters of these interactions have been fixed by fitting empirical properties of nuclear matter, properties of some selected nuclei and/or some predictions of more microscopic many-body theories.
Shown in Fig.\,\ref{FIGEsymDecom} are components of the $E_{\mathrm{sym}}(\rho)$ and $L(\rho)$ using the four RMF interactions,
it is seen that while the kinetic contributions to both the $E_{\mathrm{sym}}(\rho)$ and $L(\rho)$ are approximately identical using all interactions as one expects,
the different interactions predict significantly different values for the interaction components and the results also depend strongly on whether the isovector-scalar $\delta $ meson is included or
not. In particular, compared with the FSUGold and IU-FSU, the HA predicts very similar total
$E_{\mathrm{sym}}(\rho )$ but significantly different
$E_{\mathrm{sym}}^{\mathrm{1st,S}}(\rho )$ and
$E_{\mathrm{sym}}^{\mathrm{1st,V}}(\rho )$.
It is also interesting to note that the higher-order contribution
$L^{\mathrm{2nd}}(\rho)$ from the second-order symmetry
self-energies,  similar to the $U_{\rm{sym},2}$ and $L_5$ terms in the non-relativistic decomposition of $E_{\mathrm{sym}}(\rho)$ and $L(\rho)$ in Eqs.\,(\ref{Esymexp2}) and (\ref{Lexp2}),
generally cannot be neglected. The lack of a clear systematics in Fig.\,\ref{FIGEsymDecom} signifies the need of better understanding the isovector features of nuclear interactions in the RMF models.
In fact, it was suggested that one may try to determine each component of the $E_{\text{sym}}(\rho )$ and $L(\rho )$ from
experiments (e.g., Dirac phenomenology ) or microscopic calculations
based on nucleon-nucleon interactions derived from scattering phase
shifts.

We now turn to the neutron-proton effective mass splittings in relativistic approaches.
The isospin splitting of the Landau effective mass is given by
\begin{align}
m_{\rm{n-p}}^{\ast,\rm{Lan}}
\approx&-\frac{2M_{0,\rm{Lan}}^{\ast,2}\delta}{3Mk_{\rm{F}}{E}_{\rm{F}}^{\ast,3}}
\times\Bigg\{k_{\rm{F}}{E}_{\rm{F}}^{\ast,3}\Bigg[\frac{k_{\rm{F}}^{\ast}}{{E}_{\rm{F}}^{\ast}}
\frac{\partial^2\Sigma_{\rm{K}}^0(\rho,{k})}{\partial
{k}^2}+\frac{M_0^{\ast}}{{E}_{\rm{F}}^{\ast}}\frac{\partial^2\Sigma_{\rm{S}}^0(\rho,{k})}{\partial
{k}^2}+\frac{\partial^2\Sigma_{\rm{V}}^0(\rho,{k})}{\partial
{k}^2}\Bigg]_{k_{\rm{F}}}\notag\\
&+k_{\rm{F}}M_0^{\ast,2}\left[1+\frac{\partial\Sigma_{\rm{K}}^0(\rho,{k})}{\partial
{k}}\right]_{k_{\rm{F}}}^2+k_{\rm{F}}k_{\rm{F}}^{\ast,2}\left[\frac{\partial\Sigma_{\rm{S}}^0(\rho,{k})}{\partial
{k}}\right]_{k_{\rm{F}}}^2
-2k_{\rm{F}}k_{\rm{F}}^{\ast}M_0^{\ast}\left[1+\frac{\partial\Sigma_{\rm{K}}^0(\rho,{k})}{\partial
{k}}\right]_{k_{\rm{F}}}\left.\frac{\partial\Sigma_{\rm{S}}^0(\rho,{k})}{\partial
{k}}\right|_{k_{\rm{F}}}\notag\\
&+3\left[M_0^{\ast}\Sigma_{\rm{K}}^{\rm{sym},1}(\rho,k_{\rm{F}})
-k_{\rm{F}}^{\ast}\Sigma_{\rm{S}}^{\rm{sym},1}(\rho,k_{\rm{F}})\right]
\times\Bigg\{M_0^{\ast}\left[1+\frac{\partial\Sigma_{\rm{K}}^0(\rho,{k})}{\partial
{k}}\right]_{k_{\rm{F}}}
-k_{\rm{F}}^{\ast}\left.\frac{\partial\Sigma_{\rm{S}}^0(\rho,{k})}{\partial
{k}}\right|_{{k}=k_{\rm{F}}}\Bigg\}\notag\\
&+3{E}_{\rm{F}}^{\ast,3}\Bigg[\frac{k_{\rm{F}}^{\ast}}{{E}_{\rm{F}}^{\ast}}
\frac{\partial\Sigma_{\rm{K}}^{\rm{sym},1}(\rho,{k})}{\partial
{k}}+\frac{M_0^{\ast}}{{E}_{\rm{F}}^{\ast}}\frac{\partial\Sigma_{\rm{S}}^{\rm{sym},1}(\rho,{k})}{\partial
{k}}
\frac{\partial\Sigma_{\rm{V}}^{\rm{sym},1}(\rho,{k})}{\partial
{k}}\Bigg]_{k_{\rm{F}}}-\frac{{E}_{\rm{F}}^{\ast,3}k_{\rm{F}}}{M_{0,\rm{Lan}}^{\ast}}
\Bigg\} \label{ss_135}
\end{align}
at leading order in isospin asymmetry $\delta$. We note again here that  $M_{0,\rm{Lan}}^{\ast}$ is the Landau effective mass while the $M_0^{\ast}$
denotes the Dirac nucleon effective mass both in SNM.

In relativistic approaches neglecting the momentum dependence of self-energies, the above formula can be simplified to (also neglecting the K-part)
\begin{equation}\label{Lan1}
m_{\rm{n-p}}^{\ast,\rm{Lan}}
=-\frac{2M_{0,\rm{Lan}}^{\ast,2}\delta}{3Mk_{\rm{F}}{E}_{\rm{F}}^{\ast,3}}
\times\left[k_{\rm{F}}M_0^{\ast,2}-3M_0^{\ast}k_{\rm{F}}^{\ast}\Sigma_{\rm{S}}^{\rm{sym},1}(\rho,k_{\rm{F}})
-\frac{k_{\rm{F}}{E}_{\rm{F}}^{\ast,3}}{M_{0,\rm{Lan}}^{\ast}}
\right].
\end{equation}
Similarly, one can obtain the isospin splitting for the Dirac mass as
\begin{equation}\label{Dir1}
m_{\rm{n-p}}^{\ast,\rm{Dir}}\approx\frac{2\delta}{M}\left[\frac{k_{\rm{F}}}{3}
\left.\frac{\partial\Sigma_{\rm{S}}^0(\rho,{k})}{\partial
{k}}\right|_{k_{\rm{F}}}+\Sigma_{\rm{S}}^{\rm{sym},1}(\rho,k_{\rm{F}})\right].\end{equation}
It shows that besides the scalar-isovector meson (the second term), the momentum-dependent scalar-isoscalar interaction (characterized by the first term)
can also contribute to the Dirac neutron-proton effective mass splitting.
Applying the formulae (\ref{Lan1}) and (\ref{Dir1}) to the RMF models, one can immediately obtain
\begin{equation}
m_{\rm{n-p}}^{\ast,\rm{Lan}}=
\frac{2k_{\rm{F}}^2\delta}{3M{E}_{\rm{F}}^{\ast}}\left(1-\frac{3g_{\delta}^2M_0^{\ast,2}\rho}{k_{\rm{F}}^2
{E}_{\rm{F}}^{\ast}Q_{\delta}}\right) ,~~m_{\rm{n-p}}^{\ast,\rm{Dir}}=-\frac{2g_{\delta}^2\rho\delta}{{E}_{\rm{F}}^{\ast}Q_{\delta}}\frac{M_0^{\ast}}{M},
\end{equation}
and consequently,
\begin{equation}
s_{\rm{L}}=\frac{2k_{\rm{F}}^2}{3M{E}_{\rm{F}}^{\ast}}\left(1+\frac{3MM_0^{\ast}s_{\rm{D}}}{2k_{\rm{F}}^2}\right),
\end{equation}
where $s_{\rm{L/D}}$ is the linear isospin splitting coefficient defined through $m_{\rm{n-p}}^{\ast,\rm{Lan/Dir}}\approx s_{\rm{L/D}}\delta$.
It is obvious that the Dirac splitting coefficient $s_{\rm{D}}$ in the RMF models is always negative, indicating that the Dirac effective mass
of neutrons is smaller than that of protons. On the other hand, the Landau splitting coefficient $s_{\rm{L}}$ can take both positive and negative values,
depending on the model details.

The lack of momentum dependence in the RMF models will introduce a mismatch when the Schr\"odinger equivalent potential $U_{\rm{SEP},J}$ defined in the subsection \ref{Rmass} is used
to recalculate the EOS. Specifically, when one calculates the kinetic symmetry energy based on the $U_{\rm{SEP},J}$ constructed from the RMF models, one
obtains $E_{\rm{sym}}^{\rm{kin}}(\rho)=k_{\rm{F}}^2/6M$ (since there is no momentum dependence in the self-energies) instead of the $E_{\rm{sym}}^{\rm{kin}}(\rho)=k_{\rm{F}}^2/6M_{0,\rm{Lan}}^{\ast}=k_{\rm{F}}^2/6E_{\rm{F}}^{\ast}$ as in the original RMF models. A similar mismatch will also occur at higher orders in isospin asymmetry, e.g., in calculating the fourth-order symmetric energy and the slope parameter $L$.
In order to remedy these drawbacks, one can redefine an equivalent non-relativistic potential perturbatively in the RMF models as
\begin{equation}
U_{\rm{SEP},J}^{\rm{RMF}}(\rho,\delta)
=U_{\rm{SEP},J}(\rho,\delta)+\zeta(\rho)\tau_3^{J}\delta+\beta(\rho)\delta^2+\chi(\rho)\tau_3^{J}\delta^3+\mathcal{O}(\delta^4)
\end{equation}
where
\begin{align}
\zeta(\rho)=&2\left(\frac{k_{\rm{F}}^2}{6{E}_{\rm{F}}^{\ast}}-\frac{k_{\rm{F}}^2}{6M}\right),\\
\beta(\rho)=&\frac{k_{\rm{F}}^2M_0^{\ast,2}}{18{E}_{\rm{F}}^{\ast,3}}
-\frac{k_{\rm{F}}^2}{9{E}_{\rm{F}}^{\ast}}-\frac{k_{\rm{F}}^2}{18M},\\
\chi(\rho)=&4\left(\frac{k_{\rm{F}}^2}{648}\frac{4M_0^{\ast,4}+11M_0^{\ast,2}k_{\rm{F}}^2
+10k_{\rm{F}}^4}{{E}_{\rm{F}}^{\ast,5}}-\frac{1}{162}\frac{k_{\rm{F}}^2}{M}\right).
\end{align}
Now the $U_{\rm{SEP},J}^{\rm{RMF}}$ can correctly reproduce both the kinetic and fourth-order symmetry energy as well as the $L(\rho)$ in the original RMF models.
For example, using the non-relativistic decomposition of $L(\rho)$ only with the $U_{\rm{SEP}}$  one obtains
\begin{equation}\label{xxx}
L(\rho)=\frac{k_{\rm{F}}^2}{3M}+\frac{3}{2}U_{\rm{sym}}(\rho)+3U_{\rm{sym},2}(\rho).
\end{equation}
Its kinetic part (the first term) is not the same as $L^{\rm{kin}}(\rho)={k_{\rm{F}}^2}/{6{E}_{\rm{F}}^{\ast}}+{k_{\rm{F}}^2M_0^{\ast,2}
}/{6{E}_{\rm{F}}^{\ast,3}}$ for RMF models as in Eq.\,(\ref{RMF_Lkin}).
With the $U_{\rm{SEP},J}^{\rm{RMF}}$, however, the $L^{\rm{kin}}(\rho)$ also gains two extra parts $3\zeta/2+3\beta={k_{\rm{F}}^2}/{6{E}_{\rm{F}}^{\ast}}+{k_{\rm{F}}^2M_0^{\ast,2}
}/{6{E}_{\rm{F}}^{\ast,3}}-{k_{\rm{F}}^2}/{3M}$. Combining the latter with $k_{\rm{F}}^2/3M$ in Eq.\,(\ref{xxx}) exactly gives the $L^{\rm{kin}}(\rho)$ in the RMF models.
The kinetic and fourth-order symmetry energies thus obtained using the $U_{\rm{SEP},J}^{\rm{RMF}}(\rho,\delta)$
are also the same as in the original RMF models.

In summary of this section, the HVH theorem reveals directly and analytically the relations among the effective masses, their own momentum dependences,
symmetry potentials and symmetry energies order by order in isospin-asymmetry in both
non-relativistic and relativistic frameworks. These relations are independent of the models. Their applications help us better understand the physical origin of
the symmetry energies, neutron-proton effective mass splitting in neutron-rich matter and their uncertainties in a model independent way.

\FloatBarrier
\begin{figure}[tbh]
\centering
\includegraphics[width=8.cm,height=6cm]{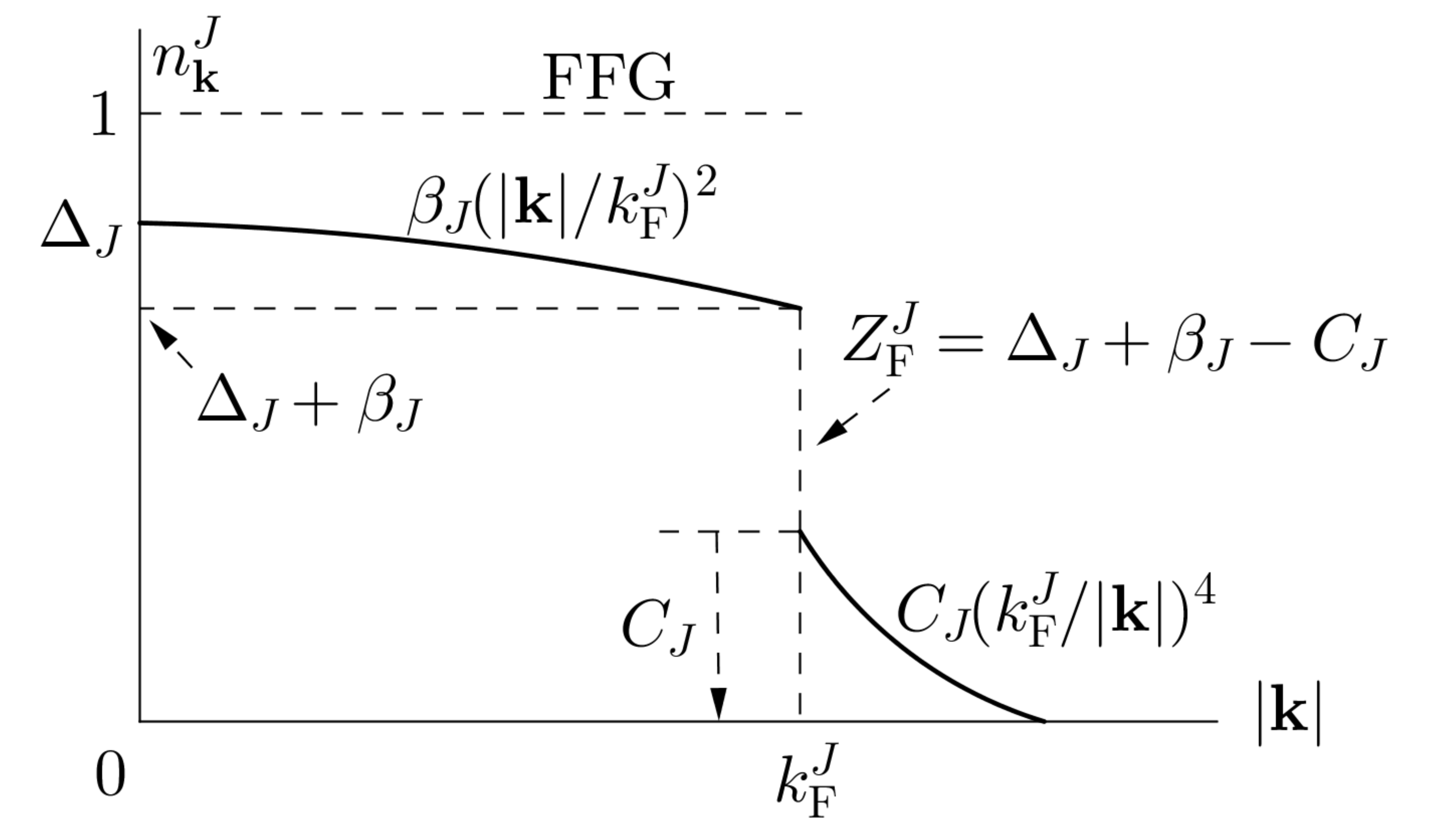}
\hspace{0.5cm}
\includegraphics[width=7.cm,height=6cm]{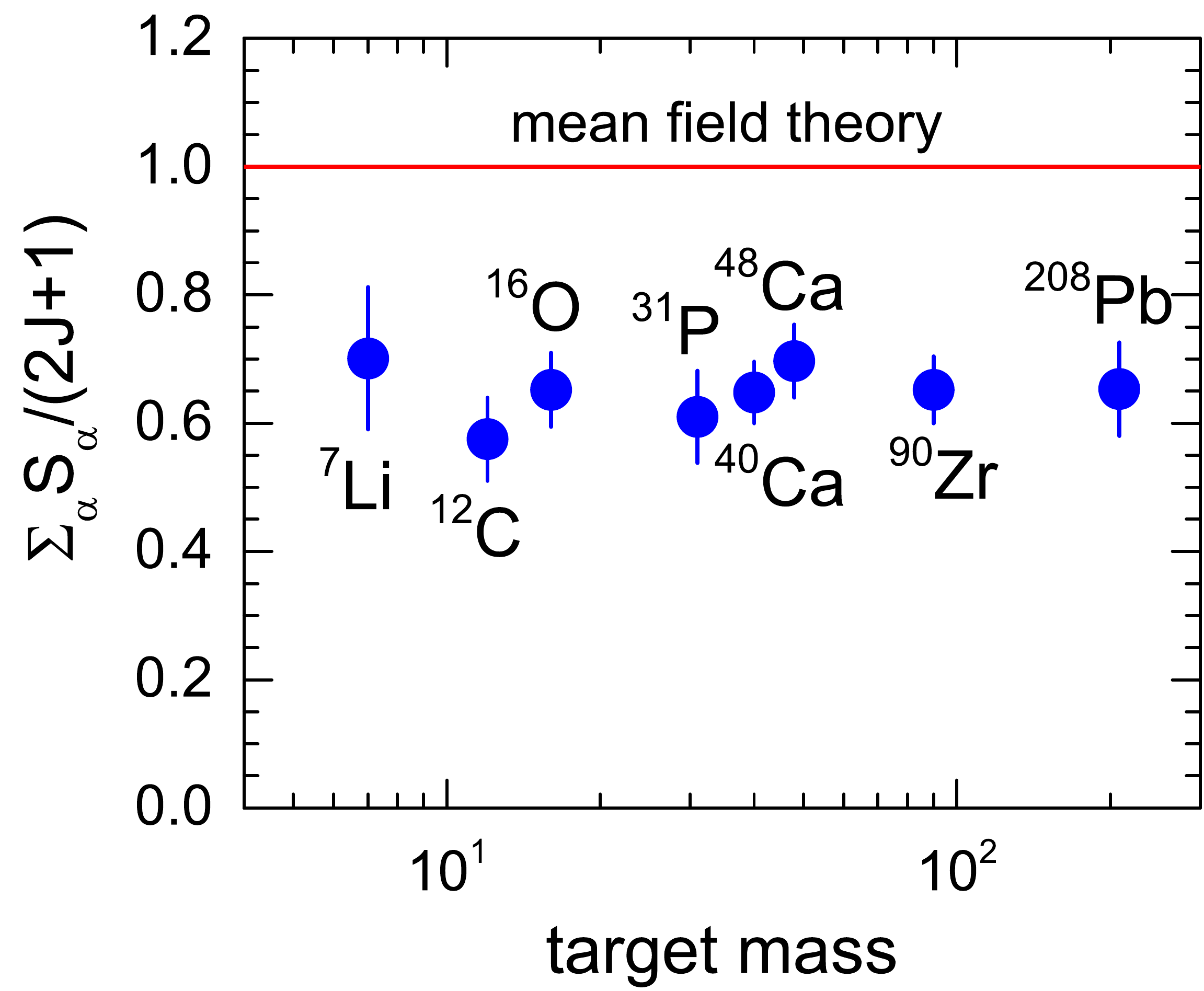}
\caption{Left: A sketch of single-nucleon momentum distribution function in ANM. Taken from ref.\,\cite{CaiLi16a}.
Right: Spectroscopic factor as a function of target mass. Taken from ref.\,\cite{Lap93}.\label{fig_SF}}
\end{figure}
\section{Nucleon E-mass, its mean free path and short-range correlations (SRC) in neutron-rich matter}
In this section, we review effects of the nucleon-nucleon short range correlations (SRC) on the isospin dependence of nucleon E-mass and mean free path (MFP) as well as the symmetry energy in neutron-rich matter.
 After briefly recalling the Migdal--Luttinger theorem connecting the E-mass with the discontinuity of single-nucleon momentum distribution $n_{\v{k}}^J$ at the Fermi momentum,
we discuss a phenomenological $n_{\v{k}}^J$ encapsulating the SRC-induced high momentum tail (HMT) with its major features constrained theoretically and/or experimentally.
As an example of direct consequences of the isospin dependence of the HMT in the $n_{\v{k}}^J$, the enhancement (reduction) of the kinetic EOS of SNM (kinetic symmetry energy) will then be discussed. The extracted constraints on the nucleon E-mass and its isospin dependence in ANM will be presented in subsection \ref{subx.4}. Finally, implications of the experimentally constrained SRC and E-mass
on the critical proton fraction for the direct URCA process to occur in protoneutron stars and the isospin dependence of nucleon MFP in neutron-rich matter will be discussed.

\subsection{Migdal--Luttinger theorem connecting the E-mass with single-nucleon momentum distributions in nuclear matter}\label{subx.1}
As discussed in Section 2, an independent determination of
either the E-mass or k-mass together with the total effective mass
will then allow us to know all three kinds of non-relativistic nucleon effective
masses. Interestingly, the Migdal--Luttinger
theorem\,\cite{Mig57,Lut60} connects the nucleon E-mass directly with
the jump (discontinuity)
\begin{equation}
Z_{\rm{F}}^J\equiv n_{\v{k}}^J(k^J_{\rm{F}-0})-n_{\v{k}}^J(k^J_{\rm{F}+0}),
\end{equation}
of the
single-nucleon momentum distribution $n_{\v{k}}^J$ at the Fermi
momentum $k^J_{\rm{F}}$ via\,\cite{CaiLi16a}
\begin{equation}
{M_{J}^{\ast,\rm{E}}}/{M}=1/Z_{\rm{F}}^J,
\end{equation}
see the left part of Fig.\,\ref{fig_SF}.
The Migdal--Luttinger jump $Z^J_{\rm{F}}$  also called the ``renormalization (strength) function",  contains rich information about the nucleon effective E-mass and its non-trival isospin dependence\,\cite{Jeu76,Mah85}.
The nuclear physics community has devoted much efforts to probing the depletion of the
nucleon Fermi sphere by using transfer, pickup and (e,e$'$p)
reactions. Results of these studies normally given in terms of the
nucleon spectroscopic factors can constrain the
$n_{\v{k}}^J(k^J_{\rm{F}-0})$ from below the Fermi surface\,\cite{Jeu76,Mah85,Jam89,Lap93,Wak17}. As illustrated in the plot of spectroscopic factors versus the target mass in the right window of Fig.\,\ref{fig_SF},  it is well known that mean field theories over-predict the spectroscopic factor by about 20-40\% due to the lack of correlations in the theories.
On the other hand, quantitative information about the size, shape and isospin dependence of
the high-momentum tail (HMT) above the Fermi surface has been
extracted recently from analyzing both inclusive
and exclusive electron-nucleus scattering cross sections\,\cite{Hen15,Hen14,Col15,Egi06,Shn07,Wei11,Kor14,Fa17}, nucleon-nucleus reactions\,\cite{Tang03,Pia06} as well as medium-energy photonuclear absorptions\,\cite{Wei15},
providing a constraint on the $n_{\v{k}}^J(k^J_{\rm{F}+0})$ from above the Fermi surface. These
experimental results together with model analyses provide a
significant empirical constraint on the Migdal--Luttinger jump, thus the E-mass in neutron-rich matter.

\begin{figure}[h!]
\centering
  \includegraphics[width=7.5cm,angle=-90]{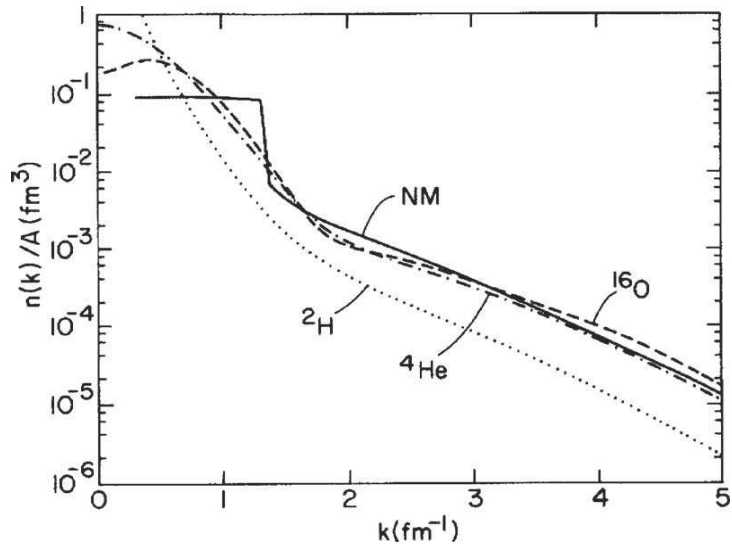}
  \caption{The nucleon momentum distribution in $^{2}\rm{H}$, $^4{\rm{He}}$, $^{16}\rm{O}$ and that in infinite isospin-symmetric nuclear matter. Taken from ref.\,\cite{Ben93}. }
  \label{fig_Benh08}
\end{figure}
Next, we discuss in more detail the single-nucleon momentum distribution function $n_{\v{k}}^J$ having a depletion (HMT) under (above) the Fermi momentum $k_{\rm{F}}^J$ in ANM.
It is well known that nuclear interactions, in particular the nucleon-nucleon short-range repulsive core (correlations) and
tensor force, usually lead to a high (low) momentum tail (depletion) in the
single-nucleon momentum distribution above (below) the nucleon Fermi
surface\,\cite{Mig57,bethe,pan92,Pan99}. During the last two decades, significant progresses
both theoretically and experimentally have been made continuously
in quantifying, for instance, the nucleon HMT in
ANM\,\cite{Hen15,Hen14,Egi06,Shn07,Wei11,Kor14,Fa17,Pia06,Wei15}.
For example, shown in Fig.\,\ref{fig_Benh08} are calculated nucleon momentum distributions in $^{2}\rm{H}$, $^4{\rm{He}}$, $^{16}\rm{O}$ and SNM\,\cite{Ben93}.
Here, the $n_{\v{k}}^0$ (the superscript ``0" is for SNM) for $^{2}\rm{H}$, $^4{\rm{He}}$, $^{16}\rm{O}$ were obtained with the Argonne $v(ij)$ and model
VII $V(ijk)$, and those for SNM\,\cite{Fan84} were obtained with the Urbanna $v(ij)$ with density-dependent terms mimicking the effects of $V(ijk)$.
One of the most important features of these momentum distribution functions is the universal scaling of the HMT in all systems considered, indicating the shared short-range nature of the HMT.

Guided by state-of-the-art recent microscopic calculations and some well-known earlier predictions from other nuclear many-body
theories, see, e.g., ref.\,\cite{Ant88,Ant93} for a historical review, as well as recent experimental
findings\,\cite{Hen15,Hen14,Wei15}, the single-nucleon
momentum distribution in ANM has often been parameterized by, e.g. ref.\,\cite{Cai15a},
\begin{equation}\label{MDGen}
n^J_{\v{k}}(\rho,\delta)\equiv n_J(\rho,|\v{k}|,\delta)=\left\{\begin{array}{ll}
\Delta_J+\beta_J{I}\left(\displaystyle\frac{|\v{k}|}{k_{\rm{F}}^J}\right),~~&0<|\v{k}|<k_{\rm{F}}^J,\\
&\\
\displaystyle{C}_J\left(\frac{k_{\rm{F}}^{J}}{|\v{k}|}\right)^4,~~&k_{\rm{F}}^J<|\v{k}|<\phi_Jk_{\rm{F}}^J.
\end{array}\right.
\end{equation}
Here, the
$\Delta_J$ is the depletion of the Fermi sphere at zero momentum
with respect to the free Fermi gas (FFG) model prediction while the $\beta_J$ characterizes the
strength of the momentum dependence
$I(|\v{k}|/k_{\rm{F}}^J)$\,\cite{Czy61,Bel61,Sar80} of the depletion
near (and at) the nucleon Fermi surface. The constraints on the function $I(x)$
will be discussed later.

Before going any further, it is necessary to address a possible drawback
of the parametrization in Eq.\,(\ref{MDGen}). To our best knowledge, the original derivations
\cite{Mig57,Lut60} of the Migdal--Luttinger theorem do not explicitly
require the derivative of the momentum distribution
 $[\partial n_{\v{k}}^J/\partial|\v{k}|]_{|\v{k}|=k^J_{\rm{F}}\pm0}$ at the Fermi momentum $k_{\rm{F}}^J$ to be $-\infty$.
 Of course, one can associate mathematically loosely the finite drop in $n^J_{\v{k}}$
over zero increase in momentum at $k^J_{\rm{F}}$ to a slope of
$-\infty$. Some later derivations using various approximation
schemes, such as the ``derivative expansion" in which the momentum
distribution is expressed in terms of energy derivatives of the mass
operator by Mahaux and Saror\,\cite{Mah92}, have shown that the slope
of the momentum distribution should have the asymptotic behavior of
$[\partial
n^J_{\v{k}}/\partial|\v{k}|]_{|\v{k}|=k_{\rm{F}}^J\pm0}=-\infty$.
Similar to many other calculations including some examples given in
refs.\,\cite{Mah85,Mah92}, the parametrization of Eq.\,(\ref{MDGen})
does not have such behavior from neither side of the discontinuity.
However, similar to what has been done in ref.\,\cite{Bal90}, in
parameterizing the $n_{\v{k}}^J$ both above and below the
$k_{\rm{F}}^J$ one can add a term that is vanishingly small in
magnitude but asymptotically singular in slope at $k_{\rm{F}}^J$.
For example, the form of the HMT of the single-nucleon momentum
distribution, i.e., $C_J(k_{\rm{F}}^J/|\v{k}|)^4$, is physical
verified by microscopic calculations and experiments. While
generally, one can always write $n_{\v{k}}^J$ as
\begin{equation}
n_{\v{k}}^J(\rho,\delta)=\delta
n_{\v{k}}^J(\rho,\delta)+C_J\left({k_{\rm{F}}^J}/{|\v{k}|}\right)^4
\end{equation}
above the Fermi surface, where $\delta n_{\v{k}}^J(\rho,\delta)$ is a
correction which has the following properties,
\begin{align}
\lim_{|\v{k}|\to k_{\rm{F}}^J}\delta n_{\v{k}}^J(\rho,\delta)=0,~~
\left|\delta
n_{\v{k}}^J(\rho,\delta)\right|\ll&C_J\left({k_{\rm{F}}^J}/{|\v{k}|}\right)^4.
\end{align}
These two properties guarantees that the correction is small, namely,
\begin{equation}
n_{\v{k}}^J(\rho,\delta)=C_J\left(\frac{k_{\rm{F}}^J}{|\v{k}|}\right)^4\cdot\left[1+\frac{1}{C_J}\left(\frac{|\v{k}|}{k_{\rm{F}}^J}\right)^4
\delta n_{\v{k}}^J(\rho,\delta)\right]\approx
C_J\left(\frac{k_{\rm{F}}^J}{|\v{k}|}\right)^4.\end{equation} Thus,
it is difficult to probe the correction $\delta n_{\v{k}}^J$
experimentally. For all physical quantities explored at the Fermi
surface directly related to the $n_{\v{k}}^J$ instead of any form of
its derivatives/integrations, the correction $\delta n_{\v{k}}^J$
has no effects. The E-mass proportional to the jump of the momentum
distribution at the Fermi surface is just one such quantity. But
the integrations/derivatives of $\delta n_{\v{k}}^J$ can probably
have singularities at $|\v{k}|=k_{\rm{F}}^J$. For instance, the
following correction has such property\,\cite{Bal90}
\begin{equation}
\delta
n_{\v{k}}^J(\rho,\delta)=\zeta_J\cdot\frac{|\v{k}|-k_{\rm{F}}^J}{\Lambda_J}\ln\left(\frac{|\v{k}|-k_{\rm{F}}^J}{\Lambda_J}\right)\end{equation}
where $ 0<\zeta_J\ll1$ and $\Lambda_J$ is a positive constant
independent of momentum. In this case, one can show that
\begin{equation}
\frac{\partial n_{\v{k}}^J(\rho,\delta)}{\partial|\v{k}|}
=\frac{\zeta_J}{\Lambda_J}\left[1+\ln\left(\frac{|\v{k}|-k_{\rm{F}}^J}{\Lambda_J}\right)\right]-\frac{4C_J}{|\v{k}|}\left(\frac{k_{\rm{F}}^J}{|\v{k}|}\right)^4,
~~ \lim_{|\v{k}|\to (k_{\rm{F}}^J)^+}\frac{\partial
n_{\v{k}}^J(\rho,\delta)}{\partial|\v{k}|}=-\infty,~~(k_{\rm{F}}^J)^+\equiv
k_{\rm{F}}^J+0^+.\end{equation} Of course, in this more general form, one then has to determine
totally 4 additional parameters ($\zeta_J$ and $\Lambda_J$ for
neutrons and protons above and below their respective Fermi
momenta) and readjust the other parameters already used in Eq.\,(\ref{MDGen}).
This has the potential of reducing the uncertainties of
the quantities one extract but requires more experimental
information that is currently very limited. While keeping in mind that the description about the
discontinuity of $n_{\v{k}}^J$ at $k_{\rm{F}}^J$ in Eq.\,(\ref{MDGen}) certainly should be
investigated further and possibly improved in the future, some of the following discussions are based on analyses using the parametrization of
Eq.\,(\ref{MDGen}).

The contact ${C}_J$ and cutoff coefficient $\phi_J$ determine the
fraction of nucleons in the HMT through the following relation
\begin{equation}\label{def_xJHMT}
x_J^{\rm{HMT}}=\left.\int_{k_{\rm{F}}^J}^{\phi_Jk_{\rm{F}}^J}
n_{\v{k}}^J\d\v{k}\right/{\displaystyle\int_0^{\phi_Jk_{\rm{F}}^J}
n_{\v{k}}^J\d\v{k}}=3C_{{J}}\left(1-\frac{1}{\phi_{{J}}}\right).
\end{equation}
The normalization
condition \begin{equation}\label{def_NC}
\frac{2}{(2\pi)^3}\int_0^{\infty}n^J_{\v{k}}(\rho,\delta)\d\v{k}=\rho_J=\frac{(k_{\rm{F}}^{J})^3}{3\pi^2}
\end{equation} requires that only three of the four parameters, i.e., $\beta_J$,
${C}_J$, $\phi_J$ and $\Delta_J$, are independent. For example, one can choose
the first three as independent and then determine the $\Delta_J$ from
\begin{equation}\label{DeltaJ}
\Delta_J=1-\frac{3\beta_J}{(k_{\rm{F}}^{J})^3}\int_0^{k_{\rm{F}}^J}{I}\left(\frac{k}{k_{\rm{F}}^J}\right)k^2\d
k-3{C}_J\left(1-\frac{1}{\phi_J}\right).
\end{equation}
Hinted by findings within the SCGF\,\cite{Rio09} and BHF theory\,\cite{Yin13} that the
depletion $\Delta_J$ has an approximately linear dependence on $\delta$ in
the opposite directions for neutrons and protons, it was assumed that all of the four
parameters in Eq.\,(\ref{MDGen}) can be expanded in isospin asymmetry in the form of $Y_J=Y_0(1+Y_1^J\delta)$\,\cite{Cai15a}.  It can be justified by considering the isospin dependence of the resulting average kinetic energy per nucleon in ANM
\begin{equation}\label{kinE}
E^{\rm{kin}}(\rho,\delta)=\frac{1}{\rho}\frac{2}{(2\pi)^3}\sum_{J=\rm{n,p}}\int_0^{\phi_Jk_{\rm{F}}^J}\frac{\v{k}^2}{2M}n_{\v{k}}^J(\rho,\delta)\d\v{k}.
\end{equation}
Using the distribution in Eq.\,(\ref{MDGen}), the $E^{\rm{kin}}(\rho,\delta)$ acquires a linear term in isospin asymmetry $\delta$ with a coefficient of
\begin{align}\label{Ekin1}
E_1^{\rm{kin}}(\rho)\equiv\left.\frac{\partial E^{\rm{kin}}(\rho,\delta)}{\partial\delta}
\right|_{\delta=0}=&\frac{3}{5}\frac{k_{\rm{F}}^2}{2M}\Bigg[\frac{5}{2}C_0\phi_0(\phi_1^{\rm{n}}+\phi_1^{\rm{p}})
+\frac{5}{2}C_0(\phi_0-1)(C_1^{\rm{n}}+C_1^{\rm{p}})\notag\\
&+\frac{1}{2}\Delta_0(\Delta_1^{\rm{n}}+\Delta_1^{\rm{p}})
+\frac{5\beta_0(\beta_1^{\rm{n}}+\beta_1^{\rm{p}})}{2k_{\rm{F}}^5}
\int_0^{k_{\rm{F}}}I\left(\frac{k}{k_{\rm{F}}}\right)k^4\d k\Bigg].
\end{align}
To enforce the neutron-proton exchange symmetry of the EOS, by setting the $E_1^{\rm{kin}}(\rho)$ to zero one naturally obtains
$\Delta_1^{\rm{n}}=-\Delta_1^{\rm{p}}$,
$\beta_1^{\rm{n}}=-\beta_1^{\rm{p}}$,
${C}_1^{\rm{n}}=-{C}_1^{\rm{p}}$ and
$\phi_{1}^{\rm{n}}=-\phi_1^{\rm{p}}$, namely, they all have the form of $Y_J=Y_0(1+Y_1^J\delta)$. In the following subsection, we will discuss how all of these parameters  are determined using some experimental data and state-of-the-art microscopic many-body theory calculations for pure neutron matter.

\begin{figure}[h!]
\centering
    \includegraphics[width=15cm,height=9cm]{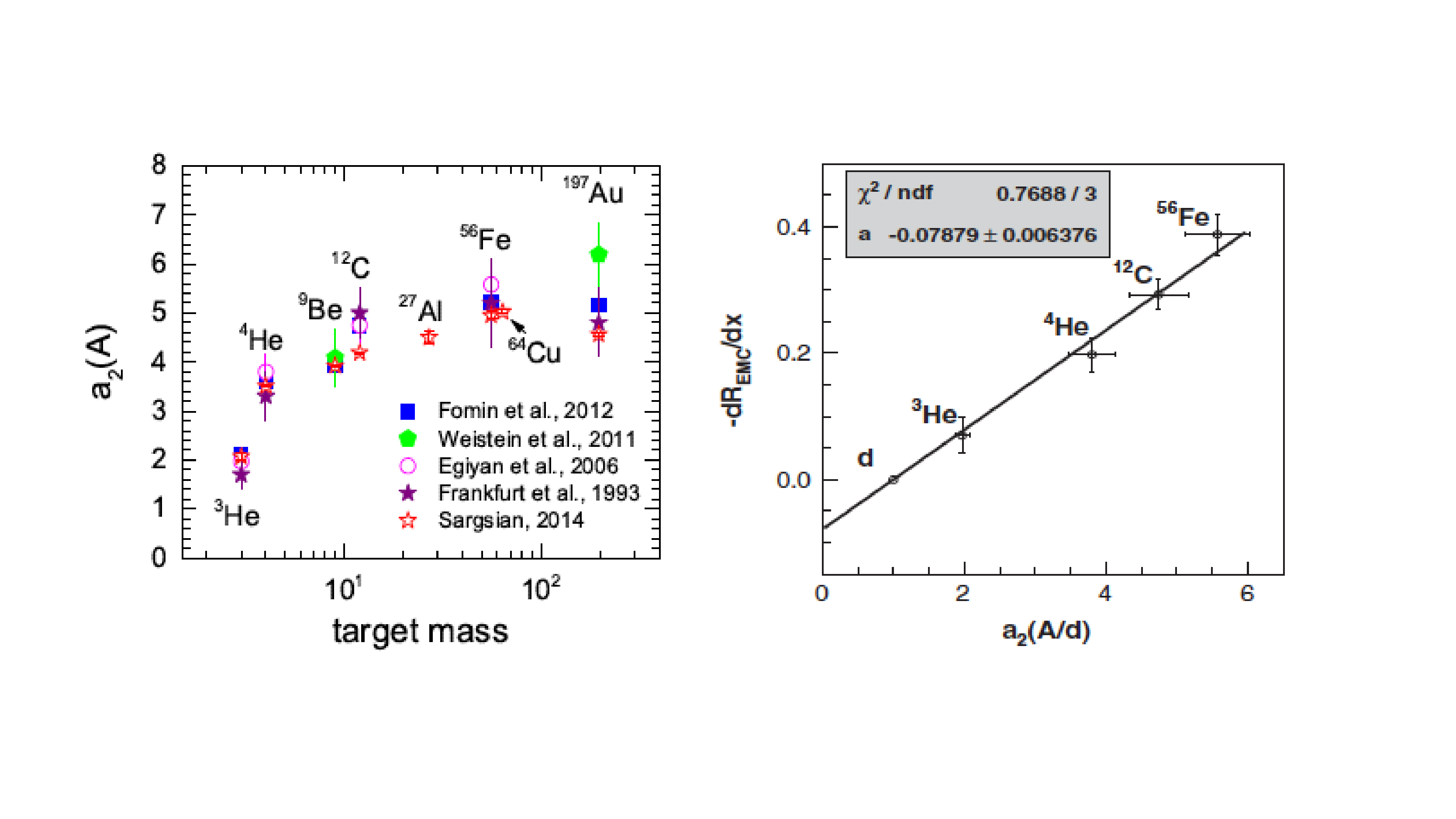}
\vspace{-2cm}
  \caption{(Left: Experimental $a_2(A)$ for several typical nuclei, i.e., $^3\rm{He}$,
 $^4\rm{He}$, $^9\rm{Be}$, $^{12}\rm{C}$, $^{56}\rm{Fe}(^{63}\rm{Cu})$ and $^{197}\rm{Au}$\,\cite{Fra93,Wei11,Egi06,Fom12,Sar14}.
 Right: the EMC-effect slope as a function of the SRC factor $a_2$. Taken from ref.\,\cite{Wei11}. }
 \label{fig_a2exp}
\end{figure}

\subsection{The size, shape and isospin dependence of the SRC-induced nucleon high-momentum tail}
As illustrated in Fig.\,\ref{fig_Benh08},  the nucleon HMT approximately scales from deuteron to infinite nuclear matter according to\,\cite{Fan84,Pie92,Cio96},
\begin{equation}
n_{\v{k}}^A=a_2(A)n_{\v{k}}^{\rm{d}},
\end{equation}
where $n_{\v{k}}^A$ and $n_{\v{k}}^{\rm{d}}$ are the high-momentum parts of
the nucleon momentum distribution for a nucleus of atomic
number $A$ and deuterium, respectively. The factor $a_2(A)$
is independent of the momentum and is the probability of finding a
high momentum neutron-proton pair in nucleus $A$ relative to the deuterium\,\cite{Fra88}.
It is extracted experimentally from the plateau in the ratio of per nucleon
inclusive $(\rm{e},\rm{e}^{\prime})$ cross sections for heavy nuclei
with respect to deuteron for the Bjorken scaling parameter
$x_{\rm{B}}$ between about 1.5 and 1.9 \cite{Arr12}, i.e., $a_2(A)=(\sigma_A/A)/(\sigma_{\rm{d}}/2)$\,\cite{Hen15}. Shown in the left window of Fig.\,\ref{fig_a2exp} are examples
of the experimental $a_2(A)$ for several typical nuclei, i.e., $^3\rm{He}$,
 $^4\rm{He}$, $^9\rm{Be}$, $^{12}\rm{C}$, $^{56}\rm{Fe}(^{63}\rm{Cu})$ and $^{197}\rm{Au}$\,\cite{Egi06,Wei11,Fom12,Fra93}, as well as
a theoretical prediction\,\cite{Sar14}.
When extrapolated to infinite nuclear matter, one then obtains the asymptotic value of\,\cite{Pia12,McG11,Att91,Hen15b}
\begin{equation}\label{a2inf}
a_2(\infty)\approx7\pm1.
\end{equation}
As pointed out in ref.\,\cite{Hen16x}, the large uncertainties
in the SRC ratios of ref.\,\cite{Fra93} are mainly due
to extrapolating data from different experiments measured
at different kinematics. The SRC ratios reported
in ref.\,\cite{Egi06} were used in the original EMC-SRC correlation analysis of ref.\,\cite{Wei11}. It is interesting to note that the observed linear correlation between the slope of the EMC effects
and the $a_2$ factor\,\cite{Wei11} shown in the right window of Fig.\,\ref{fig_a2exp} is still not completely understood and more studies have been planned\,\cite{LRP2015}.
\begin{figure}[h!]
\centering
 \vspace{-0.45cm}
    \includegraphics[width=15cm,height=9cm]{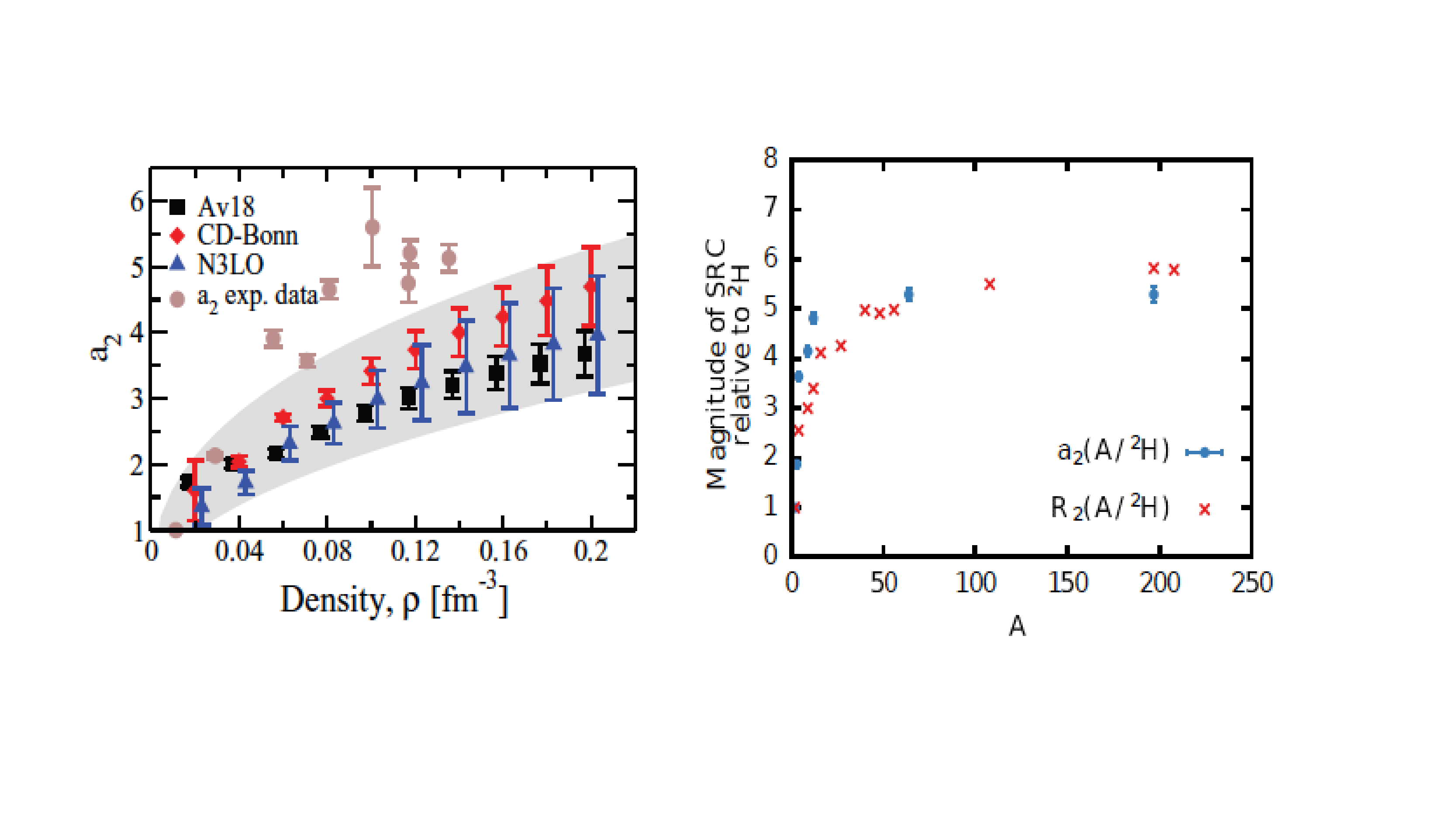}
\vspace{-2cm}
  \caption{Comparisons of the $a_2(A)$ data with predictions using the SCGF theory by Rios \textit{et al.}\,\cite{Rio14} (left) and
  an independent particle model (IPM) for nucleon pairs by Ryckebusch \textit{et al.}\,\cite{Ryc15,Van11,Van12} (right).  The symbols with error bars are the experimental data. The
  gray band (squares) are the calculations with (without) corrections for the c.m. motion of the pairs.}
  \label{fig_a2the}
\end{figure}
\begin{figure}[h!]
\centering
 \includegraphics[width=4.5cm]{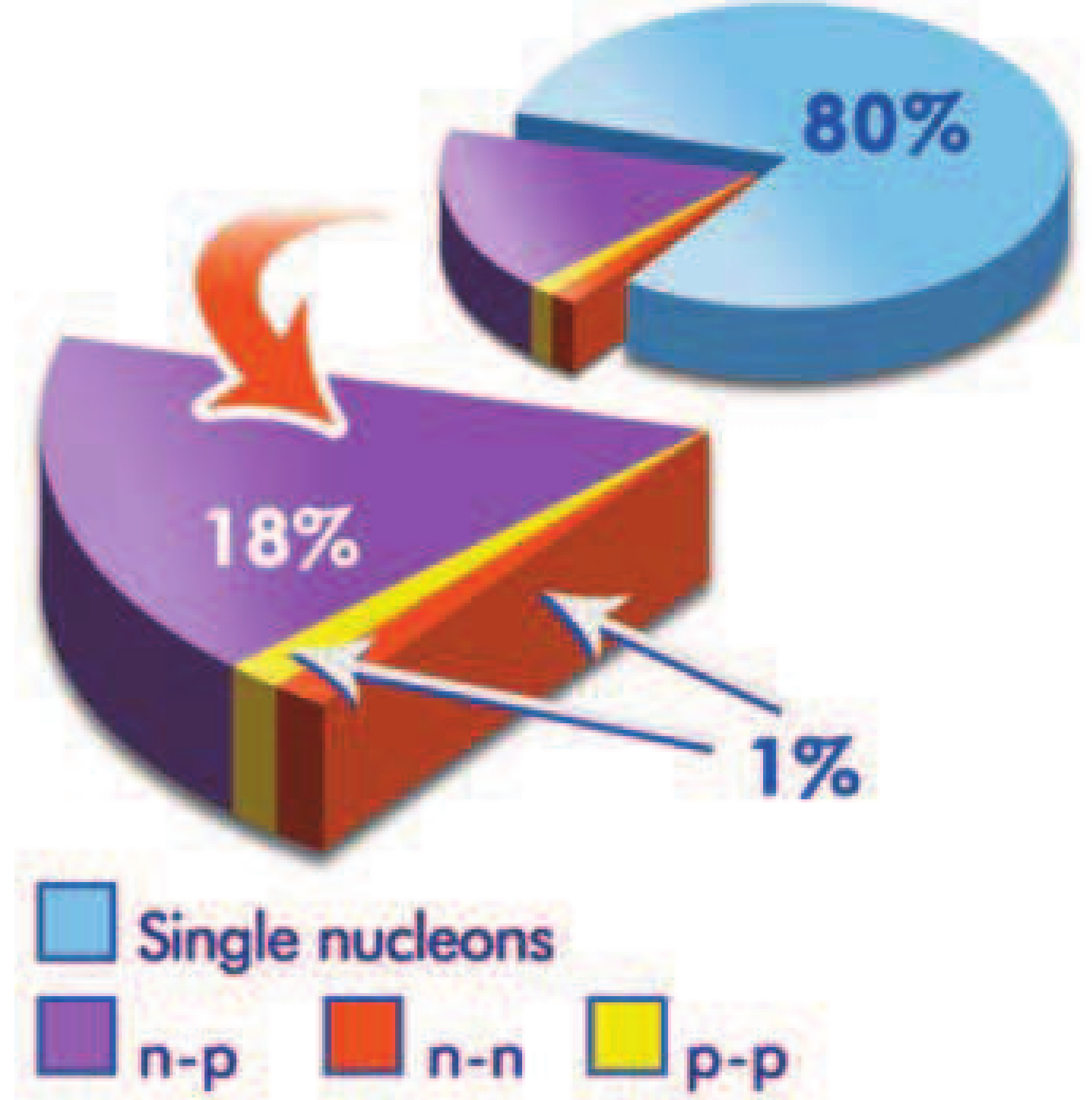}
 \hspace{1cm}
  \includegraphics[width=5.cm,angle=-90]{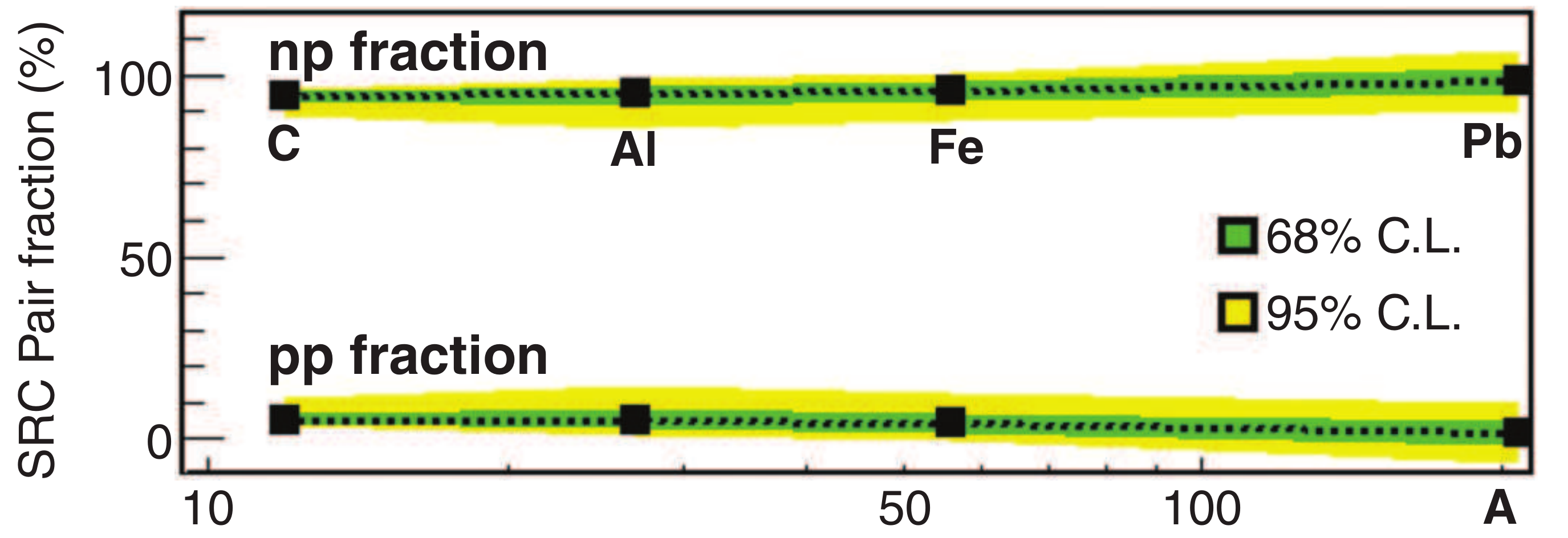}
  \caption{Upper: The average fraction of nucleons in the various initial-state configurations of $^{12}\rm{C}$. Taken from ref.\,\cite{Sub08}.
  Lower: The extracted fractions of np (top) and pp (bottom) SRC pairs from the sum of pp and np pairs in the four nuclei indicated. Taken from ref.\,\cite{Hen14}. }\label{fig_Sub}
\end{figure}
Shown in Fig.\,\ref{fig_a2the} are two examples of theoretical predictions in comparison with the $a_2(A)$ data\,\cite{Rio14,Van11,Van12}.
In ref.\,\cite{Rio14}, the $a_2$ was estimated as $a_2=\langle n_{\v{k}}^0/n_{\v{k}}^{\rm{d}}\rangle_{|\v{k}|=400\sim550\,\rm{MeV}}$,
and an empirical formula for it was given, i.e., $a_2(\rho)\approx b_1\rho^{b_2}$ with $b_1\approx7\sim10$ and $b_2\approx0.4\sim0.5$.
After considering the fact that the saturation density in the SCGF\,\cite{Rio14} is normally larger than the empirical one, at about $\rho_0^{\rm{SCGF}}\approx0.34\,\rm{fm}^{-3}$,
the extrapolation of the $a_2(A)$ to infinite SNM in SCGF at its own saturation density is found to be $a_2(\rho_0^{\rm{SCGF}})\approx5.2\pm1.0$, which
under-predicts the data but is still approximately consistent with Eq.\,(\ref{a2inf}). The (low) high momentum cutoff used in the SCGF, i.e., (400\,MeV) 550\,MeV,
is probably another factor for under-estimating the $a_2$ since the momentum range for the two-body SRC pairs used in the data analysis is about 300\,MeV to 600\,MeV\,\cite{Hen14}.
On the other hand, in refs.\,\cite{Ryc15,Van11,Van12} using correlation operators within an independent particle model (IPM) for nucleon pairs, the $a_2$ factor was calculated as
\begin{equation}
a_2(A/\rm{d})=\frac{2}{A}N_{\rm{pn}(S=1)}(A,Z)\int_{\rm{PS}}\d\v{P}_{12}F^{\rm{pn}}(P_{12}),
\end{equation}
where $N_{\rm{pn}(S=1)}(A,Z)$ is the proportionality factor counting the number
of correlated np pairs in the nucleus $A$. The integration extends over those parts of the center of mass (c.m.)
phase space (PS) included in the data and $F^{\rm{pn}}$ is the c.m. distribution of the correlated pairs.
One of the main results of this approach is shown in the right window of Fig.\,\ref{fig_a2the}. Their prediction on the $a_2(A)$ reproduces well the experimentally observed saturation of $a_2$ for heavy nuclei. 
The above typical examples of model calculations indicate that while the underlying physics of the SRC is qualitatively understood,
its size and isospin dependence are not quantitatively well reproduced by some of the most advanced nuclear many-body theories available in this field.

It is well known that to reproduce the experimentally measured electrical quadrupole moment of deuteron and the spin correlation parameter of electron scattering on polarized deuteron
about $4-5\%$ of tensor-force induced S-D wave mixing is required\,\cite{Fra88,Pas12}. Combing this information with the extrapolated $a_2(\infty)$ value in Eq.\,(\ref{a2inf}), it was
estimated that the HMT fraction in infinite SNM is $x^{\rm{HMT}}_{\rm{SNM}}\approx28\%\pm4$\%\cite{Hen16x,Hen14,Egi06,Pia06,Shn07,Wei11,Kor14,Hen15b}.
As pointed out in ref.\,\cite{Sub08} that about 80\% of the nucleons in
the $^{12}\rm{C}$ nucleus acted independently or as described
within the shell model. Whereas among the remaining
20\% correlated nucleons, $90\pm10$\% of them were found in the
form of pn SRC pairs and only $5\pm1.5$\% were in the form
of pp SRC pairs. These findings indicate that the np pairs outnumber pp pairs by a factor of about 20 \cite{Egi06,Sub08} as illustrated in Fig.\ \ref{fig_Sub}.
This result was confirmed in more experiments with heavier targets from C to $^{208}$Pb\,\cite{Hen14}, see the lower window of Fig.\,\ref{fig_Sub}.
By isospin symmetry, they inferred that there existed about $5\pm1.5$\% SRC nn pairs, leading to a HMT fraction in PNM to be about
$x_{\rm{PNM}}^{\rm{HMT}}\approx1.5\%\pm0.5\%$\,\cite{Hen15,Hen14,Egi06,Pia06,Shn07,Wei11,Kor14,Hen15b}.

\begin{figure}[tbh]
\centering
  \includegraphics[width=8.cm]{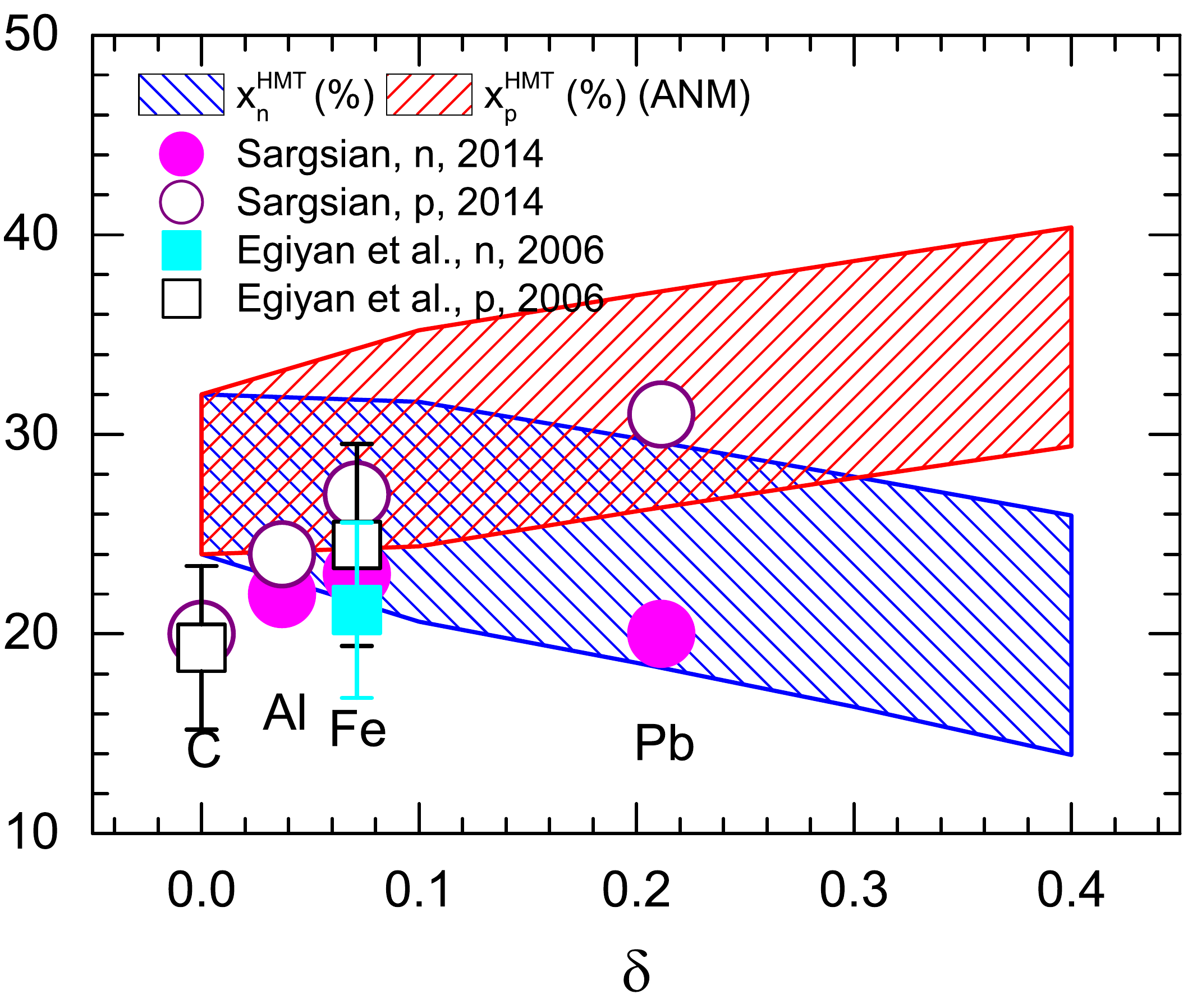}
    \caption{Fractions of high momentum neutrons (blue) and protons (red) as a function of isospin asymmetry $\delta$ in ANM at saturation density as evaluated from Eq. (\ref{def_xJHMT}) in comparison with those for
    finite nuclei extracted from analyzing the SRC data within the neutron-proton dominance model.}
  \label{fig_xnxpdelta}
\end{figure}

The observed dominance of pn over pp SRC pairs is a clear consequence of the nucleon-nucleon
tensor correlation. Based on the neutron-proton dominance picture of the tensor force induced SRC, there are as many protons as
neutrons in the HMT. Thus, one expects that the fraction of protons in the HMT with respect to the total proton number is greater than that of neutrons in neutron-rich matter\,\cite{Hen14,Sar14}.
The isospin dependence of $x_J^{\rm{HMT}}$ is shown in Fig.\,\ref{fig_xnxpdelta} as the blue (red) band for neutrons (protons) using parameters describing the $n^J_{\v{k}}(\rho,\delta)$ of Eq. ({\ref{MDGen})
given in ref.\,\cite{Cai15a}. Also shown are the high momentum nucleon fractions in several typical finite nuclei based on the analyses of Sargsian et al. in refs.\,\cite{Egi06,Sar14} where
it is estimated from
\begin{equation}\label{SarE}
x_J^{\rm{HMT}}(A,y)\approx\frac{1}{2x_J}a_2(A,y)\int_{k_{\rm{F}}}^{\infty}n_{\v{k}}^{\rm{d}}\d\v{k}
\end{equation}
with $a_2(A,y)$ given by the estimates in ref.\,\cite{Egi06,Sar14,McG11}, $x_{\rm{p}}=Z/A, x_{\rm{n}}=N/A$ and $y=|x_{\rm{p}}-x_{\rm{n}}|$.
As it follows from Fig.\,\ref{fig_xnxpdelta}, with the increase
of the isospin asymmetry the imbalance between the high momentum
fractions of protons and neutrons grows. For example, in $^{208}\rm{Pb}$,
the relative fraction of high momentum protons is about 55\% more than that of neutrons.

\begin{figure}[h!]
\centering
\includegraphics[width=17.cm]{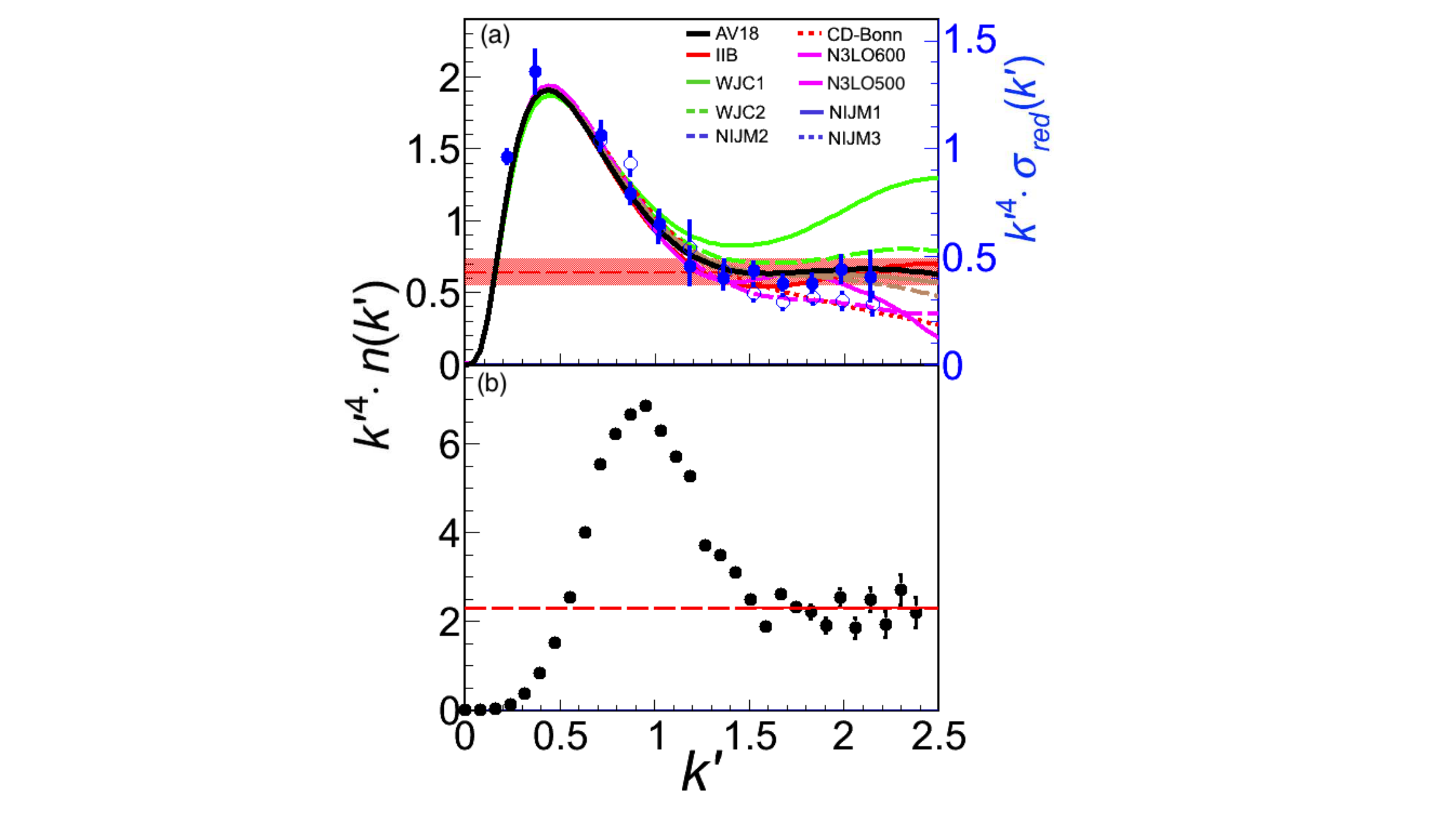}
\caption{The scaled momentum distribution, $k'^4n_{k'}$ where $k'
= |\v{k}|/k_{\rm{F}}$ for (a) deuteron and (b) ultra-cold $^{40}$K atomic systems in two-spin states with a short-range interaction. Taken from ref.\,\cite{Hen15}.}\label{fig_nk4univ}
\end{figure}

Interestingly, the $1/{|\v{k}|^4}$ shape of the nucleon HMT is a general property shared by several quantum many-body systems\,\cite{Gio08,Blo08}. In fact, the existence
of the HMT especially in SNM is widely supported by recent theoretical and/or
experimental studies.  Physically, the parameter $C$ in the expression of  $n^J_{\v{k}}(\rho,\delta)$ measures the probability of finding correlated pairs in HMT of the system considered. Indeed, the HMT for deuterons from variational many body (VMB)
calculations using several modern nuclear forces decrease as
$|\v{k}|^{-4}$ within about 10\% in quantitative agreement with
that from analyzing the $\rm{d}(\rm{e}, \rm{e}^{\prime}\rm{p})$
cross section in directions where the final state interaction suffered
by the knocked-out proton is effectively small\,\cite{Hen15}. The extracted
magnitude $C_{\rm{SNM}}=C_0$ of the HMT in  SNM at $\rho_0$ is $C_0=R_{\rm{d}}a_2(\infty)k_{\rm{F}}\rho/k_{\rm{F}}^4$, where the $R_{\rm{d}}$ is a factor introduced through
$n_{|\v{k}|/k_{\rm{F}}}^A(\v{k}/k_{\rm{F}})^4=R_{\rm{d}}a_2(A)$, see Fig.\,\ref{fig_nk4univ}\,\cite{Hen15} where VMB predictions for the HMT using several interactions are also shown.
Here the $R_{\rm{d}}$ is extracted from the deuteron momentum
distribution and it characterizes the plateau of the HMT as shown in Fig.\,\ref{fig_nk4univ}. At
even higher momenta, the momentum distribution drops much
more rapidly. Interestingly, the experimentally measured $\rm{d(e,e'p)}$ cross sections (blue dots with error bars) in the region of $1.3k_{\rm{F}} \lesssim |\v{k}|\lesssim 2.5k_{\rm{F}}$ also appears to scale as $|\v{k}|^{-4}$.
Using results from these theoretical and experimental investigations, it was found that
${C}_0\approx 0.15\pm0.03$\,\cite{Hen15} (properly rescaled
considering the factor of 2 difference in the adopted normalizations
of $n_{\v{k}}$ here and that in refs.\,\cite{Hen15,Hen15b}).\footnote{In refs.\,\cite{Hen15,Hen15b}, the momentum distribution is normalized as
$[2/(2\pi)^3]\int_0^{\infty}n_{\v{k}}^0\d\v{k}=\rho$, differs by a factor 2 from the definition of Eq. (\ref{def_NC}),
from which the corresponding relation reads $[4/(2\pi)^3]\int_0^{\infty}n_{\v{k}}^0\d\v{k}=\rho$ for SNM.}
The lower window of Fig.\,\ref{fig_nk4univ}\,\cite{Hen15} is an example of the well-known $1/{|\v{k}|^4}$ type HMT of the momentum distributions of cold atoms in two-spin states with a short-range interaction between the different spin-states, see e.g., refs. \cite{Ste10,Kuh10,Hu11,Sag15} and more discussions in subsection \ref{simls}.

Rather remarkably, a recent evaluation of medium-energy photonuclear
absorption cross sections has also presented independent and clear
evidence for the ${C}/{|\v{k}|^4}$ behavior of the HMT. Moreover, a relation between the contact $C_0$ and the Levinger constant was established,
i.e, $C_0=\rho\pi L/a_{\rm{t}}k_{\rm{F}}^4=2L/3\pi k_{\rm{F}}a_{\rm{t}}$, here $L$ is the Levinger constant and $a_{\rm{t}}$ is the scattering length for the $^3\rm{S}_1$ channel.
Using the value $1/k_{\rm{F}}a_{\rm{t}}\approx0.15$ and the experimental value for $L\approx5.50\pm0.21$\,\cite{Wei15}, a value of ${C}_0\approx 0.172\pm0.007$\,\cite{Wei15} for SNM at
$\rho_0$ was extracted\,\cite{Cai15a}. It is in very good agreement with that found in ref.\,\cite{Hen15}.
Combining the above two independent studies on the $C_0$, an average value of $C_0\approx0.161\pm0.015$ was used in refs.\,\cite{CaiLi16a,Cai15a}. With this $C_0$ and the value
of $x^{\rm{HMT}}_{\rm{SNM}}$ given earlier, the HMT cutoff parameter
in SNM is then $\phi_0=(1-x_{\rm{SNM}}^{\rm{HMT}}/3{C}_0)^{-1}\approx2.38\pm0.56$ well within the range of the SRC pairs, i.e., from about 300\,MeV to 600\,MeV, indicated by the SRC experiments.

\begin{figure}[tbh]
\centering
  \includegraphics[width=12.cm]{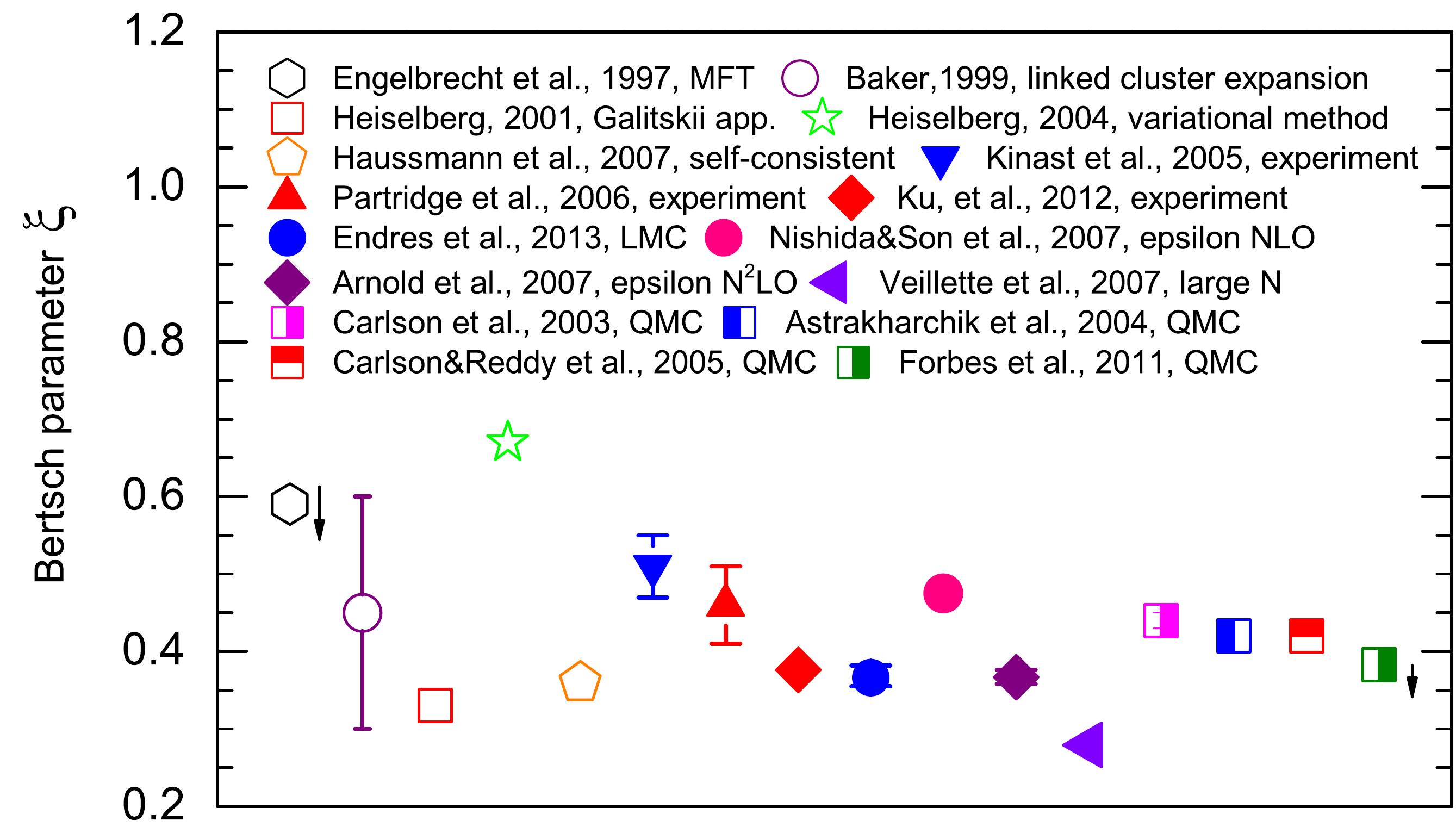}
\caption{Status of the constraints on the Bertsch parameter $\xi$ from both
experimental and theoretical studies. See the text for details.}
\label{xiRange}
\end{figure}
\subsection{The EOS and single-neutron momentum distribution in pure neutron matter (PNM)}
To fix all isospin-dependent parameters in Eq.\,(\ref{MDGen}), as boundary conditions one also needs information about the EOS and single-neutron momentum distribution in PNM.
Some of the information can be obtained from the state-of-the-art microscopic many-body calculations at sub-saturation densities.
Since the neutron-neutron s-wave scattering length in PNM is unnaturally large, i.e.,
$a_{\rm{nn}}(^1\rm{S}_0)\approx-18.8$\,fm, it is generally accepted that the
PNM is closer to the unitary limit at low densities\,\cite{Sch05}. It has been shown that the EOS of PNM can
be expressed as\,\cite{Bak99}
\begin{equation}\label{EPNMTan}
E_{\rm{PNM}}(\rho)\approx\frac{3}{5}\frac{(k^{\textrm{PNM}}_{\textrm{F}})^2}{2M}\left[\xi-\frac{\zeta}{k^{\textrm{PNM}}_{\textrm{F}}a_{\textrm{nn}}}
-\frac{5\nu}{3(k^{\textrm{PNM}}_{\textrm{F}}a_{\textrm{nn}})^2}\right]
\end{equation}
where $k_{\rm{F}}^{\rm{PNM}}=2^{1/3}k_{\rm{F}}$ is the Fermi
momentum in PNM, $\xi$ is the Bertsch
parameter\,\cite{BertPara}, $\zeta\approx\nu\approx1$ are two
universal constants\,\cite{Bul05}. Shown in Fig.\,\ref{xiRange} are the Bertsch parameter $\xi$ from
both experimental and theoretical studies over the last 20 years \cite{Kin05,Par06,Ku12}. They include predictions from the mean field theory (MFT)\,\cite{Eng97}, linked cluster expansion
method\,\cite{Bak99}, Galitskii approximation\,\cite{Hei01}, variational
method\,\cite{Hei04}, self-consistent method\,\cite{Hau07}, lattice
Monte Carlo simulation\,\cite{End13}, effective field theory based on
$\epsilon$-expansion at next-leading-order (NLO)\,\cite{Nis06,Nis06a}
as well as at next-next-leading-order (N$^2$LO)\,\cite{Arn07}, large-$N$
approach\,\cite{Vei07} and quantum Monte Carlo
simulation\,\cite{Car03,Ast04,Car05,For11}.
While historically the Bertsch parameter $\xi$ has scattered around $\xi\approx0.4\pm0.1$ as illustrated in Fig.\,\ref{xiRange}, it has converged to around $\xi = 0.376\pm0.005$ over the last 5 years.
Currently, the best estimate from the lattice Monte Carlo studies is $\xi = 0.372(5)$ consistent with the
most accurate experimental value of $\xi = 0.376(4)$.

Shown in the left window of Fig.\,\ref{TanPNMLow} is a comparison of the EOS of PNM
obtained from Eq.\,(\ref{EPNMTan}) (dashed red band with $\xi\approx0.4\pm0.1$) with several
state-of-the-art calculations using modern microscopic many-body
theories. At densities less than about 0.01\,$\textrm{fm}^{-3}$, as
shown in the inset, the Eq.\,(\ref{EPNMTan}) is found to be fairly consistent with the
prediction by the effective field theory\,\cite{Sch05}. In the range
of 0.01\,$\textrm{fm}^{-3}$ to about $0.02\,\textrm{fm}^{-3}$, it
has some deviations from predictions in ref.\,\cite{Sch05} but agrees
very well with the NLO lattice simulations\,\cite{Epe09a}. At higher
densities up to about $\rho_0$, it overlaps largely with predictions
by the chiral perturbation theories (ChPT)\,\cite{Tew13} and the quantum
Monte Carlo (QMC) simulations\,\cite{Gez13,Gez10}. 
\begin{figure}[h!]
\centering
\includegraphics[width=17.cm,height=9cm]{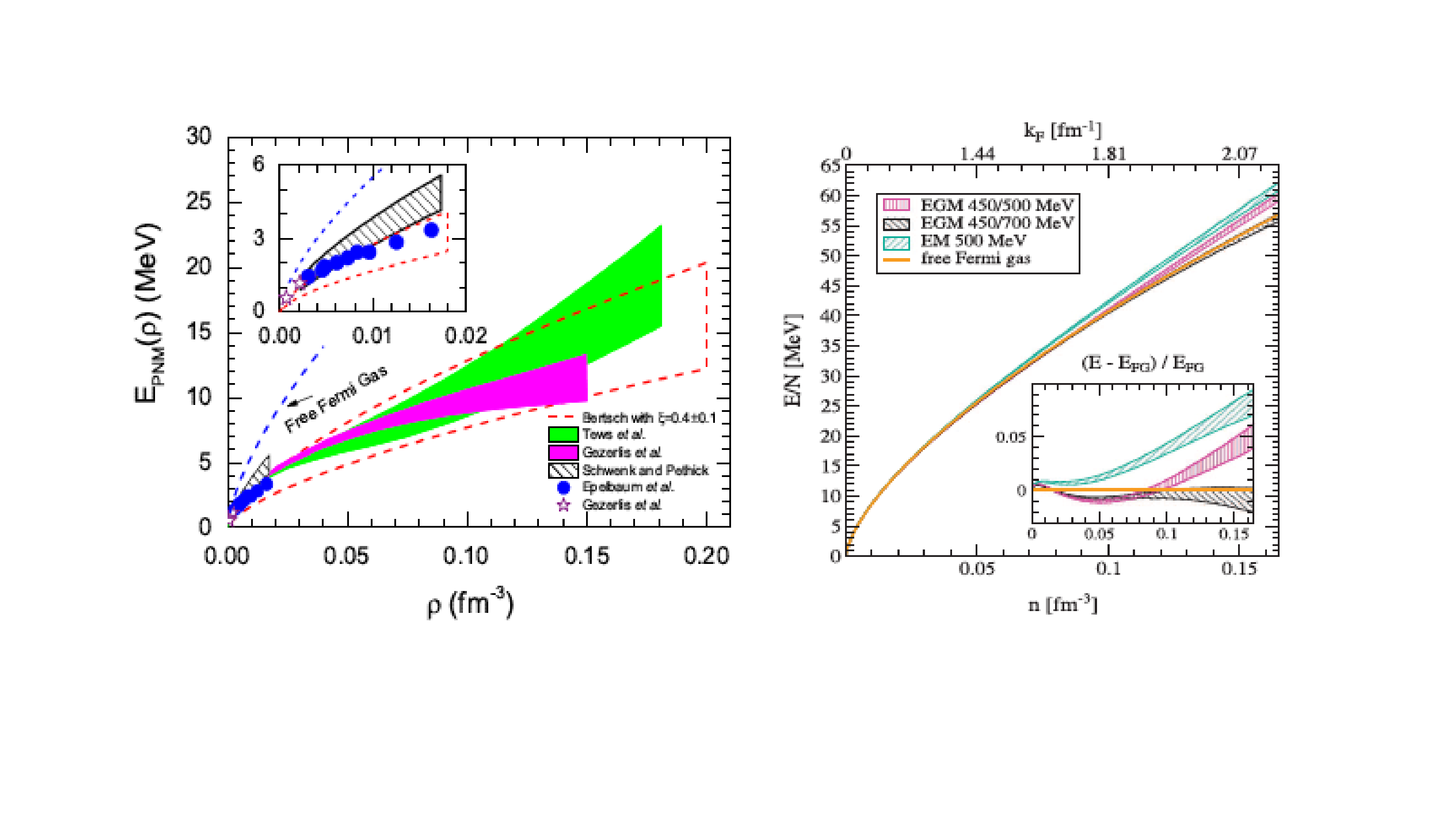}
\vspace{-2.3cm}
\caption{Left: EOS of PNM obtained from Eq.\,(\ref{EPNMTan}) (dashed red band) and that from next-leading-order
(NLO) lattice calculation\,\cite{Epe09a} (blue solid points), chiral
perturbative theories\,\cite{Tew13} (green band), quantum Monte Carlo
simulations (QMC)\,\cite{Gez13,Gez10} (magenta band and purple stars),
and effective field theory\,\cite{Sch05}. Taken from ref.\,\cite{Cai15a}.
Right: Energy per particle of spin-polarized neutron matter as a function of density.
The solid (orange) line is the energy of a FFG. Taken from ref.\,\cite{Kru15}.}\label{TanPNMLow}
\end{figure}
In addition, recent studies on the spin-polarized neutron matter within the chiral
effective field theory including two-, three-, and four-neutron
interactions show that the properties of PNM is very similar to the
unitary Fermi gas at least up to about $\rho_0$, which is far beyond the
scattering-length regime of $\rho\lesssim\rho_0/100$\,\cite{Kru15}, see the right window of Fig.\,\ref{TanPNMLow}.
Overall, these comparisons and studies clearly justify the use of
Eq.\,(\ref{EPNMTan}) to calculate the EOS of PNM up to about the saturation density.
In fact,  the EOS of PNM obtained through Eq.\,(\ref{EPNMTan}) near the unitary limit was recently used to constrain
several quantities characterising the EOS of neutron-rich matter\,\cite{Tew17,Zha17} and properties of neutron stars\,\cite{Tew17}.

\begin{figure}[h!]
\centering
\includegraphics[width=6.cm,height=8cm,angle=-90]{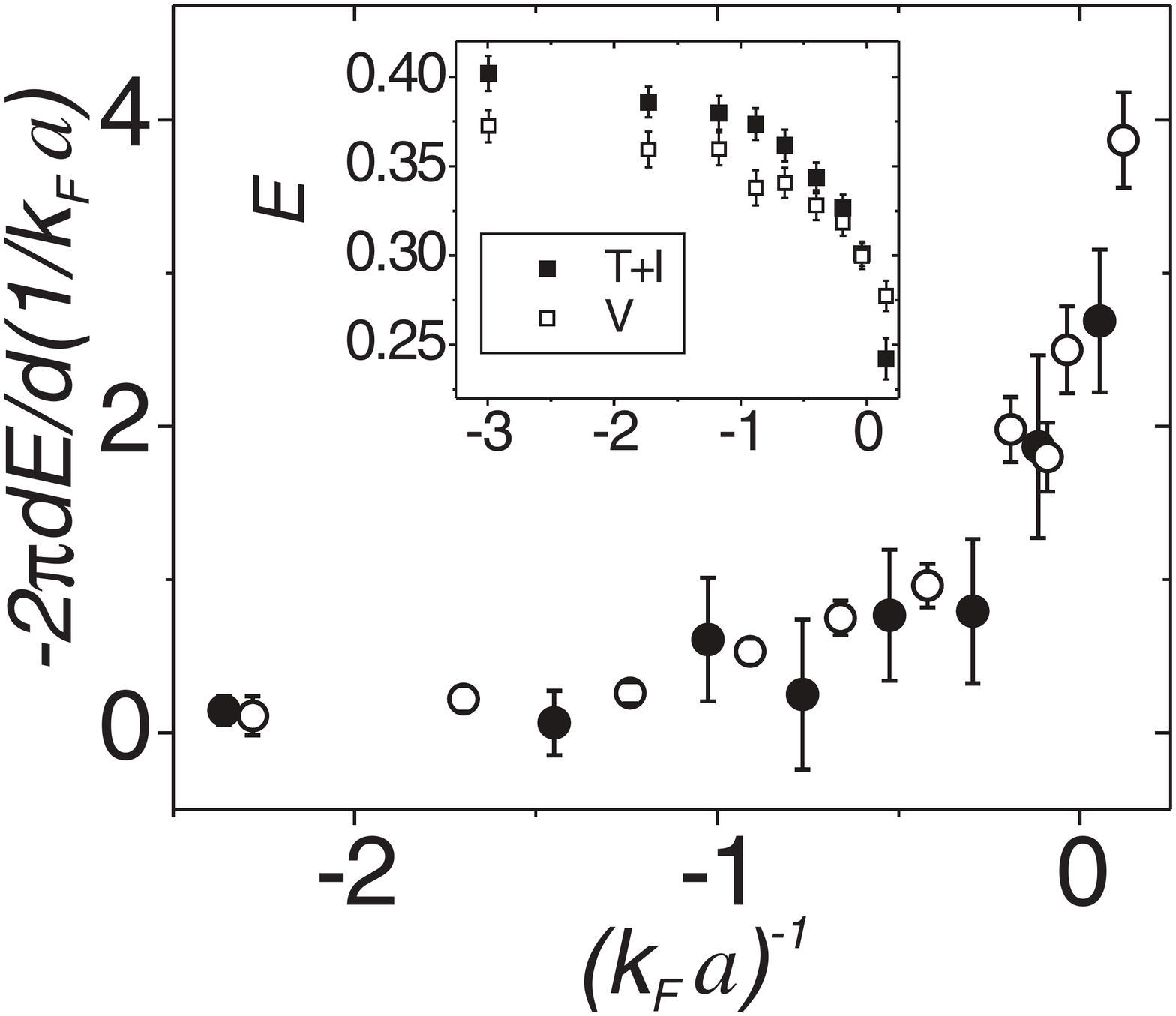}
\hspace{0.2cm}
\includegraphics[width=6.cm,height=8cm,angle=-90]{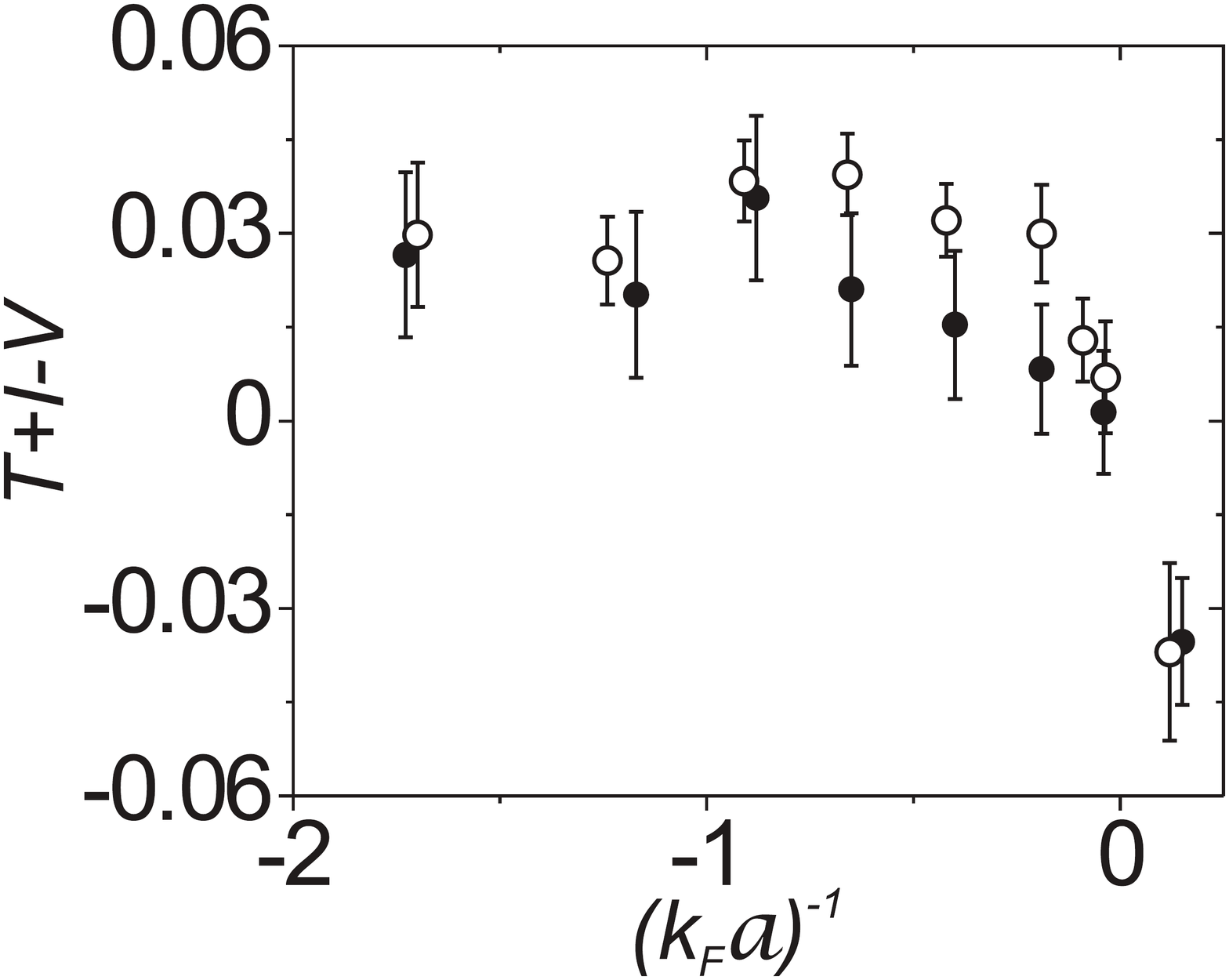}
\caption{Testing of the adiabatic sweep theorem in ultra-cold atomic gases (left) and the generalized virial theorem (right).
The solid circle in the left panel is for the value $-2\pi\d E/\d(1/k_{\rm{F}}a)$ and the harrow circle for the one by measuring the contact experimentally.
Taken from ref.\,\cite{Ste10}.}\label{fig_Stew10}
\end{figure}

The HMT of single-particle momentum distributions and the EOS of a many-body system can be experimentally measured independently
and calculated simultaneously within the same model. Interestingly, Tan has proven an adiabatic sweep theorem directly relating the contact parameter $C$ with the EOS\,\cite{Tan08}. It is valid for any
two-component Fermi systems under the same conditions as the
Eq.\,(\ref{EPNMTan}) near the unitary limit. Moreover, it was accurately verified in several experiments using ultra-cold atomic gases.
For instance, shown in Fig.\,\ref{fig_Stew10} are the values obtained by the adiabatic sweep theorem, i.e., $-2\pi\d E/\d(1/k_{\rm{F}}a)$ (solid circle)
in comparison with the experimentally measured contacts (hollow circle) in the ultra-cold $^{40}\rm{K}$ system. Here the relation connecting the energy and the contact $\mathcal{C}$ of the $^{40}\rm{K}$
system is given by\,\cite{Ste10}
\begin{equation}\label{def_gvt}
E-2V=T+I-V=-{\mathcal{C}}/{4\pi k_{\rm{F}}a}
\end{equation}
where $\mathcal{C}$ is the contact ($\mathcal{C}\leftrightarrow 3\pi^2C_{\rm{n}}^{\rm{PNM}}$ for the PNM case) and $I$ is the interaction energy\,\cite{Tan08}.
It is obvious that the independent measurements
of the left hand side and the right hand side of Eq.\,(\ref{def_gvt}) agree to
within the error bars. It is also interesting to note that the measured energy difference $T+I-V$ is so small (in
units of Fermi energy) that even a Fermi gas with a strongly attractive contact interaction nearly obeys the noninteracting
virial equation, i.e., $E-2V=T+I-V\approx 0$\,\cite{Ste10}.

For PNM, the adiabatic sweep theorem states
\begin{equation}\label{ast}
C_{\rm{n}}^{\rm{PNM}}\cdot(k_{\rm{F}}^{\rm{PNM}})^4=-4\pi
M\cdot{\d (\rho E_{\rm{PNM}})}/{\d(a^{-1})}.
\end{equation}
While the results shown in Fig.\,\ref{TanPNMLow} justify the use of
Eq.\,(\ref{EPNMTan}) for the EOS of PNM up to about $\rho_0$,
there is currently no proof that the
Eq.\,(\ref{ast}) is valid in the same density range as the EOS of
Eq.\,(\ref{EPNMTan}). Thus, it would be very interesting to examine
the validity range of Eq.\,(\ref{ast}) using the same models as those
used to calculate the EOS. Assuming the Eqs.\,(\ref{EPNMTan}) and (\ref{ast}) are both valid in the same
density range, the strength of the HMT in PNM can be consequently obtained as
\begin{equation}\label{C_Tan}
C_{\rm{n}}^{\rm{PNM}}\approx 2\zeta/5\pi+4\nu/(3\pi
k_{\rm{F}}^{\rm{PNM}}a_{\rm{nn}}(^1\rm{S}_0))\approx0.12.
\end{equation}
After obtaining the contact $C_{\rm{n}}^{\rm{PNM}}$ in PNM and using the relation $C_{\rm{n}}^{\rm{PNM}}=C_0(1+C_1)$, a value of
$C_1\approx-0.25\pm0.07$ is obtained. Similarly, by
inserting the $x_{\rm{PNM}}^{\rm{HMT}}$ and
$C_{\rm{n}}^{\rm{PNM}}$ found earlier into $x_{\rm{n}}^{\rm{PNM}}$, the high momentum
cutoff parameter for PNM is found to be about
$\phi_{\rm{n}}^{\rm{PNM}}\equiv
\phi_0(1+\phi_1)=(1-x_{\rm{PNM}}^{\rm{HMT}}/3C_{\rm{n}}^{\rm{PNM}})^{-1}\approx1.04\pm0.02$.
It is easy to understand that the $\phi_{\rm{n}}^{\rm{PNM}}$ is very
close to unity, since the high momentum neutron fraction in PNM is only about 1.5\%.
Finally, inserting the $\phi_0$ obtained earlier,
$\phi_1\approx-0.56\pm0.10$ is found. The two parameters $\beta_0$ and
$\beta_1$ in $\beta_J=\beta_0(1+\beta_1\tau_3^J\delta)$ depend on
the function $I(|\v{k}|/k_{\rm{F}}^J)$ which is model
dependent\,\cite{Cai15a}, and the effects of $\beta_0$ and $\beta_1$ on the kinetic EOS will be explored more in later subsections.
\begin{figure}[h!]
\centering
\includegraphics[width=15cm]{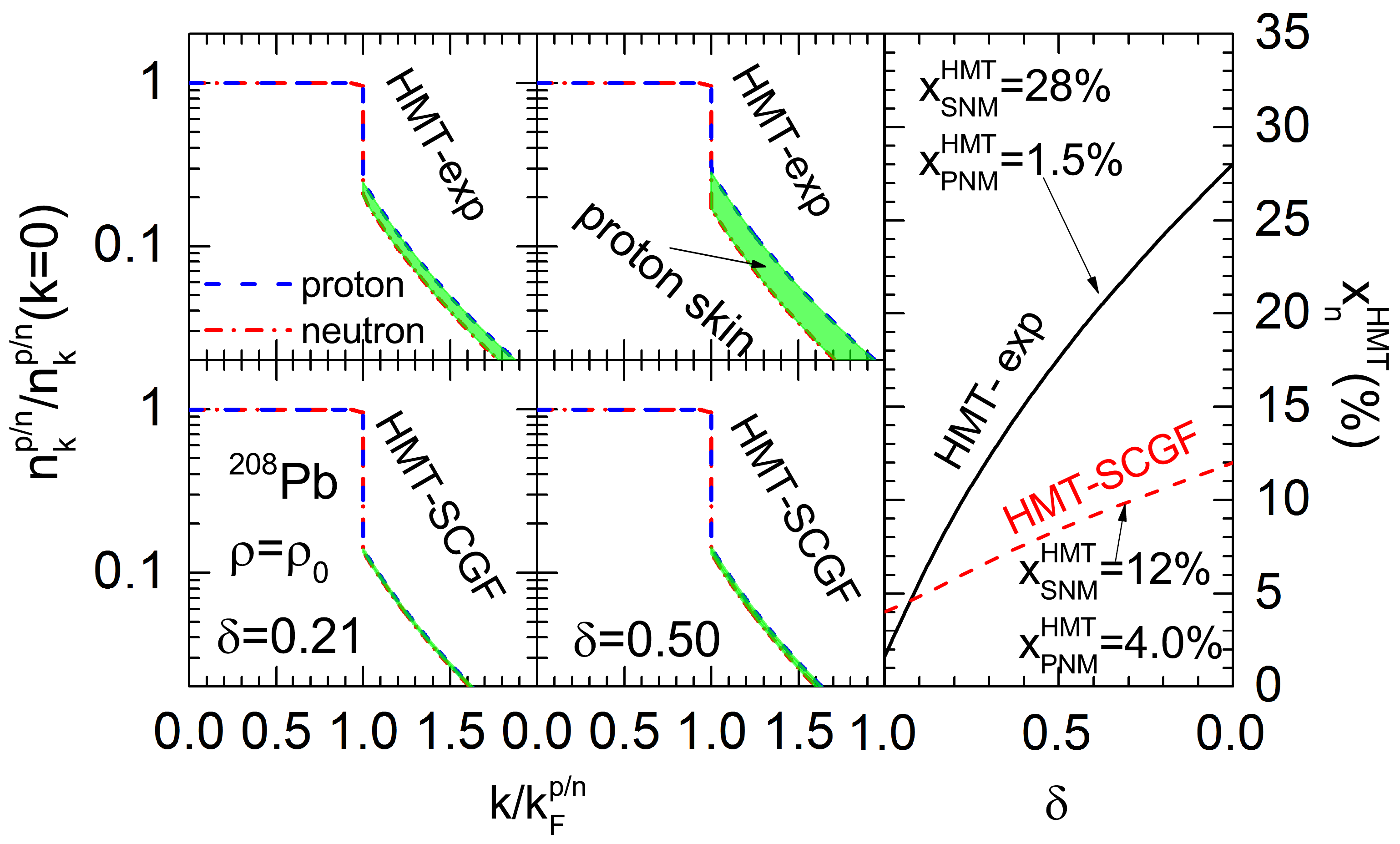}
\caption{Reduced nucleon  momentum distribution of neutron-rich nucleonic matter with an isospin asymmetry of $\delta=0.21$ (left) and 0.50 (middle),
and the fraction of neutrons in the HMT as a function of $\delta$ using the HMT-exp and HMT-SCGF parameter set (right). Taken from ref.\,\cite{Cai16b}.}\label{fig_xCaiFig1}
\end{figure}

To this end, some comments about the uncertainty of the HMT fractions especially in PNM should be given. As illustrated in Fig. \ref{fig_a2the},
although different many-body theories have consistently predicted
the effects of the SRC on the HMT qualitatively consistent with the experimental findings, the predicted size of the HMT still depends on the model and
interaction used. For example,  the SCGF theory with the Av18 interaction predicts a 11-13\% HMT for
SNM at saturation density $\rho_0$\,\cite{Rio09,Rio14}. While the latest BHF calculations
predict a HMT ranging from about 10\% using the N$^3$LO450 to
over 20\% using the Av18, Paris or Nij93 interactions\,\cite{ZHLi16},
the latest variational Monte Carlo (VMC) calculations for $^{12}$C gives a 21\% HMT in SNM\,\cite{Wir14,Wir16}.
On the other hand, very little information about the isospin dependence of the HMT has been extracted directly from experiments so far.
As we discussed earlier, the SRC between a neutron-proton pair is about 18-20 times that of two protons\,\cite{Hen14,Sub08}, the HMT in PNM was then estimated to be about 1-2\% \cite{Hen15b}.
However, some recent calculations such as the SCGF indicates a significantly higher HMT in PNM of approximately 4-5\% \cite{Rio09,Rio14}.
This uncertainty about the size of HMT in PNM affects some quantitative results of model calculations. Hopefully, more calculations of the HMT in PNM can be done soon also with those models and interactions used to calculate the EOS of PNM. More precise knowledge about the HMT in PNM is important. For example, similar to the parameter set described above (abbreviated as the HMT-exp), the HMT-SCGF parameter
set contains the following values, i.e.,  $ x_{\rm{SNM}}^{\rm{HMT}}=12\%$,
$x_{\rm{PNM}}^{\rm{HMT}}=4\% $, $\phi_0=1.49,
\phi_1=-0.25$, $C_0=0.121$ and $C_1=-0.01$\,\cite{Cai16b}. The resulting neutron
fractions in the HMT versus $\delta$ at $\rho_0$ using the two parameter sets are compared in the right panel of Fig.\,\ref{fig_xCaiFig1}. The reduced nucleon momentum distributions
(normalized to unity at zero momentum) with $\delta=0.21$ and $0.5$ for both cases are shown in the left panel. Clearly, relative to the center in momentum-space, nucleonic matter has a distinct proton-skin in momentum-space
and its thickness grows with the isospin asymmetry $\delta$ at a rate depending on the sizes of the HMT parameters adopted\,\cite{Cai16b}. The neutron-skins in coordinate-space and proton-skins in momentum-space coexist in heavy nuclei and their correlation is governed by Liouville's theorem and Heisenberg's uncertainty principle\,\cite{Cai16b}. A careful study of both skins simultaneously may help better constrain the underlying isovector interaction.
In the following discussions, we will use the HMT-exp parameter set unless specified otherwise.

\begin{figure}[h!]
\centering
\includegraphics[width=16.cm,height=11.cm]{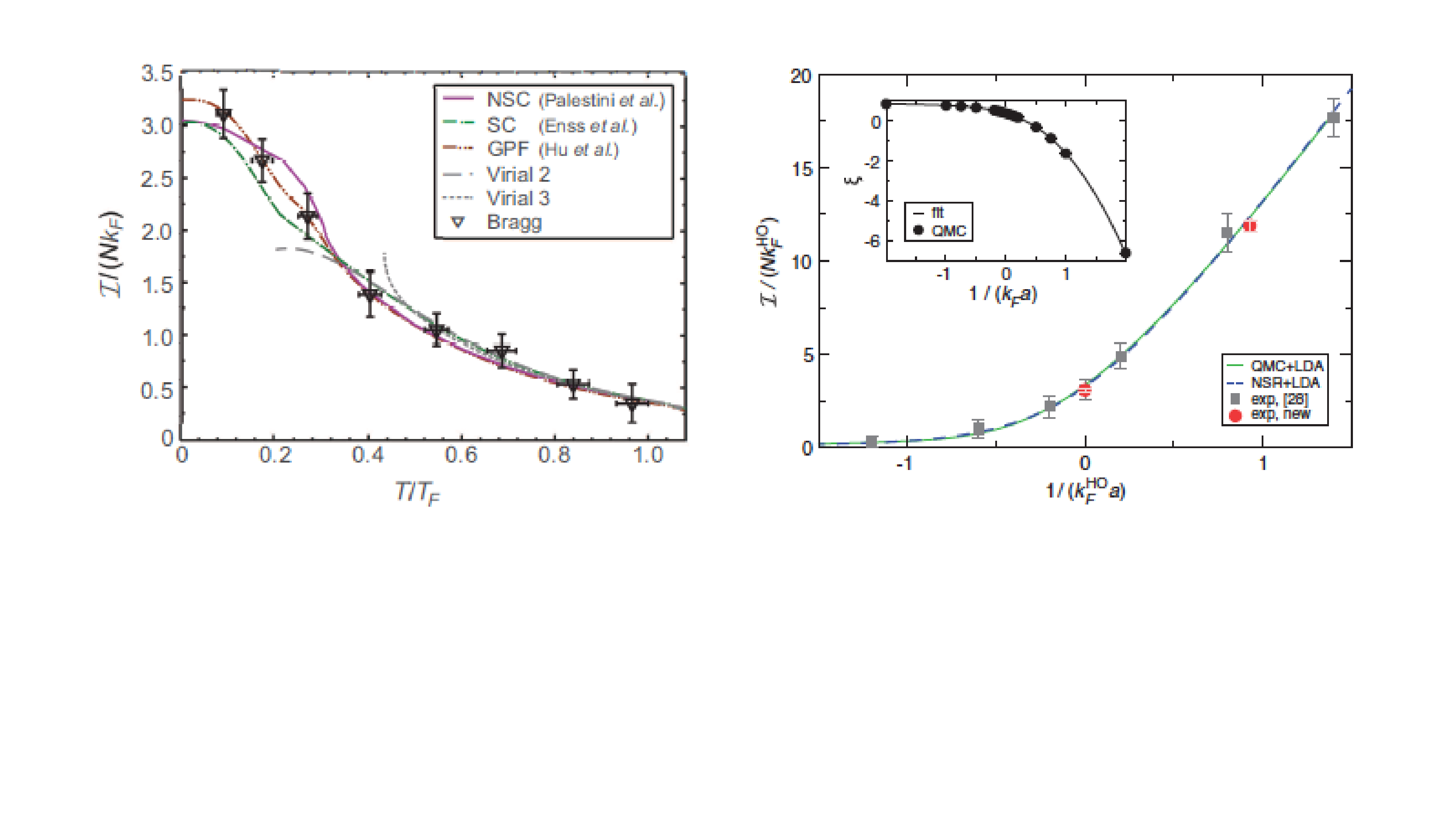}
\vspace{-4cm}
\caption{Left: Measurements of the temperature dependence of the contact in a unitary Fermi
gas ($^6\rm{Li}$) using Bragg spectroscopy techniques. Taken from ref.\,\cite{Kuh11}.
Right: Measurement of the contact in a strongly interacting trapped gas of $^6\rm{Li}$
by an equal mixture of the states $|f=1/2, m_f=\pm1/2\rangle$. Taken from ref.\,\cite{Hoi13}.}\label{fig_Kuhn11}
\end{figure}

\begin{figure}[h!]
\centering
\includegraphics[width=16cm,height=10.cm]{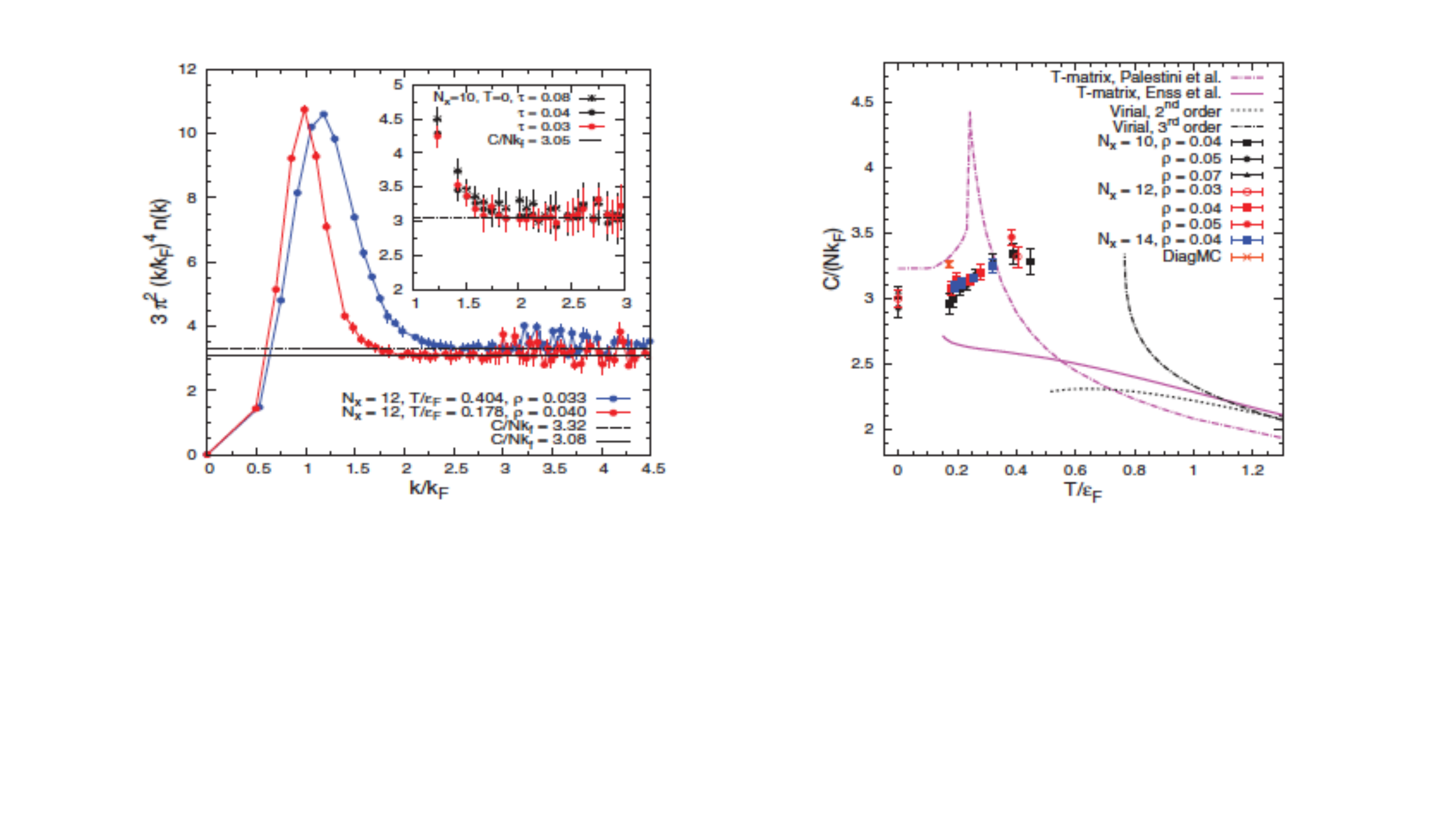}
\vspace{-4cm}
\caption{Reduced momentum distribution for the unitary Fermi gas (left) and the temperature
dependence of the contact (right) obtained by QMC. Taken from ref.\,\cite{Dru11}.}\label{fig_Drut11}
\end{figure}

\subsection{Exploring the similar shapes of the high-momentum tails in nuclei and cold atoms}\label{simls}
As shown already in Fig.\,\ref{fig_nk4univ}, the ${1}/{|\v{k}|^4}$ shape of the HMT for nucleons is almost identical to that in two-component (spin-up and -down)
cold atomic Fermi gases.  For the latter, it was first predicted by Tan\,\cite{Tan08} and then
quickly verified experimentally mainly using the $^6\rm{Li}$ and $^{40}\rm{K}$
systems\,\cite{Ste10,Kuh10}. It is necessary to recall
that Tan's general prediction is for all two-component Fermion systems having a
contact interaction with a s-wave scattering length $a$ much larger than
the inter-particle distance $d\sim\rho^{-1/3}$ which meanwhile has to be much longer than the
effective interaction range $r_{\rm{eff}}$, i.e., $r_{\rm{eff}}\ll d\ll a$.
At the unitary limit, i.e.,
$k_{\rm{F}}a\rightarrow \pm\infty$, Tan's prediction is universal for
all Fermion systems. Since the HMT in nuclei and SNM is known to be
dominated by the tensor force induced neutron-proton pairs with
$a\approx 5.4$ fm and $d\approx 1.8$ fm at $\rho_0$, as noted in
refs.\,\cite{Hen15,Wei15}, Tan's stringent conditions for unitary
Fermions is obviously not well satisfied in normal nuclei and SNM. Thus the
observed almost-identical ${1}/{|\v{k}|^4}$ behavior of the HMT in nuclei
and cold atomic Fermi gases may have some deeper physical meanings deserving further investigations both theoretically and experimentally\,\cite{Hen15}.

The $|\v{k}|^{-4}$ behavior in nuclei can be understood as arising from the
importance of the one-pion-exchange (OPE) contribution
to the tensor potential $V_{\rm{T}}$, starting from the second order\,\cite{Hen15,Col15}.
Based on the Schr$\ddot{\rm{o}}$dinger equation for the spin-1 two-nucleon system involving S- and D-state components, one can obtain an effective S-state potential as
$V_{00}=V_{\rm{T}}(-B-H_0)^{-1}V_{\rm{T}}$, where $V_{\rm{T}}$ connects the S and D
states and $B$ is the binding energy of the system.
The intermediate Hamiltonian $H_0$ dominated by the
effects of the centrifugal barrier can be approximated by
the kinetic energy operator\,\cite{Hen15}. Then evaluating the S-state potential, and neglecting
the small effects of the central potential in the intermediate D
state, one obtains\,\cite{Hen15}
\begin{equation}
V_{00}(|\v{k}|,|\v{k}'|)=-\frac{16M^4f}{m_{\pi}^4\pi^4}\int\frac{p^2p\d p}{MB+p^2}I_{02}(|\v{k}|,p)I_{20}(p,|\v{k}'|)
\end{equation}
where M is the nucleon static mass, $f^2\approx0.08$ is the coupling
constant, $m_{\pi}$ is the pion mass, and $I_{LL'}$'s are Fourier transformations
of the OPE tensor potential. At large momenta, one then has $\lim_{p\to\infty}I_{02}(p,|\v{k}|)\approx1 - (|\v{k}|^2 + m_{\pi}^2)/p^2+\cdots$.
Thus, the integrand of $V_{00}(|\v{k}|,|\v{k}'|)$ is dominated by large values of the momentum $p$ and
diverges unless there is a high momentum cutoff. This means that the $V_{00}(|\v{k}|,|\v{k}'|)$ is approximately a constant independent of $|\v{k}|$ and $|\v{k}'|$. As pointed out in refs.\,\cite{Hen15,Col15},
this is just the signature of a short-range correlation. It is actually the necessary and sufficient condition
to obtain an asymptotic two-nucleon wave function scaling as $|\v{k}|^{-2}$ and consequently a momentum probability density proportional to $|\v{k}|^{-4}$.

At large momenta, the static structure factor of the ultra-cold atomic gases is given by $S_{\uparrow\downarrow}=(\mathcal{I}/Nk_{\rm{F}})(k_{\rm{F}}/4|\v{k}|)$,
which can be quantitatively measured by using the Bragg spectroscopy\,\cite{Kuh10,Kuh11}. Here the $\mathcal{I}/Nk_{\rm{F}}$ is the dimensionless contact (equivalent to $3\pi^2C_{\rm{n}}^{\rm{PNM}}$ studied in nuclei).
In the left window of Fig.\,\ref{fig_Kuhn11}, the temperature dependence of the contact is shown for the cold $^6\rm{Li}$ gas. It is seen that an extrapolation to zero temperature gives approximately $\mathcal{I}/Nk_{\rm{F}}\approx3.11\pm0.23$ (or equivalently $C_{\rm{n}}^{\rm{PNM}}\approx0.11\pm0.01$)\,\cite{Kuh11}.
Similar experiments studied in ref.\,\cite{Hoi13} gives $\mathcal{I}/Nk_{\rm{F}}^{\rm{HO}}\approx3.15\pm0.09$ at zero temperature. In ref.\,\cite{Hoi13}, the Bertsch parameter $\xi$ was also studied and a value about 0.3899 was obtained (inset of the right window in Fig.\,\ref{fig_Kuhn11}), which is slightly larger than the measurement\,\cite{Ku12}.
Moreover, a recent calculation based on the QMC algorithm also confirmed the ${1}/{|\v{k}|^4}$ behavior (left window of Fig.\,\ref{fig_Drut11}) and found that the dimensionless contact of the unitary Fermi gas
is about $2.95\pm0.01$\,\cite{Dru11}, see Fig.\,\ref{fig_Drut11} (right) for the temperature dependence of the contact.
All of the above experimental and theoretical studies on the contact of the unitary Fermi gas point to a value of $C_{\rm{n}}^{\rm{PNM}}=0.12$ consistent with the one from applying Tan's formula in Eq. (\ref{C_Tan}).

\begin{figure}[h!]
\centering
\hspace*{-0.5cm}\includegraphics[width=12.cm]{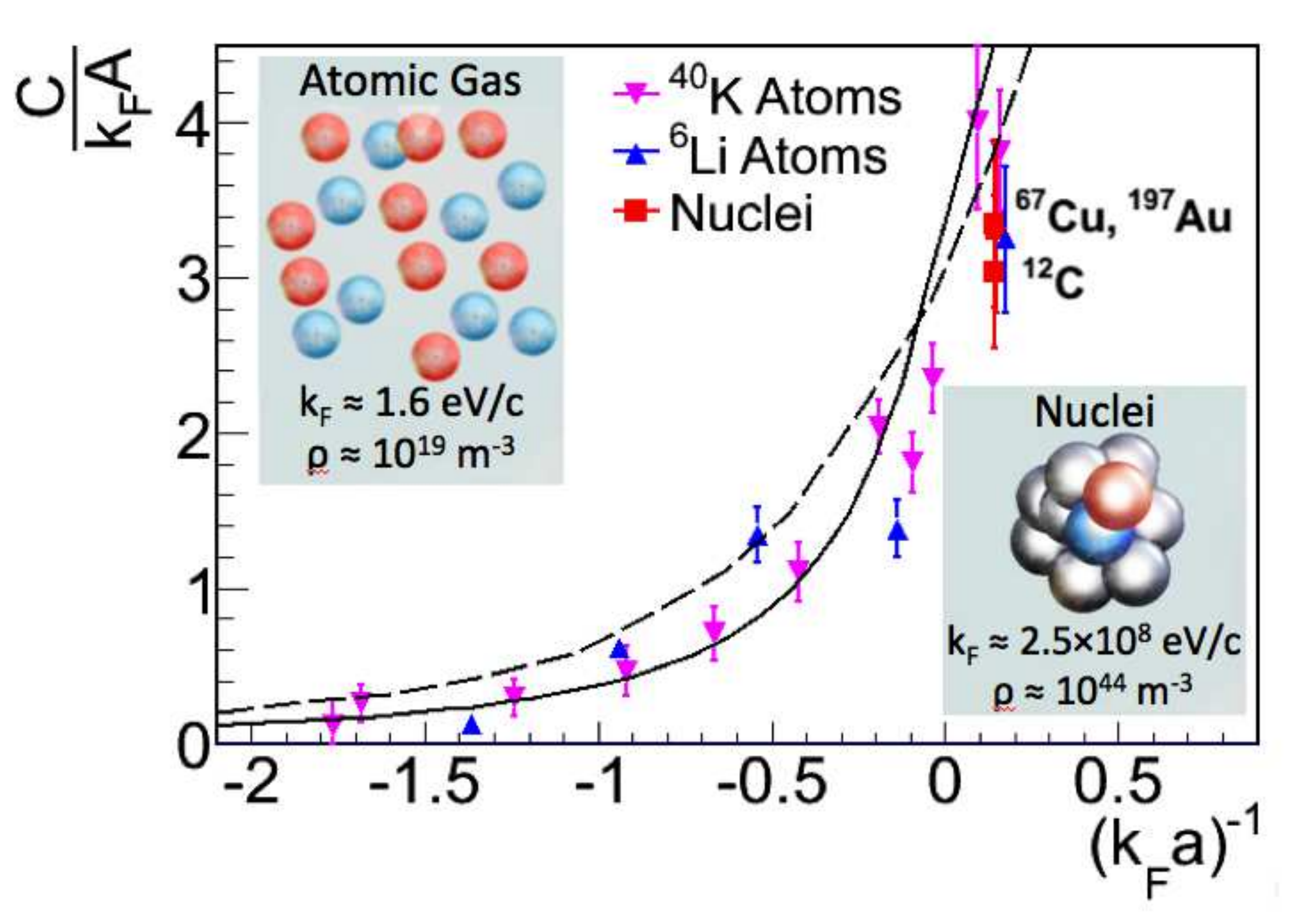}
\caption{Comparison of contacts of atomic nuclei and ultra-cold Fermi gas. Taken from ref.\,\cite{Hen15}.}\label{fig_Contact}
\end{figure}
As it was discussed by Hen et al. in ref.\,\cite{Hen15}, although the density of the ANM and that of the ultra-cold Fermi gas differs by about 25 orders of magnitude, the contact describing the probability of finding particle pairs is very similar as shown in  Fig.\,\ref{fig_Contact} (also see the lower panel of Fig.\,\ref{fig_nk4univ} and the left panel of Fig.\,\ref{fig_Drut11}). The two body
correlation function in a two-component Fermi gas has
the general structure of $\langle\psi_1^{\dag}(\v{r}')\psi_2^{\dag}(\v{0})
\psi_2(\v{0})\psi_1(\v{r})\rangle=\sum_i\gamma_i\phi_i(\v{r})\phi_i^{\ast}(\v{r}')$\,\cite{Yu13},
i.e., it regards the correlation function as a Hermitian
operator in $\v{r}$ and $\v{r}'$. For $r\ll d$, the inter-particle distance,
the functions $\phi_i$, essentially s-wave Jastrow factors in the
many-body wave function at short inter-particle distance are
determined by two-body physics. They have the form of $\sin[kr+\delta_0(k)]/r$, where $\delta_0(k)$
is the s-wave phase shift and $k$ is the wave vector of the scattering wave.
The s-wave scattering amplitude $f_0(k)=[-k\cot\delta_0(k)+ik]^{-1}$ is independent of the scattering angle.
In the limit $k\to0$, it tends to a constant value, i.e., $f_0(k\to0)=-a$, where $a$ is the s-wave scattering
length, playing a crucial role in the scattering processes at low energies\,\cite{Gio08}.
After including terms to order $k^2$ in the perturbative expansion of the phase shift $\delta_0(k)$ at small momenta $k$, the
scattering amplitude takes the form of
\begin{equation}
f_0(k)=-\frac{1}{a^{-1}-k^2r_{\rm{eff}}/2+ik}
\end{equation}
where $r_{\rm{eff}}$ is the effective range of the interaction. Now in the limit $a\to\infty$, i.e., the unitary limit,
the above scattering amplitude with wave vectors $k\ll r_{\rm{eff}}^{-1}$ obeys the universal law $f_0(k)=i/k$,
independent of the interaction. Under these conditions, the wave function $\sin(kr+\delta_0)/r\approx\sin\delta_0\chi(r)/r$,
where $\chi(r)=1-r/a$, and thus
\begin{equation}
\langle\psi_1^{\dag}(\v{r}')\psi_2^{\dag}(\v{0})
\psi_2(\v{0})\psi_1(\v{r})\rangle=C\left(\frac{\chi(r)}{r}\right)^2=C\left(\frac{1}{r}-\frac{1}{a}\right)^2,
\end{equation}
where $C$ is the contact parameter. At the unitary limit, the Fourier transformation of the above correlation gives a term $\sim|\v{k}|^{-4}
$ for the high-momentum part of the momentum distribution.
Furthermore, the $|\v{k}|^{-4}$ form of the HMT was also found recently in a Bose system both theoretically\,\cite{Bra11,Wer12,Smi14} and
experimentally\,\cite{Mak14,Fle17}, indicating again the very generality
of the HMT shared by some quantum many-body systems.

\begin{figure}
\centering
\vspace{-0cm}
\includegraphics[width=8cm,angle=90]{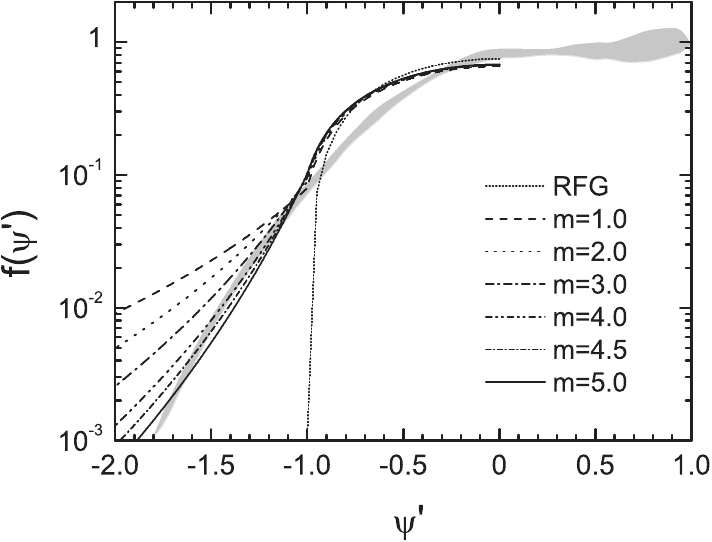}
\vspace{-0cm}
\caption{\label{fig5ant} The super scaling function in a dilute Fermi gas
calculated for different values of $m$ in
the asymptotics of the momentum distribution $n(k) \sim 1/k^{4+m}$
in comparison with the Relativistic Fermi Gas (RFG) model result and the experimental QE scaling function (grey area)~\cite{DS99}. Taken from ref. \cite{Ant07}.}
\end{figure}
While it is exciting that the similar $|\v{k}|^{-4}$ shapes of the HMT are found in both nuclei and cold atoms near the unitary limit, we notice that the evidences for the nuclear case are mostly based on theoretical models. There are only few indirect experimental indications so far unlike the case for cold atoms. As we mentioned earlier, there is clearly a need to further investigate the shape of the HMT in nuclei and nuclear matter both theoretically and experimentally. For this purpose, we make a few more observations and comments in the following. In fact, the study on the shape and range of the HMT for Fermions at $k>k_F$ has a long history, see e.g., refs. \cite{Mig57,Gal58,Lut60,Czy61,Bel61,Sar80}. However, these earlier studies are mostly done for SNM only. The HMT of $n(k)$ is often obtained by expanding quantities of interest in a dilute gas of hard spheres in terms of the $k_F\cdot a$ where a is the scattering length or hard-core radius. The asymptotic behavior of the momentum distribution $n(k)$ is generally related to the Fourier transform $\widetilde{V}_\text{NN}(k)$ of the nucleon-nucleon (NN) force via ~\cite{Amado76a,Amado76b,Amado77,Ant07}
$
n(k)\xrightarrow[k\rightarrow
\infty]{}[\widetilde{V}_\text{NN}(k)/k^2]^2.
$
However, it is still unclear if $k$ or $k/A$ where $A$ is the mass number must be large for this asymptotic behavior to be valid \cite{Ant07}. Moreover, the spin-isospin dependence of the short-range interaction and the resulting isospin dependence of the HMT are unclear. In the case of using a contact force, the asymptotic HMT naturally reduces to the $n(k) \sim1/k^4$ as for the ultra cold atoms near the unitary limit or the nuclear HMT in SNM within the neutron-proton dominance model assuming a contact nuclear force.  Extensive and systematic analyses of scaling functions, e.g., y-scaling and superscaling in the quasi-elastic (QE) region of inclusive electron-nucleus scatterings \cite{Ant07,AGK+04,AGK+05,AIG+06,AIG+06a} found that the $n(k)$ scales as $n(k) \sim1/k^{4+m}$ with $m\approx 4-4.5$ for momenta $k$ up to $(1.59\sim1.97)k_F$ with $k_F=250$ MeV/c. The corresponding NN force from an inverse Fourier transform for $m=4$ and $m=5$ behaves as ${V}_\text{NN}(r)\sim 1/r$ and ${V}_\text{NN}(r)\sim (1/r)^{1/2}$, respectively.
For example, shown in Fig.~\ref{fig5ant} are the scaling functions for different values of $m$ compared with the Relativistic Fermi Gas (RFG) model result and the experimental QE scaling function from ref.~\cite{DS99}.
It is seen that a good agreement with the experimental QE data is achieved with $m \approx  4.5$, indicating a power-law of $n(k)\approx 1/k^{8.5}$. This example also illustrates clearly how the study of HMT can help us probe the
short-range behavior of nuclear forces. It is worth noting that the fraction of HMT nucleons with $n(k) \sim1/k^{8.5}$ and how it compares with the recent data from JLAB as well as the various model calculations we mentioned earlier deserve further investigations.

\begin{figure}[th]
\centering
\includegraphics[width=0.5\textwidth]{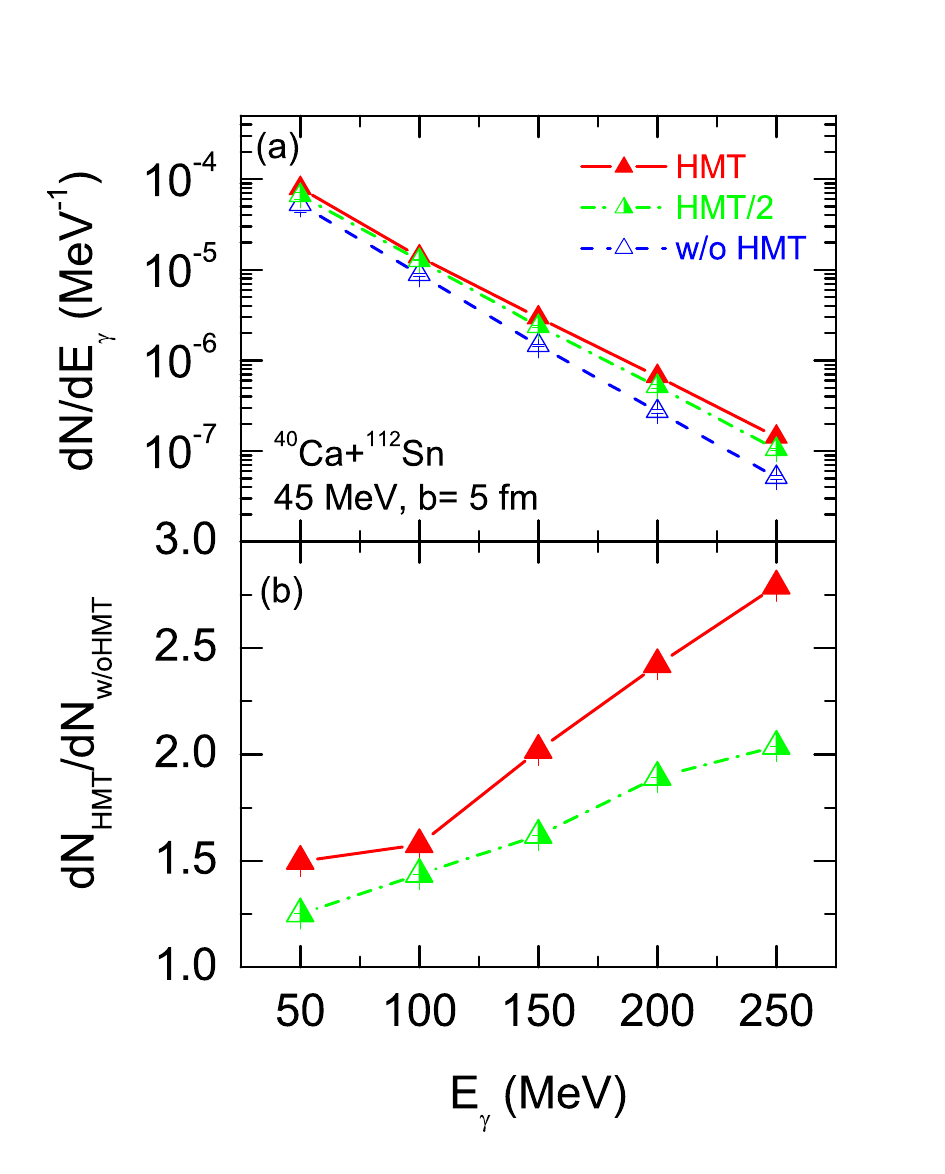}
\caption{Nucleon momentum distribution n(k) of $^{56}_{26}$Fe with normalization condition $\int_{0}^{\lambda k_{F}}n(k)k^{2}dk$ = 1 with different fractions (HMT: $20\%, 10\%$ and 0) of nucleons in the high-momentum tail.
Right upper-window: HMT effects on hard photon productions in $^{40}$Ca + $^{112}$Sn reactions at a beam energy of 45 MeV/nucleon and an impact parameter of 5 fm. Taken from ref. \cite{YongLi}.} \label{YongFig}
\end{figure}
Based on our observations as illustrated above, there appears to be inconsistent findings regarding the shape and range of the HMT in nuclei from analyzing some electron-nucleus scattering data.  Obviously, the discrepancy should be further explored. One interesting proposal put forward recently is to use hard photons from first-chance $p+n\rightarrow p+n+\gamma$ scatterings in heavy-ion reactions at low-intermediate energies to probe the size, shape, range and isospin dependence of HMT in heavy nuclei \cite{YongLi}. Earlier experiments at GSI by the TAPES collaboration have demonstrated that high-energy photons can be measured to a very high precision \cite{TAPS}. Moreover, theoretical studies have shown that the first chance $p+n\rightarrow p+n+\gamma$ scatterings are the main source of hard photons \cite{nif90,cassrp}. The proposed approach has two major advantages. First of all, the FSI suffered by outgoing hadrons has always been a major issue in interpreting accurately most SRC experimental data especially the exclusive ones where two outgoing nucleons are measured in coincidence to study the isospin dependence of the SRC. Since photons only interact with nucleons electromagnetically, it is the most FSI free probe and may thus carry the most reliable information about the HMT in the target and projectile involved in the reaction. Secondly, compared to the reactions induced by a single nucleon, electron or photon on a heavy target, collisions between two heavy nuclei can use constructively the two HMTs existing in both the target and projectile, e.g., two identical HMTs in the $^{208}$Pb target and projectile. More specifically, the center-of-mass energy of two colliding nucleons from the two HMTs will be much higher than those involving nucleons all from below the Fermi surfaces of the two colliding nuclei. These higher energies available will make sub-threshold productions of various particles, such as high-energy photons, pions, kaons, nucleon-antinucleon pairs, etc, possible. In particular, according to the proton-neutron dominance model of the tensor force induced SRC, there are equal numbers of neutrons and protons in the HMT \cite{Hen14}. Since the hard photons are mostly from the first-chance collisions of protons from one nucleus with neutrons from the other nucleus and the elementary cross section for the $p+n\rightarrow p+n+\gamma$ process depends strongly on the neutron-proton relative velocity, their yield is expected to be increased significantly by the two HMTs in the target and projectile. As an example of a preliminary study using an isospin- and momentum-dependent Boltzmann-Uehling-Uhlenbeck transport model (IBUU) \cite{LCK08}, shown in the right upper-panel of Fig.~\ref{YongFig} are hard photon spectra calculated by using the $|\v{k}|^{-4}$ shape but different fractions ($20\%, 10\%$ and 0) of nucleons in the HMT. It is clearly seen that with the HMT, there is a clear increase of hard photon production. More quantitatively, reducing the fraction of HMT nucleons from $20\%$ to $10\%$ leads to a reduction of hard photons around 200 MeV by a factor of 0.7.
These results indicate clearly that nucleon-nucleon collisions involving HMT nucleons play a major role in producing hard photons with energies above 50 MeV. More detailed studies by varying the shape, size, range and isospin dependence of the HMT are necessary to further test the feasibility of using hard photons in heavy-ion reactions at radioactive beam facilities to probe the HMT in neutron-rich matter.

\subsection{SRC-induced reduction of kinetic symmetry energy and enhancement of the isospin-quartic term in the EOS of neutron-rich matter}\label{subx.3}
The modification of the single-nucleon momentum distribution $n_{\v{k}}^J$ due to the SRC-induced HMT has direct consequences on the kinetic EOS of ANM\,\cite{Cai15a} as we shall
discuss in this subsection. The kinetic EOS can be expanded in $\delta$ as
\begin{equation}
E^{\rm{kin}}(\rho,\delta)\approx E_0^{\rm{kin}}(\rho)+E_{\rm{sym}}^{\rm{kin}}(\rho)\delta^2+E_{\rm{sym,4}}^{\rm{kin}}(\rho)\delta^4+\mathcal{O}(\delta^6)
.
\end{equation}
The coefficients evaluated from Eq.\,(\ref{kinE}) using the
$n^J_{\v{k}}(\rho,\delta)$ in Eq.\,(\ref{MDGen}) with $\beta_J=0$ are
\begin{align}
E^{\rm{kin}}_0(\rho)=&\frac{3}{5}E_{\rm{F}}(\rho)\left[
1+{C}_0\left(5\phi_0+\frac{3}{\phi_0}-8\right)\right],\label{E0kin}\\
E_{\rm{sym}}^{\rm{kin}}(\rho)=&\frac{1}{3}E_{\rm{F}}(\rho)\Bigg[1+{C}_0\left(1+3{C}_1\right)\left(5\phi_0+\frac{3}{\phi_0}-8\right)\notag\\
&\hspace*{2cm}+3{C}_0\phi_1\left(1+\frac{3}{5}{C}_1\right)\left(5\phi_0-\frac{3}{\phi_0}\right)+\frac{27{C}_0\phi_1^2}{5\phi_0}\Bigg],\label{Esymkin}\\
E_{\rm{sym,4}}^{\rm{kin}}(\rho)=&\frac{1}{81}E_{\rm{F}}(\rho)\Bigg[1+{C}_0(1-3{C}_1)\left(5\phi_0+\frac{3}{\phi_0}
-8\right)+3{C}_0\phi_1(9{C}_1-1)\left(5\phi_0-\frac{3}{\phi_0}\right)\notag\\
&\hspace*{2.cm}+\frac{81{C}_0\phi_1^2(9\phi_1^2-9{C}_1\phi_1-15\phi_1+15{C}_1+5)}{5\phi_0}
\Bigg].\label{Esymkin4}
\end{align}
Since in the presence of the HMT, $\phi_0>1$ and thus $5\phi_0+3/\phi_0-8>0$,
the kinetic EOS of the SNM is expected to be enhanced compared to the FFG prediction\,\cite{Hen14,CXu11,CXu13}.
This is easy to understand as the neutron-proton SRC is most important in SNM, the HMT enhances very efficiently its kinetic energy through the $k^4$ factor in Eq.\,(\ref{kinE}).
In the FFG where there is no HMT, $\phi_0=1$, $\phi_1=0$ and thus
$5\phi_0+3/\phi_0-8=0$, the above expressions reduce naturally to
the well known results
\begin{equation}E^{\rm{kin}}_0(\rho)=3E_{\rm{F}}(\rho)/5,~~
E_{\rm{sym}}^{\rm{kin}}(\rho)=E_{\rm{F}}(\rho)/3,~~
E_{\rm{sym,4}}^{\rm{kin}}(\rho)=E_{\rm{F}}(\rho)/81
\end{equation}
where $E_{\rm{F}}(\rho)=k_{\rm{F}}^2/2M$ is the Fermi energy.

For the interacting nucleons in ANM with the momentum distribution and
its parameters given earlier, it was found that\,\cite{CaiLi16a}
$E_0^{\rm{kin}}(\rho_0)\approx40.45\pm8.15\,\rm{MeV}$,
$E_{\rm{sym}}^{\rm{kin}}(\rho_0)\approx-13.90\pm11.54\,\rm{MeV}$ and
$E_{\rm{sym,4}}^{\rm{kin}}(\rho_0)\approx7.19\pm2.52\,\rm{MeV}$,
respectively. Relativistic corrections to these values were found small\,\cite{Cai17-rel}.
Compared to the corresponding values for the FFG, it is seen that
the isospin-dependent HMT increases significantly the average
kinetic energy $E_0^{\rm{kin}}(\rho_0)$ of SNM but decreases the
kinetic symmetry energy $E_{\rm{sym}}^{\rm{kin}}(\rho_0)$ of ANM to
a negative value qualitatively consistent with findings of several
recent studies of the kinetic EOS considering short-range
nucleon-nucleon correlations using both phenomenological models and
microscopic many body
theories\,\cite{Rio14,CXu11,CXu13,Vid11,Lov11,Car12,Car14}. However,
it was completely unknown before whether the empirical isospin parabolic
law is still valid for the kinetic EOS of ANM when the
isospin-dependent HMTs are considered. Very surprisingly and
interestingly, it was shown clearly broken seriously\,\cite{CaiLi16a}.
It is very difficult to expect that the corrections in the square brackets for
$E_0^{\rm{kin}}(\rho)$, $E_{\rm{sym}}^{\rm{kin}}$ and $E_{\rm{sym,4}}^{\rm{kin}}$
are small quantities, i.e., the usual
parabolic approximation for the EOS of ANM, even at
the kinetic level, is expected to be broken.
More quantitatively, the ratio
$|E_{\rm{sym,4}}^{\rm{kin}}(\rho_0)/E_{\rm{sym}}^{\rm{kin}}(\rho_0)|$
is about $52\%\pm26$\% that is much larger than the FFG value of
$3.7$\%. It was also found that the large quartic term is mainly due to
the strong isospin dependence of the HMT cutoff described by the $\phi_1$
parameter. For instance, by artificially setting $\phi_1=0$, one shall
obtain $E^{\rm{kin}}_{\rm{sym}}(\rho_0)\approx14.68\pm2.80$\,MeV and
$E^{\rm{kin}}_{\rm{sym},4}(\rho_0)\approx1.12\pm0.27$\,MeV, which are
very close to the corresponding FFG values. In Fig.\,\ref{fig_E2E4withFraction},
the dependence of the kinetic symmetry energy and the isospin-quartic term
on the fractions $x_{\rm{SNM}}^{\rm{HMT}}$ and
$x_{\rm{PNM}}^{\rm{HMT}}$ is shown. It is obvious that a large quartic term is
generated by a large difference between the $x_{\rm{SNM}}^{\rm{HMT}}$ and
$x_{\rm{PNM}}^{\rm{HMT}}$. Considering short-range nucleon-nucleon
correlations but assuming that the isospin parabolic approximation
is still valid, some previous studies have evaluated the kinetic
symmetry energy $E_{\rm{sym}}^{\rm{kin}}$ by simply taking the difference
between the kinetic energy of the PNM and that of the SNM, i.e., subtracting the
$E_{\rm{PNM}}^{\rm{kin}}$ by $E_0^{\rm{kin}}$. This approximation actually
takes the sum of the kinetic symmetry energy and the quartic kinetic energy as the
symmetry energy, i.e.,
$E_{\rm{sym}}^{\rm{kin,appro}}(\rho_0)\approx E_{\rm{sym}}^{\rm{kin}}(\rho_0)+E_{\rm{sym,4}}^{\rm{kin}}(\rho_0)\approx-6.71\pm9.11$\,MeV.
There is thus no surprising that this value is consistent quantitatively with
the $E_{\rm{sym}}^{\rm{kin}}(\rho_0)$ found in, e.g., ref.\,\cite{Hen15b},
under the parabolic approximation.

\begin{figure}[h!]
\centering
  \includegraphics[width=13.cm]{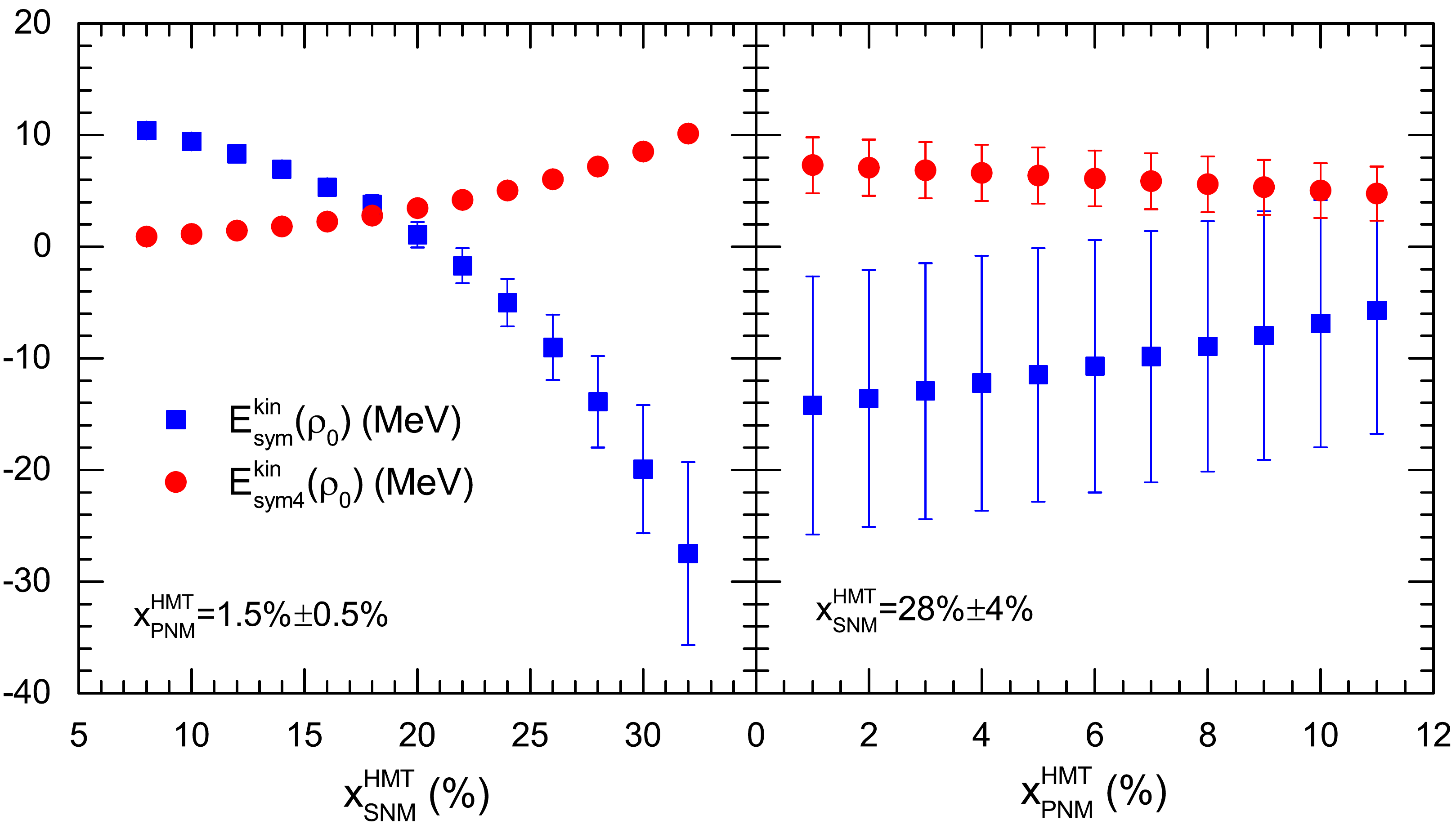}
  \caption{Dependence of the kinetic symmetry energy and the isospin-quartic term on the fractions of high-momentum nucleons in symmetric nuclear matter ($x_{\rm{SNM}}^{\rm{HMT}}$) and
  pure neutron matter  ($x_{\rm{PNM}}^{\rm{HMT}}$).}
  \label{fig_E2E4withFraction}
\end{figure}

As we have discussed earlier, using the HVH theorem valid at the mean-field level, an appreciable isospin-quartic term in the EOS may appear depending on the high-order derivatives of the momentum-dependent symmetry potential or the difference between the nucleon isoscalar and isovector effective masses. Here we have seen that the SRC modifies the momentum distribution of nucleons such that these quasi-particles have a new momentum distribution that is different from the usual step function of the FFG. As the SRC is induced mainly by the tensor force, one can also attribute the reduction of kinetic symmetry energy and the enhancement of isospin-quartic term for quasi-particles to tensor force which does not contribute to the mean-field of spin saturated systems. Thus, the SRC-induced modifications of the kinetic symmetry energy and isospin-quartic term seem to be additions to the EOS evaluated at the mean-field level. However, more quantitative studies remain to be done.

\begin{figure}[h!]
\centering
  \includegraphics[width=9.cm]{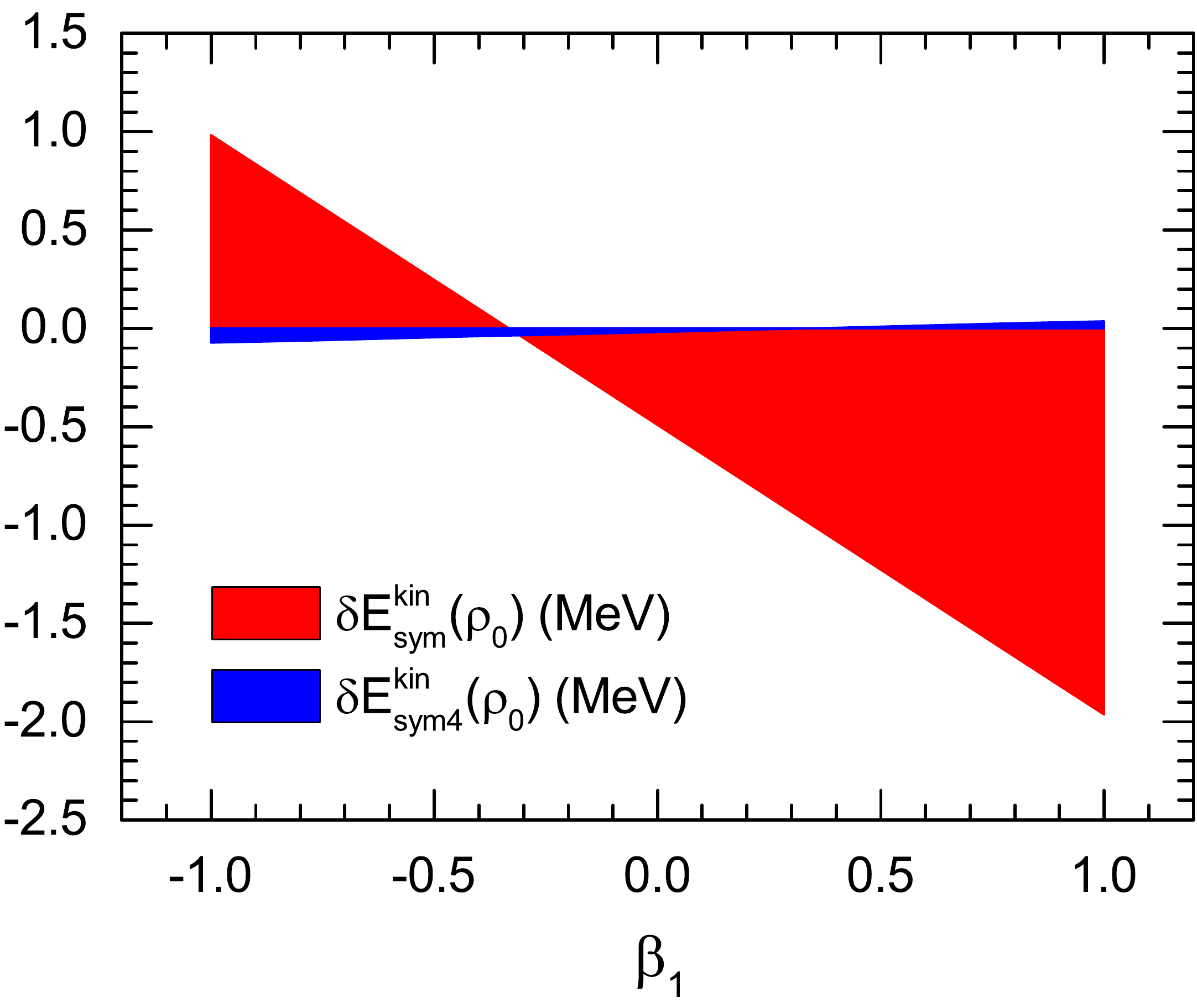}\hspace*{1.cm}
  \caption{Corrections to the $E_{\rm{sym}}^{\rm{kin}}(\rho_0)$ and $E_{\rm{sym,4}}^{\rm{kin}}(\rho_0)
  $ as functions of $\beta_1$ with $\beta_0=-0.35$. Taken from ref.\,\cite{Cai15a}.}
  \label{fig_betaJeffect}
\end{figure}

Most of the phenomenological models describe the depletion of the Fermi sea with a constant reduction factor $\Delta_J$ from 1 as illustrated in Fig.\,\ref{fig_SF}.
In the following, we briefly discuss effects of the two parameters in the $\beta_J=\beta_0(1+\tau_3^J\beta_1\delta)$ used in the
$n^J_{\v{k}}(\rho,\delta)$ of Eq.\,(\ref{MDGen}).  They were used to characterize the depletion of the Fermi sea at finite momenta below/around the Fermi surface and can affect the E-mass to be extracted using the
Migdal--Luttinger theorem. To estimate their effects,
one can consider a widely used single-nucleon momentum distribution
parameterized in ref.\,\cite{Cio96} based on calculations using many-body theories.
More specifically, for $|\v{k}|\lesssim2\,\rm{fm}^{-1}$,
the function $I$ goes like $\sim e^{-\alpha|\v{k}|^2}$ with $\alpha\approx 0.12\,\rm{fm}^2$.
At $\rho_0$ since $\alpha k_{\rm{F}}^2\approx0.21\ll1$, it is good enough to use the approximation $e^{-\alpha|\v{k}|^2}\approx1-\alpha|\v{k}|^2
+\mathcal{O}(|\v{k}|^4)$, in the range of $0<|\v{k}|<k_{\rm{F}}^J$.
Thus, a quadratic function
\begin{equation}
{I}\left(\frac{|\v{k}|}{k_{\rm{F}}^J}\right)=\left(\frac{|\v{k}|}{k_{\rm{F}}^J}\right)^2
\end{equation} is effective and reasonable to
describe the depletion of the Fermi sea at finite momenta. The constants
in the parametrization of ref.\,\cite{Cio96} are absorbed here into the
parameters $\Delta_J$ and $\beta_J$. Then, the Eq.\,(\ref{DeltaJ}) gives
$\Delta_J=1-3\beta_J/5-3{C}_J\left(1-1/\phi_J\right)$. Specifically,
one can obtain
$\beta_0=(5/3)[1-\Delta_0-3C_0(1-\phi_0^{-1})]=(5/3)[1-\Delta_0-x_{\rm{SNM}}^{\rm{HMT}}]$
for SNM. Then, using the predicted value  of $\Delta_0\approx
0.88\pm0.03$\,\cite{Pan99,Fan84,Yin13} and the experimental value of
$x_{\rm{SNM}}^{\rm{HMT}}\approx 0.28\pm 0.04$, the value of
$\beta_0$ is estimated to be about $-0.27\pm0.08$. Similarly, the
condition $\beta_J=\beta_0(1+\beta_1\tau_3^J\delta)<0$, i.e.,
$n_{\v{k}}^J$ is a decreasing function of momentum towards
$k_{\rm{F}}^J$, indicates that $|\beta_1|\leq1$.

A finite value of $\beta_J$ is expected to affect the
``renormalization function"  $Z^J_{\rm{F}}$. For SNM, one has
$Z_{\rm{F}}^0=1+2\beta_0/5-C_0-x_{\rm{SNM}}^{\rm{HMT}}\approx0.45\pm0.07$
($0.56\pm0.04$) in the presence (absence) of $\beta_0$. For ANM,
however, the $Z^J_{\rm{F}}$ depends on the less constrained value of
$\beta_1$. It is worth noting that $\beta_1$ also strongly determines the
neutron-proton effective E-mass splitting. Contributions from a finite $\beta_J$ to the first three
terms of the kinetic EOS are
\begin{align}
\delta
E_{0}^{\rm{kin}}(\rho)=&\frac{3}{5}E_{\rm{F}}(\rho_0)\cdot\frac{4\beta_0}{35},\\
\delta
E_{\rm{sym}}^{\rm{kin}}(\rho)=&\frac{1}{3}E_{\rm{F}}(\rho_0)\cdot\frac{4\beta_0(1+3\beta_1)}{35},\\
\delta
E_{\rm{sym,4}}^{\rm{kin}}(\rho)=&\frac{1}{81}E_{\rm{F}}(\rho_0)\cdot\frac{4\beta_0(1-3\beta_1)}{35}.
\end{align}
With the largest magnitude of $\beta_0\approx-0.35$, the corrections to the
$E_{\rm{sym}}^{\rm{kin}}(\rho_0)$ and
$E_{\rm{sym,4}}^{\rm{kin}}(\rho_0)$ as functions of $\beta_1$ in its
full range allowed are shown in the left window of
Fig.\,\ref{fig_betaJeffect}. Here the maximum effects of the finite
$\beta_J$ are explored. It is seen that the correction on the
$E_{\rm{sym,4}}^{\rm{kin}}(\rho_0)$ is negligible while that on the $E_{\rm{sym}}^{\rm{kin}}(\rho_0)$ is less than
2\,MeV.
Considering the corrections due to the finite $\beta_0$ and
$\beta_1$ and their uncertainties, one finally obtain
$E_0^{\rm{kin}}(\rho_0)\approx39.77\pm8.13\,\rm{MeV}$,
$E_{\rm{sym}}^{\rm{kin}}(\rho_0)\approx-14.28\pm11.59\,\rm{MeV}$ and
$E_{\rm{sym,4}}^{\rm{kin}}(\rho_0)\approx7.18\pm2.52\,\rm{MeV}$,
respectively.  Moreover, the $\delta^6$ term was also
consistently evaluated and found egligible at
the saturation density, i.e., including the quartic term is good
enough to describe the kinetic EOS of ANM perturbatively.
It is necessary to point out the limitations of the above approach. Since the parameters of the
nucleon momentum distribution (Eq.\,(\ref{MDGen})) were fixed by using
experimental data and/or model calculations at the saturation density, the possible density dependence of these parameters has not been explored yet. The density dependence of the various terms in the
kinetic EOS is thus only due to that of the Fermi energy as shown in
Eqs.\,(\ref{E0kin})-(\ref{Esymkin4}). In this limiting case, the slope parameter
of the kinetic symmetry energy itself, i.e., $L^{\rm{kin}}=3\rho_0dE_{\rm{sym}}^{\rm{kin}}(\rho)/d\rho|_{\rho=\rho_0}\approx-27.81\pm
23.08$\,MeV while that of the FFG is about 25.04\,MeV.

\begin{figure}[h!]
\centering
  \includegraphics[width=9.cm]{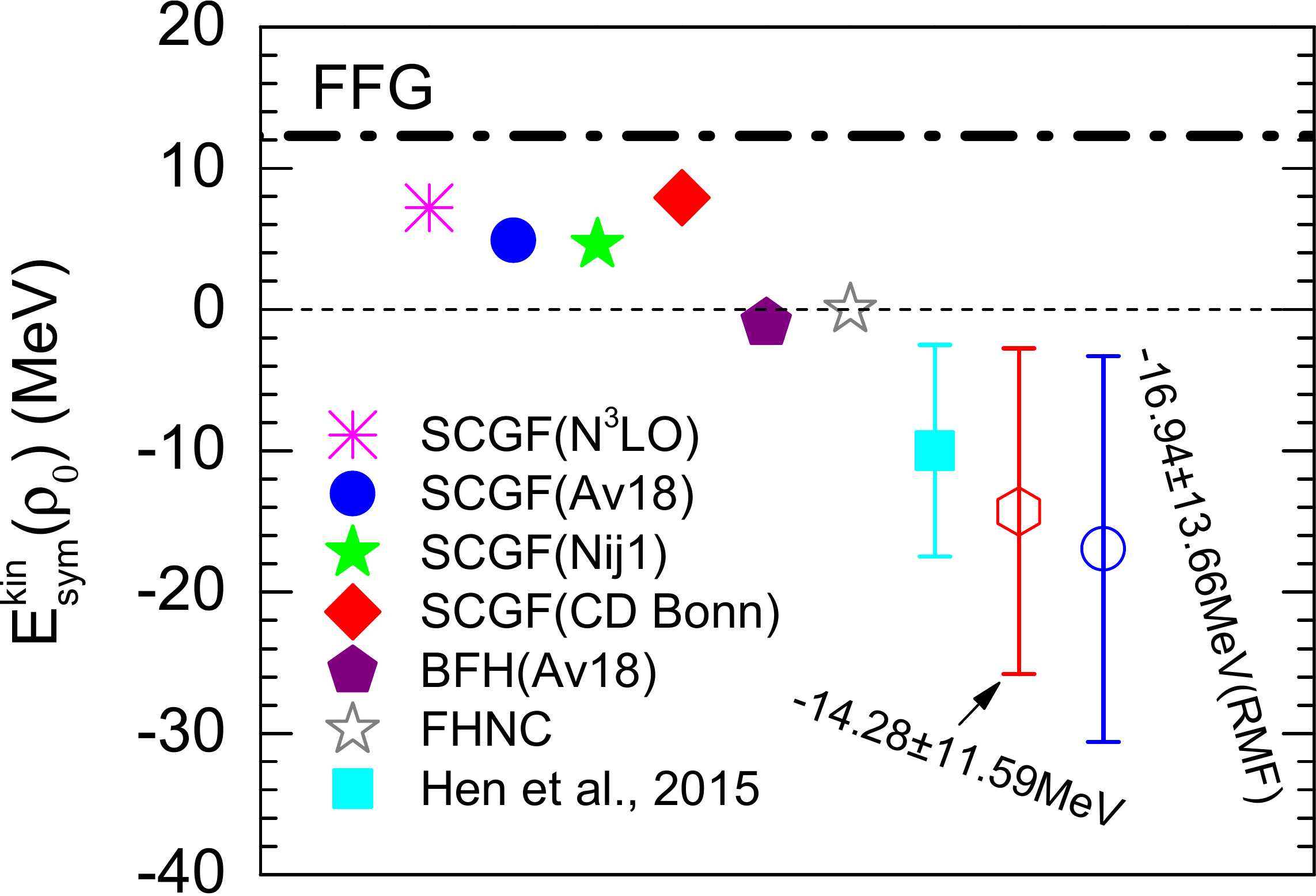}
  \caption{SRC-induced reductions of nucleon kinetic symmetry energy from several models with respect to the value (12.3 MeV) for the free Fermi gas (FFG).}
  \label{fig_esym0}
\end{figure}

It is important to emphasize that the SRC-induced reduction of the kinetic symmetry energy with respect to the FFG value of about 12.3 MeV at saturation density is commonly seen in many-body theories where the SRC effects
are considered. Shown in Fig.\,\ref{fig_esym0} are comparisons of the kinetic symmetry energy from the above analysis and those predicted by several microscopic many-body
theories and phenomenological models including the SCGF using different interactions (N$^3$LO, Av18, Nij1, and CD Bonn)\,\cite{Car12}, BHF approach using the Av18 potential plus the Urbana IX three-body force (TBF)\,\cite{Vid11}, the Fermi hypernetted chain (FHNC) method\,\cite{Lov11}, and a phenomenological parametrization of ref.\,\cite{Hen15b}.
As mentioned earlier, the kinetic symmetry energy obtained
in ref.\,\cite{Car12} and refs.\,\cite{Hen15b,Vid11,Lov11} are actually approximated by the difference between the $E_{\rm{PNM}}^{\rm{kin}}$
and $E_0^{\rm{kin}}$ using the parabolic approximation of the EOS, or, equivalently, the sum of the kinetic symmetry
energy and the kinetic quartic term from ref.\,\cite{Cai15a}. They all agree qualitatively that the kinetic symmetry energy is reduced by the SRC but disagree quantitatively.

\begin{figure}[h!]
\centering
  \includegraphics[width=9.cm]{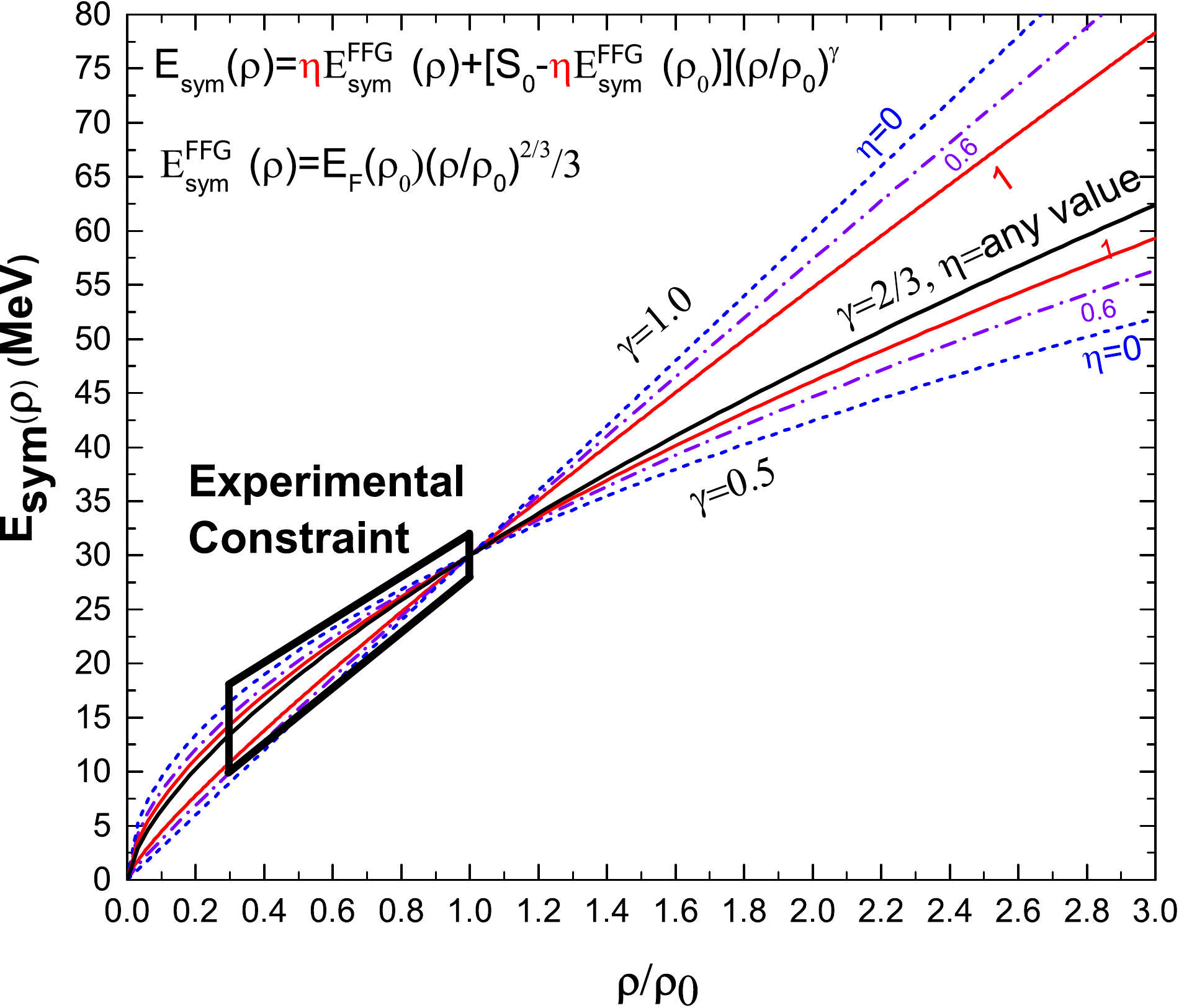}
  \caption{Freedom of diving total symmetry energy into its kinetic and potential parts within the existing constraints. Taken from ref. \cite{Li15}.}
  \label{Esym-eta}
\end{figure}
The SRC-reduced kinetic symmetry energy with respect to the FFG
prediction has been found to affect significantly not only the
understanding about the origin of the symmetry energy but also
several isovector observables, such as the free neutron/proton and
$\pi^-/\pi^+$ ratios in heavy-ion collisions\,\cite{Hen15b,Li15,Yong1,Yong2}, as well
as some properties of neutron stars \cite{HenSteiner}. In both transport model simulations of heavy-ion reactions and constructing phenomenological EOS for astrophysical studies, it is customary to parameterize the symmetry
energy as a sum of the kinetic symmetry energy of a FFG and an interaction contribution as
\begin{equation}\label{esym}
E_{\rm{sym}}(\rho)=E_{\rm{sym}}^{\rm{kin}}(\textrm{FFG})(\rho)+\left[S_0-E_{\rm{sym}}^{\rm{kin}}(\textrm{FFG})(\rho_0)\right]\left(\frac{\rho}{\rho_0}\right)^{\gamma}
\end{equation}
where $E_{\rm{sym}}^{\rm{kin}}(\textrm{FFG})(\rho)\equiv(2^{{2}/{3}}-1)({3}/{5})E_{\rm{F}}(\rho)\approx 12.3(\rho/\rho_0)^{2/3}\equiv E^{\rm{kin}}_{\rm{sym}}(\rho)$,
$S_0$ is the total symmetry energy at $\rho_0$ and $\gamma$ is a free parameter that one hopes to extract from fitting isospin-sensitive observables.
Normally, the standard value of $E_{\rm{sym}}^{\rm{kin}}(\textrm{FG})(\rho_0)=12.3$ MeV is used. Our discussions above put this practice into serious question especially in simulating heavy-ion reactions. For calculating global properties of neutron stars, it is the total pressure and energy density that go into solving the TOV equation, it thus does not matter much how they are divided. For solving the transport equation of quasi-nucleons for heavy-ion reactions, however, only the potential part of the symmetry energy is a direct input needed. Under the condition that the $S_0$ is a constant consistent with existing experimental constraints, what value one uses for the kinetic symmetry energy is important. It affects what one extracts for the $\gamma$ parameter from fitting experimental data. Unfortunately, the currently existing constraints on the symmetry energy around and below
$\rho_0$ do not limit the division of the total symmetry energy as illustrated in Fig. \ref{Esym-eta}. To consider the possibility that the kinetic symmetry energy of quasi-nucleons is reduced with respect to the FFG value, one can introduce a reduction factor $\eta$ and rewrite the $E_{\rm{sym}}(\rho)$ as
\begin{equation}\label{esym}
E_{\rm{sym}}(\rho)=\eta\cdot E_{\rm{sym}}^{\rm{kin}}(\textrm{FFG})(\rho)+\left[S_0-\eta\cdot E_{\rm{sym}}^{\rm{kin}}(\textrm{FFG})(\rho_0)\right]\left(\frac{\rho}{\rho_0}\right)^{\gamma}.
\end{equation}
In fact, since $E_{\rm{sym}}^{\rm{kin}}(\textrm{FFG})(\rho)$ goes as $(\rho/\rho_0)^{2/3}$, $\eta$ can be anything when $\gamma=2/3$ with a total symmetry energy of $E_{\rm{sym}}(\rho)=S_0\cdot (\rho/\rho_0)^{2/3}$. For other values of $\gamma$ between 0.5 to 1, all resulting $E_{\rm{sym}}(\rho)$ functions with $\eta$ between 0 and 1 considered in ref. \cite{Li15} pass the existing constraints around and below $\rho_0$. Thus, to our best knowledge, nothing forbids
the reduction of kinetic symmetry energy of quasi-nucleons from the value for a FFG.

So far, no investigation on possible effects of a large isospin quartic term in the EOS of ANM on heavy-ion collisions has been done yet while its effects on properties of neutron stars, especially the crust-core transition density and pressure, have been studied extensively. This is mainly because of the very small isospin asymmetry normally reached in heavy-ion collisions.
Of course, effects of the quartic and quadratic
terms should be studied together within the same approach. To
extract from nuclear reactions and neutron stars information about
the EOS of neutron-rich nucleonic matter, one often parameterizes the EOS as
a sum of the kinetic energy of a FFG and a potential energy
involving unknown parameters up to the isospin-quadratic term only.
The findings obtained in ref.\,\cite{Cai15a} and the discussions above indicate that it is important to
include the isospin-quartic term in both the kinetic and potential
parts of the EOS. Moreover, to accurately extract the completely
unknown isospin-quartic term $E^{\rm{pot}}_{\rm{sym,4}}(\rho)\delta^4$ in the potential EOS, it is
important to use the kinetic EOS of quasi-particles with reduced
kinetic symmetry energy and an enhanced quartic term due to the
isospin-dependence of the HMT. Most relevant to the isovector
observables in heavy-ion collisions, such as the neutron-proton
ratio and differential flow, is the nucleon isovector potential.
Besides the Lane potential $\pm 2\rho
E^{\rm{pot}}_{\rm{sym}}(\rho)\delta$ where the
$E^{\rm{pot}}_{\rm{sym}}(\rho)$ is the potential part of the
symmetry energy and the $\pm$ sign is for neutrons/protons, the
$E^{\rm{pot}}_{\rm{sym,4}}(\rho)\delta^4$ term contributes an
additional isovector term  $\pm 4 \rho
E^{\rm{pot}}_{\rm{sym,4}}(\rho)\delta^3$. In neutron-rich systems
besides neutron stars, such as nuclear reactions induced by rare
isotopes and peripheral collisions between two heavy nuclei having
thick neutron-skins, the latter may play a significant role in
understanding the isovector observables or extracting the sizes of
neutron-skins of the neutron-rich nuclei involved.

\subsection{SRC effects on the density dependence of nuclear symmetry energy within energy density functionals}
Here we examine how the SRC can modify simultaneously both the kinetic and potential parts of the EOS of ANM, especially the density dependence of nuclear symmetry energy within a nonlinear RMF model and
then a modified-Gogny energy functional. To include the SRC-induced HMT in RMF models,  the kinetic energy density and the corresponding pressure were modified according to\,\cite{Cai16c}
\begin{align}
\varepsilon^{\rm{kin}}_{J} =&\frac{2}{(2\pi )^{3}}\int_{0}^{k_{\textrm{F}}^{J}}\Delta_J\textrm{d}%
\textbf{k}\sqrt{|\textbf{k}|^{2}+{M_J^{\ast
2}}}+\frac{2}{(2\pi )^{3}}\int_{k_{\rm{F}}^J}^{\phi_Jk_{\textrm{F}}^{J}}C_J\left(\frac{k_{\rm{F}}^J}{|\v{k}|}\right)^4\textrm{d}%
\textbf{k}\sqrt{|\textbf{k}|^{2}+{M_J^{\ast 2}}},\label{EnDenKin}\\
P_{\rm{kin}}^{J}=&\frac{1}{3\pi ^{2}}\int_{0}^{k_{\textrm{F}}^{J}}\Delta_J\textrm{d}k\frac{k^{4}}{%
\sqrt{k^{2}+{M_J^{\ast}}^{2}}}+\frac{1}{3\pi ^{2}}\int_{k_{\rm{F}}^J}^{\phi_Jk_{\textrm{F}}^{J}}C_J\left(\frac{k_{\rm{F}}^J}{k}\right)^4\textrm{d}k\frac{k^{4}}{%
\sqrt{k^{2}+{M_J^{\ast}}^{2}}}. \label{pressureKin}
\end{align}
Generally, for any function $f$ at zero temperature when the step function for the momentum distribution in FFG is replaced by the two parts of the $n_{\v{k}}^J\,(\rm{HMT})$ below and above the Fermi surface, one has to
make the following replacement
\begin{equation}
\int_0^{k_{\rm{F}}^J}(\rm{FFG step
function})f\d\v{k}\longrightarrow\int_0^{\phi_Jk_{\rm{F}}^J}n_{\v{k}}^J\,(\rm{HMT})f\d\v{k}.
\end{equation}
It is necessary to point out here an inconsistency of this hybrid approach, namely,
the RMF model itself can not self-consistently reproduce
the phenomenological $n_{\v{k}}^J$ in Eq.(\ref{MDGen}) constrained by the experiments we discussed earlier. In fact, this
inconsistency exists in almost all phenomenological mean-field
models. Ideally, one should first reproduce quantitatively and self-consistently the
experimentally constrained $n_{\v{k}}^J$ by adjusting parameters in
the model Lagrangian. Unfortunately, as shown by the strong model
dependence in predicting the $n_{\v{k}}^J$ using various models and
interactions, our poor knowledge on the isospin dependence of SRC
still hinders reproducing quantitatively the experimentally constrained
$n_{\v{k}}^J$ within some models. Thus, the hybrid approach of using directly the
phenomenologically constrained $n_{\v{k}}^J$ and readjusting only the interaction parts of the EOS can give
us some useful perspectives on the effects of the SRC on the EOS of
ANM while keeping in mind the shortcomings of the approach.

The kinetic symmetry energy $E_{\rm{sym}}^{\rm{kin}}(\rho)$ in the RMF model with HMT is\,\cite{Cai16c}
\begin{align}
E_{\rm{sym}}^{\rm{kin}}(\rho)=&\frac{k_{\rm{F}}^2}{6E_{\rm{F}}^{\ast}}\left[1-3C_0\left(1-\frac{1}{\phi_0}\right)\right]
-3E_{\rm{F}}^{\ast}C_0\left[C_1\left(1-\frac{1}{\phi_0}\right)+\frac{\phi_1}{\phi_0}\right]\notag\\
&-\frac{9M_0^{\ast,4}}{8k_{\rm{F}}^3}\frac{C_0\phi_1(C_1-\phi_1)}{\phi_0}
\left[\frac{2k_{\rm{F}}}{M_0^{\ast}}\left(\left(\frac{k_{\rm{F}}}{M_0^{\ast}}\right)^2+1\right)^{3/2}\right.
\left.-\frac{k_{\rm{F}}}{M_0^{\ast}}\left(\left(\frac{k_{\rm{F}}}{M_0^{\ast}}\right)^2+1\right)^{1/2}-\rm{arcsinh}\left(\frac{k_{\rm{F}}}{M_0^{\ast}}\right)\right]\notag\\
&+\frac{2k_{\rm{F}}C_0(6C_1+1)}{3}\left[\rm{arcsinh}\left(\frac{\phi_0k_{\rm{F}}}{M_0^{\ast}}\right)-
\sqrt{1+\left(\frac{M_0^{\ast}}{\phi_0k_{\rm{F}}}\right)^2}\right.
\left.-\rm{arcsinh}\left(\frac{k_{\rm{F}}}{M_0^{\ast}}\right)+
\sqrt{1+\left(\frac{M_0^{\ast}}{k_{\rm{F}}}\right)^2}\right]\notag\\
&+\frac{3k_{\rm{F}}C_0}{2}\Bigg[\frac{(1+3\phi_1)^2}{9}\left(\frac{\phi_0k_{\rm{F}}}{F_{\rm{F}}^{\ast}}
-\frac{2F_{\rm{F}}^{\ast}}{\phi_0k_{\rm{F}}}\right)
+\frac{2F_{\rm{F}}^{\ast}(3\phi_1-1)}{9\phi_0k_{\rm{F}}}
-\frac{1}{9}\frac{k_{\rm{F}}}{E_{\rm{F}}^{\ast}}+\frac{4E_{\rm{F}}^{\ast}}{9k_{\rm{F}}}\Bigg]\notag\\
&+\frac{C_0(4+3C_1)}{3}\left[\frac{F_{\rm{F}}^{\ast}(1+3\phi_1)}{\phi_0}-E_{\rm{F}}^{\ast}\right]\label{EsymkinHMT}
\end{align}
where $E_{\rm{F}}^{\ast}=(M_0^{\ast,2}+k_{\rm{F}}^2)^{1/2}$ and
$F_{\rm{F}}^{\ast}=(M_0^{\ast,2}+(\phi_0k_{\rm{F}})^2)^{1/2}$, $M_0^{\ast}=M-g_{\sigma}\sigma$ is the Dirac effective mass of a nucleon in SNM with $g_{\sigma}
$ the coupling constant between the $\sigma$ meson and the nucleon. 
\begin{figure}[h!]
\centering
  \includegraphics[width=8.cm]{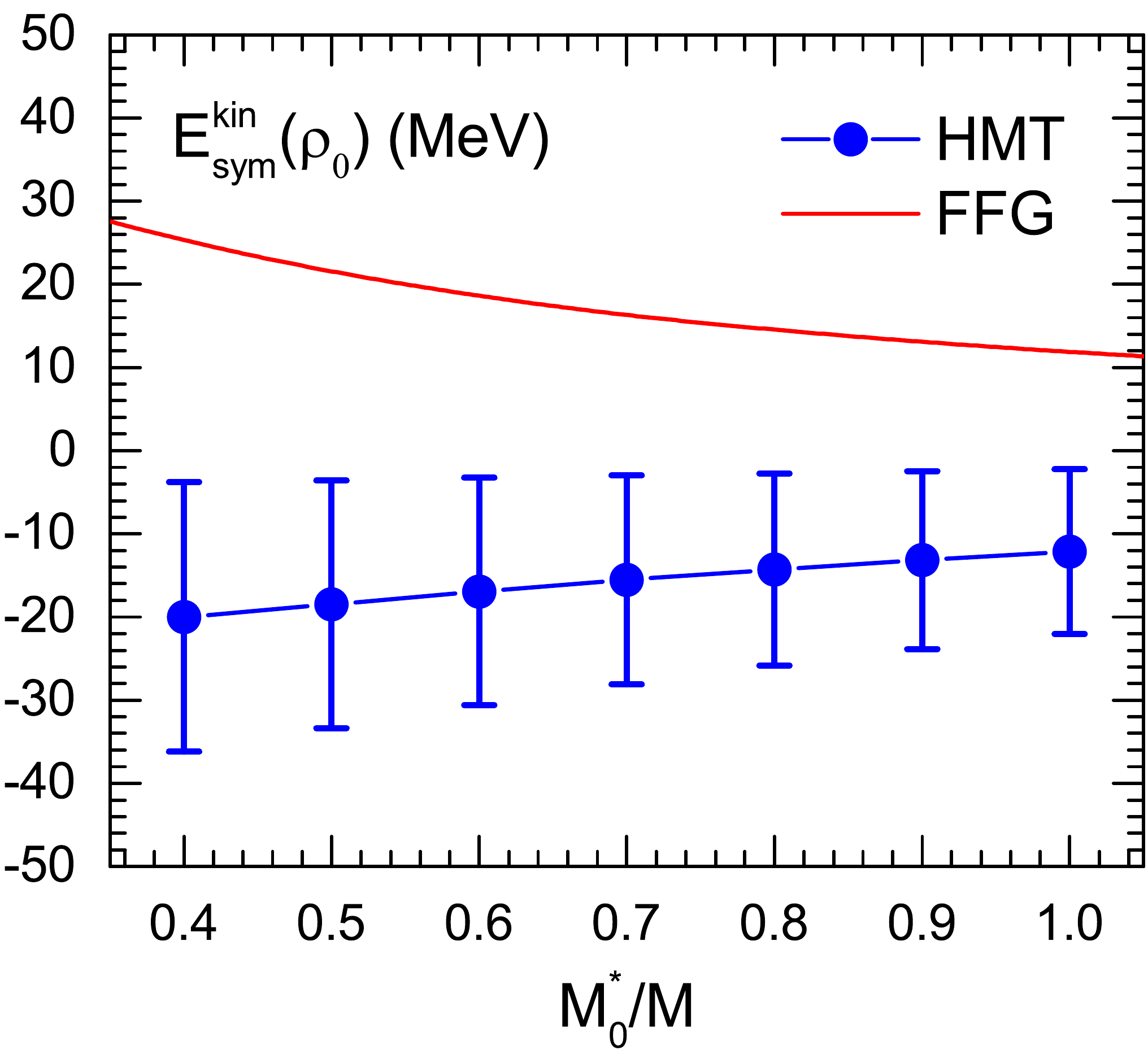}
  \caption{Kinetic symmetry energy as a function of Dirac effective mass of nucleons in SNM both in the RMF-FFG model and RMF-HMT model at saturation density.
 Taken from ref.\,\cite{Cai16c}.}\label{fig_xCaiFig2}
\end{figure}
In the FFG limit, $\phi_0=1,\phi_1=0$, only the first term of the above
expression survives and leads to $ E_{\rm{sym}}^{\rm{kin}}(\rho)\to
E_{\rm{sym}}^{\rm{kin}}(\rho)\equiv k_{\rm{F}}^2/6E_{\rm{F}}^{\ast}
$ as in traditional RMF models. The kinetic symmetry energy in the
presence of HMT is a function only of the Dirac effective mass
$M_0^{\ast}$. Using the values of $C_0,C_1,\phi_0$ and $\phi_1$
given earlier, as shown in Fig.\,\ref{fig_xCaiFig2}, the kinetic symmetry energy with HMT is significantly smaller than that of the FFG in the whole range of $M_0^{\ast}$ considered as
reasonable. For example, with $M_0^{\ast}/M=0.6$ the kinetic
symmetry energy is $E_{\rm{sym}}^{\rm{kin}}(\rho_0)\approx-16.94\pm13.66\,\rm{MeV}$.
This value is close to the non-relativistic result of
$E_{\rm{sym}}^{\rm{kin}}(\rho_0)\approx-13.90\pm11.54\,\rm{MeV}$ (and $-14.28\pm11.59$ including the effects of $\beta_0$ and $\beta_1$). Thus, the reduction of the kinetic symmetry
energy in the presence of HMT is general in both relativistic and
non-relativistic calculations.

\begin{figure}[h!]
\centering
  \includegraphics[width=7.5cm]{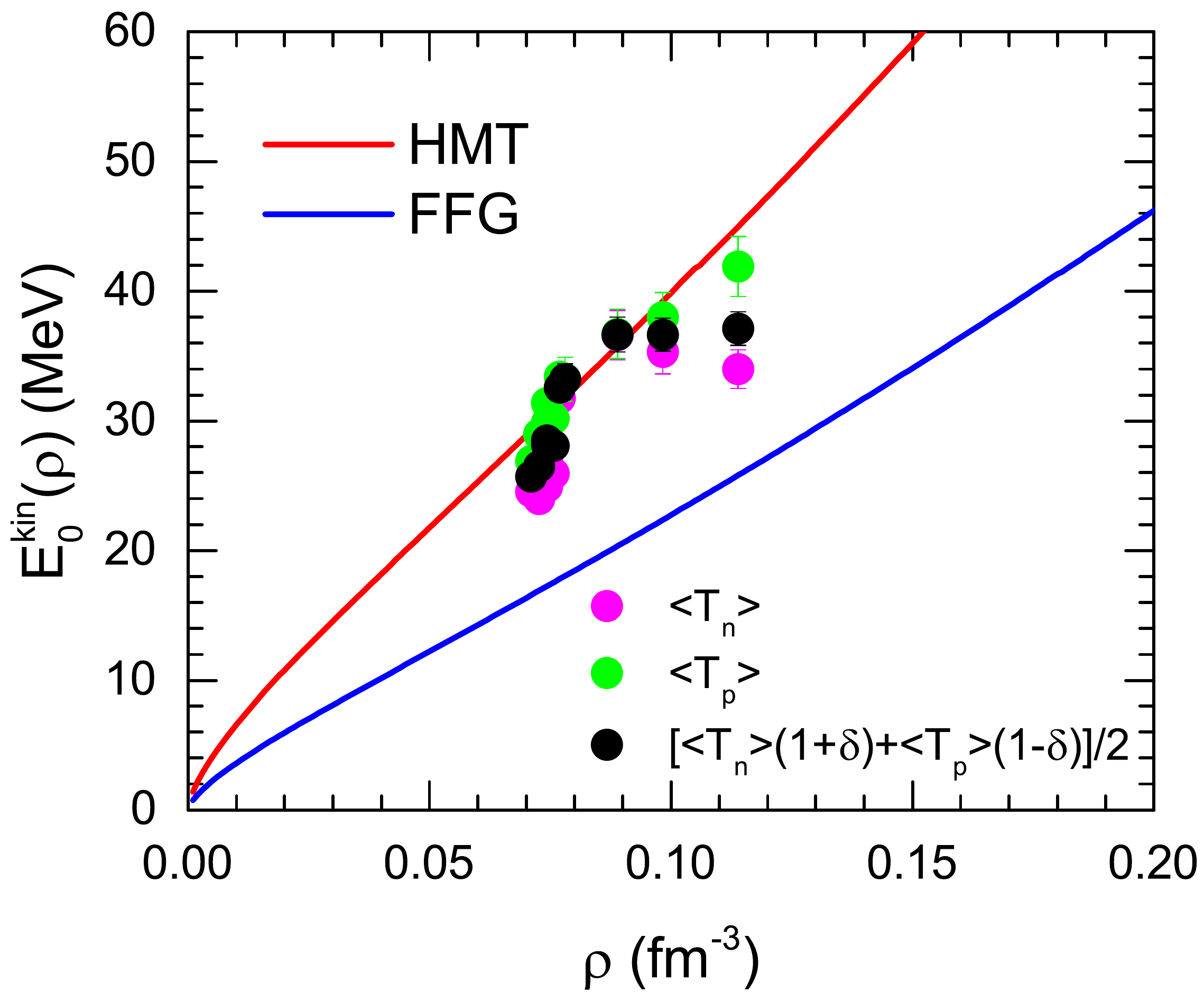}
  \hspace{1cm}
   \includegraphics[width=6.5cm,angle=-90]{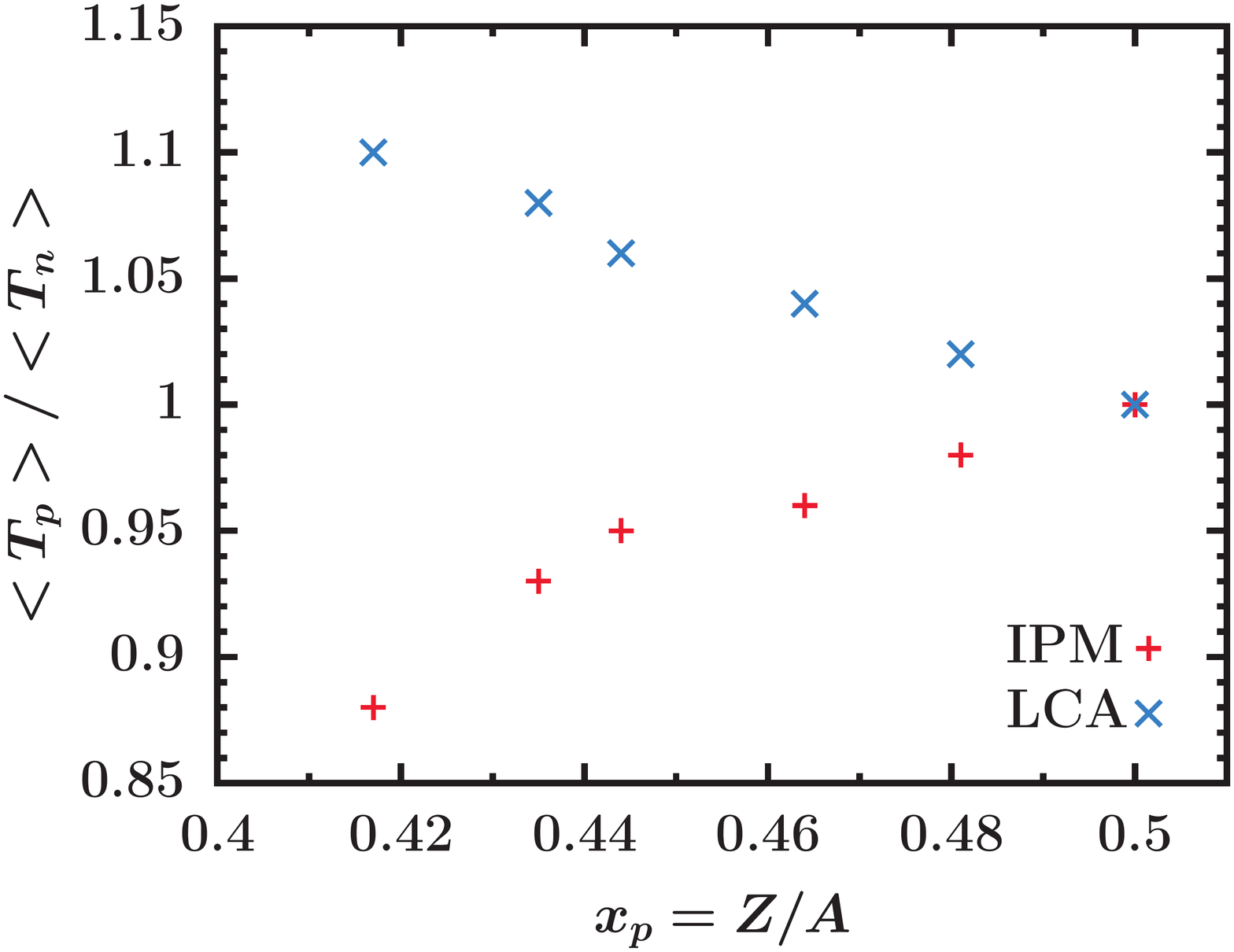}
  \caption{Upper: Kinetic EOS of SNM defined by Eq. (\ref{E0kin-RMF}).
The average kinetic energy of neutrons and protons in C, Al, Fe
and Pb with error bars\,\cite{Hen14} and $^{7,8,9}\textrm{Li}$,
$^{9,10}\textrm{Be}$ and $^{11}\textrm{B}$\,\cite{Sar14} were extracted from data analysis using the neutron-proton dominance model. The lines are results of the RMF-HMT and RMF-FFG calculations\,\cite{Cai16c}.
Lower: Ratio of the average kinetic energy of protons to that of neutrons as a function of proton fraction $x_{\rm{p}}=Z/A$ within the independent particle model (IPM) and
the low order correlation operator approximation (LCA). Taken from ref.\,\cite{Ryc15}.}
  \label{fig-Ekin0RMFHMT}
\end{figure}

One can also gain some confidence about the hybrid approach used here by comparing its prediction for the kinetic EOS with those by other models\,\cite{Ryc15,Sar14}.
In the nonlinear RMF model, the kinetic EOS of SNM is defined as
\begin{equation}
E_0^{\rm{kin}}(\rho)\equiv\frac{1}{\rho}\frac{2}{(2\pi)^3}\int_0^{\phi_0k_{\rm{F}}}n_{\v{k}}^0\sqrt{\v{k}^2+M_0^{\ast,2}}\d\v{k}-M_0^{\ast},\label{E0kin-RMF}
\end{equation}
where $n_{\v{k}}^0$ is the momentum distribution of nucleons in SNM.
The HMT and FFG model predictions for the $E_0^{\rm{kin}}(\rho)$ are
shown in the upper window of Fig.\,\ref{fig-Ekin0RMFHMT}. Recently, the average kinetic
energy of neutrons and protons in C, Al, Fe and Pb with error bars
as well as $^{7,8,9}\textrm{Li}$, $^{9,10}\textrm{Be}$ and
$^{11}\textrm{B}$ without error bars were extracted from analyzing several
electron-nucleus scattering experiments using a neutron-proton
dominance model\,\cite{Hen14,Sar14}. The mass number $A$-dependence of the nucleon kinetic energy can be translated into its density
dependence through a well-established empirical
relationship\,\cite{Cen09,Che11a,Maz13,Mye69,Dan03,Dan09,Dan14}
\begin{equation}
\rho_{A}\approx\frac{\rho_0}{1+\alpha/A^{1/3}}
\end{equation}
where $\alpha$ reflects the balance between the volume and surface
symmetry energies. Using $\alpha\approx2.8$\,\cite{Dan03} appropriate for the mass range
considered, the average kinetic energy per nucleon, i.e., $\langle T\rangle=[\langle
T_{\rm{n}}\rangle(1+\delta)+\langle T_{\rm{p}}\rangle(1-\delta)]/2$, calculated with the HMT and FFG are compared with that from the neutron-proton dominance model in the upper window of Fig.\,\ref{fig-Ekin0RMFHMT}.
According to the parabolic approximation for the EOS of ANM, i.e.,
$E^{\rm{kin}}_{\rm{ANM}}(\rho)\approx
E^{\rm{kin}}_{0}(\rho)+\delta^2E^{\rm{kin}}_{\rm{sym}}(\rho)$, even
for the most neutron-rich nucleus considered $^{208}\rm{Pb}$ with an
isospin asymmetry $\delta^2\approx0.045$, we still have
$E^{\rm{kin}}_{\rm{ANM}}(\rho)\approx E_{0}^{\rm{kin}}(\rho)$. This
means that the results shown in the upper window of Fig.\,\ref{fig-Ekin0RMFHMT} are approximately
equal to the kinetic EOS $E_{0}^{\rm{kin}}(\rho)$ of SNM.
It is interesting to see that the RMF-HMT result is consistent with the predictions of the neutron-proton dominance model
while the FFG prediction falls about 40\% below.
As indicated in right window of Fig. \ref{fig_SF}, pure mean-field models fail to describe the
spectroscopic factors extracted from electron scatterings on nuclei
from $^{7}$Li to $^{208}$Pb by about 30-40\% due to the lack of
occupations of energetic orbitals in these models where the
SRC effects are not considered\,\cite{Lap93}. The fact that the RMF-FFG model under-predicts the average
nucleon kinetic energy compared to the RMF-HMT is due to the same physical reason.

There is also an interesting common feature\,\cite{Ryc15,Hen14} indicating that protons move faster than neutrons in neutron-rich nuclei especially in neutron-skins of heavy nuclei\,\cite{Cai16b}.
As shown in the upper window of Fig.\,\ref{fig-Ekin0RMFHMT}, the proton-neutron dominance model predicts that the average kinetic energy of protons is higher than that of neutrons and their difference increases with the
isospin asymmetry of the nucleus considered. In ref.\,\cite{Ryc15}, the average kinetic energy of nucleons were calculated via a low order correlation operator approximation (LCA) where one can separate contributions of the central, spin-isospin and tensor correlations and study how they affect the relative strength of nn, pp and pn pairs in the HMT.
For instance, for $1.5\lesssim|\v{k}|\lesssim3\,\rm{fm}^{-1}$, it was shown that the momentum distribution is dominated by tensor-induced
pn correlations. Moreover, the prediction on the relative strength of pp and pn pairs in the tail part
is consistent with observations in the exclusive two-nucleon knockout studies, i.e., a
strong dominance of the np SRC pairs over the pp SRC pairs\,\cite{Ryc15}. As shown in the lower window of Fig.\,\ref{fig-Ekin0RMFHMT}, the average kinetic energy of neutrons
in the independent particle model (IPM) is generally larger than that of protons, but the inclusion of the correlations makes the prediction flipped.
Moreover, they also found that the difference between the $\langle T_{\rm{p}}\rangle$ and $\langle T_{\rm{n}}\rangle$ increases
roughly linearly with the decreasing proton fraction $x_{\rm{p}}$. For the most asymmetric nucleus considered there, i.e., $^{48}\rm{Ca}$, the $\langle T_{\rm{p}}\rangle$ is about 10\% larger than the $\langle T_{\rm{n}}\rangle$,
qualitatively consistent with the results of ref.\,\cite{Hen14}, see the magenta
and the green solid circles in the left window of Fig.\,\ref{fig_xCaiFig2}.

\begin{figure}[h!]
\centering
  \includegraphics[width=7.5cm]{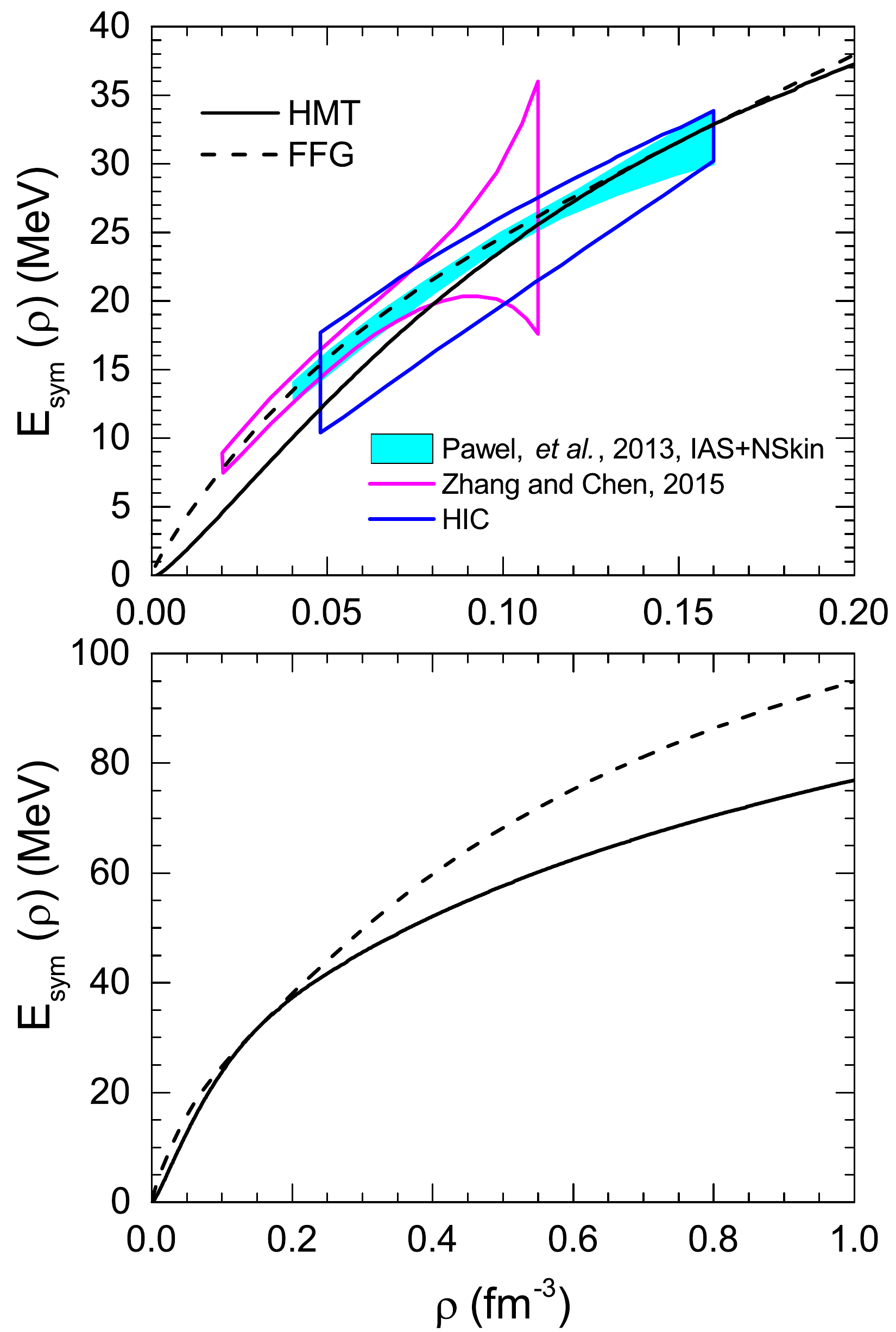}
  \hspace{0.5cm}
   \includegraphics[width=8cm]{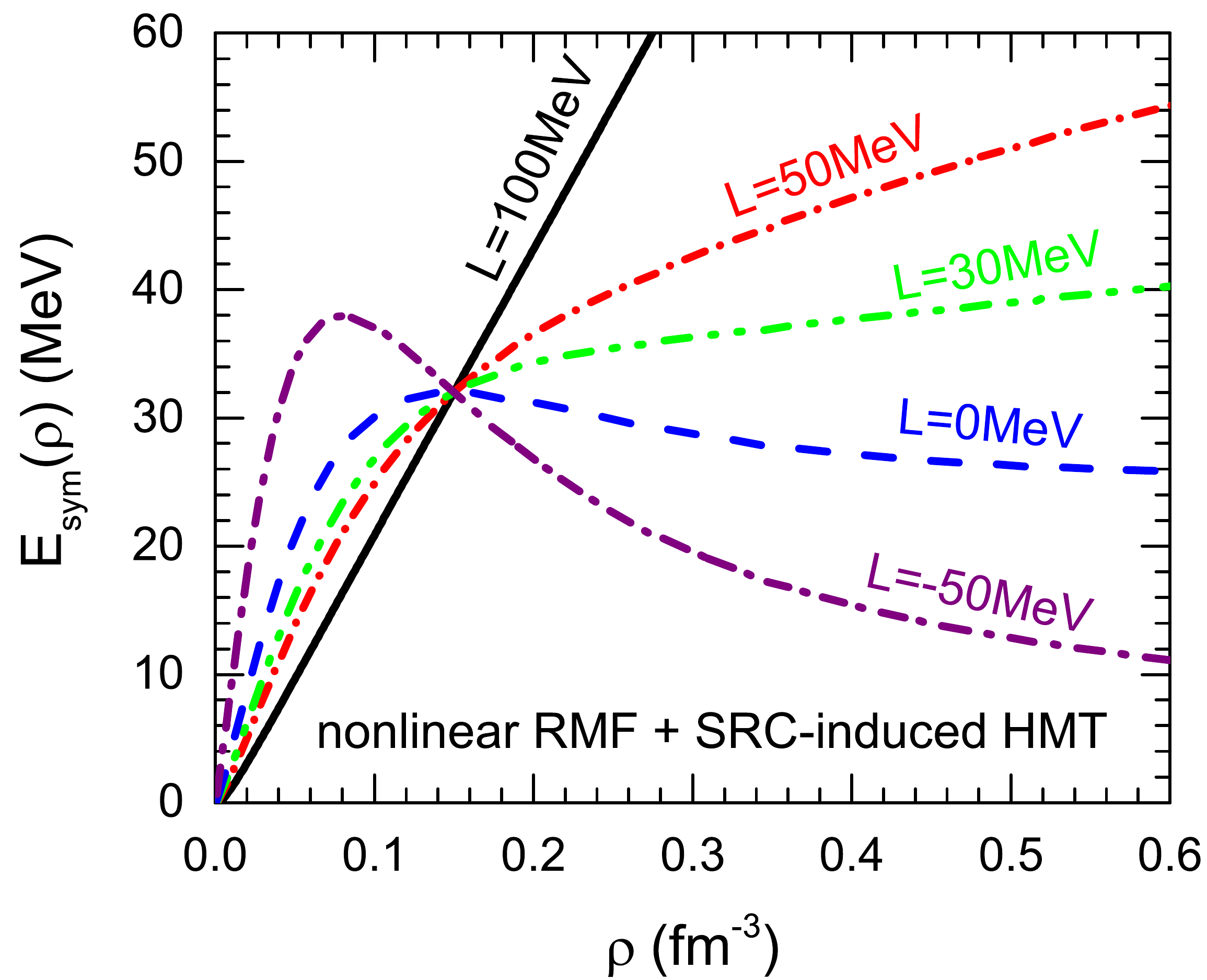}
 \caption{Left: Total symmetry energy as a function of density in the HMT and FFG models in comparison with constraints
  from several recent studies\,\cite{Dan14,Tsa12,Zha15}. Taken from ref.\,\cite{Cai16c}.
  Right: Symmetry energy with different $L$ as a function of density in the nonlinear RMF models with SRC-induced HMT.
  The symmetry energy at normal density is fixed.}
  \label{fig-EsymRMFHMT}
\end{figure}

\begin{figure}[h!]
\centering
  \includegraphics[width=9.cm]{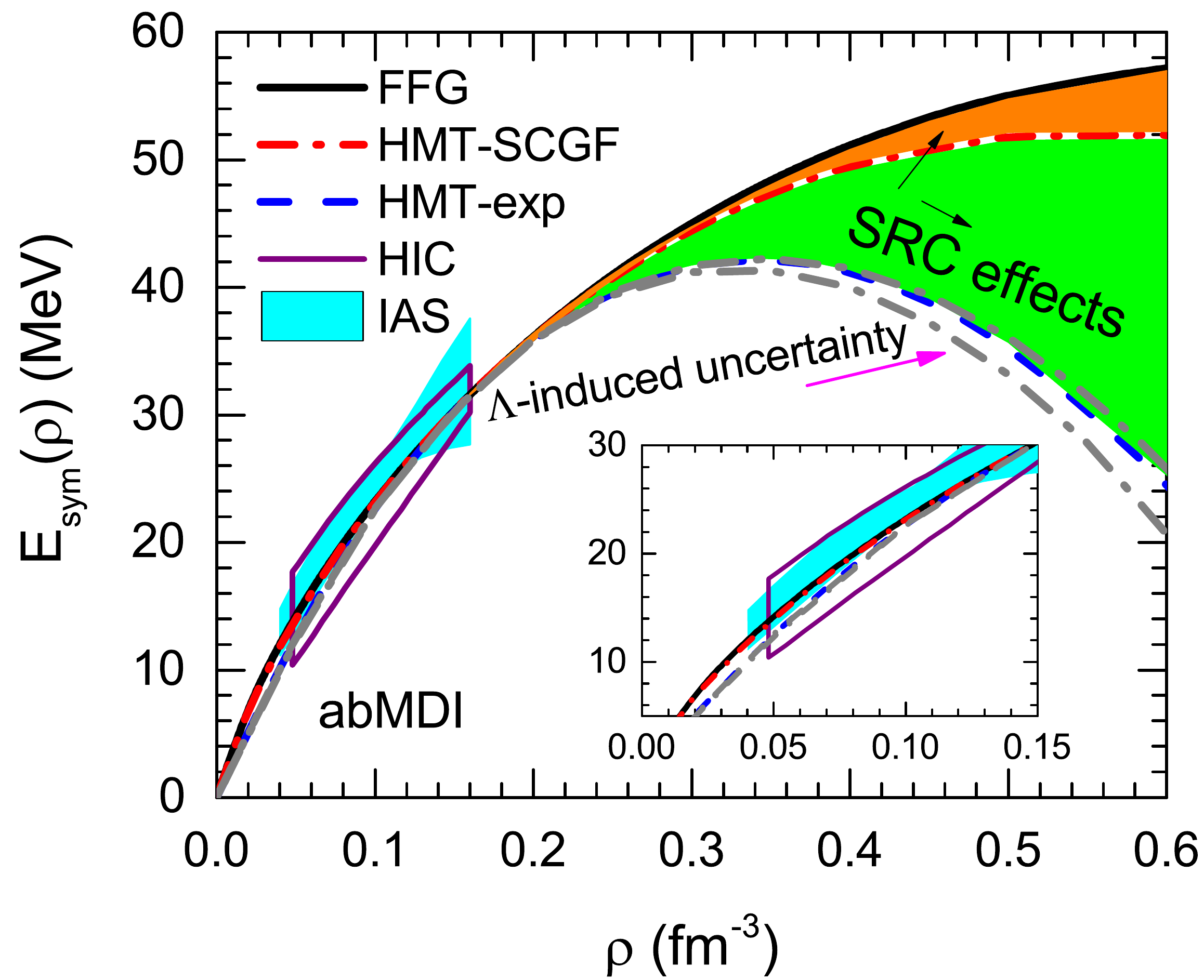}
  \caption{Comparisons of nuclear symmetry energy obtained within a modified Gogny energy density functional using the FFG, HMT-SCGF and HMT-exp parameter sets.
  Constraints on the symmetry energy from analyzing heavy-ion collisions (HIC)\,\cite{Tsa12} and the Isobaric Analog States (IAS)\,\cite{Dan14} are also shown. Taken from ref.\,\cite{Cai17}.}
  \label{fig_xFig1}
\end{figure}
Going back to the RMF calculations, we now discuss effects of the SRC-reduced kinetic symmetry energy on the calculations of the density dependence of the total symmetry energy $E_{\rm{sym}}(\rho)$.
As the total symmetry energy $E_{\rm{sym}}(\rho_0)$  and its slope parameter $L(\rho_0)$ at the saturation density are both relatively well constrained as we discussed earlier,
the reduced kinetic $E^{\rm{kin}}_{\rm{sym}}(\rho_0)$ requires an enhanced potential $E^{\rm{pot}}_{\rm{sym}}(\rho_0)$ such that their sum is still consistent with the known empirical constraints. Results of a recent study\,\cite{Cai16c} by fixing six quantities, i.e., $M_0^{\ast}$, $E_0(\rho_0)$, $\rho_0$, $K_0$, $E_{\rm{sym}}(\rho_0)$
and $L(\rho_0)$ at their empirical values using the RMF-HMT and RMF-FFG parameters are compared in Fig.\,\ref{fig-EsymRMFHMT}. Results of several other predictions\,\cite{Dan14,Zha15,Tsa12} for the $E_{\rm{sym}}(\rho)$
near $\rho_0$ are also shown. It is seen from the upper panel that
the HMT reduces the magnitude but hardens the slope of $E_{\rm{sym}}(\rho)$ at sub-saturation
densities. For instance, at densities around 0.04\,fm$^{-3}$, the
effect is about 30\% which is larger than the width of the existing
constraint\,\cite{Zha15}. On the other hand, at supra-saturation densities as shown in the lower panel, the symmetry energy is also
significantly reduced by the HMT. For instance, the effect is about
25\% at densities around 0.5\,fm$^{-3}$. Overall, while the  $E_{\rm{sym}}(\rho_0)$  and its slope parameter $L(\rho_0)$ are fixed, the HMT significantly reduces the curvature, i.e, the $E_{\rm{sym}}(\rho)$ becomes more concave, at saturation density. It is interesting to note that in the original nonlinear RMF model, the high density
$E_{\rm{sym}}(\rho)$ can not be made too small because of the limitation of the RMF model structure itself. In the presence of HMT, however,
mainly owing to the negative kinetic symmetry energy it is possible to
make the total $E_{\rm{sym}}(\rho)$ very soft and even
decreases at high densities by using different $L(\rho_0)$ values but a fixed $E_{\rm{sym}}(\rho_0)$, see the right window of Fig.\,\ref{fig-EsymRMFHMT}.
We notice that there is indeed some circumstantial evidences supporting a super-soft $E_{\rm{sym}}(\rho)$\,\cite{Xia09}.

Since the HMT and FFG models are designed to have the same values of
symmetry energy $E_{\rm{sym}}(\rho_0)$ and its slope $L(\rho_0)$, it is
useful to use the curvature of the symmetry energy, $K_{\rm{sym}}$ to measure the HMT effects on the total symmetry energy.  More quantitatively, the values of
$K_{\rm{sym}}^{\rm{FFG}}\approx-37\,\rm{MeV}$ and $
K_{\rm{sym}}^{\rm{HMT}}\approx-274\,\rm{MeV}$ were found.
The corresponding isospin-coefficients of the incompressibility are
$K_{\tau}^{\rm{FFG}}\approx-174\,\rm{MeV}$ and
$K_{\tau}^{\rm{HMT}}\approx-470\,\rm{MeV}$, respectively. The latter is in
very good agreement with the best estimate of
$K_{\tau}=-550\pm 100$\,MeV from analyzing many different
kinds of experimental data currently available\,\cite{Col14}. Overall,
the HMT is to make the symmetry energy significantly more concave
around the saturation density, leading to a stronger isospin
dependence in the incompressibility of ANM compared to the FFG
model.

As an example of non-relativistic calculations, shown in Fig.\,\ref{fig_xFig1} are the results of calculations for the $E_{\rm{sym}}(\rho)$ within a modified Gogny (with the MDI interaction) energy density functional (EDF) using the FFG, HMT-SCGF and HMT-exp parameter sets\,\cite{Cai17}. In this study, several quantities were fixed at their empirical values, i.e., $M_0^{\ast}$, $E_0(\rho_0)$, $\rho_0$, $K_0$, $E_{\rm{sym}}(\rho_0)$, $L$, $U_0(\rho_0,0)$
and $U_{\rm{sym}}(\rho_0,1\,\rm{GeV})$, where $U_0(\rho,|\v{k}|)$ and $U_{\rm{sym}}(\rho,|\v{k}|)$ is the single-nucleon isoscalar and isovector potential, respectively\,\cite{Cai17}.
It is interesting to notice that the SRC-induced reduction of
$E_{\rm{sym}}(\rho)$ within the non-relativistic EDF approach is qualitatively
consistent with the RMF-HMT result shown in Fig.\,\ref{fig-EsymRMFHMT}. Nevertheless,
since there is no explicit momentum dependence in the RMF EDF, the
corresponding reduction of $E_{\rm{sym}}(\rho)$ is smaller. Obviously, the
momentum-dependent interaction in the modified Gogny EDF makes the SRC-induced softening of the symmetry
energy at supra-saturation densities more evident.

The above two examples have shown that the tensor force induced SRC can soften the symmetry energy at supra-saturation densities, leading to a decreasing \esym that may even become negative at high densities. 
In fact, such kind of \esym has also been found in many other calculations in the literature, see, e.g., refs. \cite{Pan72,WFF,Kut94,Kub99,Szm06,Fri81,eft,Kra06,Cha97,Dec80,MS,Kho96,Bas07,Ban00,Ch09,HKLee}. 
 A summary of more predictions of super-soft high-density symmetry energy can be found in Ref. \cite{Kut06}. Not surprisingly, one of the possible origins of the
super-soft \esym at supra-saturation densities is the isospin dependence of nuclear tensor force.  Examples of detailed studies on the role of tensor force on the high-density behavior of nuclear symmetry energy can be found 
in refs. \cite{Kuo65,Eng97,Xu-tensor,HKLee11,AngLi}. At very high densities when the short-range repulsive tensor force due to the $\rho$-meson exchange makes the EOS of symmetric matter increase faster with density than that of pure neutron matter where the tensor force is much weaker, the \esym decreases or even becomes negative at high densities \cite{Pan72,WFF}. To our best knowledge, such kind of unusual \esym is not excluded by any known fundamental physics principle. It is also not clear if there are experimental/observational evidences excluding the super-soft \esym in either nuclear physics or astrophysics.
It is actually an interesting topic that has attracted considerable attention since it was first predicted by Pandharipande in 1972 \cite{Pan72}. The super-soft symmetry energy may lead to new phenomena, such as the formation of proton polarons \cite{Kut93} or isospin separation instability (i.e., symmetric matter is unstable against being separated into clusters of pure neutrons and protons) \cite{BALI02} in neutron stars. Moreover, there is a degeneracy between the high-density EOS and strong-field gravity in determining properties of massive neutron stars \cite{Ded03,Psa08,Kri09,Wen09,Wen12,Wlin14,Jiang15,He15}. A better understanding of the high-density 
\esym will help test theories of strong-field gravity that is not completely understood yet. In the era of gravitational wave astronomy just begun with the recent detection of GW170817 \cite{LIGO-PRL}, reliable information about the high-density \esym is becoming even more important \cite{Farrooh,Plamen18}. As to supporting massive neutron stars heavier than two solar masses with a super-soft \esym, similar to situations where various possible phase transitions and/or new particles are considered, there are probably many options. For example, since the EOS of super-dense isospin-symmetric nucleonic matter is not determined either \cite{Cai-Chen-J0}, a super-stiff EOS of SNM can be constructed by varying the skewness parameter $J_0$ of SNM within its uncertainty range and staying causal to support massive neutron stars when the \esym is super-soft \cite{NBZhang2}. Moreover, several seemingly extreme mechanisms, such as the modified gravity or new light mesons have also been proposed to support massive neutron stars when the EOS at high densities is super-soft,
see, e.g., refs. \cite{Kri09,Wen09,Wen12,Wlin14,Jiang15}.

\begin{figure}[h!]
\centering
   \includegraphics[width=18cm,height=10cm]{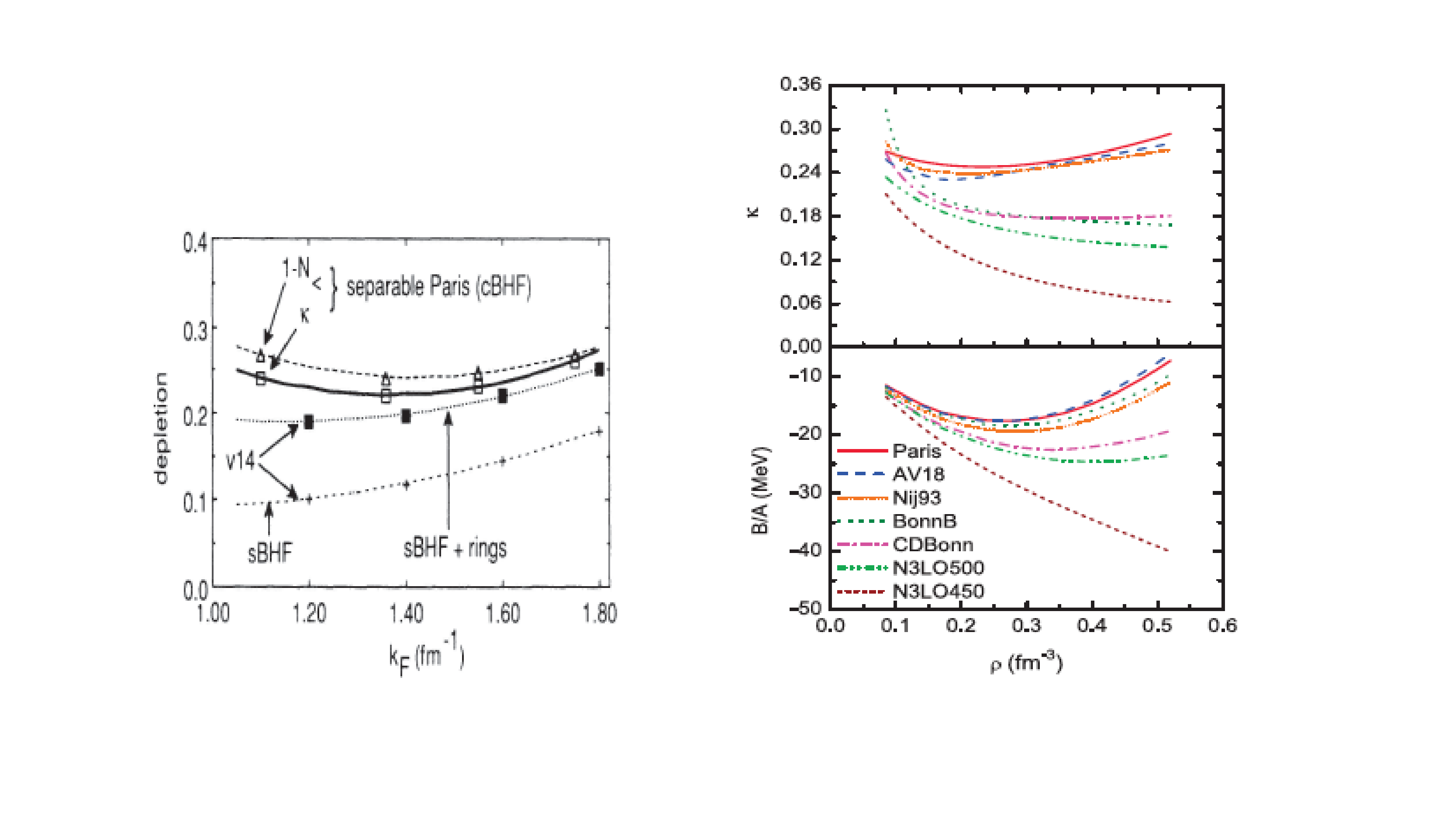}
   \vspace{-2cm}
  \caption{Left:  The depletion parameter $\kappa$ as a function of Fermi momentum $k_{\rm{F}}$ in SNM from the BHF
approach with a separable representation of the Paris interaction. Taken from ref.\,\cite{Bal90}.
   Right: Average depletion in SNM (upper) as a function of density and the EOS of SNM (lower) from BHF calculations using several modern interactions. Taken from ref.\,\cite{ZHLi16}. }
  \label{Fig-kappa}
\end{figure}
\FloatBarrier

\subsection{SRC effects on the isospin-dependent depletion of nucleon Fermi sea and E-mass in neutron-rich matter}\label{subx.4}
In this subsection, we discuss constraints and SRC effects on the nucleon E-mass and its isospin splitting via the Migdal--Luttinger theorem\,\cite{Mig57,Lut60}. For this purpose, one needs reliable information about the nucleon momentum distributions from both right below and above the Fermi surface. In the previous sections, we have focused on the contact and the HMT above the Fermi surface. Here we shall first discuss the depletion below the Fermi surface, then combine all relevant information together to extract the E-mass from applying the Migdal--Luttinger theorem. Of course, the isospin-dependent depletion of the nucleon Fermi sea and various experimental efforts probing it are extremely interesting in their own rights. For an overview of earlier studies on effects of both the SRC and long-range correlations on the depletion of the nucleon Fermi sea, we refer the reader to the comprehensive review by Dickhoff and Barbieri \cite{Wim04}.

\begin{figure}[h!]
\centering
  \includegraphics[width=7.5cm]{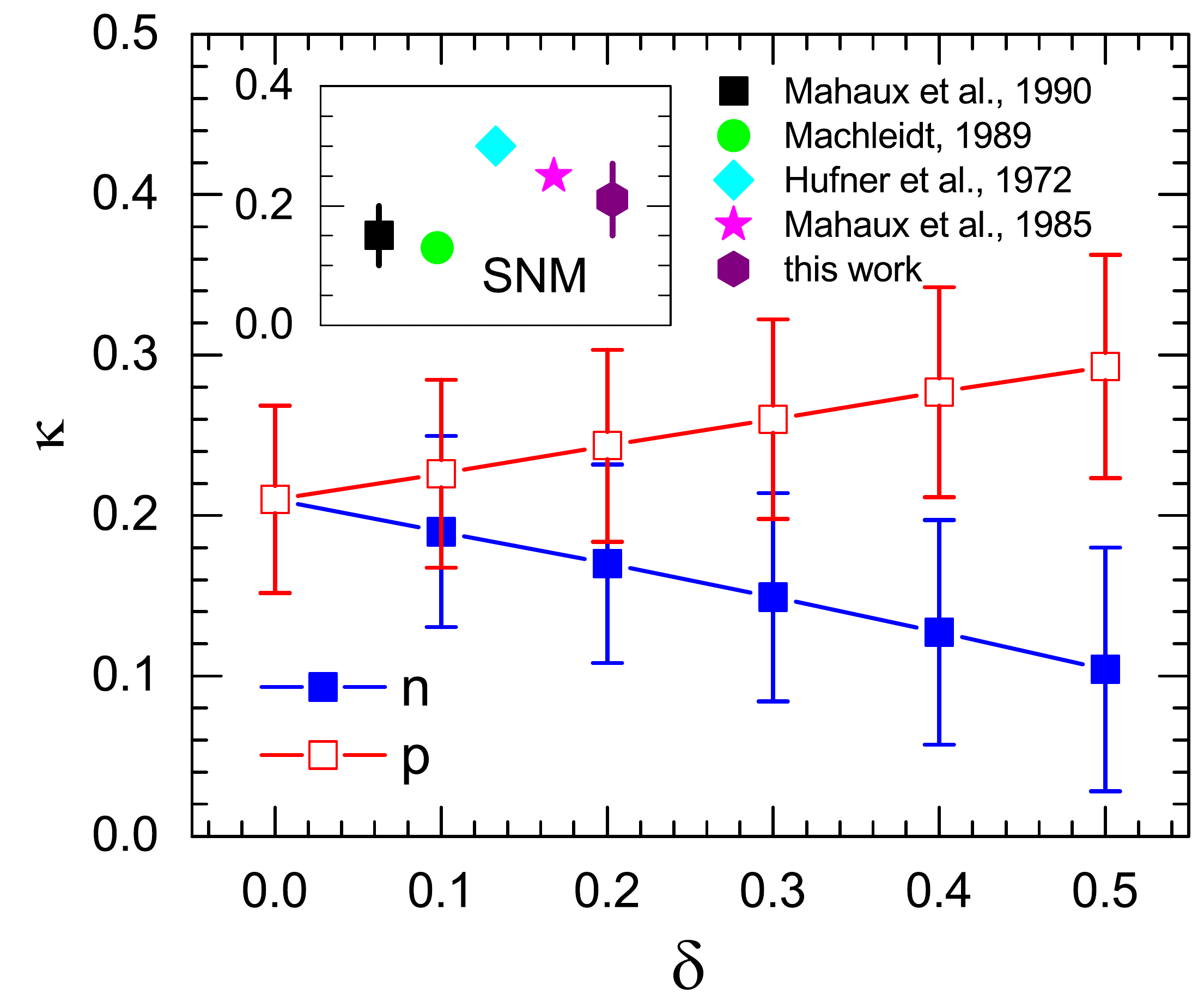}
  \caption{Average depletion of the neutron and proton Fermi surface as a function of isospin asymmetry in neutron-rich matter.
  The inset compares the average nucleon depletion in symmetric matter from several studies\,\cite{Mah85,Mac89,Mah90,Huf72}. Taken from ref.\,\cite{CaiLi16a}.}
  \label{Cai-kappa}
\end{figure}

The depletion $\kappa_J$ of the nucleon momentum distribution below the Fermi surface
provides a quantitative measure of the validity of the  HVH theorem\,\cite{Hug58} and more generally
independent particle models. A deeper depletion indicates a more
serious violation of the HVH theorem\,\cite{Mac89,Ram90,Mah90,Huf72,Jon91}.
Experimentally, it can be measured by using the nucleon
spectroscopic factor from transfer, pickup and (e,e$'$p)
reactions\,\cite{Mah90}. As illustrated already in Fig.\,\ref{fig_SF}, a well-known finding is that
mean-field models over-predict the occupation of low-momentum nucleon
orbitals compared to data of electron scatterings on nuclei from
$^{7}$Li to $^{208}$Pb by about 30-40\% due to the neglect of
correlations\,\cite{Lap93}. The $\kappa_J$ is also believed to
determine the rate of convergence of the hole-line expansion of the
nuclear potential\,\cite{Mah85,bethe,Mah90,Huf72}. It was shown earlier that the $\kappa_J$ depends strongly on the tensor part of the nucleon-nucleon
interaction\,\cite{Fan84,Ram90}. Now let's look at a few examples. Shown in the left window of Fig.\,\ref{Fig-kappa} are the density dependence of the depletion parameter $\kappa$ in SNM (black line) obtained using the BHF
approach with a separable representation of the Paris interaction\,\cite{Bal90}. The most likely value of the density inside the Fermi surface
was found to be $3k_{\rm{F}}/4$ based on the potential form used. The $\kappa$ reached about 0.27 in the range of densities considered there.
Since the density dependence of the $\kappa$ parameter is rather weak, the studies on the $\kappa$ at a fixed density (e.g., the saturation density)
is already useful. In the right window of Fig.\,\ref{Fig-kappa}, the average depletion $\kappa$ (upper) and the EOS of SNM (lower) from BHF calculations using
several interactions are shown\,\cite{ZHLi16}. One of the main features is that at high densities, e.g.,
$\rho>\rho_0\approx0.17\,\rm{fm}^{-3}$, the depletions from the calculations using the r-space potentials Paris, Av18, Nij93 remain
fairly high, whereas those using the k-space and in particular the chiral
potentials feature much lower values. In the latter case, there
is a strong dependence on the chiral cutoff parameter used. As pointed out in ref.\,\cite{ZHLi16},
the depletion at high densities is determined by the repulsive
core of the nucleon-nucleon interaction. The latter is very strong in the r-space
potentials used but very weak in the chiral potentials with small momentum cutoff parameters. On the other hand, even around the saturation density, the $\kappa$ obtained using different interactions shown in Fig.\,\ref{Fig-kappa}
takes value from about 0.12 to 0.27, leading to a relative variation of about 37\% in the same BHF model. This represents the current level of the interaction-dependence in predicting the depletion of the nucleon Fermi sea.

Corresponding to the $n_{\v{k}}^J$ in Eq.\,(\ref{MDGen}),
the average depletion of the Fermi sphere in ANM at $\rho_0$ is
\begin{align}
\kappa_J=&1-\Delta_J-\frac{1}{k_{\rm{F}}^J}\int_0^{k_{\rm{F}}^J}\beta_J\left(\frac{|\v{k}|}{k_{\rm{F}}^J}\right)^2\d
k=\frac{4}{15}\beta_J+3C_J\left(1-\frac{1}{\phi_J}\right).
\end{align}
More quantitatively, for SNM, $\kappa=4\beta_0/15+x_{\rm{SNM}}^{\rm{HMT}}\approx0.21\pm0.06$ is comparable with results obtained earlier from other studies
\cite{Mah85,Mac89,Mah90,Huf72}, as shown in the inset of Fig.\,\ref{Cai-kappa} and those shown in Fig.\,\ref{Fig-kappa}.
The corresponding neutron-proton depletion splitting is approximately
\begin{equation}\label{ksplitting}
\kappa_\rm{n}-\kappa_\rm{p}\approx[8\beta_0\beta_1/15+6C_0\phi_1/\phi_0+6C_0C_1(1-\phi_0^{-1})]\delta\approx(-0.37\pm0.16)\delta.
\end{equation}
Shown in Fig.\,\ref{Cai-kappa} are the $\kappa$ values for neutrons and protons separately as functions of isospin asymmetry $\delta$. It is interesting to see that the neutron/proton
depletion decreases/increases with $\delta$ approximately linearly as shown analytically in Eq. (\ref{ksplitting}), indicating that protons with energies near
the Fermi surface experience larger correlations with increasing
isospin asymmetry. Within the tensor force induced neutron-proton dominance model used in deriving the Eq. (\ref{ksplitting}), it is simply because protons have more neutrons to be paired with in more neutron-rich matter.
This feature is in qualitative agreement with findings from studies using both microscopic many-body theories\,\cite{Yin13,Ram90,Frick} and
phenomenological models\,\cite{Sar14}.

\begin{figure}[h!]
\centering
   \includegraphics[width=15cm,height=8cm]{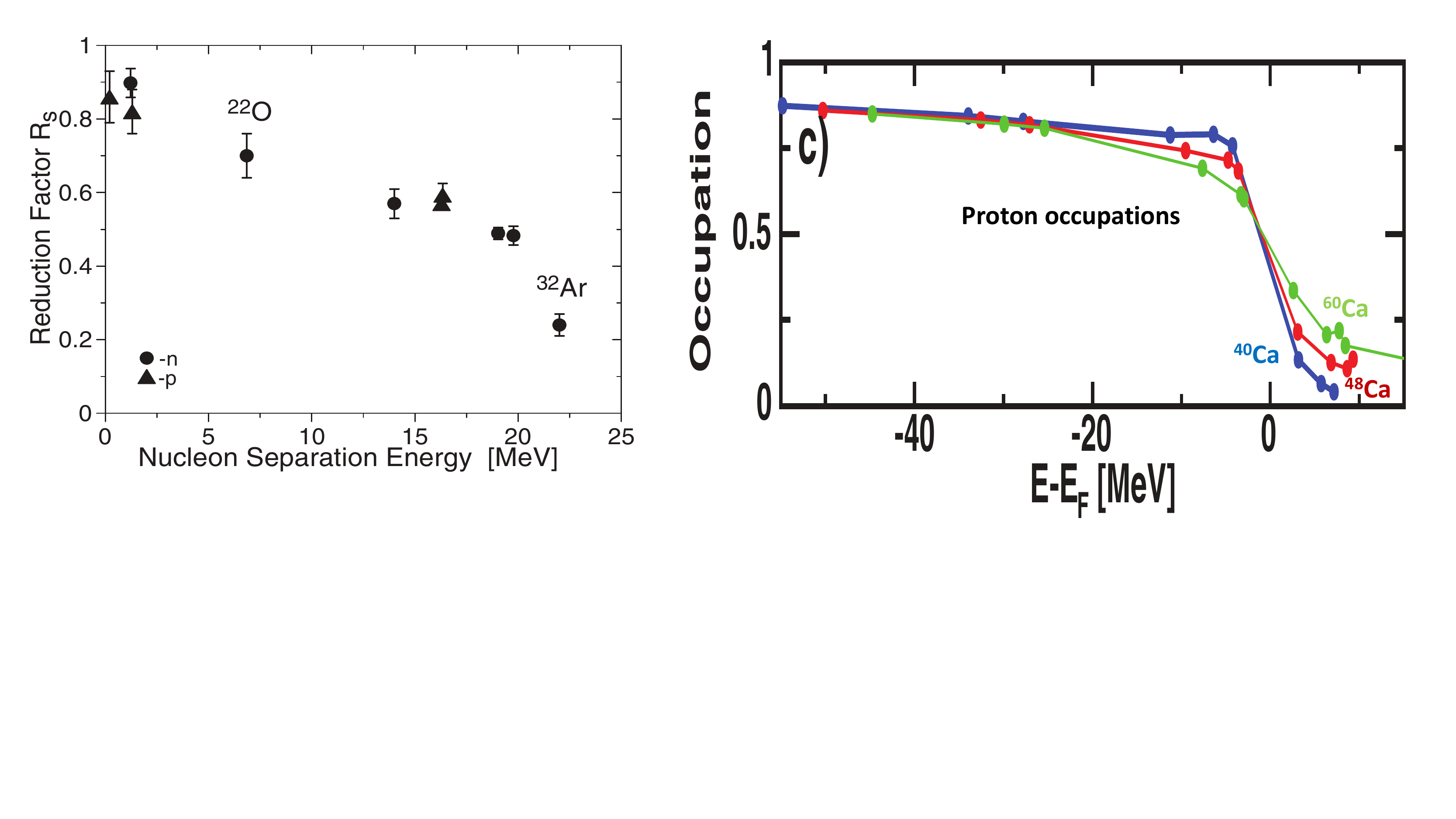}
 \vspace{-3cm}
  \caption{Left:  Inferred reduction factors $R_s$ as a function of nucleon separation energy. Taken from ref.\,\cite{Gade}.
 Right: Proton occupation probability as a function of Fermi energy for $^{40}\rm{Ca}$ (blue) and $^{48}\rm{Ca}$ (red) from a dispersive optical model analysis of p+Ca scattering data and $^{60}\rm{Ca}$ (green) from model calculations. Taken from ref.\,\cite{Bob06}}\label{Gade-Bob}
\end{figure}
The indication that the minority particles are more correlated and deeply depleted from its Fermi sea in imbalanced 2-component systems is also consistent with
some experimental observations from studying nucleon spectroscopic
factors in radioactive nuclei\,\cite{Gade}, dispersive optical model analyses of
proton-nucleus scatterings\,\cite{Bob06} and the neutron-proton
dominance model analyses of electron-nucleus scatterings\,\cite{Hen14}.
For example, from neutron knock-out reactions, Gade et al. \cite{Gade} deduced the occupancy of the $0d_{5/2}$ neutron
hole state in the proton-rich $^{32}$Ar nucleus from studying the reduction factor $R_s$ defined as the ratio of the
experimental and theoretical value for the spectroscopic factor \cite{Gade}.
The latter measures the overlap of the initial and final states with the same orbital and total angular momenta.
Shown in the left window of Fig.\,\ref{Gade-Bob} is their compilation of the reduction factor $R_{\rm{s}}$ as a function of nucleon separation energy\,\cite{Gade}.
Most interestingly in the context of the discussions here, the neutron reduction factor for the proton-rich $^{32}$Ar is much lower than that for the neutron-rich $^{22}$O both having 14 neutrons,
indicting clearly that the minority neutrons in the proton-rich environment is more correlated/depleted.
On the other hand, Charity et al. \cite{Bob06} found that protons in neutron-rich systems are more strongly correlated from a dispersive optical model analysis of p+$^{40}$Ca and p+$^{48}$Ca scattering data.
More quantitatively, the inferred spectroscopic factors are 65\%, 56\% and 50\% for the $0d_{3}/2$ proton level in $^{40}$Ca, $^{48}$Ca and $^{60}$Ca, respectively.
Shown in the right window of Fig.\,\ref{Gade-Bob} are the proton occupation probabilities they inferred from the experimental data of proton scattering on $^{40}$Ca and $^{48}$Ca as well as model calculations for $^{60}$Ca.
Their results show that for proton hole states just below the Fermi energy, the occupation probabilities have decreased
for the more neutron-rich $^{48}$Ca while the opposite is true
for the particle states. This trend is even stronger in the extrapolation to $^{60}$Ca based on their model calculations.
These results clearly imply that protons with energies near the Fermi surface experience stronger correlations with increasing isospin-asymmetry in neutron-rich matter.
The above experimental observations are qualitatively consistent with indications of the Eq. (\ref{ksplitting}).
It is, however, necessary to emphasize that the particle occupancy is not directly observable and the reduction factor is by definition model dependent. All reported results thus have some intrinsic model dependences,
see discussions in, e.g., ref. \cite{Tsang-sf}. For example, it was reported by Lee et al. \cite{Lee-Tsang} that spectroscopic factors extracted for proton-rich $^{34}$Ar and neutron-rich $^{46}$Ar using the (p,d) neutron transfer reactions show little reduction of the ground state neutron spectroscopic factor in the proton-rich nucleus $^{34}$Ar compared to that of $^{46}$Ar, inconsistent with the trends observed in the knockout reaction measurements by Gade et al. \cite{Gade}.

\begin{figure}[h!]
\centering
  \includegraphics[width=8.5cm]{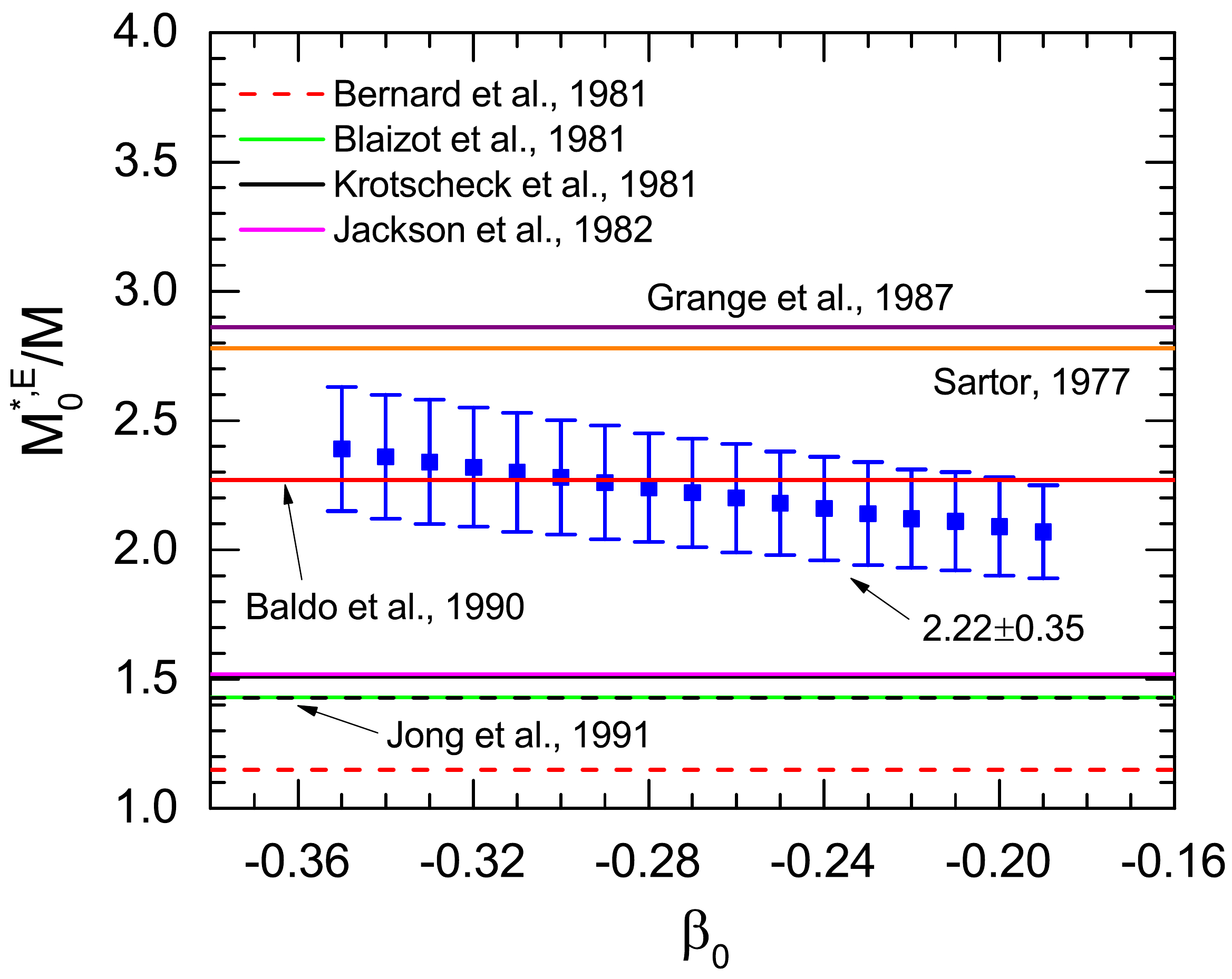}
  \hspace{0.cm}
   \includegraphics[width=7.5cm,height=8cm]{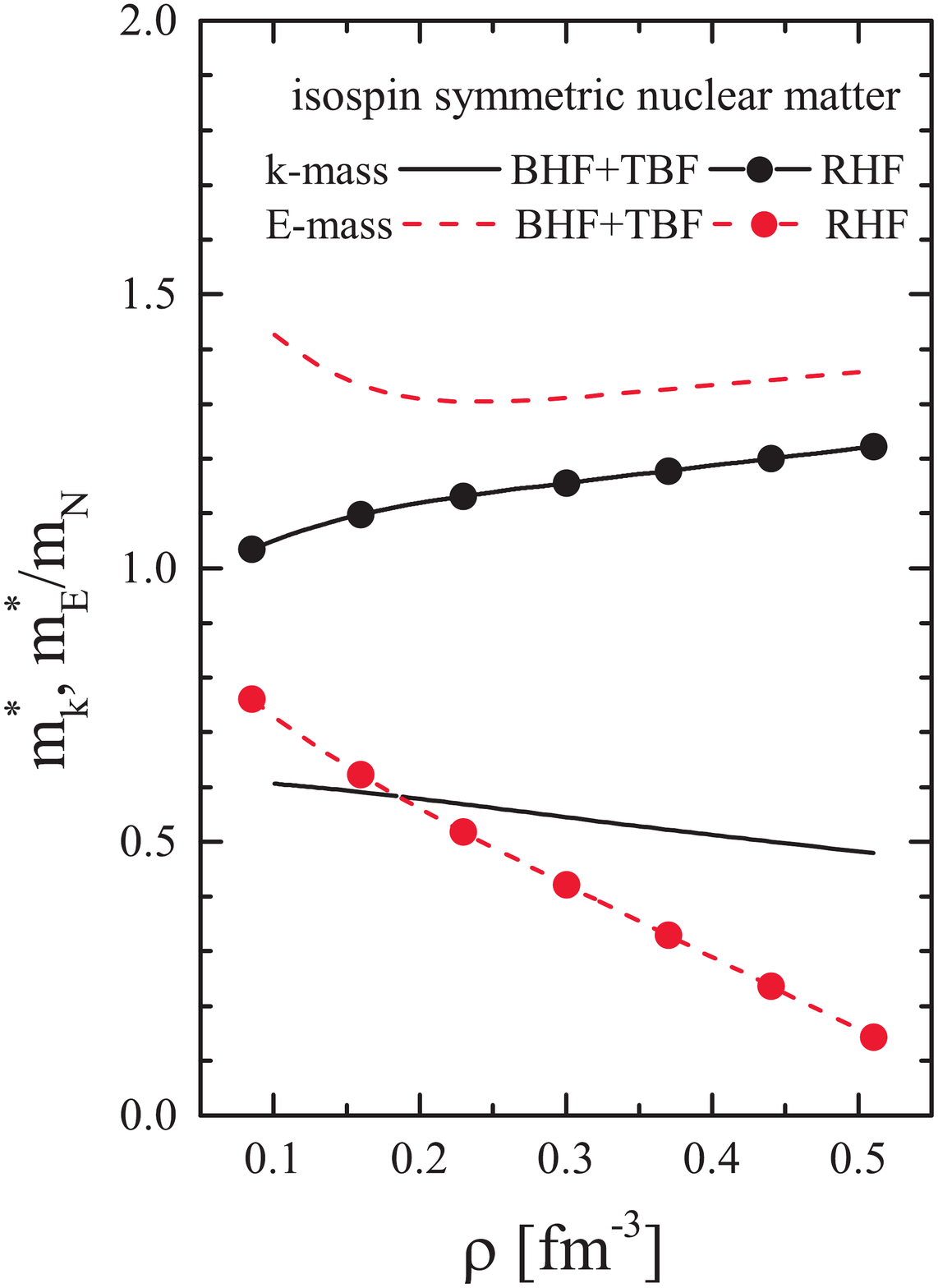}
  \caption{Left: Nucleon effective E-mass in SNM (blue lines with error bars) at normal density
  within the uncertainty range of the shape parameter $\beta_0$ of the nucleon momentum distribution extracted using the Migdal--Luttinger theorem
  in comparison with predictions of earlier studies\,\cite{Bla81,Kro81,Jac82,Bal90,Jon91,Ber81,Gra87,Sar77}. Taken from ref.\,\cite{CaiLi16a}.
  Right: Density dependence of both the k-mass and E-mass in the BHF + TBF and RHF model for SNM. Taken from ref.\,\cite{ALi16}.}
  \label{Fig-Emass0}
\end{figure}

We now turn to the nucleon E-mass obtained through the
Migdal--Luttinger theorem. In terms of the parameters describing the
single-nucleon momentum distribution $n_{\v{k}}^J$ in Eq.\,(\ref{MDGen}), one has
$Z_{\rm{F}}^J=\Delta_J+\beta_J-C_J=1+2\beta_J/5-4C_J+3C_J\phi_J^{-1}$.
For SNM, it is given by
$Z_{\rm{F}}^0=1+2\beta_0/5-4C_0+3C_0\phi_0^{-1}=1+2\beta_0/5-C_0-x_{\rm{SNM}}^{\rm{HMT}}$,
then using the values for $\beta_0$, $\phi_0$, $C_0$ and
$x_{\rm{SNM}}^{\rm{HMT}}$ given above, one obtains a value of
$M_0^{\ast,\rm{E}}/M\approx2.22\pm0.35$. Shown in the left window of Fig.\,\ref{Fig-Emass0} with the filled squares are the extracted nucleon
E-mass in SNM within the uncertain range of the $\beta_0$ parameter.
It is seen that the variation of $M_0^{\ast,\rm{E}}/M$ with
$\beta_0$ is rather small. For comparisons, also shown are earlier
predictions based on (1) a semi-realistic parametrization through a
relative s-wave exponential nucleon-nucleon interaction potential
(red dash line)\,\cite{Ber81}, (2) a Green's function method
considering collective effects due to the coupling of nucleons with
the low-lying particle-hole excitations of the medium (green solid
line)\,\cite{Bla81}, (3) a correlated basis function (CBF) method
using the Reid and Bethe-Johnson potentials (black and magenta solid
lines)\,\cite{Kro81,Jac82}, (4) two non-relativistic models with the
Paris nuclear potential (purple and red solid
line)\,\cite{Gra87,Bal90}, (5) a low density expansion of the optical
potential (orange solid line)\,\cite{Sar77} and (6) a relativistic
Dirac-Brueckner approach (dash black line)\,\cite{Jon91}.

\begin{figure}[h!]
\centering
  \includegraphics[width=8.cm,height=9.5cm,angle=-90]{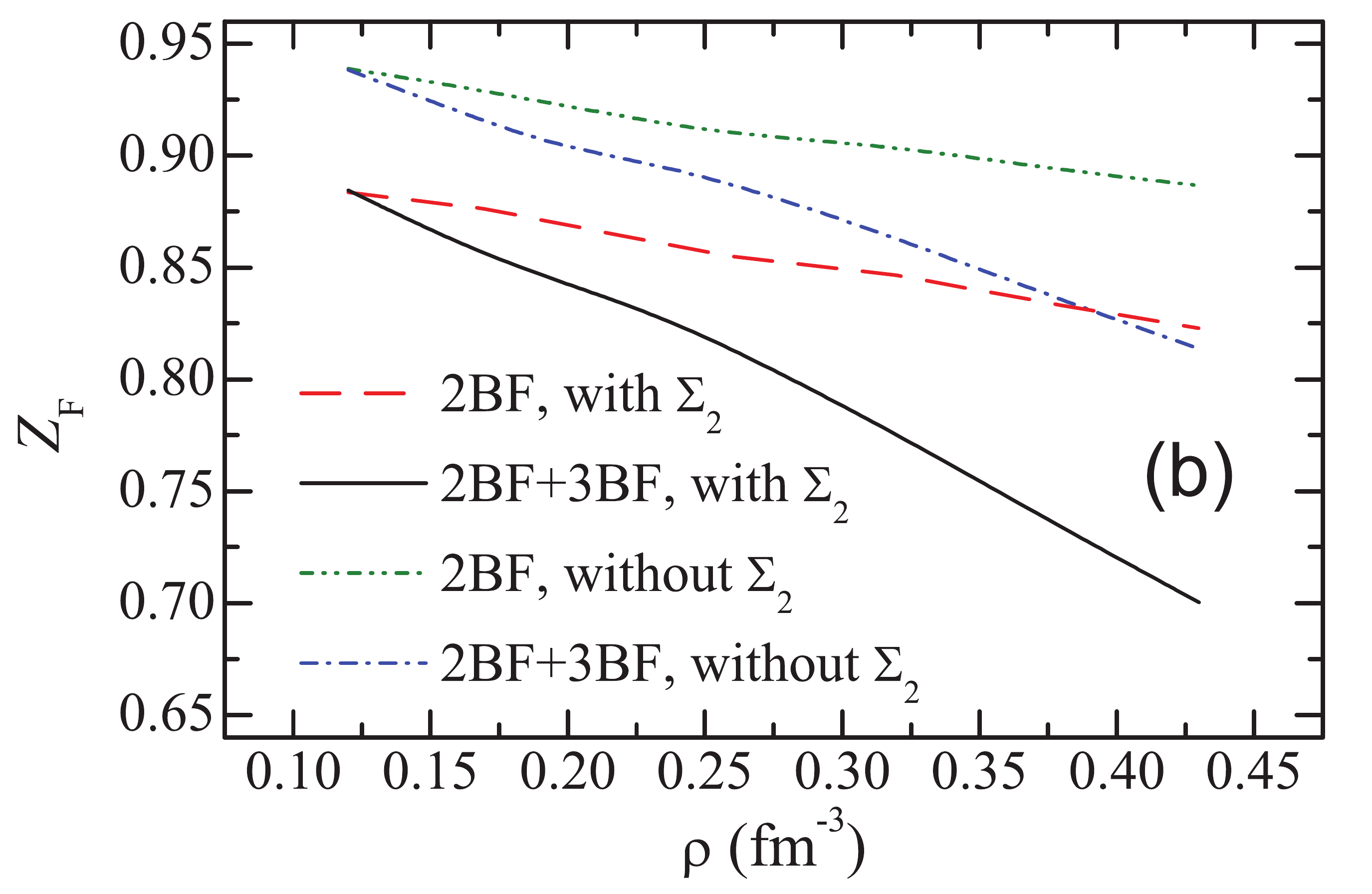}
  \caption{Density dependence of the Migdal--Luttinger jump in PNM within the BCS framework. Taken from ref.\,\cite{Don13}. }
  \label{fig_Dong13}
\end{figure}
The Migdal--Luttinger jump in SNM is $Z_{\rm{F}}^0=1+2\beta_0/5-C_0-x_{\rm{SNM}}^{\rm{HMT}}$ where the last term (fraction of nucleons in the HMT) is a measure of the SRC strength. It is clear that a stronger SRC
makes the nucleon E-mass larger. This conclusion was also reached by comparing predictions of the BHF and RHF which does not consider the SRC effects as can be seen immediately in
the right window of Fig.\,\ref{Fig-Emass0}\,\cite{ALi16}. It was pointed out that the ladder diagram/three-body force (TBF) considered in the Brueckner pairs included strong SRC effects.
The latter become increasingly important in the high-density region, i.e., the SRC can generate a strong enhancement of the E mass at high densities.
On the other hand, one has $Z_{\rm{F}}^{\rm{PNM}}=1+2\beta_{\rm{n}}^{\rm{PNM}}/5-x_{\rm{HMT}}^{\rm{PNM}}-C_{\rm{n}}^{\rm{PNM}}
$ for the jump in PNM. The $Z_{\rm{F}}^{\rm{PNM}}$ (neutron E-mass in PNM) is still being reduced (enhanced) mainly owing to the
contact $C_{\rm{n}}^{\rm{PNM}}\approx0.12$ since the high momentum neutron fraction in PNM $x_{\rm{HMT}}^{\rm{PNM}}\approx1.5\%$
is even smaller. Consequently, the enhancement of neutron E-mass in PNM is smaller than that in SNM. Shown in Fig.\,\ref{fig_Dong13} is the Migdal--Luttinger jump in PNM
as a function of density within the BCS framework including both two-body and three-body forces (TBF)\,\cite{Don13}. Similar to the findings in ref.\,\cite{ALi16},  the TBF is found to reduce (enhance) the jump (E-mass) in PNM\,\cite{Don13}. For comparisons, it is interesting to note that  the tensor force mainly acting between neutron-proton pairs in SNM,
makes the E-mass (k-mass) in SNM higher (smaller) than that in PNM around the Fermi surface as demonstrated in the HF and BHF calculations with only two-body forces\,\cite{Heb09}.

\begin{figure}[h!]
\centering
  \includegraphics[width=8.cm,height=8cm]{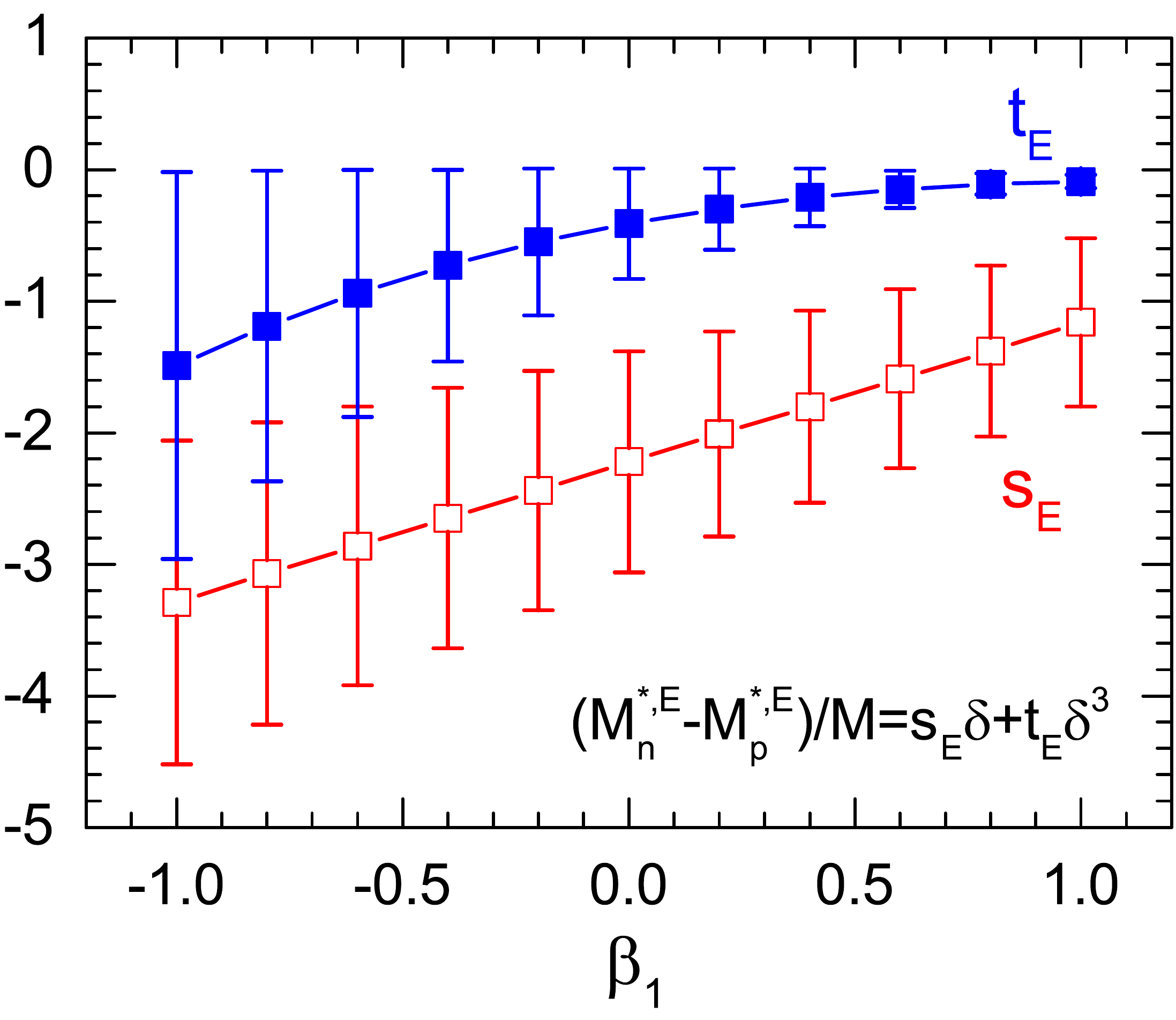}
  \caption{Linear and cubic splitting coefficients $s_{\rm{E}}$ and $t_{\rm{E}}$ at normal density within the uncertain range of the $\beta_1$-parameter characterizing the isospin-dependence of
  the nucleon momentum distribution near the Fermi surface. Taken from ref.\,\cite{CaiLi16a}.}
  \label{Fig-Splitting}
\end{figure}
In neutron-rich nucleonic matter, an interesting quantity is the
neutron-proton E-mass splitting that can be generally expressed as
\begin{equation}
({M_{\rm{n}}^{\ast,\rm{E}}-M_{\rm{p}}^{\ast,\rm{E}}})/{M}=s_{\rm{E}}\delta+t_{\rm{E}}\delta^3
+\mathcal{O}(\delta^5)
\end{equation}
where $s_{\rm{E}}$ and $t_{\rm{E}}$ are the linear and cubic
splitting coefficients, respectively. Shown in Fig.\,\ref{Fig-Splitting} are the values of $s_{\rm{E}}$ and $t_{\rm{E}}$
at the nucleon Fermi momentum in nuclear matter at $\rho_0$ within
the uncertainty range of the $\beta_1$-parameter. More
quantitatively, at the lower limit, mid-value and the upper limit of the
$\beta_1$-parameter, one has
$s_{\rm{E}}(\beta_1=-1)\approx-3.29\pm1.23$, $t_{\rm{E}}(\beta_1=-1)
\approx-1.49\pm1.47$, $s_{\rm{E}}(\beta_1=0)\approx-2.22\pm0.84$,
$t_{\rm{E}}(\beta_1=0)\approx-0.41\pm0.42$,
$s_{\rm{E}}(\beta_1=1)\approx-1.16\pm0.64$ and
$t_{\rm{E}}(\beta_1=1)\approx-0.09\pm0.05$, respectively. An important feature shown in Fig.\,\ref{Fig-Splitting} is that in neutron-rich nucleonic matter, the
E-mass of neutrons is smaller than that of protons, i.e.,
$M_{\rm{n}}^{\ast,\rm{E}}<M_{\rm{p}}^{\ast,\rm{E}}$. As we discussed earlier in this review, such a splitting was predicted by most microscopic many-body theories although
its magnitude is still model and interaction dependent. The results shown in Fig.\,\ref{Fig-Splitting} indicate that the neutron-proton E-mass splitting in ANM has an appreciable dependence
on the largely uncertain $\beta_1$-parameter characterizing the
isospin-dependence of the nucleon momentum distribution near the
Fermi surface. Since the $\beta_1$-parameter is currently unconstrained but important for determining the neutron-proton E-mass splitting in ANM, its experimental constraints are desired.
Hopefully, experiments measuring the isospin dependence of SRC using proton-nucleus scatterings involving rare isotopes in inverse kinematics and/or electron-nucleus scatterings on a series of isotopes
as well as the nucleon spectroscopic factors from direct reactions with various radioactive beams will help constrain the $\beta_1$ in the near future.

\subsection{SRC effects on the critical proton fraction for the direct URCA process to occur in protoneutron stars}\label{subx.5}
Besides several indirect astrophysical consequences through the EOS of ANM especially the density dependence of nuclear symmetry energy, the SRC-induced HMT may also have some direct impacts
on some astrophysical objects and/or processes. One such process is the cooling of protoneutron stars through neutrino emissions. In particular, the SRC may affect both the critical proton fraction for the fast direct URCA
to occur and the neutrino emissivity itself. There were some works on both in the literature, see, e.g., refs.\,\cite{Fra08,Don15}. However, to our best knowledge, there is no community consensus yet, especially since the limited observational data available seem to favor the slow modified URCA process while some earlier studies considering SRC effects have indicated that the critical density for the direct URCA is very low. Thus, much more efforts are
needed in this direction of research. Here we discuss only effects of the SRC-induced HMT on the critical proton fraction $x_{\rm{p}}^{\rm{cric}}$ kinematically. A consistent analysis of both the $x_{\rm{p}}^{\rm{cric}}$ and emissivity may be better made by using the nucleon spectral functions in neutron-rich matter\,\cite{Yak01}.

In the presence of HMT within the simplest model for protoneutron stars consisting of ${\it npe}$ matter only, there are not only opening space below the Fermi surface for newcomers to occupy but those nucleons in the HMT also have more momentum and energy to share with other particles. The critical momentum conservation condition for the direct URCA process $\rm{n}\to\rm{p}+\rm{e}^{-}+\overline{\nu}_{\rm{e}^-}$ to occur then generally requires $
k_{\rm{F}}^{\rm{n}}+\varepsilon_{\rm{F}}^{\rm{n}}\leq
k_{\rm{F}}^{\rm{p}}+\varepsilon_{\rm{F}}^{\rm{p}}
+k_{\rm{F}}^{\rm{e}}$, where $\varepsilon_{\rm{F}}^J$ is a measure of nucleon momentum relative to its Fermi momentum when the SRC is considered.
Including the HMT, the relation between the density and Fermi momentum can still be approximated by $k_{\rm{f}}^j=(3\pi^2\rho_j)^{1/3}$ with
$j$=p/e, thus the neutrality $\rho_p=\rho_e$ still requires $k_{\rm{F}}^{\rm{p}}=k_{\rm{F}}^{\rm{e}}$. One then has $
k_{\rm{F}}^{\rm{n}}(1+t_{\rm{n}})\leq
k_{\rm{F}}^{\rm{p}}(2+t_{\rm{p}})$, where $t_J=\varepsilon_{\rm{F}}^J/k_{\rm{F}}^J$ (with $-1\leq t_J\leq\phi_J$).
Thus, considering the SRC-induced HMT and using $k_{\rm{F}}^J=\rm{const.}\times\rho^{1/3}(1+\tau_3^J\delta)^{1/3}$, the general condition for the direct URCA can be written as
\begin{equation}\label{E1}
x_{\rm{p}}\geq
x_{\rm{p}}^{\rm{cric}}(t_{\rm{p}},t_{\rm{n}})\equiv\frac{(t_{\rm{n}}+1)^3}{(t_{\rm{n}}+1)^3+(t_{\rm{p}}+2)^3}.
\end{equation}
In the FFG where $t_{\rm{p}}=t_{\rm{n}}=0$, the above condition then gives the standard value of
$x_{\rm{p}}^{\rm{cric}}(0,0)=1/9\approx11\%$ widely used in the literature.
While in the presence of HMT and the corresponding low-momentum depletion around the Fermi surface,
there exist some probabilities for the $x_{\textmd{p}}^{\textmd{cric}}$ to be smaller than $1/9$ as now particles involved can be from the HMT and dive deep into the Fermi sea.
The specific probability for a certain direct URCA process is proportional to the product of nucleon momentum distributions in the initial and final state
\begin{align}\label{Pnp-2}
\mathcal{P}_{\rm{pn}}\equiv&\mathcal{P}(t_{\rm{p}},t_{\rm{n}}) =
\left\{\begin{array}{ll}
\Delta_{\rm{n}}\cdot(1-\Delta_{\rm{p}}),&~~-1\leq
t_{\rm{n}}\leq0,~-1\leq
t_{\rm{p}}\leq0,\\
\Delta_{\rm{n}}\cdot\left[1-C_{\rm{p}}/(t_{\rm{p}}+1)^4\right],&~~-1\leq
t_{\rm{n}}\leq0,~0<t_{\rm{p}}\leq
\phi_{\rm{p}}-1,\\
0,&~~-1\leq t_{\rm{n}}\leq0,~t_{\rm{p}}>\phi_{\rm{p}}-1;\\
\left[C_{\rm{n}}/(t_{\rm{n}}+1)^4\right]\cdot(1-\Delta_{\rm{p}}),&~~0<t_{\rm{n}}\leq\phi_{\rm{n}}-1,~-1\leq
t_{\rm{p}}\leq0,\\
\left[C_{\rm{n}}/(t_{\rm{n}}+1)^4\right]\cdot\left[1-C_{\rm{p}}/(t_{\rm{p}}+1)^4\right],&~~0<t_{\rm{n}}\leq\phi_{\rm{n}}-1,~0<t_{\rm{p}}\leq\phi_{\rm{p}}-1,\\
0,&~~0<t_{\rm{n}}\leq\phi_{\rm{n}}-1,~t_{\rm{p}}>\phi_{\rm{p}}-1;\\
0,&~~\rm{others}.
 \end{array}\right.
\end{align}
For example, the first one corresponds to the process where a neutron below its Fermi surface decays into a proton
below its own Fermi surface. It is obvious that the process $-1\leq
t_{\rm{n}}\leq0,~0<t_{\rm{p}}\leq \phi_{\rm{p}}-1$ (the neutron is
below the Fermi surface while the available phase space for the proton is above its Fermi surface)
has the largest probability. The corresponding critical proton
fraction of this channel is expected to be much smaller than $1/9$ according to Eq.\,(\ref{E1}).

\begin{figure}[tbh!]
\centering
  \includegraphics[width=7.5cm]{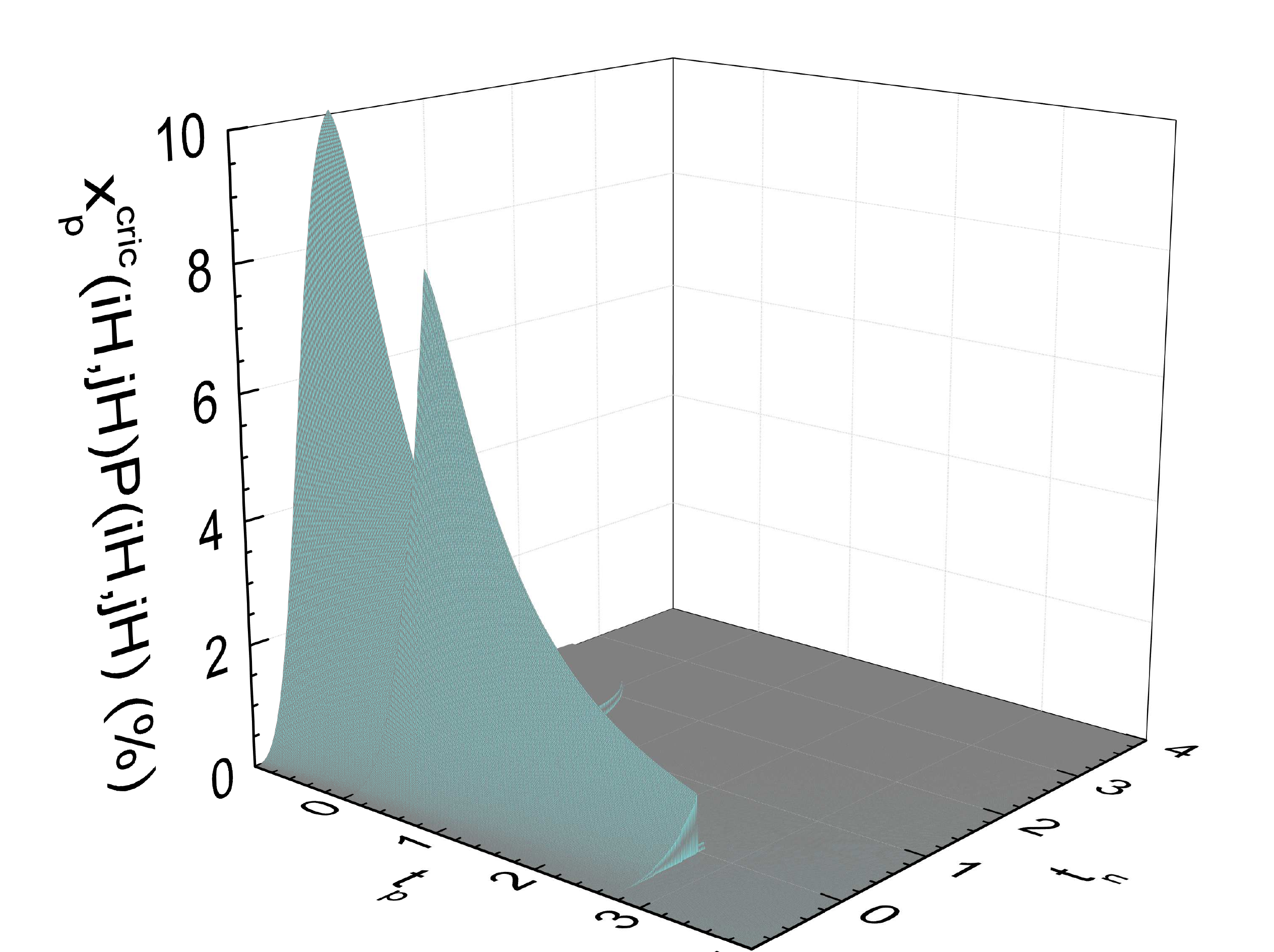}
  \hspace{0.5cm}
   \includegraphics[width=7.5cm]{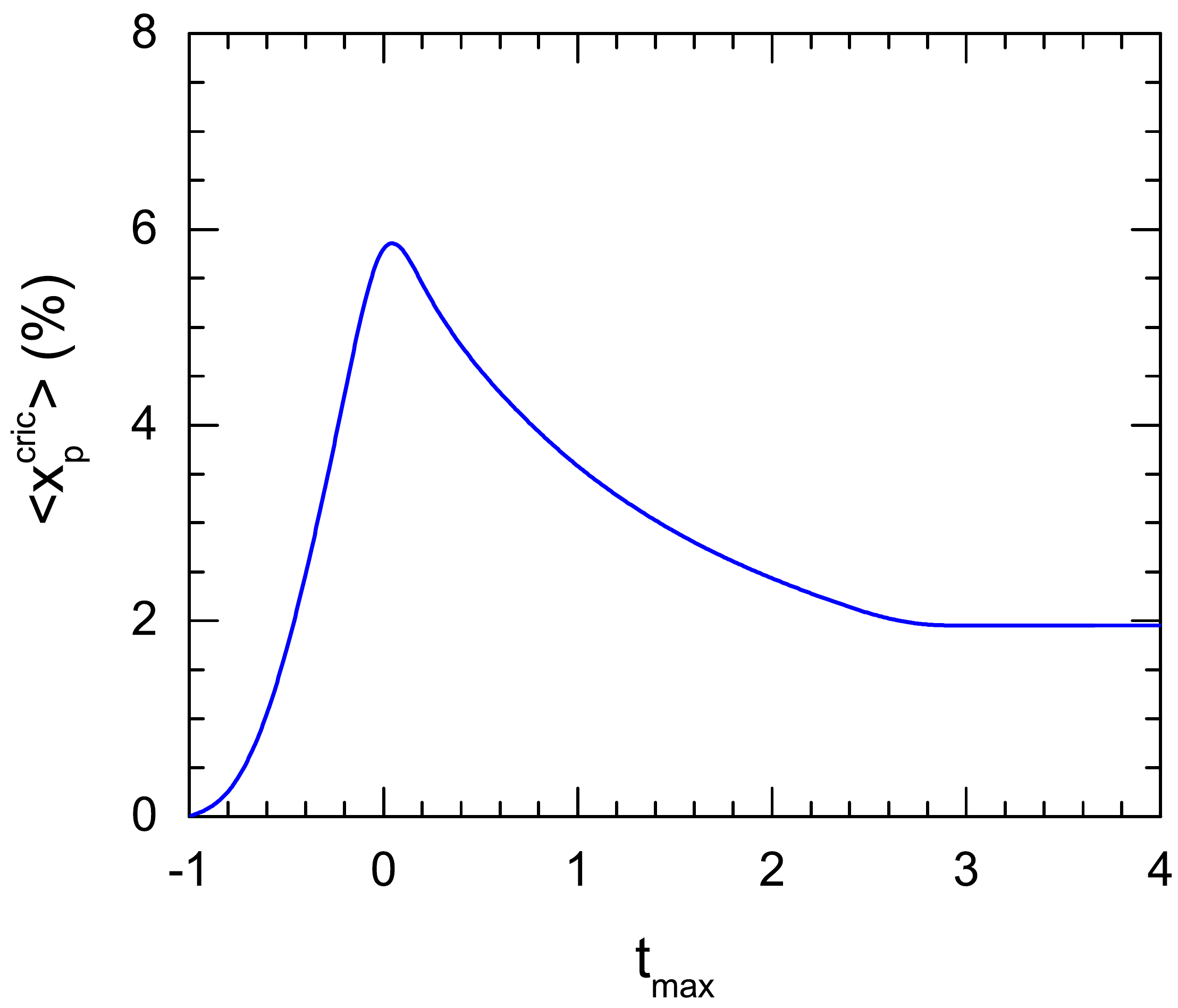}
 \caption{Left: Product of the critical proton fraction and the relative probability at each ``site" (relative contribution of each site in momentum-space of neutrons and protons to the average critical proton fraction for the direct URCA to occur). Right: Critical proton fraction as a function of $t_{\max}$.}\label{xpcProb1}
\end{figure}\par

Considering the possibilities that both nucleons involved can be either above or below their own Fermi surfaces, a statistical average of the critical proton fraction can be evaluated from
\begin{equation}
\langle
x_{\rm{p}}^{\rm{cric}}\rangle\equiv\frac{\sum_{i,j}\mathcal{P}(iH,jH)x_{\rm{p}}^{\rm{cric}}(iH,jH)}{\sum_{i,j}\mathcal{P}(iH,jH)},\end{equation}
where $iH=t_{\rm{p}}^i$ and $jH=t_{\rm{n}}^j$ running from $-1$ to a
maximum value of
$t_{\max}\lesssim t_{\max}^{\rm{th}}\equiv\phi_{\max}-1=\phi_0(1-\phi_1)-1\approx2.71$, $H$
is the corresponding step size on a two-dimensional lattice in momentum-space.
For instance, if $-1\leq t_{\max}\leq0$, then one artificially only takes the processes in
which protons and neutrons are both under their Fermi surfaces into consideration, indicating the probability $\Delta_{\rm{n}}(1-\Delta_{\rm{p}})$
is dominant. Combining both the values of $t_{\rm{n}}$ and $t_{\rm{p}}$ (and thus $x_{\rm{p}}^{\rm{cric}}(t_{\rm{p}},t_{\rm{n}})$) in this region (i.e., $-1\leq t_{\rm{p/n}}\leq0$)
and the probability $\mathcal{P}_{\rm{pn}}$, there is no surprise that the averaged $\langle x_{\rm{p}}^{\rm{cric}}\rangle$ will be smaller than 1/9.
The $x_{\rm{p}}^{\rm{cric}}(iH,jH)$ is the critical proton fraction at the ``site" $(t_{\rm{p}}^i,t_{\rm{n}}^j)$,
see Fig.\,\ref{xpcProb1} for the product of the critical proton faction and the relative probability at each ``site".
The averaged critical proton fraction $\langle x_{\rm{p}}^{\rm{cric}}\rangle$ should be
obtained by including all the possible processes with different $t_{\rm{p}}$ and $t_{\rm{n}}$ values.
In the right window of Fig.\,\ref{xpcProb1}, the average
critical proton fraction as a function of $t_{\max}$ is shown. It is interesting to see that the direct URCA processes with low threshold
proton fractions even approaching zero can occur. Generally speaking, the HMT and the low momentum depletion makes the averaged
critical proton fraction much smaller than 11\%. Considering that the SRC experiments indicate a value of $t_{\max}\approx 2.71$,
one thus expects that $\langle x_{\rm{p}}^{\rm{cric}}\rangle\approx2\%$ after including all the processes with possible $t_{\rm{p}}$ and $t_{\rm{n}}$, i.e.,
the direct URCA process is statistically much easier to occur. This finding is consistent with an earlier study based on different arguments\,\cite{Fra08}.
To better understand the neutron star cooling data\,\cite{Yak01,Pag06}, obviously the SRC effects on the cooling mechanism deserve further investigations.

It was suggested recently in ref.\,\cite{Don15} that the neutrino emissivity of the direct URCA
process is reduced by a factor $\eta=Z_{\rm{F}}^{\rm{p}}Z_{\rm{F}}^{\rm{n}}$ compared to the FFG
model with $Z_{\rm{F}}^J$ being the discontinuity of nucleon $J$'s momentum distribution at the Fermi momentum. In models with SRC-induced HMT, the
depletion of the Fermi sphere together with the sizable value of
$C_J$ makes the factor $\eta$ much smaller than unity. However,
effects of HMT nucleons not considered in ref.\,\cite{Don15}
may enhance the emissivity of neutrinos\,\cite{Fra08}. Thus, to our
best knowledge, the net effects of the entire single-nucleon
momentum distribution modified by the SRC on both the critical
density for the direct URCA process to occur and the associated cooling rate of protoneutron stars are still unclear.

\subsection{SRC effects on nucleon mean free path in neutron-rich matter}\label{subx.6}
The nucleon mean free path (MFP) in neutron-rich matter is an important
quantity useful for simulating heavy-ion collisions (HICs)\,\cite{LCK08} and transmutations of nuclear wastes as well as understanding the dynamics in nuclear reactors and collisions between two neutron stars, just to name a few.
It is related to the in-medium nucleon-nucleon (NN) cross sections via $\lambda_{\rm{p}}^{-1}=\rho_{\rm{p}}\sigma_{\rm{pp}}^{\ast}+\rho_{\rm{n}}\sigma_{\rm{pn}}^{\ast}$
and
$\lambda_{\rm{n}}^{-1}=\rho_{\rm{n}}\sigma_{\rm{nn}}^{\ast}+\rho_{\rm{p}}\sigma_{\rm{np}}^{\ast}$,
in the two-body collision picture. Here, $\lambda_{\rm{n}}$
($\lambda_{\rm{p}}$) is the MFP of a neutron (proton) in ANM given by\,\cite{Neg81}
\begin{equation}\label{def_lambda}
\lambda_J=\frac{k_{\rm{R}}^J}{2M_{J}^{\ast,\rm{k}}|W_J|}
\end{equation}
where $k_{\rm{R}}^J=[2M(E-U_J)]^{1/2}$, $U_J/W_J$ is the real/imaginary part
of the nucleon optical potential. Since the SRC-induced HMT affect either directly or indirectly all three factors in the above expression, it is interesting to investigate effects of the SRC on the MFP.
Since the total nucleon effective mass in SNM at saturation density is empirically constrained, the SRC-induced enhancement of  E-mass will naturally reduce the k-mass $M_0^{\ast,\rm{k}}$
via $M_0^{\ast,\rm{k}}=M^*_0/M_0^{\ast,\rm{E}}$ relation, leading to a longer nucleon MFP according to Eq.\,(\ref{def_lambda}).
From a collisional picture, if nucleon-nucleon SRC pairs in the HMT with tensor correlations can be considered as effective
``dimers"\,\cite{Nis12}, then the two-body scattering picture between nucleons and dimers will hold with the average nuclear density unchanged. However, to get quantitative information about the nucleon MFP in this clustered matter one needs more information about nucleon-dimer collision cross sections.
In the following, we only discuss SRC effects on the nucleon MFP through the nucleon k-mass in Eq. (\ref{def_lambda}).
This simplified approach is incomplete but can still help us get some useful insight.  Similar to the Lane form for the real potential, i.e.,
$U_J\approx U_0+U_{\rm{sym}}\tau_3^J\delta$, the imaginary potential
can also be decomposed as $W_J\approx
W_0+W_{\rm{sym}}\tau_3^J\delta$ to leading order in isospin asymmetry. As $W_0$ is generally negative\,\cite{Hol16}, we have $|W_J|=-W_J$ at small $\delta$. The nucleon MFP can further be written in the Lane form of
$\lambda_J\approx\lambda_0+\lambda_{\rm{sym}}\tau_3^J\delta$ with
\begin{equation}
\lambda_0=-\frac{k_{\rm{R}}^0}{2M_0^{\ast,\rm{k}}W_0}
\end{equation}
being the MFP in SNM. Since $k_{\rm{R}}^0=[2M(E-U_0)]^{1/2}$, the $\lambda_0$ can be expressed with the parameters describing the single-nucleon momentum distribution as
\begin{align}\label{es_l0}
\lambda_0=\left.-\frac{k_{\rm{R}}^0}{2W_0M_0^{\ast}}\right/\left(1+\frac{2}{5}\beta_0-C_0-x_{\rm{SNM}}^{\rm{HMT}}\right).
\end{align}
Using the empirical values of the HMT parameters discussed earlier, i.e, $\beta_0\approx-0.27$, $C_0\approx0.161$ and $x_{\rm{SNM}}^{\rm{HMT}}\approx28\%$,
the value of the parenthesis in the denominator of Eq.\,(\ref{es_l0}) is about 0.45 leading to an enhance of the MFP in SNM at saturation density by a factor of about 2.2.

In neutron-rich matter, an interesting question is whether neutrons or protons have a longer MFP and how the answer may depend on the nucleon energy as well as the density and isospin asymmetry of the medium.
As we shall see, answers to these questions also depend on our knowledge about the isospin-dependence of the SRC. For illustrations, let's consider the isospin-dependent part of the MFP
\begin{align}\label{def_lsym}
\lambda_{\rm{sym}}
=&-\frac{M}{4\lambda_0M_{0}^{\ast,\rm{k},3}W_0^3}\cdot\bigg[\left[M(E-U_0)s_{\rm{k}}+M_{0}^{\ast,\rm{k}}U_{\rm{sym}}\right]W_0+2(E-U_0)M_{0,\rm{k}}^{\ast}W_{\rm{sym}}\bigg],
\end{align}
where $s_{\rm{k}}$ is the isospin splitting coefficient of the nucleon
k-mass, i.e., $m_{\rm{n}}^{\ast,\rm{k}}\equiv
M_{\rm{n}}^{\ast,\rm{k}}/M\approx
m_0^{\ast,\rm{k}}+2^{-1}s_{\rm{k}}\delta$ and
$m_{\rm{p}}^{\ast,\rm{k}}\equiv M_{\rm{p}}^{\ast,\rm{k}}/M\approx
m_0^{\ast,\rm{k}}+2^{-1}s_{\rm{k}}\delta$, thus
$(m_{\rm{n}}^{\ast,\rm{k}}-m_{\rm{p}}^{\ast,\rm{k}})/m_{0}^{\ast,\rm{k}}\approx
s_{\rm{k}}\delta$, $m_0^{\ast,\rm{k}}\equiv M_0^{\ast,\rm{k}}/M$ is
the reduced nucleon k-mass in SNM. The front-factor in
(\ref{def_lsym}), i.e., $
-{M}/{4\lambda_0M_{0}^{\ast,\rm{k},3}W_0^3}$ is positive since
$W_0<0$. Thus the sign of $\lambda_{\rm{sym}}$ is determined by that of the bracket, i.e.,
$f_{\rm{sym}}=[M(E-U_0)s_{\rm{k}}+M_{0}^{\ast,\rm{k}}U_{\rm{sym}}]W_0+2(E-U_0)M_{0}^{\ast,\rm{k}}W_{\rm{sym}}
$. Obviously, the dependence of $f_{\rm{sym}}$ (thus
$\lambda_{\rm{sym}}$) on the $M_0^{\ast,\rm{k}}$ and $s_{\rm{k}}$ is
nontrivial. Two useful features of the $f_{\rm{sym}}$ can be obtained though:
a) since $W_{\rm{sym}}>0$ and $W_0<0$\,\cite{Hol16}, a reduction of the $M_{0}^{\ast,\rm{k}}$
makes the $\lambda_{\rm{sym}}$ smaller at low energies, even to negative values (if $s_{\rm{k}}$ is large enough), i.e., the
MFP of neutrons may be smaller than that of protons.
b) the $U_{\rm{sym}}$ becomes negative above certain critical energy $E_{\rm{c}}$ as we discussed extensively in earlier sections of this review, and the term
$M(E-U_0)s_{\rm{k}}W_0$ wins the competition with
$2(E-U_0)M_{0}^{\ast,\rm{k}}W_{\rm{sym}}$ since $W_0$ generally
becomes more negative while $W_{\rm{sym}}$ becomes less positive,
indicating the MFP of neutrons is smaller than that of proton at high energies\,\cite{Jia07}.

\begin{figure}[h!]
\centering
\includegraphics[width=8.5cm]{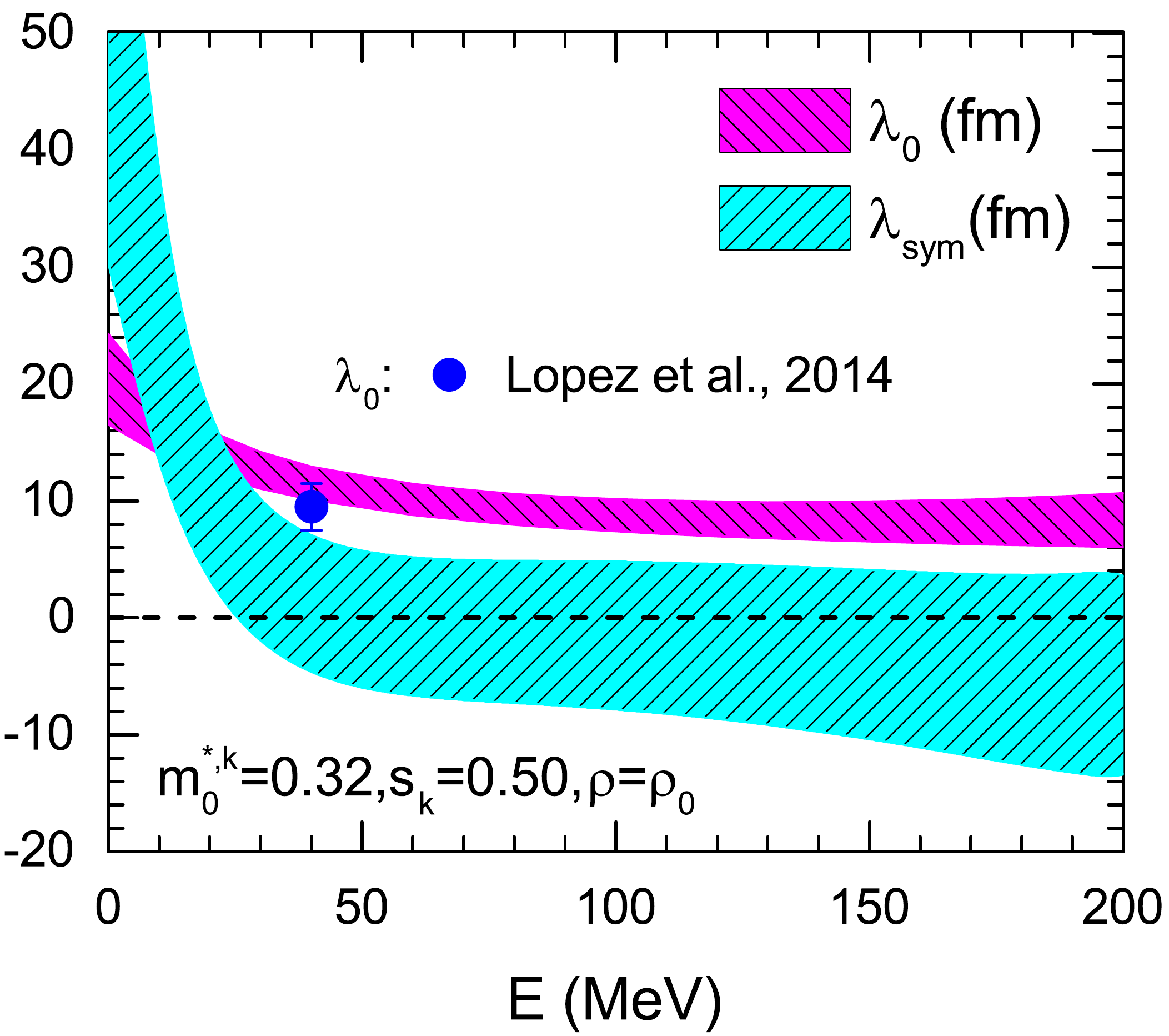}\caption{Energy dependence of the isoscalar $\lambda_0$ and isovector $\lambda_{\rm{sym}}$ nucleon mean free path at saturation density.
The blue solid circle with an error bar represents
the experimental data reported in ref.\,\cite{Lop14}.} \label{fig_Holtl0lsym-2}
\end{figure}

From the relations discussed above, it is clear that the isovector part $\lambda_{\rm{sym}}$ of the nucleon MFP in neutron-rich matter depends sensitively on the isospin splittings of nucleon effective masses as well as the isospin dependence of the SRC. However, as we discussed in the previous sections, our knowledge about these isospin splittings and dependences are quantitatively still rather poor, although qualitatively clear. Numerical evaluations of the
$\lambda_{\rm{sym}}$ will thus be rather model dependent. Here we make an estimate to get some guidance with selected inputs that could be model dependent.
For the total nucleon effective mass in SNM we use $M_0^{\ast}/M\approx0.70\pm0.10$\,\cite{LiChen15}. For the isospin splitting of the total effective mass, we take
$s\approx0.41\pm0.15$ from analyzing the nucleon-nucleus scatterings discussed in section \ref{Exp-cs}. We also take the nucleon k-mass splitting coefficient $s_{\rm{k}}=({1}/{m_0^{\ast,\rm{E}}})(s-{s_{\rm{E}}m_0^{\ast}}/{m_0^{\ast,\rm{E}}})\approx0.50\pm0.24$ and $m_0^{\ast,\rm{k}}={m_0^{\ast}}/{m_0^{\ast,\rm{E}}}\approx0.32\pm0.07
$\,\cite{LiBA16}. Moreover, we assume that the effective masses are independent of energy.  In addition, we adopt the
nucleon optical potential $U_0$, $U_{\rm{sym}}$, $W_0$ and $W_{\rm{sym}}$ obtained by Holt \textit{et al.} in ref.\,\cite{Hol16} using the chiral effective field theory incorporating the
realistic chiral two- and three-body interactions over a range of resolution scales about 400-500\,MeV.
Shown in Fig.\,\ref{fig_Holtl0lsym-2} are the resulting isoscalar $\lambda_0$ and isovector $\lambda_{\rm{sym}}$ parts of the MFP at $\rho_0$ as functions of nucleon energy at saturation density.
The uncertainties of $\lambda_0$ and $\lambda_{\rm{sym}}$ shown in Fig.\,\ref{fig_Holtl0lsym-2} are mainly due to the
uncertainties of those potentials\,\cite{Hol16}.
A recent experimental study on the $\lambda_0$ using nuclear stopping power in heavy-ion reactions is also
shown for a comparison (blue solid circle), i.e., $\lambda_0\approx9.5\pm2\,\rm{fm}$ with energy between about
35\,MeV to 40\,MeV\,\cite{Lop14}. The estimate indicates that at energies around/above 50\,MeV nucleons have a MFP around 10\,fm in SNM at saturation density, e.g., more quantitatively,
at $E=50\,$MeV, $\lambda_0\approx10.7\pm1.3\,\rm{fm}$ consistent with the experimental finding. The isospin-dependent part of the MFP $\lambda_{\rm{sym}}$ changes sign at energies as low as 30 MeV, but
its uncertainty is still too big to tell its sign at higher energies, implying that one can not tell
whether neutrons or protons have a longer MFP at energies above 30\,MeV based on the estimated results alone.
It should be stressed that the imaginary part of the single-nucleon potential used here is not experimentally constrained. In principle, to do a consistent analysis, both the real and imaginary parts of the
single-nucleon optical potential should be constrained by the same sets of nucleon-nucleus scatterings data.
We notice that an earlier study on the nucleon MFP and its isospin dependence based on the DBHF theory\,\cite{Sam08} found that at kinetic energies below about 60\,MeV,
the MFP of neutrons is longer than that of protons. At higher energies, however, the DBHF theory predicted an approximately identical MFP for neutrons and protons, consistent with the approximately
zero value of $\lambda_{\rm{sym}}$ from the estimate albeit within a large error band.
As discussed in some details in ref.\,\cite{LiChen15}, calculations of the isospin-dependence of MFP using various microscopic theories
and phenomenological models have given rather diverse predictions especially for high-energy nucleons. Our discussions above indicate that the reason may be traced back again to
our poor knowledge about the isospin dependence of the SRC and the isospin splittings of nucleon effective masses..

In summary of this section, the study of nucleon E-mass using the Migdal--Luttinger theorem and its connections with many interesting issues regarding the isospin-dependence of SRC in neutron-rich matter are truelly exciting.
The empirically constrained single-nucleon momentum distribution with a HMT can already provide some useful insight about several interesting issues. The EOS of neutron-rich matter especially the density dependence and kinetic/potential composition of nuclear symmetry energy, the isospin dependence of the nucleon E-mass and its MFP in neutron-rich matter all depend on the isospin dependence of the SRC.
In particular, the SRC makes the symmetry energy more concave around the saturation density, leading to an isospin-dependent incompressibility of neutron-rich matter in better agreement with the experimental results.
Moreover, it enhances the nucleon MFP in SNM at saturation density by a factor of about 2. The SRC may also affect significantly the critical proton fraction for the direct URCA process to happen in protoneutron stars. Needless to say, there are many interesting questions regarding the E-mass and SRC in neutron-rich matter remain to be investigated both theoretically and experimentally.

\FloatBarrier

\section{Impacts of isovector nucleon effective mass on thermal and transport properties of neutron-rich matter}
\label{thermal}
The momentum dependence of nuclear interactions are expected to affect dynamical processes of nuclear reactions and thermal equilibrium properties of various systems including hot nuclei and nuclear matter.
In this section, using several examples we discuss a few specific effects of the isovector nucleon effective mass on the isospin-dependent in-medium NN cross sections, symmetry energy of hot ANM as well as the liquid-gas phase transition and viscosity of neutron-rich matter. While the studies of these topics are important in their own rights, results of these studies also have broad impacts in several areas of both nuclear physics and astrophysics including nuclear reactions, supernovae explosions, neutron stars and gravitational wave emissions from oscillations of isolated neutron stars as well as their collisions.
\begin{figure}
\centering
\vspace{-0.3cm}
\includegraphics[width=16cm,height=10cm]{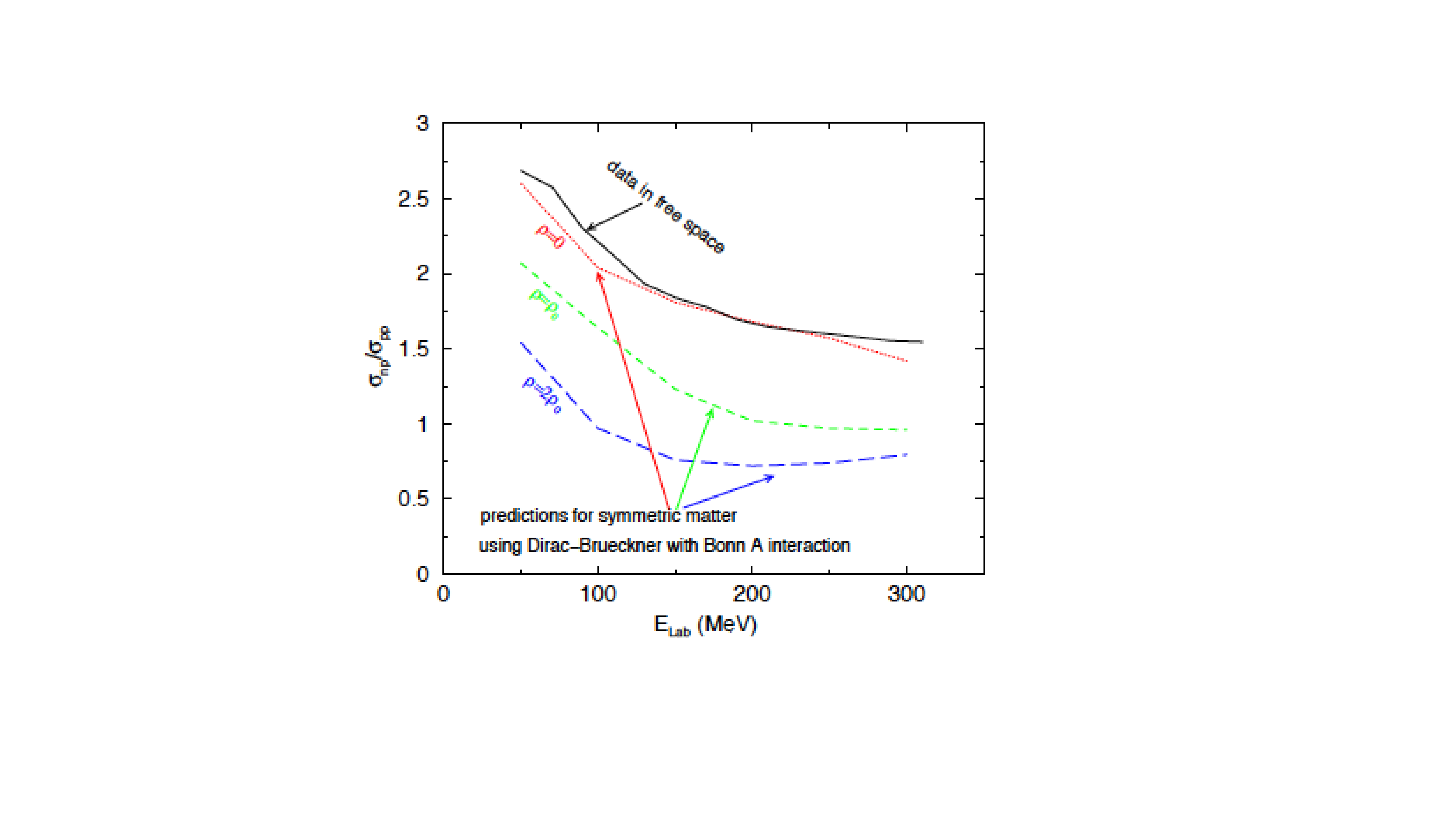}
\vspace{-2cm}
\caption{The ratio of np over pp elastic scattering cross sections
as a function of nucleon kinetic energy at various densities of nuclear medium.
The solid line is the experimental data while the dashed lines are
extracted from calculations in SNM using the Bonn A potential within the
Dirac-Bruckner approach in ref.\,\cite{gqli}. Taken from ref.\,\cite{LLD}. }
\label{xsc0}
\end{figure}

\subsection{Effects of neutron-proton effective mass splitting on the isospin-dependence of in-medium nucleon-nucleon cross sections}
In-medium NN cross sections determine transport properties, e.g., the degree and speed of reaching thermal and chemical equilibrium of nuclear reactions at various beam energies. They are also
useful for nuclear waste transmutations and nuclear stockpile stewardship. In free-space, as shown in Fig.\,\ref{xsc0} the neutron-proton cross section is about three times that of proton-proton (neutron-neutron) collisions below the pion production threshold. This is partially because both the iso-singlet and iso-triplet channels contribute to
neutron-proton (np) scatterings, while in proton-proton (pp) or neutron-neutron (nn) scatterings ($\sigma_{\rm{pp}}^{\rm{free}}$) only iso-triplet channels are involved.
How does this strong isospin-dependence
behave in neutron-rich matter? An accurate answer to this question is useful for understanding properties of neutron stars and nuclear reactions especially those induced by rare isotopes.
The mean-field potential and the NN scattering cross sections are
two basic inputs in transport model simulations of heavy ion collisions. In principle, they should be determined self-consistently from the same models using the same interactions.
However, due to the complexity of the problems and still rather model and interaction dependent predictions as we have discussed to some extent in previous sections, the single-nucleon potentials and NN scattering cross sections used in most transport models are obtained separately. In particular, the experimental free-space NN scattering cross sections are often used as default inputs.
However, it is theoretically expected and there are reliable experimental evidences indicating that the in-medium NN cross sections are not only different from their values in free-space but their dependencies on the
isospin asymmetry and density of the medium are also different for nn (pp) and np nucleon pairs. These changes and dependences are reflections of the isovector nuclear
effective interactions and the isospin-dependent Pauli blocking for the intermediate states of NN scatterings in ANM. In SNM, as shown in Fig.\,\ref{xsc0}, based on the DBHF theory the $\sigma_{\rm{np}}/\sigma_{\rm{pp}}$ decreases with increasing density\,\cite{gqli}. However, several other microscopic studies have reported the opposite, i.e., the $\sigma_{\rm{np}}/\sigma_{\rm{pp}}$ ratio increases with density in SNM, see, e.g.,\,\cite{gg,mk,qli}.
This situation requires more efforts both theoretically and experimentally to better understand the isospin dependence of in-medium NN cross sections. Indeed, several promising probes of the isospin-dependence of in-medium NN cross sections using reactions with radioactive beams were proposed\,\cite{LLD}.
\begin{figure}[h]
\centering
\vspace{1cm}
\includegraphics[height=0.3\textheight]{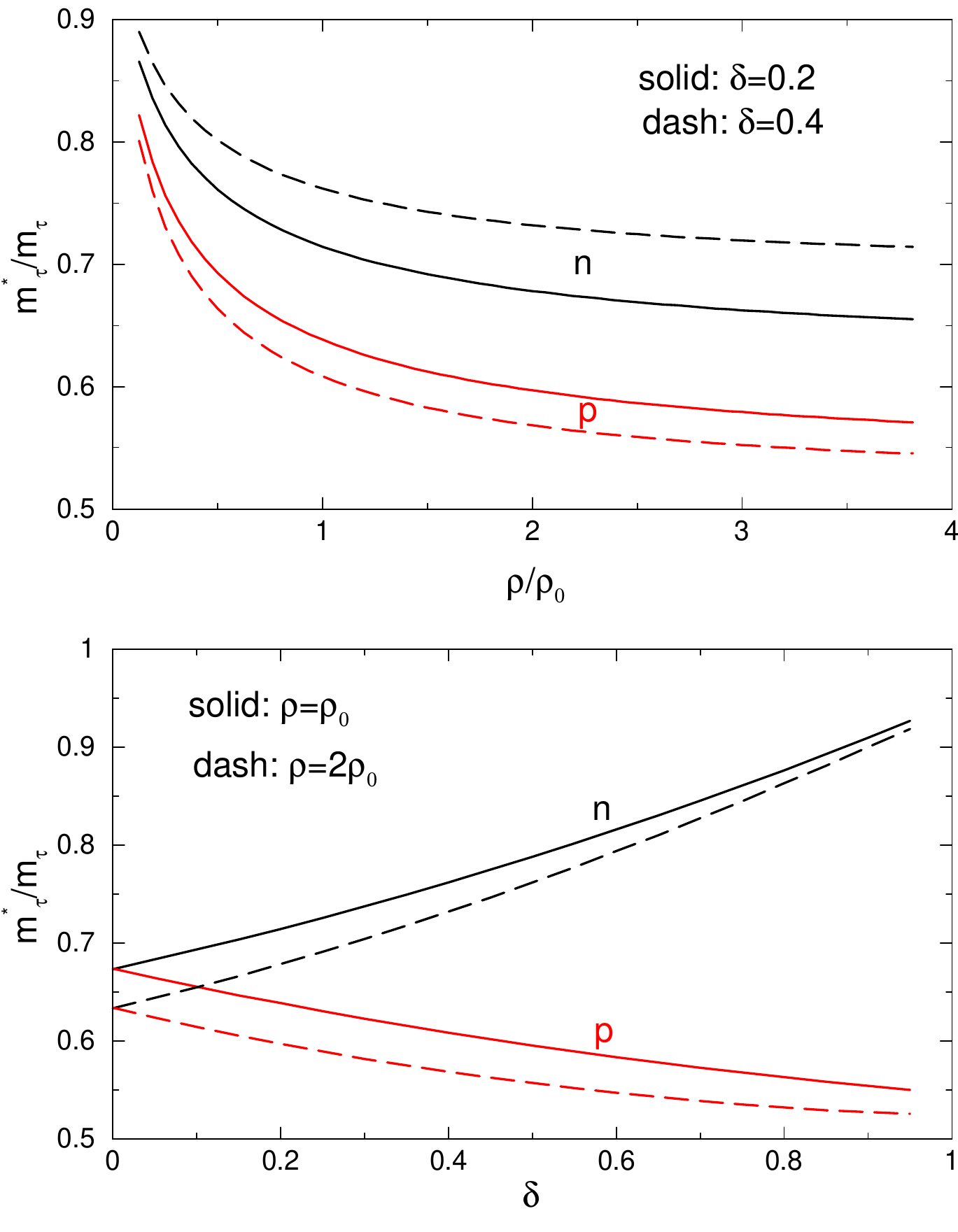}
\hspace{1cm}
\includegraphics[height=0.35\textheight]{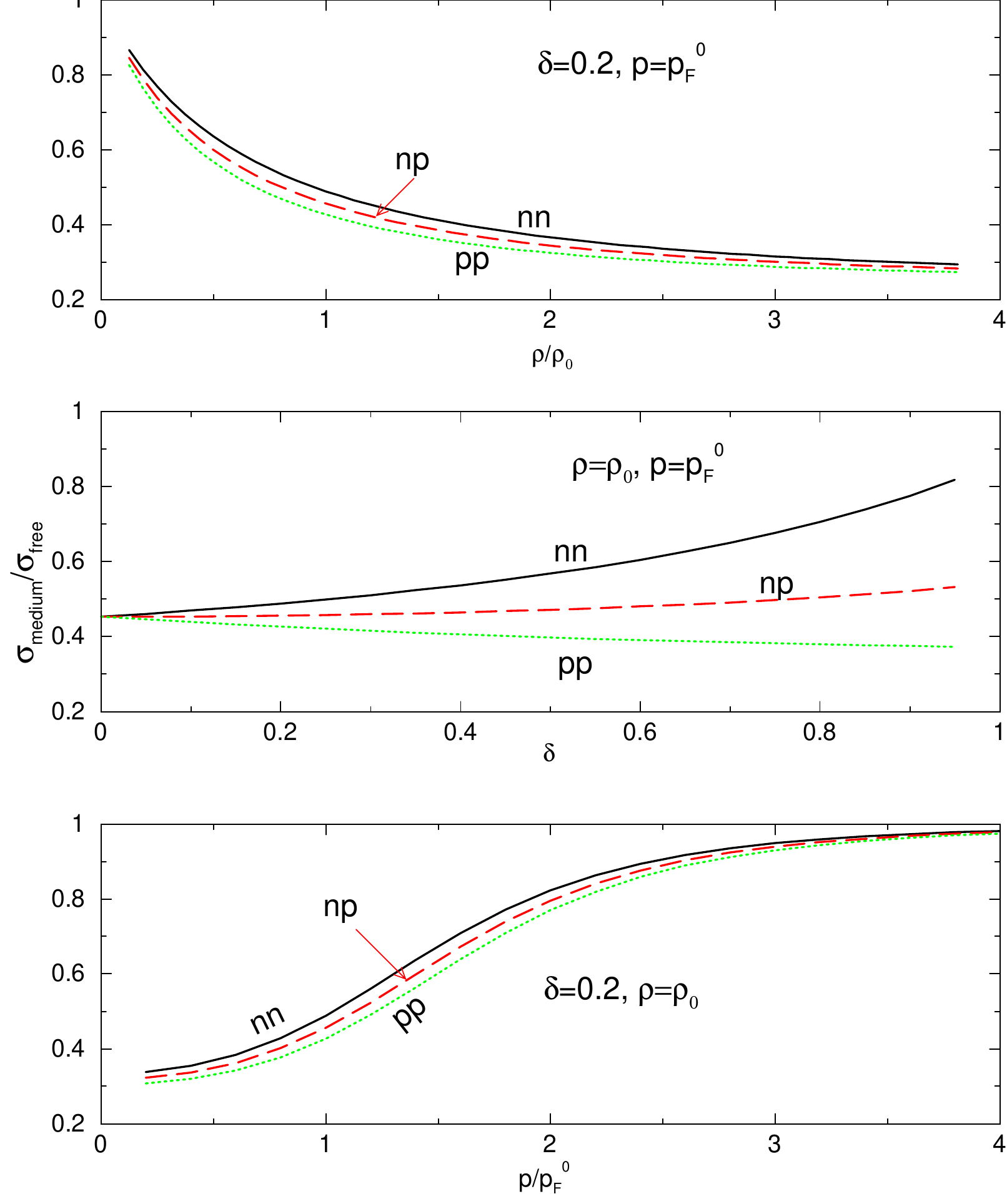}
\caption{Left panels: Neutron and proton effective masses in
asymmetric nuclear matter as functions of density (upper window) and
isospin asymmetry (lower window) from the MDI interaction.
Right panels: Ratio of nucleon-nucleon cross sections in nuclear medium
to their free-space values as a function of density (top window),
isospin asymmetry (middle window), and momentum (bottom window).
Taken from ref.\,\protect\cite{LiChen05}.}
\label{EffMassCrsc}
\end{figure}

\begin{figure}[h!]
\centering
\vspace{-0.cm}
\includegraphics[width=16cm,height=12cm]{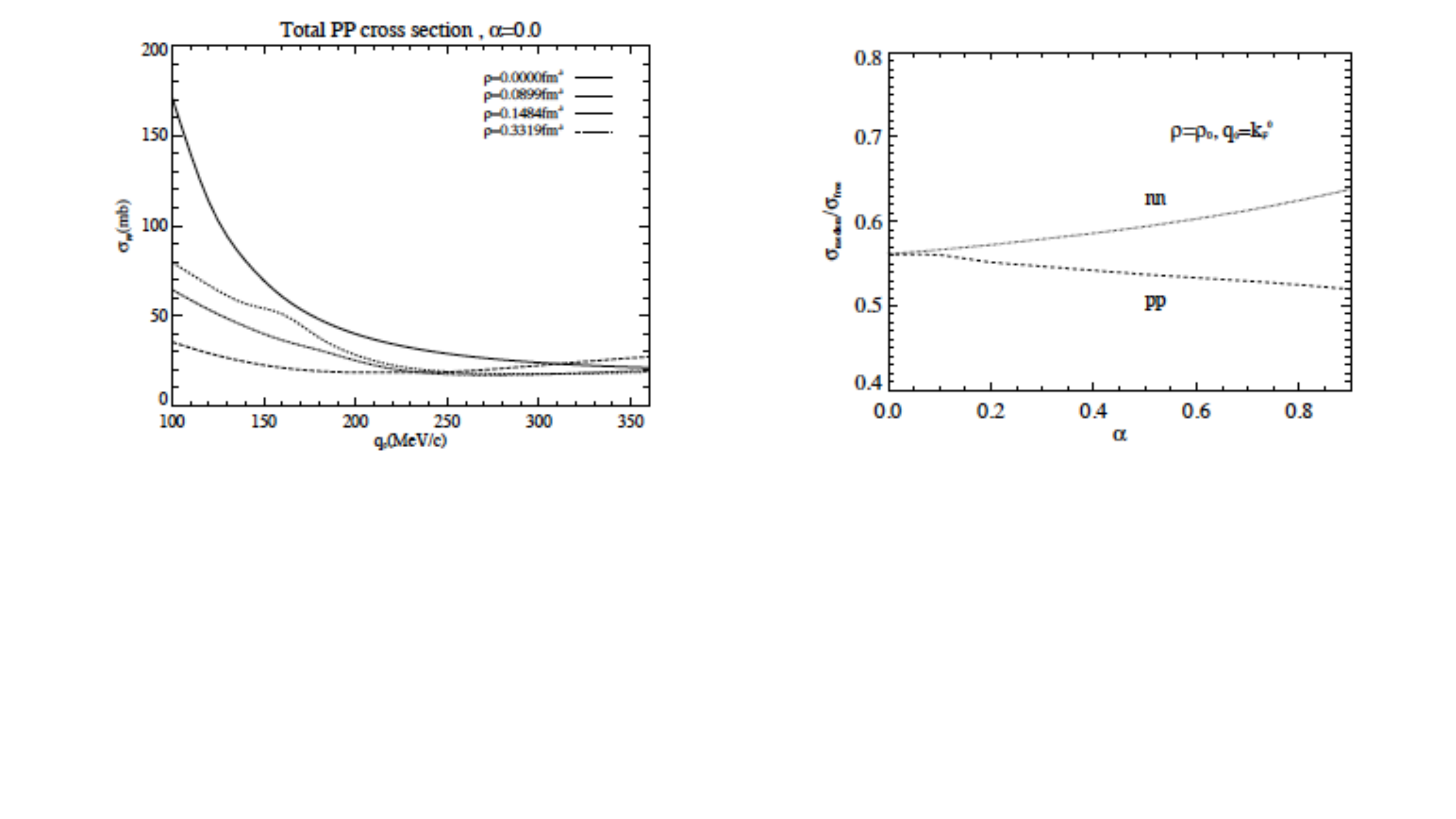}
\vspace{-5cm}
\caption{Left: Proton-proton cross sections in free space and in SNM at saturation density and in-medium cross sections at several abnormal densities
as functions of the nucleon momentum in the 2-nucleon c.m. frame.
Right: Ratios of the pp and nn cross sections to their free-space values as a function of the isospin asymmetry $\alpha$
near the saturation density and fixed 2N relative momentum. Taken from ref.\,\protect\cite{Sam05b}.}
\label{Plamen}
\end{figure}

In recent years, in a number of transport models, e.g., all versions and extensions of the IBUU transport model\,\cite{LCK08}, one uses the isospin-dependent in-medium NN cross sections obtained by extending the effective mass
scaling model for in-medium NN cross sections from SNM\,\cite{Neg81,Pan91,Gale02} to ANM. In such models, the NN interaction matrix elements
in the nuclear medium and free space are assumed to be identical. The in-medium NN cross sections ($\sigma _{\rm{NN}}^{\rm{medium}}$) thus differ
from their free-space values ($\sigma_{\rm{NN}}^{\rm{free}}$) only due to the variation in the
incoming current and the density of final states. Since both factors depend on the effective masses of the colliding nucleon pair, the in-medium NN cross sections are reduced by the factor
\begin{equation}
R_{\rm{medium}}\equiv \sigma _{\rm{NN}}^{\rm{medium}}/\sigma _{\rm{NN}}^{\rm{free}}=(\mu _{\rm{NN}}^{\ast
}/\mu _{\rm{NN}})^{2},
\label{xmedium}
\end{equation}%
where $\mu _{\rm{NN}}$ and $\mu _{\rm{NN}}^{\ast }$ are, respectively, the free-space and in-medium
reduced masses of the colliding nucleon pair.

\begin{figure}[h!]
\centering
\vspace*{1.cm} 
\includegraphics[height=0.3\textheight]{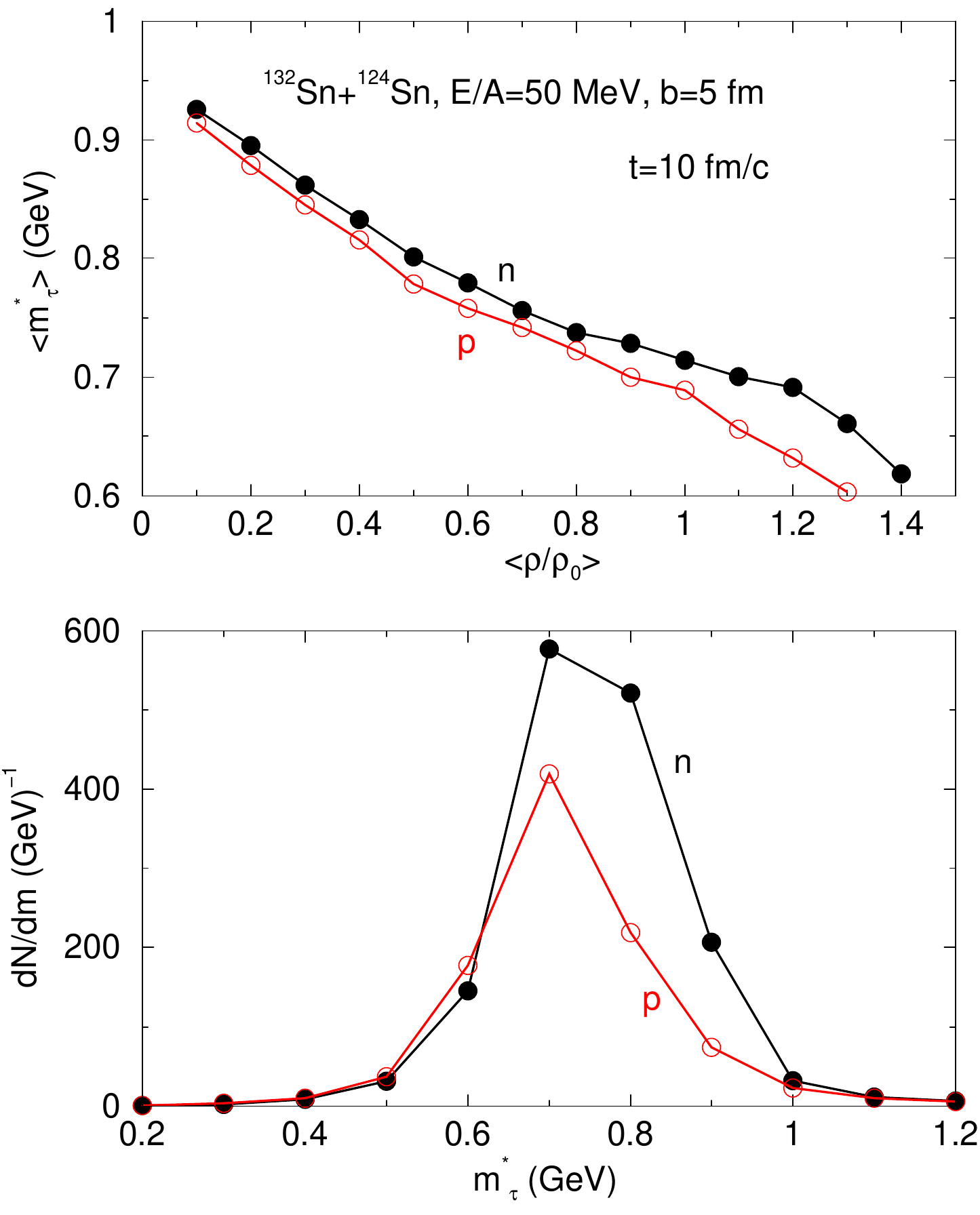}
\hspace{1.6cm}
\includegraphics[height=0.31\textheight]{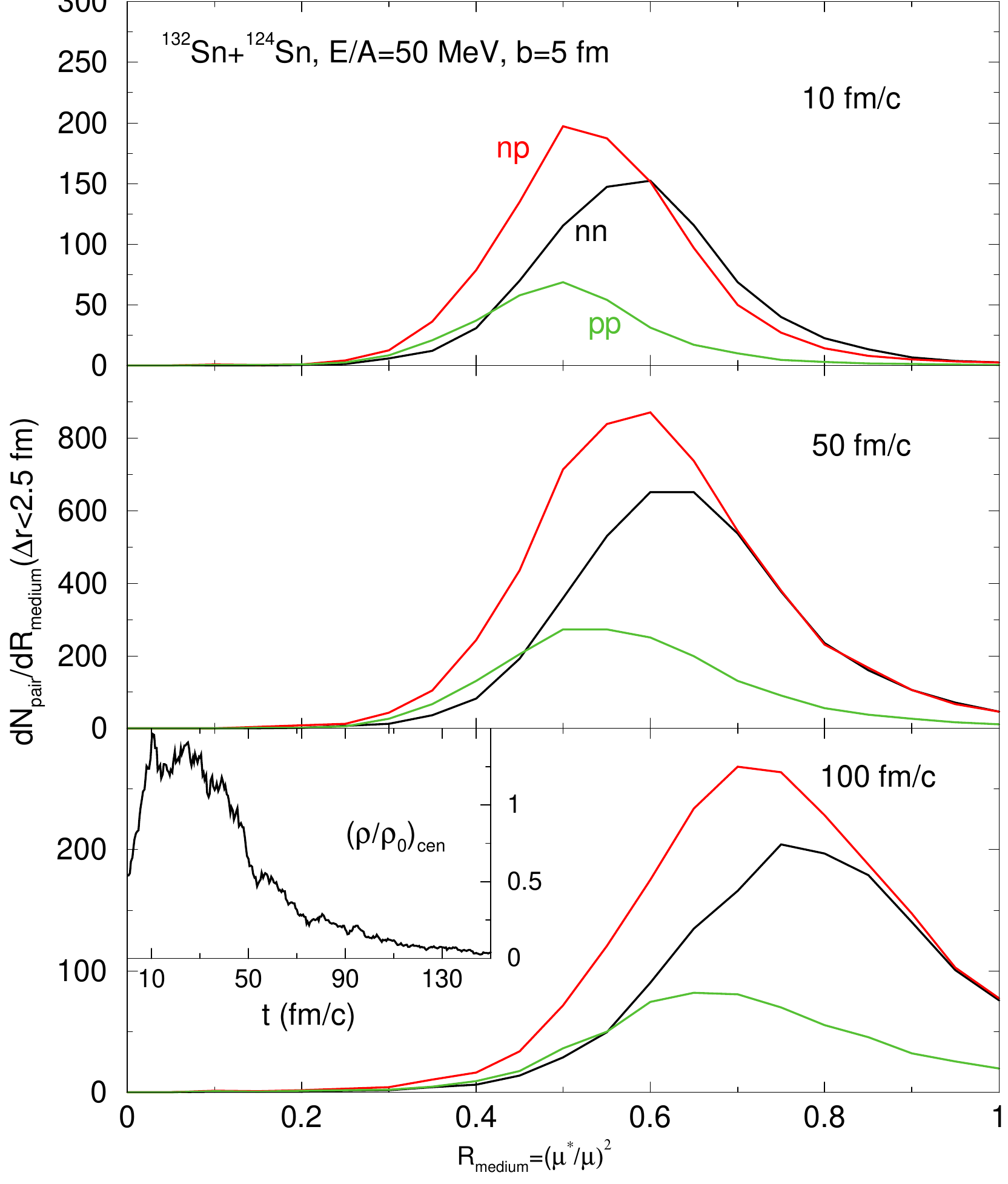}
\hspace*{1cm}
\caption{{\protect\small Left: The correlation between the average
nucleon effective mass and the average nucleon density (top), and the
distribution of nucleon effective masses (bottom) in the reaction of }$^{132}
${\protect\small Sn+}$^{124}${\protect\small Sn at a beam energy of 50 MeV/A
and an impact parameter of 5 fm.
Right: The distribution of the reduction
factor of in-medium NN cross sections in the reaction of }$^{132}$%
{\protect\small Sn+}$^{124}${\protect\small Sn at a beam energy of 50 MeV/A
and an impact parameter of 5 fm at instants $t=10$, 50 and 100 fm/c, respectively. The
insert is the time evolution of the central density in the reaction. Taken from ref.\,\cite{LiChen05}.}}
\label{xx}
\end{figure}

As an example, we cite here the main results obtained in ref.\,\cite{LiChen05} using the mass scaling model with the momentum-dependent interaction (MDI)\,\cite{LiBA04,Das03,Che05a}
\begin{align}
U_{\rm{MDI}}(\rho ,\delta ,\vec{p},\tau) =&A_{u}\frac{\rho _{-\tau }}{\rho _{0}}%
+A_{l}\frac{\rho _{\tau }}{\rho _{0}}
+B\left(\frac{\rho }{\rho _{0}}\right)^{\sigma }(1-x\delta ^{2})-4\tau x\frac{B}{%
\sigma +1}\frac{\rho ^{\sigma -1}}{\rho _{0}^{\sigma }}\delta \rho _{-\tau }
\notag \\
&+\frac{2C_l}{\rho _{0}}\int d^{3}p^{\prime }\frac{f_{\tau }(%
\vec{r},\vec{p}^{\prime })}{1+(\vec{p}-\vec{p}^{\prime })^{2}/\Lambda ^{2}}
+\frac{2C_u}{\rho _{0}}\int d^{3}p^{\prime }\frac{f_{-\tau }(%
\vec{r},\vec{p}^{\prime })}{1+(\vec{p}-\vec{p}^{\prime })^{2}/\Lambda ^{2}},
\label{MDIU}
\end{align}%
where  $\tau=1/2(-1/2)$ is the isospin for neutrons (protons); $f_{\tau }(\vec{r},\vec{p})$ is the phase-space distribution function at positive $\vec{r}$ and momentum $\vec{p}$; $x$ is the parameter controlling the relative contributions of the spin-singlet and spin-triplet channel in the 3-body force term of the MDI energy density function. The parameters $A_l$, $A_u$, $B$, $C_l$, $C_u$, $\Lambda$, and $\sigma$ are determined by the empirical properties of SNM,  symmetry energy $E_{\rm{sym}}(\rho_0)$, the isoscalar effective mass $m_{\rm{s}}^*$ as well as the asymptotic values of the isoscalar $U_{0,\infty}$ and isovector $U_{\rm{sym},\infty}$ potentials at infinite momentum all at saturation density\,\cite{Das03,Che05a}. Shown in the left panels of Fig.\,\ref{EffMassCrsc} are the effective masses of neutrons and protons in cold ANM at their respective Fermi surfaces
as functions of density (upper window) and isospin asymmetry
(lower window) using the Eq.\,(\ref{xmedium})\,\cite{LiChen05}. It is seen that neutrons have a larger effective mass than
protons in neutron-rich matter. We notice that these are all total effective masses as the on-shell dispersion relation has been used.
Moreover, the neutron-proton effective mass splitting
increases with both the density and isospin asymmetry of the nuclear medium.
Shown in the right panels of Fig.\,\ref{EffMassCrsc} is the reduction factor $R_{\rm{medium}}$
for two colliding nucleons having the same magnitude of momentum $p$ as a function of the density (upper window), the isospin asymmetry (middle window), and the nucleon momentum (bottom window). Interestingly, one can see
that not only the NN cross sections are reduced compared to their values in free space, but the cross sections for nn are
larger than those for pp pairs in neutron-rich matter although their free-space values are identical. Moreover, their difference becomes larger in more neutron-rich matter due to the positive neutron-proton effective mass splitting in neutron-rich matter with the MDI interaction.

It is interesting to note that the above results are qualitatively consistent with calculations using the DBHF theory\,\cite{Sam05b} for
colliding nucleon pairs with relative momenta less than about $240$ MeV/c at densities less than about $2\rho _{0}$. For example, shown in the left window of
Fig.\,\ref{Plamen} are in-medium pp cross sections in SNM at several densities in comparison with its free-space values
as functions of the p-p relative momentum in their CMS. The reduction of the pp cross section with increasing density is obvious and consistent with predictions by other models.
It is seen in the right window that cross sections for pp and nn split in ANM, and their difference increases with the increasing isospin-asymmetry $\alpha$, qualitatively consistent with results from
the scaling model discussed above.

As one can imagine, in heavy-ion collisions, the density and isospin asymmetry evolve with time.
One thus needs to evaluate nucleon effective masses and the corresponding in-medium
NN cross sections dynamically. To illustrate these, we present an example from the IBUU04 calculations in ref.\,\cite{LiChen05}.
Shown in the left window of Fig.\ \ref{xx} are the correlation between the average nucleon effective mass and the
average nucleon density (top), and the distribution of nucleon effective
masses (bottom) at the instant of $10$ fm/c in the reaction of $^{132}$Sn+$
^{124}$Sn at a beam energy of $50$ MeV/A and an impact parameter of $5$ fm.
The maximum density reached at $10$ fm/c is about $1.4\rho _{0}$.
Indeed, as expected, the nucleon effective masses decrease with increasing density and the neutron-proton effective mass
splitting increases slightly at supra-normal densities reached in the reaction.
Shown on the right of Fig.\ \ref{xx} are the distributions of the NN cross section reduction factor $R_{\rm{medium}}$ in the reaction
considered at three instants approximately corresponding to the compression, expansion and
freeze-out stages of the reaction. The $N_{pair}$($\Delta r<2.5$ fm)
is the number of nucleon pairs with spatial separations less than $2.5$ fm, representing potential colliding pairs whose scattering cross section will
be reduced by the factor $R_{\rm{medium}}$. As shown, as much as $50\%$ reduction occurs for NN scatterings in the early
stage of the reaction while as the system expands the average density decreases,
the reduction factor $R_{\rm{medium}}$ then gradually goes towards $1$ in the
later stage of the reaction.

\subsection{Effects of neutron-proton effective mass splitting on nuclear symmetry energy at finite temperatures}
\begin{figure}[h!]
\centering
\vspace{-1cm}
\includegraphics[scale=0.9,clip]{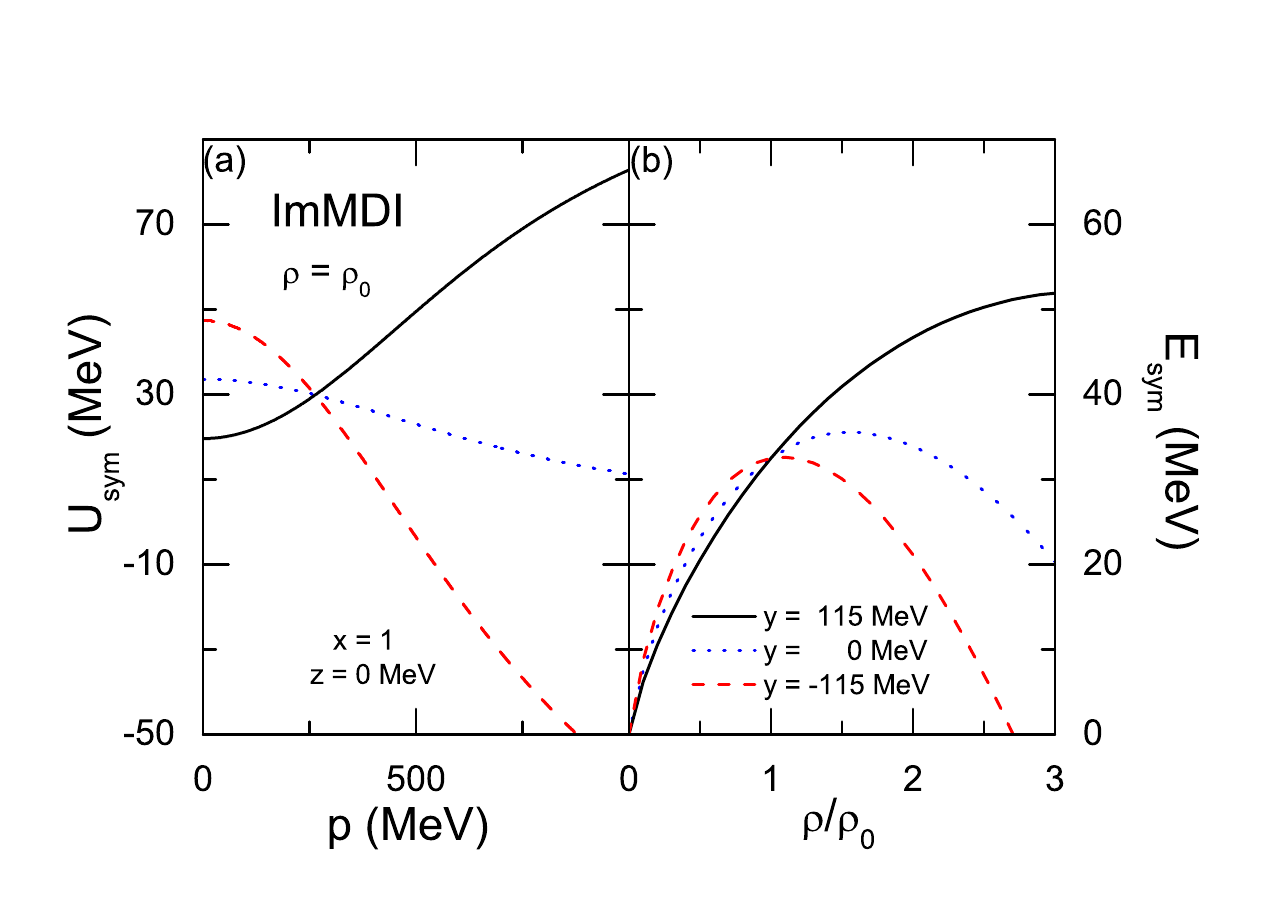}
\caption{Momentum dependence of the symmetry potential (left) and the density dependence of the nuclear symmetry energy (right) from the improved momentum-dependent interaction (ImMDI) with different $y$ parameters but fixed $x$ and $z$ parameters. Taken from ref.\,\cite{Xu15prc}.}
\label{fig:y}
\end{figure}
The nucleon effective mass, or equivalently, the momentum dependence of the nuclear mean-field potential, affects the thermodynamic properties of nuclear matter through the occupation probability in momentum-space at finite temperatures. Here we focus on discussing its effects, especially those of the isovector nucleon effective mass, on the symmetry energy of hot neutron-rich matter using two examples.
By comparing calculations using three interactions with characteristically different momentum dependences leading to neutron-proton effective mass splitting to be zero, positive and negative, it was first shown in ref.\,\cite{Xu08prc}
that the resulting kinetic and potential contribution to the total symmetry energy at finite temperature can both be rather different. This conclusion was later confirmed using other interactions\,\cite{Ou11,Xu15prc}.

As an example, let's first discuss results from a self-consistent thermodynamical calculation using an Improved Momentum-Dependent Interaction (ImMDI), see, e.g., ref.\,\cite{Xu15prc} and references therein.
The MDI interaction of Eq.\,(\ref{MDIU}) was improved by refitting the momentum dependence of the nucleon isoscalar optical potential up to nucleon kinetic energy of 1 GeV and introducing an additional parameters $y$ to adjust the asymptotic behavior of the symmetry potential at high momenta. As we discussed extensively earlier, the high-momentum behavior of the
symmetry potential and whether/where it may become negative are very uncertain. The $y$ parameter facilitates probing the high-momentum behavior of the symmetry potential, thus the neutron-proton effective mass splitting.
Fig.\,\ref{fig:y} compares the momentum dependence of the symmetry potential and the density dependence of the nuclear symmetry energy with three different $y$ values but a fixed value for the $x$ parameter at zero temperature. It is seen that with $y=-115$ (115) MeV one obtains a decreasing (increasing) symmetry potential with increasing nucleon momentum, and thus a larger (smaller) effective mass for neutrons than protons in neutron-rich medium. Moreover, since the momentum dependence of the symmetry potential affects the nuclear symmetry energy according to the HVH theorem as we discussed earlier, changing the $y$ value also modifies the density dependence of the symmetry energy as shown in the right window.

\begin{figure}[h!]
\centering
\includegraphics[width=6.5cm,height=7cm]{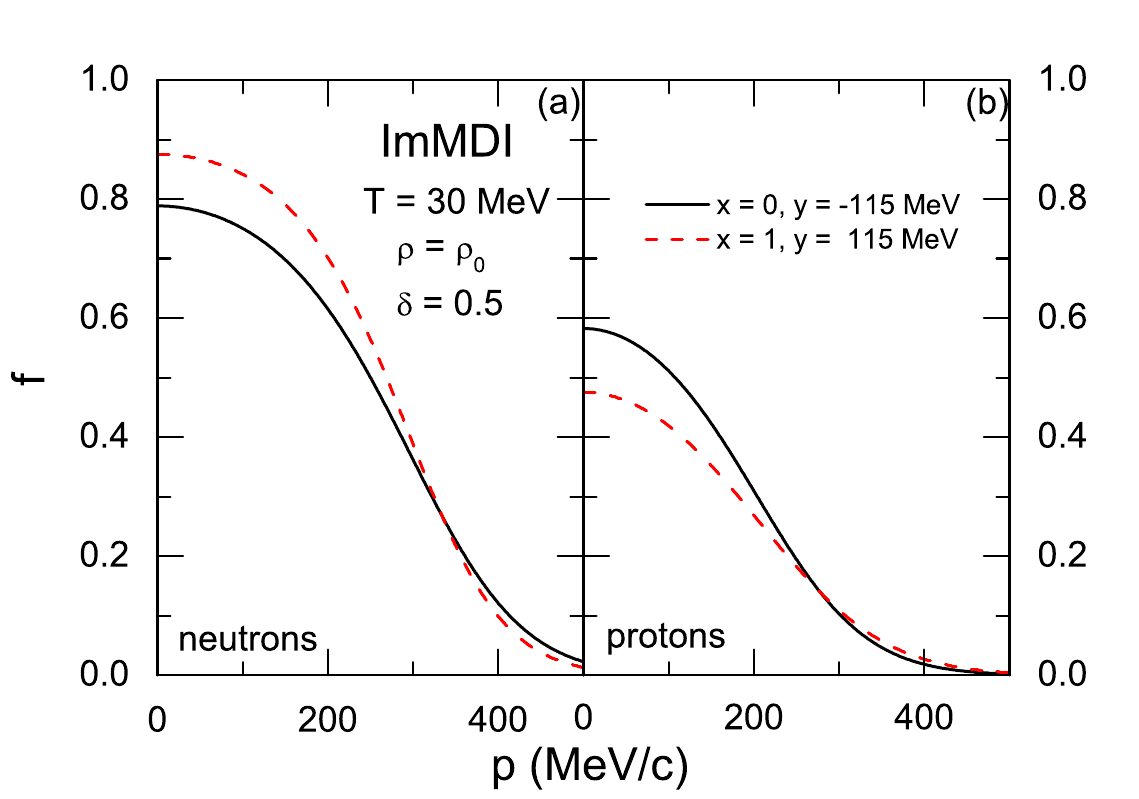}
\hspace{0.5cm}
\includegraphics[width=8.5cm,height=7cm]{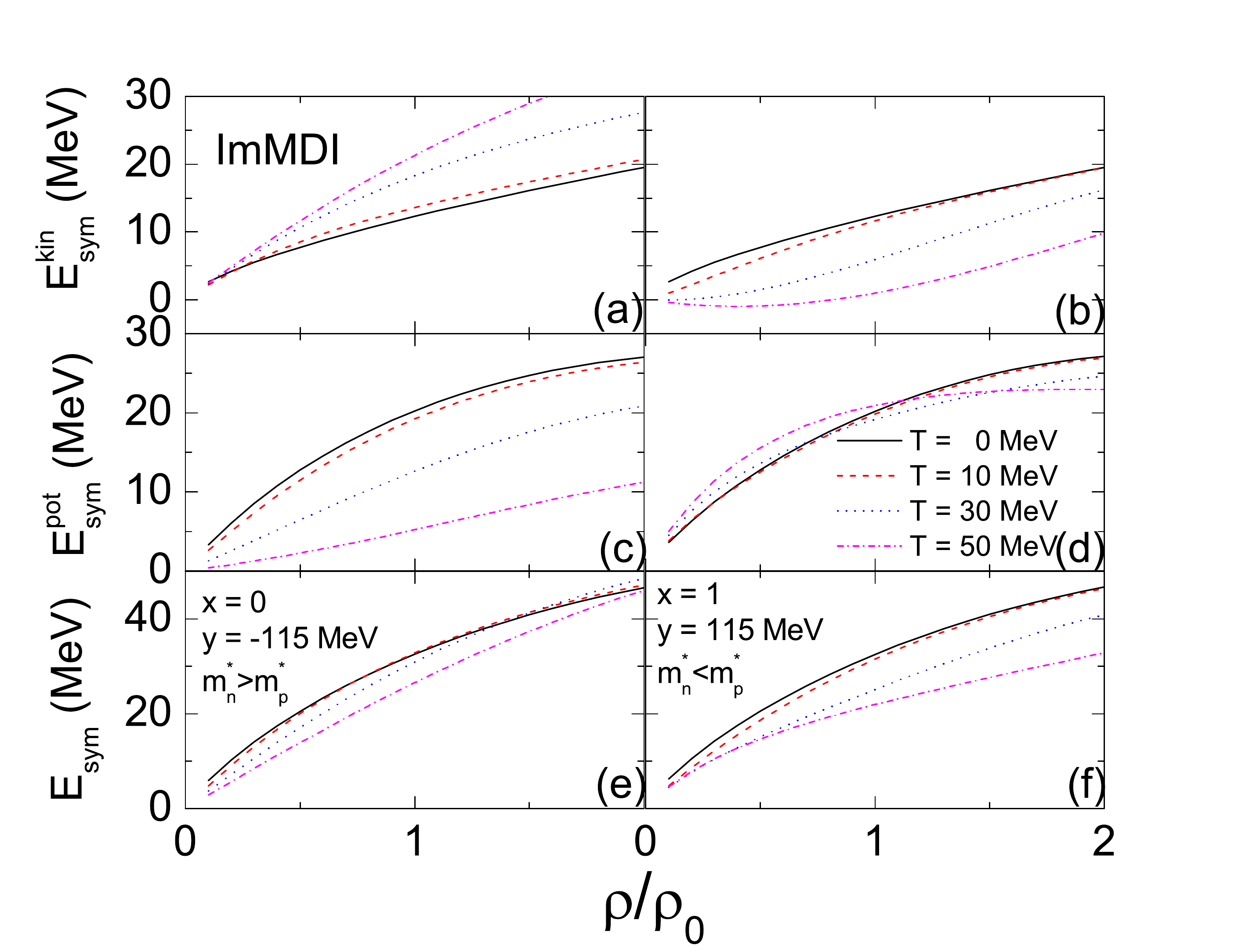}
\caption{Left: Neutron and proton occupation probability in momentum space with different isospin splittings of nucleon effective mass in neutron-rich matter from the ImMDI interaction.
Right: Density dependence of the kinetic (top panels) and potential (medium panels) contribution to the symmetry energy as well as the total symmetry energy (bottom panels) at various temperatures with different neutron-proton effective mass splittings from the ImMDI interaction. Taken/modified from ref.\,\cite{Xu15prc}.}
\label{fig:f}
\end{figure}

Using the ImMDI parametrization, it was found that the parameter sets $(x=0,y=-115~\text{MeV})$ and $(x=1,y=115~\text{MeV})$ lead to almost identical nuclear symmetry energy but different neutron-proton effective mass splittings. The two parameter sets were then used to investigate effects of the neutron-proton effective mass splitting on thermodynamic properties of neutron-rich nuclear matter. The left window of Fig.\,\ref{fig:f} compares the different neutron and proton occupation probabilities in ANM with the two parameter sets. It is seen that the $(x=0,y=-115~\text{MeV})$ parameter set leads to a more diffusive occupation probability for neutrons than protons. This is understandable since this parameter set leads to a larger effective mass for neutrons than protons especially at low nucleon momenta. After the self-consistent iteration, the momentum distribution in Fig.\,\ref{fig:f} is then favored to reach a lower energy of the system.
\begin{figure}[h!]
\centering
\includegraphics[scale=0.35,clip]{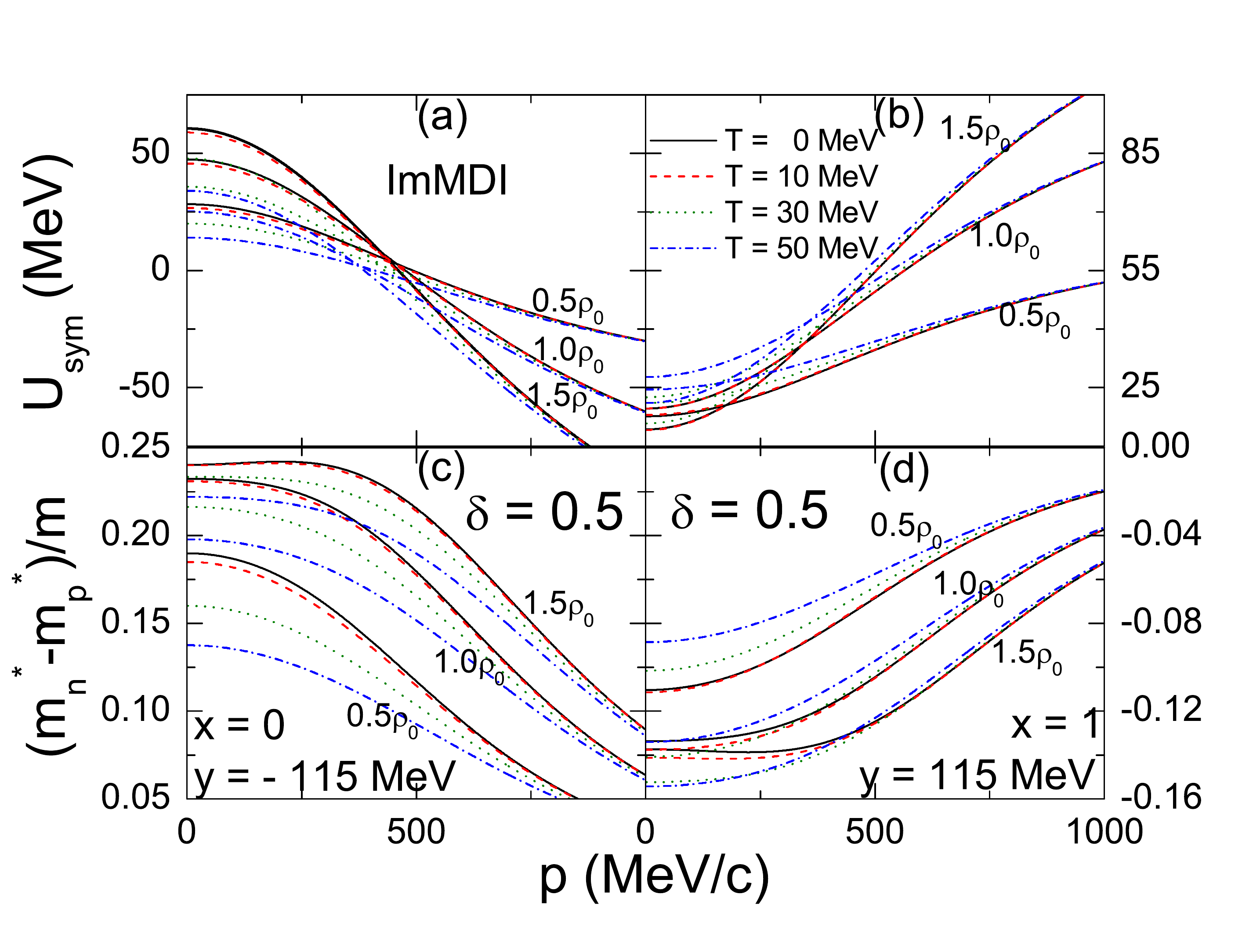}
\caption{Momentum dependence of the symmetry potential and the neutron-proton effective mass splitting at various densities and temperatures. Taken from ref.\,\cite{Xu15prc}.}
\label{fig:Usymmstar}
\end{figure}

The different neutron and proton momentum distributions from using different neutron-proton effective mass splittings will impact thermodynamic properties of neutron-rich matter. The right window of Fig.\,\ref{fig:f} displays the kinetic and potential contribution to the symmetry energy as well as the total symmetry energy at various temperatures with $m_{\rm{n}}^*>m_{\rm{p}}^*$ and $m_{\rm{n}}^*<m_{\rm{p}}^*$ obtained by using the two ImMDI parameter sets. It is seen that the kinetic contribution for $m_{\rm{n}}^*>m_{\rm{p}}^*$ increases with increasing temperature, while that for $m_{\rm{n}}^*<m_{\rm{p}}^*$ decreases with increasing temperature. This can be understood from their different momentum distributions as shown in Fig.\,\ref{fig:f}, since a more diffusive neutron momentum distribution leads to a larger kinetic contribution to the nuclear symmetry energy. On the other hand, the potential contribution decreases with increasing temperature for $m_{\rm{n}}^*>m_{\rm{p}}^*$ but remains almost unchanged for $m_{\rm{n}}^*<m_{\rm{p}}^*$, from a self-consistent thermodynamic calculation. Their combined effect leads to the different temperature dependence of the total symmetry energy, with $m_{\rm{n}}^*<m_{\rm{p}}^*$ resulting in a decreasing symmetry energy with increasing temperature, while $m_{\rm{n}}^*<m_{\rm{p}}^*$ resulting in a weaker temperature dependence and the symmetry energy may increase with increasing temperature above a certain density. Displayed in the right window of Fig.\,\ref{fig:Usymmstar} are the symmetry potential and the neutron-proton effective mass splitting at different densities and temperatures.
It is seen that the symmetry potential becomes flatter at lower densities while steeper at higher densities. In addition, the symmetry potential decreases with increasing temperature for $m_{\rm{n}}^*>m_{\rm{p}}^*$ using the $(x=0,y=-115~\text{MeV})$ parameter set while increases with increasing temperature for $m_{\rm{n}}^*<m_{\rm{p}}^*$ with the $(x=1,y=115~\text{MeV})$ parameter set, especially at lower densities. It is seen that the neutron-proton effective splitting is generally larger at lower momenta and/or higher densities, and the splitting is generally weaker at higher temperatures.

\begin{figure}[h!]
\centering
\includegraphics[width=12cm]{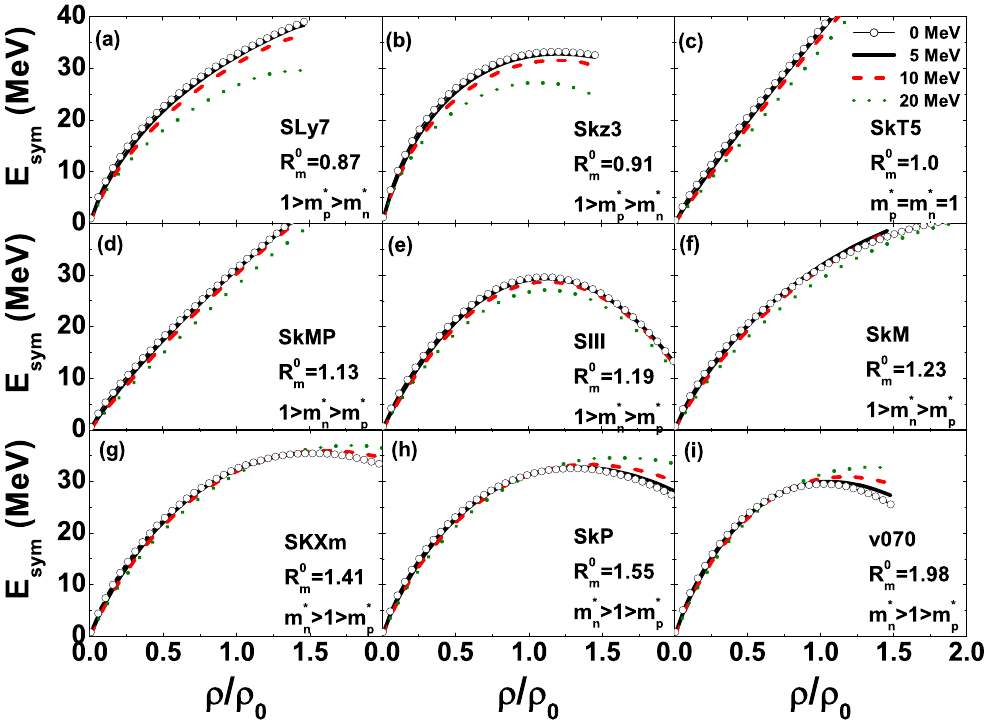}
\caption{Density dependence of the symmetry energy at various temperatures from different groups of Skyrme-Hartree-Fock interactions. Here $m^*$ is the reduced nucleon effective mass $m^*/m$. Modified from ref.\,\cite{Ou11}.}
\label{fig:EsymSkyrme}
\end{figure}

The effects of the neutron-proton effective mass splitting on the nucleon momentum distribution and the nuclear symmetry energy have also been investigated using the Skyrme-Hartree-Fock functionals corresponding to 7 groups of interactions with $m_{\rm{p}}^*=m_{\rm{n}}^*>m$, $m_{\rm{p}}^*=m_{\rm{n}}^*=m$, $m_{\rm{p}}^*=m_{\rm{n}}^*<m$, $m_{\rm{p}}^*<m<m_{\rm{n}}^*$, $m_{\rm{p}}^*<m_{\rm{n}}^*<m$, $m>m_{\rm{p}}^*>m_{\rm{n}}^*$, and $m_{\rm{p}}^*>m_{\rm{n}}^*>m$\,\cite{Ou11}.
Fig.\,\ref{fig:EsymSkyrme} compares the temperature dependence of the symmetry energy with typical Skyrme parametrizations in each group, with the subfigures ordered by the ratio $R_{\rm{m}}$ of the effective mass in pure neutron matter and that in symmetric matter, and $R_{\rm{m}}^0$ is the ratio at saturation density. It is seen that for the groups with $m_{\rm{n}}^*>m_{\rm{p}}^*$, the symmetry energy can increase with increasing temperature when the density is above a certain value. This phenomenon, called the transition of the temperature dependence of the symmetry energy in ref.\,\cite{Ou11}, has also been observed in Fig.\,\ref{fig:f}.

\begin{figure}[h!]
\centering
\includegraphics[width=12cm]{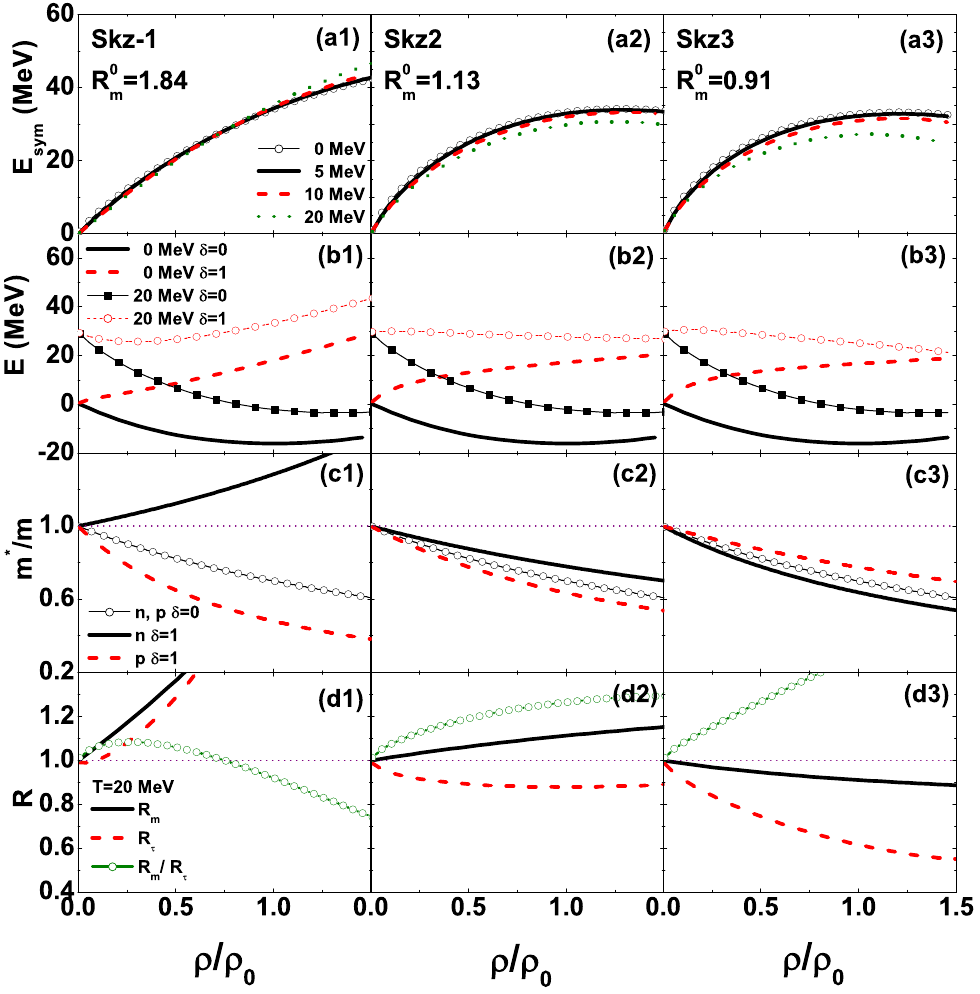}
\caption{Symmetry energy (first row), total binding energy (second row), nucleon effective mass (third row), and the ratios $R_{\rm{m}}$ and $R_\tau$ (bottom row) from three Skyrme interactions in the Skz-series. Taken from ref.\,\cite{Ou11}.}
\label{fig:EsymSkz}
\end{figure}

Fig.\,\ref{fig:EsymSkz} further displays the correlation between the neutron-proton effective mass and the temperature dependence of the symmetry energy for the Skz-series of the Skyrme forces. The authors of ref.\,\cite{Ou11} defined $R_\tau$ as the ratio between the increase in the kinetic energy density from zero temperature to finite temperature in pure neutron matter and that in symmetric matter. $R_{\rm{m}}$ reflects explicitly the neutron-proton effective mass splitting, while the symmetry energy decreases with increasing temperature if $R_{\rm{m}}/R_\tau>1$, and vice versa. Obviously, one can see a strong correlation between the temperature dependence of the symmetry energy and the ratios $R_{\rm{m}}$ as well as $R_\tau$. Similar to that observed in Fig.\,\ref{fig:f}, the different temperature dependence with different neutron-proton effective masses can be attributed to the occupation probability in momentum space.

\subsection{Effects of nucleon effective masses on the liquid-gas phase transition and differential isospin fractionation in neutron-rich matter}
The liquid-gas phase transition in SNM and the role of isoscalar nucleon effective mass on the level density and the limiting temperature extracted from nuclear reactions have been extensively studies over a long time, see, e.g.,
refs.\,\cite{Sh90,Joe02,Joe3,Lee04}. It is well known that nuclear liquid-gas phase transition in neutron-rich matter has some interesting new features, such as the order of phase transition, compared to that in SNM, see e.g., refs.\,\cite{Bes89,Mul95,LiCK97}.
It is also well known that these new features depend sensitively on the isovector interactions\,\cite{baran05,LCK08}. However, to our best knowledge, effects of the isovector nucleon effective mass on the liquid-gas phase transition have not been studied as extensively as those of the isoscalar one yet. Nevertheless, it is interesting to review a few useful results here. As we have discussed earlier, the MDI has the characteristics that it gives a positive
neutron-proton effective mass splitting while the latter is controlled by the $y$ parameter in the ImMDI. To investigate effects of the neutron-proton effective mass splitting in the MDI, the following Momentum-Independent Interaction (MID) giving the same density dependence of symmetry energy as the MDI for a given value of the $x$ parameter was constructed\,\cite{Che05a}
\begin{equation}
U_{\text{MID}}(\rho ,\delta ,\tau )=\alpha \frac{\rho }{\rho _{0}}+\beta\left(
\frac{\rho }{\rho _{0}}\right)^{\gamma }+U^{\text{asy}}(\rho ,\delta ,\tau),
\end{equation}%
with
\begin{eqnarray}
U^{\text{asy}}(\rho ,\delta ,\tau ) =\left[ 4F(x)\frac{\rho }{\rho _{0}}+4(18.6-F(x))\left(\frac{\rho }{\rho _{0}}\right)^{G(x)}\right] {\tau }{\delta}
+(18.6-F(x))(G(x)-1)\left(\frac{\rho }{\rho _{0}}\right)^{G(x)}{\delta }^{2}.
\label{Uasy}
\end{eqnarray}
In the above, the parameter $\alpha$, $\beta$, and $\gamma$ were fixed by the binding energy $E_0$ and the incompressibility $K_0$ at the saturation density $\rho_0$, and $F(x)$ and $G(x)$ are fitted to reproduce the same density dependence of symmetry energy as the MDI for a given value of the $x$ parameter\,\cite{Che05a}.

\begin{figure}[h!]
\centering
\includegraphics[scale=0.72,clip]{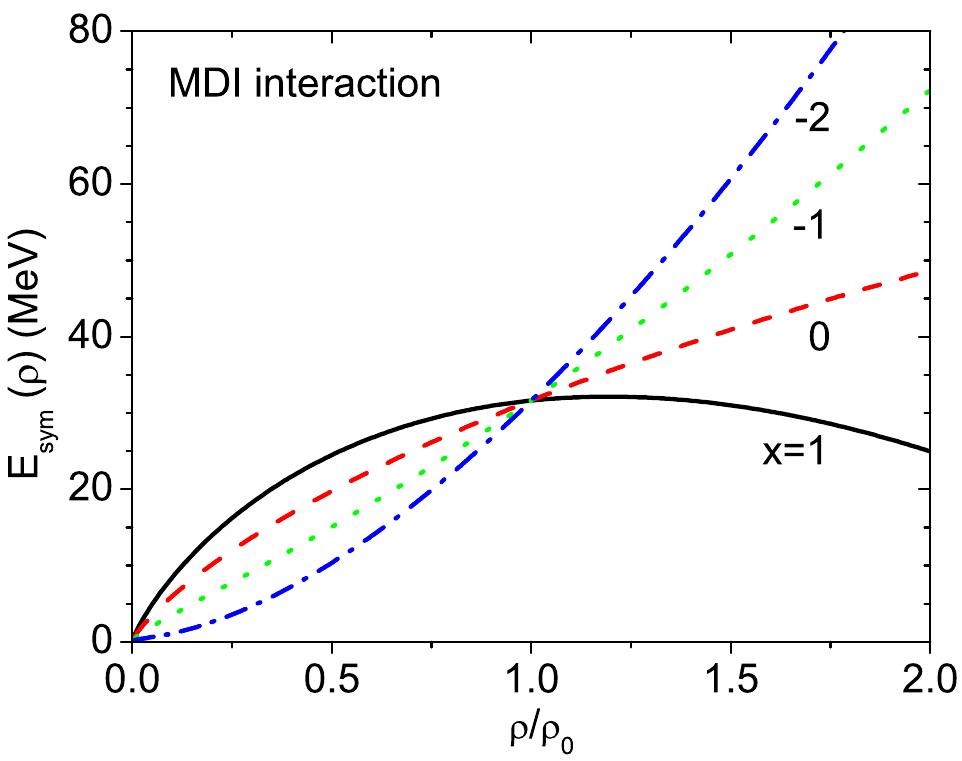}
\hspace{1cm}
\includegraphics[scale=0.8,clip]{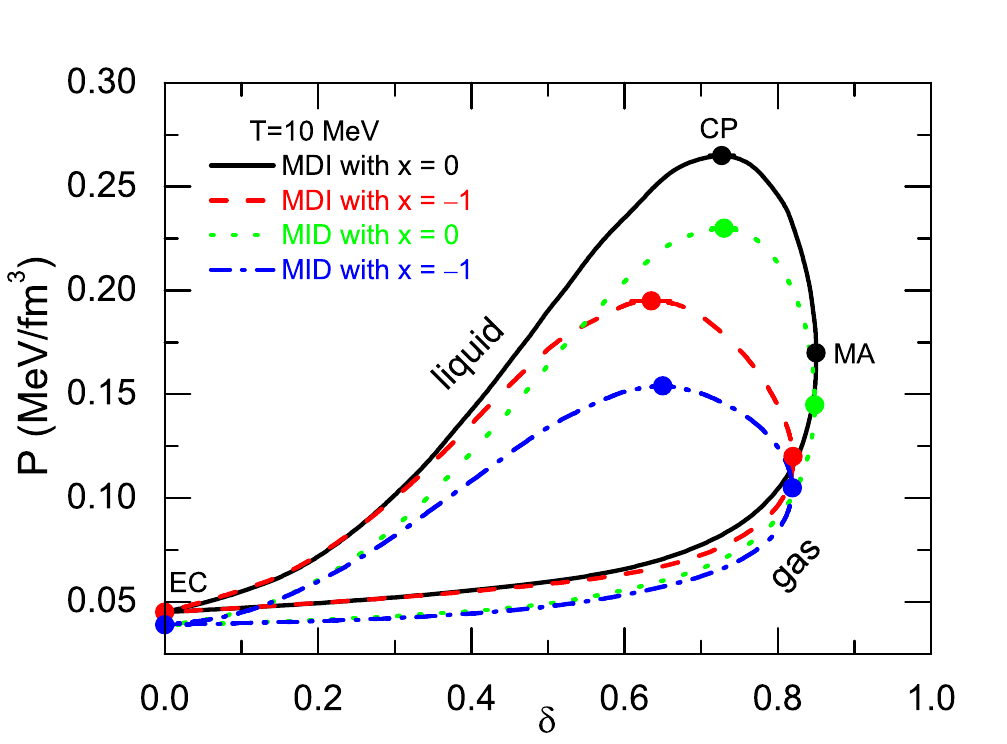}
\caption{Left: The density dependence of nuclear symmetry energy with different vales of the $x$ parameter used in the momentum-dependent interaction (MDI). Taken from ref.\,\cite{Che05a}.
Right: The binodal surface for nuclear liquid-gas phase transition at a temperature $T=10$ MeV from MDI and MID with different parameter $x$. CP, EC, and MA represent the critical point, the equal condensation, and the maximum asymmetry point of the binodal surface. Taken from ref.\,\cite{Xu07plb}. Note the different color codes used in the two windows.}
\label{fig:lgpt_2model}
\end{figure}

In the left window of Fig.\,\ref{fig:lgpt_2model} the density dependence of nuclear symmetry energy in the MDI with different $x$ parameters are shown. The right window compares the binodal surfaces of nuclear liquid-gas phase transition from the MDI and MID with different parameter $x$ in the pressure-isospin asymmetry plane (please note the different color codes used in the two windows). The binodal surface was obtained by plotting rectangles in the chemical potential isobars\,\cite{Xu07plb,Xu08ijmpe}, such that the vertexes angle of the rectangles representing the liquid and the gas phase satisfy the Gibbs condition, i.e., the same pressure, chemical potential and temperature. Inside the binodal surface is the mixed phase of the liquid and the gas, the left-hand side of it is the liquid phase, and the right-hand side is the gas phase. The parameter $x=0$ leads to a softer symmetry energy and a larger area of the mixed phase region, compared to the results with $x=-1$. On the other hand, the momentum-dependent interaction leads to a higher critical point (CP), a slightly higher equal condensation (EC) point, and also a larger area of the mix phase region, although the isospin asymmetry $\delta$ at the maximum asymmetry (MA) point is not affected. The different liquid-gas binodal surfaces from the MDI and MID can be intuitively understood from their different occupation probabilities\,\cite{Xu07plb}. For the MID, the more diffusive momentum distribution can be understood as an effectively higher temperature. Since the area of the mixed phase region generally decreases at higher temperatures, this explains why the MDI has a larger mixed phase region in the $P-\delta$ plane.
\begin{figure}[h!]
\centering
\includegraphics[scale=0.9,clip]{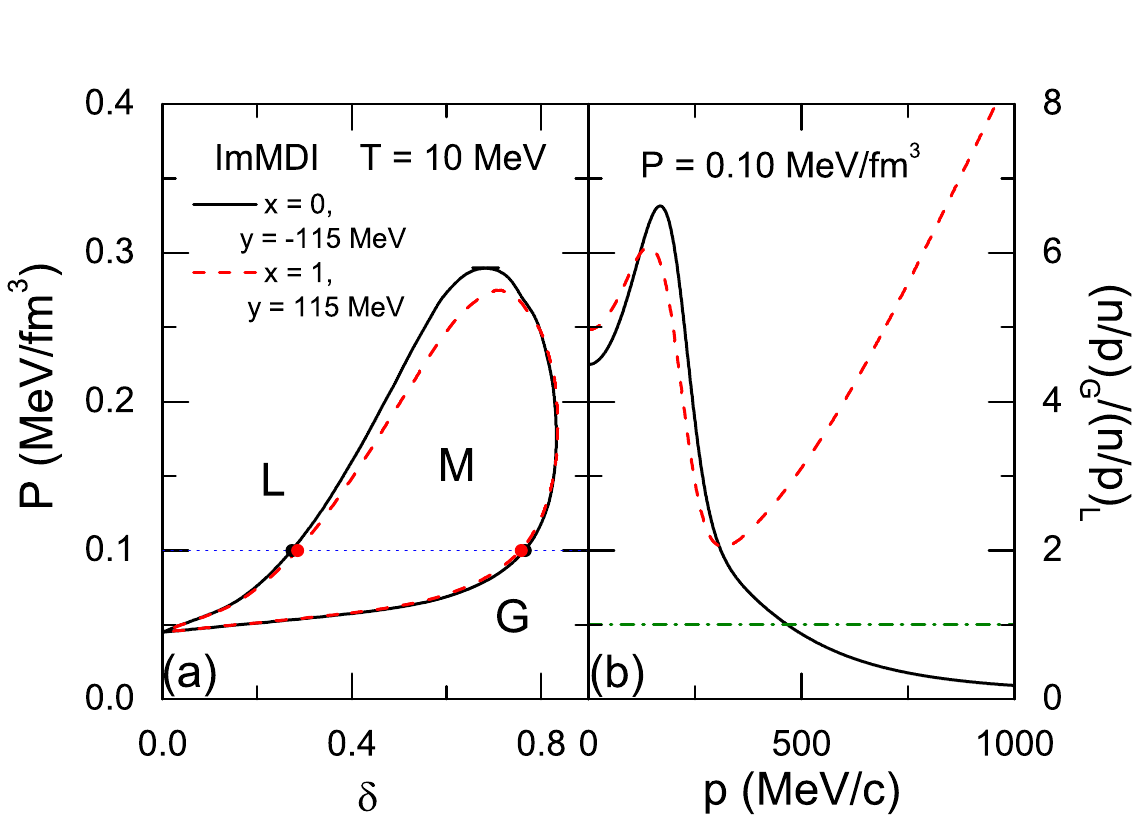}
\caption{The binodal surface of nuclear liquid-gas phase transition (left) and the ratio of neutron/proton in the gas phase with respect to that in the liquid phase (right). Taken from ref.\,\cite{Xu15prc}.}
\label{fig:isofrac}
\end{figure}

While the results and discussions above are useful, we notice that the comparisons presented there can't tell the respective effects of the isoscalar and isovector nucleon effective mass. In this regard, studies with the ImMDI interaction are more useful as the underlying neutron-proton effective mass splitting in the ImMDI can be varied under controlled conditions. As shown in the left panel of Fig.\,\ref{fig:isofrac}, the parameter sets $(x=0,y=-115~\text{MeV})$ and $(x=1,y=115~\text{MeV})$ with the same equation of states but different neutron-proton effective masses give similar liquid-gas phase boundaries, although there is a visible difference near the CP.
Thus, the difference in the MDI and MID phase boundaries observed in the right window of Fig.\,\ref{fig:lgpt_2model} is mainly due to the isoscarlar nucleon effective mass.
Another interesting phenomenon related to the neutron-proton effective mass splitting is the differential isospin fractionation in the liquid and gas phase. In ANM, to minimize its total energy the high-density liquid phase is generally less neutron-rich than the low-density gas phase due to the $E_{\rm{sym}}(\rho)\cdot \delta^2$ term in its EOS. However, it is found in ref.\,\cite{Li07prc} that the liquid phase can be more neutron-rich for energetic nucleons. It was later understood that this is related to the neutron-proton effective mass splitting or the momentum dependence of the nuclear symmetry potential. On the other hand, the energetic nucleons for $(x=0,y=-115~\text{MeV})$ with $m_{\rm{n}}^*>m_{\rm{p}}^*$ can be more neutron-rich in the liquid phase, while $(x=1,y=115~\text{MeV})$ with $m_{\rm{n}}^*<m_{\rm{p}}^*$ gives the opposite results. As found in ref.\,\cite{Li07prc}, the double ratio of neutron and proton in the gas (G) and liquid (L) phase is related to the momentum dependence of the symmetry potential through the relation
\begin{equation}
\frac{(n/p)_G}{(n/p)_L}(p) \approx \exp[-2(\delta_G \cdot U_{\rm{sym}}(p,\rho_G)-\delta_L \cdot U_{\rm{sym}}(p,\rho_L))],
\end{equation}
where $\delta_{G(L)}$ and $\rho_{G(L)}$ are respectively the isospin asymmetry and the density of the gas (liquid) phase.
The results in Fig.\,\ref{fig:isofrac} can thus be understood from Fig.\,\ref{fig:Usymmstar}, since the symmetry potential at high momenta becomes negative for $m_{\rm{n}}^*>m_{\rm{p}}^*$ but remains positive for $m_{\rm{n}}^*<m_{\rm{p}}^*$.

\begin{figure}[h!]
\centering
\vspace{-0cm}
\includegraphics[width=8cm,height=10cm]{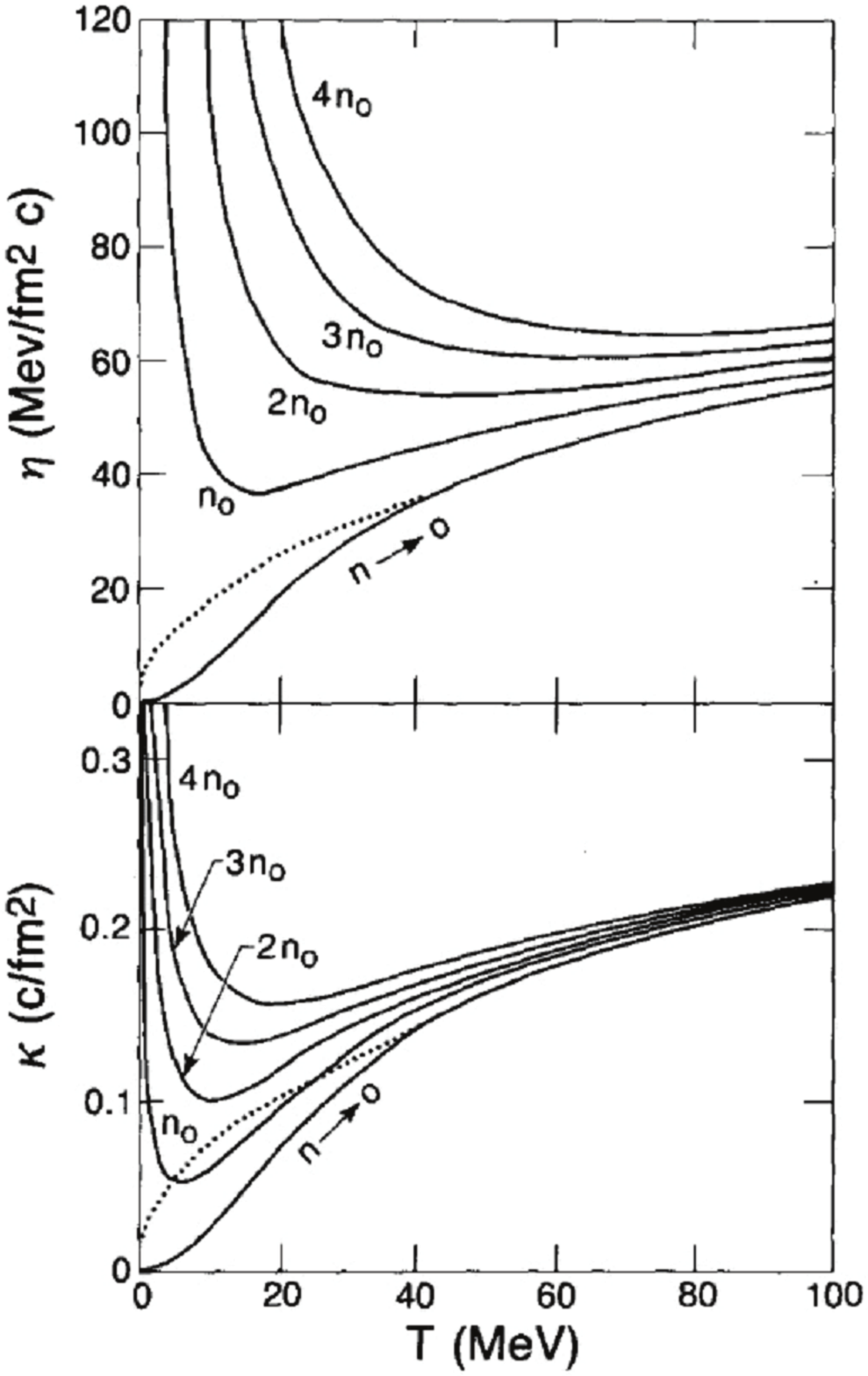}
\caption{Temperature dependence of the shear viscosity (upper panel) and thermal conductivity (lower panel) of nuclear matter at different densities. Dotted lines are Chapman-Enskog results with an effective cross section of 30 mb. Taken from ref.\,\cite{Dan84plb}.}
\label{fig:pawel}
\end{figure}
\subsection{Effects of nucleon effective masses on the viscosity and thermal conductivity of neutron-rich matter}
While the general concepts of various transport coefficients can be found in many standard textbooks and their values have been calculated for many systems in various fields, transport properties of hot and dense nuclear matter were systematically studied by Danielewicz in ref.\,\cite{Dan84plb} by linearizing the Boltzmann-Uehling-Uhlenbeck (BUU) equation. Shown in Fig.\,\ref{fig:pawel} are the shear viscosity and thermal conductivity as functions of density and temperature from this approach.
The large transport coefficients at lower temperatures are due to the Pauli blocking which is stronger at higher densities. The dotted lines are the first-order Chapman-Enskog coefficients from
\begin{align}
\eta &=\frac{5}{16}\sqrt{\pi}(mT)^{1/2}/\widetilde{\sigma},\\
\kappa &= \frac{75}{64}\sqrt{\pi}(mT)^{1/2}/\widetilde{\sigma}
\end{align}
with an effective cross section $\widetilde{\sigma} \approx 30$ mb. The density and temperature dependences of the shear viscosity and thermal conductivity were further parameterized
as\,\cite{Dan84plb}
\begin{align}
\eta \approx& (1700/T^2)(n/n_0)^2+[22/(1+T^2\times 10^{-3})](n/n_0)^{0.7}+5.8T^{1/2}/(1+160/T^2),\\
\kappa \approx& (0.15/T)(n/n_0)^{1.4}+[0.02/(1+T^4/7\times 10^{6})](n/n_0)^{0.4}+0.0225T^{1/2}/(1+160/T^2).
\end{align}
These parameterizations have been used in recent years in studying the viscosity of hot and dense matter formed in heavy-ion collisions\,\cite{Zhou13prc,Zhou14prc} by assuming that a local thermal equilibrium has been reached
in the reactions.

The transport coefficients are expected to be affected by the nucleon effective mass in two ways. First, the nucleon effective mass affects the flux since the velocity is $p/m^*$. Second, the nucleon effective mass affects the NN scattering cross sections through the effective mass scaling as we discussed earlier in this section, i.e., the in-medium scattering cross section between nucleon-1 and nucleon-2 is proportional to $[m_1^*m_2^*/(m_1^*+m_2^*)]^2$.
The isovector nucleon effective mass is expected to play a special role in determining the transport properties of neutron-rich matter. For example, a recent study of effects of the neutron-proton effective mass splitting
on the shear viscosity of neutron-rich nuclear matter through the relaxation time approach using the ImMDI interaction was found very fruitful.
As mentioned above, the parameter sets of $(x=0, y=-115~\text{MeV})$ and $(x=1, y=115~\text{MeV})$ lead respectively to $m_{\rm{n}}^*>m_{\rm{p}}^*$ and $m_{\rm{n}}^*<m_{\rm{p}}^*$ but almost the same symmetry energy, thus providing a convenient way to explore effects of the neutron-proton effective mass splitting from model comparisons. Here we outline the major steps and results of the study in ref.\,\cite{Jxu15b}.
In neutron-rich nuclear matter, the relaxation time $\tau_\tau(p_1)$, i.e., the average
collision time for a nucleon with isospin $\tau$ and momentum $p_1$, can be expressed as
\begin{equation}
\frac{1}{\tau_\tau(p_1)} = \frac{1}{\tau_\tau^{\rm{same}}(p_1)} +
\frac{1}{\tau_\tau^{\rm{diff}}(p_1)},
\end{equation}
with $\tau_\tau^{\rm{same}\,(\rm{diff})}(p_1)$ being the average time for this nucleon to collide with other nucleons of the same (different) isospin. They can
be calculated from
\begin{align}
\frac{1}{\tau_\tau^{\rm{same}}(p_1)} =&
\left(d-\frac{1}{2}\right)\frac{(2\pi)^2}{(2\pi)^3} \int p_2^2 dp_2
d\cos\theta_{12} d\cos\theta
\frac{d\sigma_{\tau,\tau}}{d\Omega}\left|\frac{\vec{p}_1}{m_\tau^\star(p_1)}-\frac{\vec{p}_2}{m_\tau^\star(p_2)}\right|
\notag\\
&\times\left[n^0_{\tau}(p_2)-n^0_{\tau}(p_2)n^0_\tau(p_1^\prime)-n^0_{\tau}(p_2)n^0_{\tau}(p_2^\prime)+n^0_\tau(p_1^\prime)n^0_{\tau}(p_2^\prime)\right],\label{tausame}\\
\frac{1}{\tau_\tau^{\rm{diff}}(p_1)} =& d\frac{(2\pi)^2}{(2\pi)^3} \int
p_2^2 dp_2 d\cos\theta_{12} d\cos\theta
\frac{d\sigma_{\tau,-\tau}}{d\Omega}\left|\frac{\vec{p}_1}{m_\tau^\star(p_1)}-\frac{\vec{p}_2}{m_{-\tau}^\star(p_2)}\right|
\notag\\
&\times\left[n^0_{-\tau}(p_2)-n^0_{-\tau}(p_2)n^0_\tau(p_1^\prime)-n^0_{-\tau}(p_2)n^0_{-\tau}(p_2^\prime)+n^0_\tau(p_1^\prime)n^0_{-\tau}(p_2^\prime)\right]\label{taudiff}
\end{align}
where $d=2$ is the spin degeneracy, $n^0$ is the equilibrium occupation probability, i.e., the Fermi-Dirac distribution, and $\sigma_{\tau,\tau}$ and $\sigma_{\tau,-\tau}$ are the collision cross sections for nucleon pairs with the same or different isospin, respectively. The NN scattering cross sections in free space are taken from ref.\,\cite{Cha90prc} while their in-medium values are calculated using the effective mass scaling model discussed earlier. In addition, the cross sections are assumed to be isotropic.

\begin{figure}[h!]
\centering
\includegraphics[scale=1.2,clip]{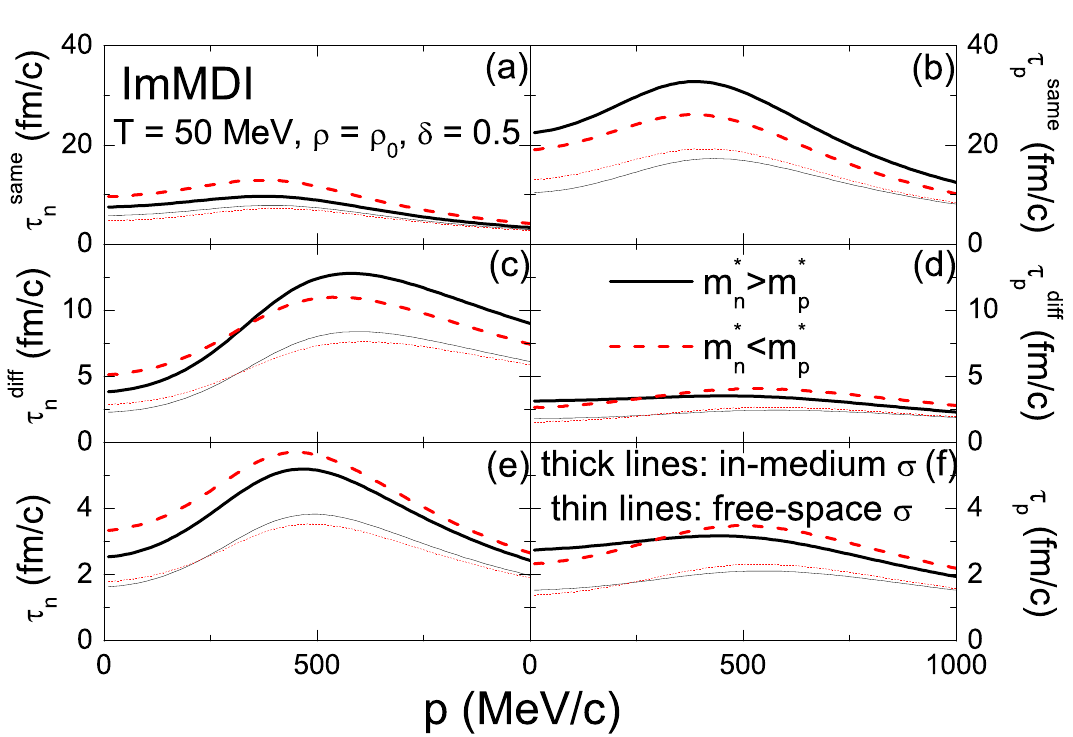}
\caption{Total relaxation times for neutrons and protons and those for colliding with another nucleon of the same or a different isospin. Taken from ref.\,\cite{Jxu15b}.}
\label{fig:tau}
\end{figure}

Fig.\,\ref{fig:tau} demonstrates effects of the neutron-proton effective mass splittings and NN scattering cross sections on the relaxation times for neutrons and protons. It is seen that the relaxation time is always smaller for colliding with nucleons of a different isospin. This is due to the larger degeneracy factor in Eq.\,(\ref{taudiff}) and the fact that np has a larger cross section than nn (pp). On the other hand, protons have a smaller total relaxation time and thus experience more frequently scatterings than neutrons in neutron-rich nuclear matter, since the total relaxation time is dominated by the $\tau^{\rm{diff}}$ which is smaller than $\tau^{\rm{same}}$. The in-medium NN scattering cross sections are generally smaller than those in free space, enhancing significantly the relaxation time. It will be shown later that the shear viscosity is mostly determined by the neutron relaxation time. More quantitatively, it is similar for $m_{\rm{n}}^*>m_{\rm{p}}^*$ and $m_{\rm{n}}^*<m_{\rm{p}}^*$ using the free-space cross sections, but is larger for $m_{\rm{n}}^*<m_{\rm{p}}^*$ than for $m_{\rm{n}}^*>m_{\rm{p}}^*$ when the isospin-dependent in-medium cross sections from the effective mass scaling are used.

\begin{figure}[h!]
\centering
\includegraphics[scale=1.2,clip]{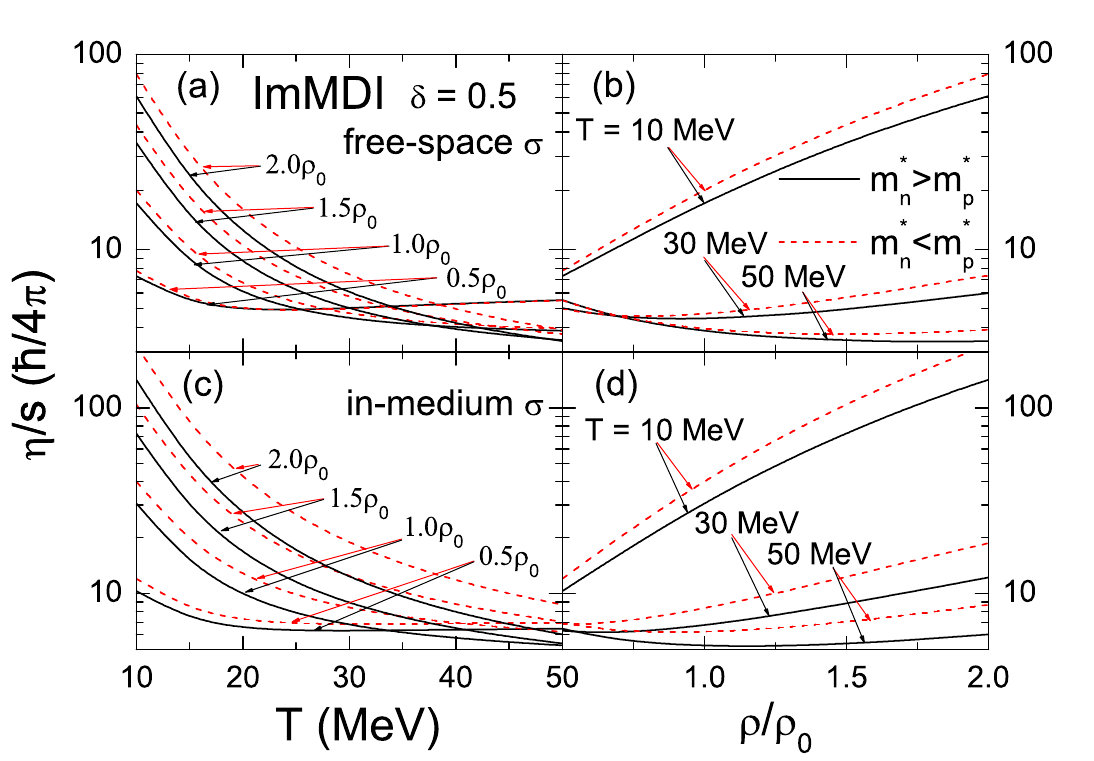}
\caption{Specific shear viscosity $\eta/s$ as a function of temperature (left) and density (right) from the free-space (upper) and in-medium (lower) nucleon-nucleon scattering cross sections from the relaxation time approach. Taken from ref.\,\cite{Jxu15b}.}
\label{fig:extensive}
\end{figure}

In the relaxation time approach, the shear
viscosity is calculated by assuming that in the uniform nuclear
system there exists a static flow field in the $z$ direction with a
flow gradient in the $x$ direction. The shear force, which is
related to the nucleon flux as well as the momentum exchange between
flow layers, is proportional to the flow gradient, and the
proportionality coefficient, i.e., the shear viscosity, turns out to
be\,\cite{Xu11prc}
\begin{eqnarray}\label{eta}
\eta = \sum_\tau -\frac{d}{(2\pi)^3} \int \tau_\tau(p) \frac{p_z^2
p_x^2}{pm^\star_\tau} \frac{d n_\tau}{dp} dp_x dp_y dp_z.
\end{eqnarray}
It is clear that the flux between flow layers is affected by the effective mass in the denominator as well. Moreover, the shear viscosity is proportional to the gradient of the occupation probability which is larger near the Fermi surface and dominated by the neutron relaxation time, since the Fermi surface of neutrons is sharper in neutron-rich nuclear matter.

The shear viscosity is often scaled by the entropy density $s$, which is approximated by that for free nucleons from the occupation probability as
\begin{equation}
s =-\sum_\tau d\int [n_{\tau }\ln n_{\tau }+(1-n_{\tau })\ln
(1-n_{\tau })]\frac{d^3p}{(2\pi)^3}. \label{S}
\end{equation}
Although the neutron-proton effective mass splitting affects the occupation probability, it is found that the effect on the entropy density is small as a result of the momentum integral. The temperature and the density dependence of the specific shear viscosity, i.e., the ratio of the shear viscosity to the entropy density $\eta/s$, are displayed in Fig.\,\ref{fig:extensive}. Since $\eta/s$ was proposed to have a lower boundary of $\hbar/4\pi$, the unit of the specific shear viscosity is taken as $\hbar/4\pi$. Similar to that observed in Fig.\,\ref{fig:pawel}, the specific shear viscosity generally increases with increasing density and/or decreasing temperature, as a result of the Pauli blocking effect. The specific shear viscosity is larger for $m_{\rm{n}}^*<m_{\rm{p}}^*$ than for $m_{\rm{n}}^*>m_{\rm{p}}^*$ even using the free-space cross sections. This is because that the neutron flux between flow layers is enhanced for $m_{\rm{n}}^*<m_{\rm{p}}^*$, and the neutron dynamics dominates the shear viscosity of the neutron-rich medium. With the isospin-dependent modifications of the in-medium NN scattering cross section, the difference of $\eta/s$ between results with $m_{\rm{n}}^*<m_{\rm{p}}^*$ and $m_{\rm{n}}^*>m_{\rm{p}}^*$ is further enhanced.

\begin{figure}[htbp]
\centering
\vspace{-0.5cm}
\includegraphics[scale=1.,clip]{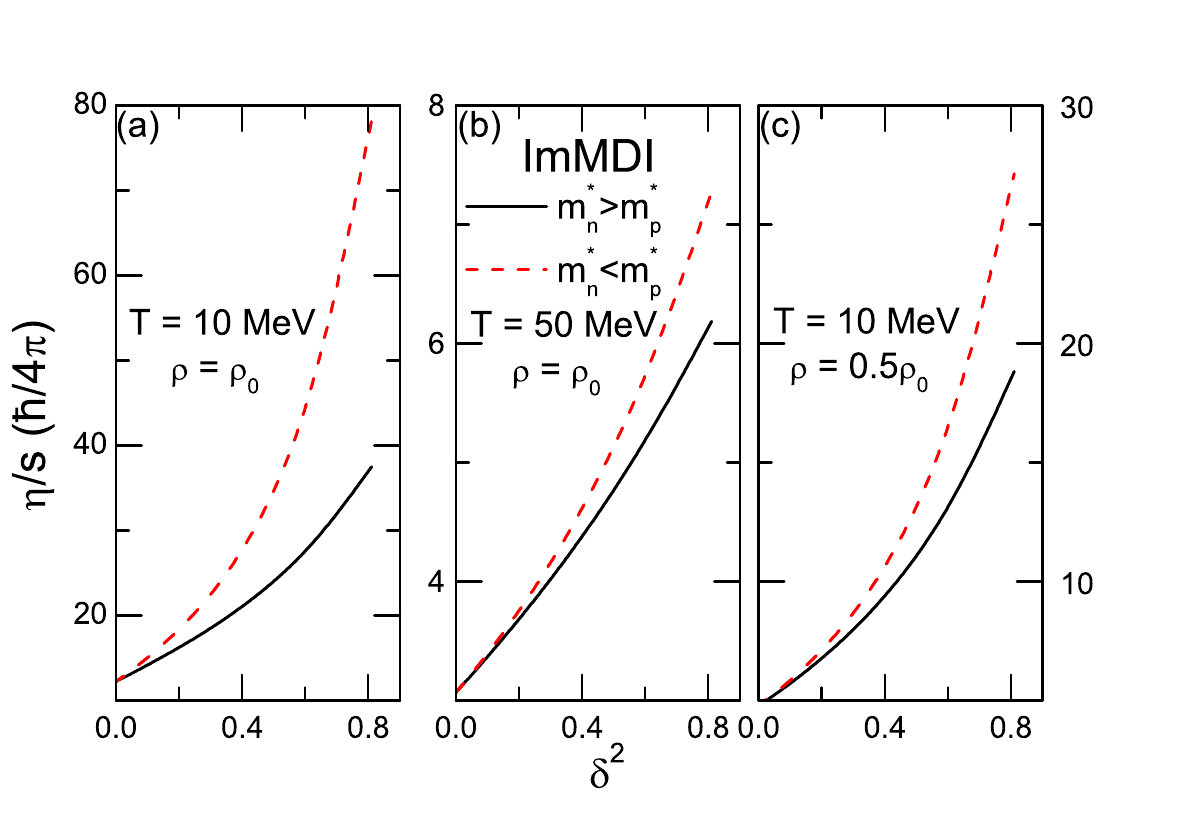}
\caption{Dependence of the specific shear viscosity $\eta/s$ on the isospin asymmetry $\delta$ of neutron-rich nuclear matter. Taken from ref.\,\cite{Jxu15b}.}
\label{fig:PA}
\end{figure}

The dependence of the specific shear viscosity on the isospin asymmetry of neutron-rich nuclear matter is further displayed in Fig.\,\ref{fig:PA} using the in-medium cross sections and different neutron-proton effective mass splittings. As expected, the specific viscosity increases with increasing isospin asymmetry of the nuclear medium. It is seen that generally the $\eta/s$ increases faster than the parabolic relation, i.e., $\eta/s \propto \delta^2$, especially at higher densities or low temperatures. The calculation with $m_{\rm{n}}^*<m_{\rm{p}}^*$ leads to a larger $\eta/s$ than that with $m_{\rm{n}}^*<m_{\rm{p}}^*$, and the effect is larger at higher isospin asymmetries.

\begin{figure}[htbp]
\centering
\vspace{-1.cm}
\includegraphics[scale=1.,clip]{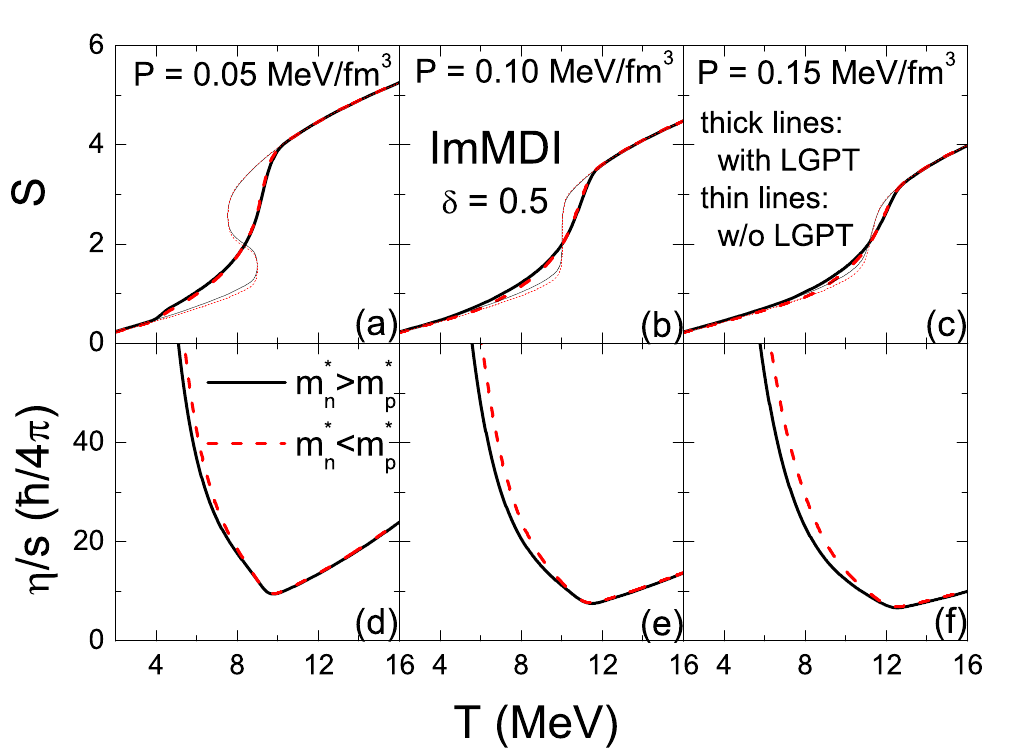}
\caption{Entropy per nucleon (upper panel) and the specific shear viscosity (lower panel) as a function of the temperature in nuclear matter at a fixed external pressure near the nuclear liquid-gas phase transition (LGPT). Taken from ref.\,\cite{Jxu15b}.}
\label{fig:etas_LGPT}
\end{figure}

It is well known that the specific shear viscosity generally has a minimum near phase transitions. It is thus interesting to examine the specific shear viscosity near the nuclear liquid-gas phase transition and its dependence on the neutron-proton effective mass splitting. As shown in the upper panels of Fig.\,\ref{fig:etas_LGPT}, the original $S$-shaped entropy per nucleon is modified to a smooth curve from the Gibbs construction during a nuclear liquid-gas phase transition, when the nuclear system is heated at a fixed pressure of $P=0.05$, 0.10, and 0.15 MeV/fm$^3$. As shown in the lower panels of Fig.\,\ref{fig:etas_LGPT}, the specific shear viscosity indeed has a minimum near the phase transition. As shown in ref.\,\cite{Xu13plb}, the minimum value of $\eta/s$ depends on the temperature or the pressure of the system, and the lowest value can be as small as $(4-5) \hbar/4\pi$. The neutron-proton effective mass splitting affects the specific shear viscosity at the low-temperature liquid side, but has almost no effect on the high-temperature gas side. This shows that the nuclear liquid-gas phase transition indeed
follows approximately the established universal pattern of having a minimum of $\eta/s$ near phase transitions, not much affected by the uncertainty in neutron-proton effective mass splitting.

\begin{figure}[h!]
\centering
\includegraphics[scale=0.25,clip]{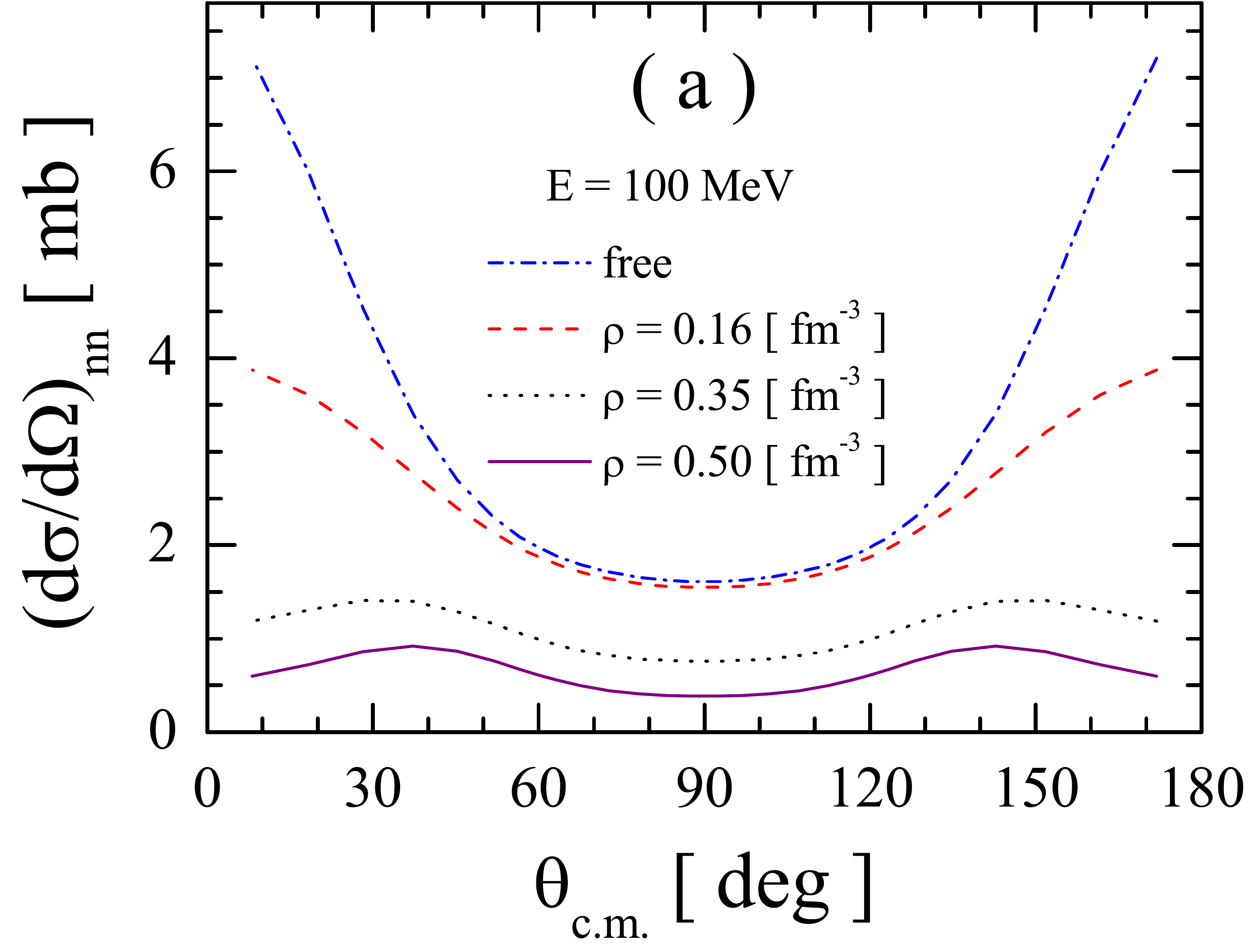}
\includegraphics[scale=0.25,clip]{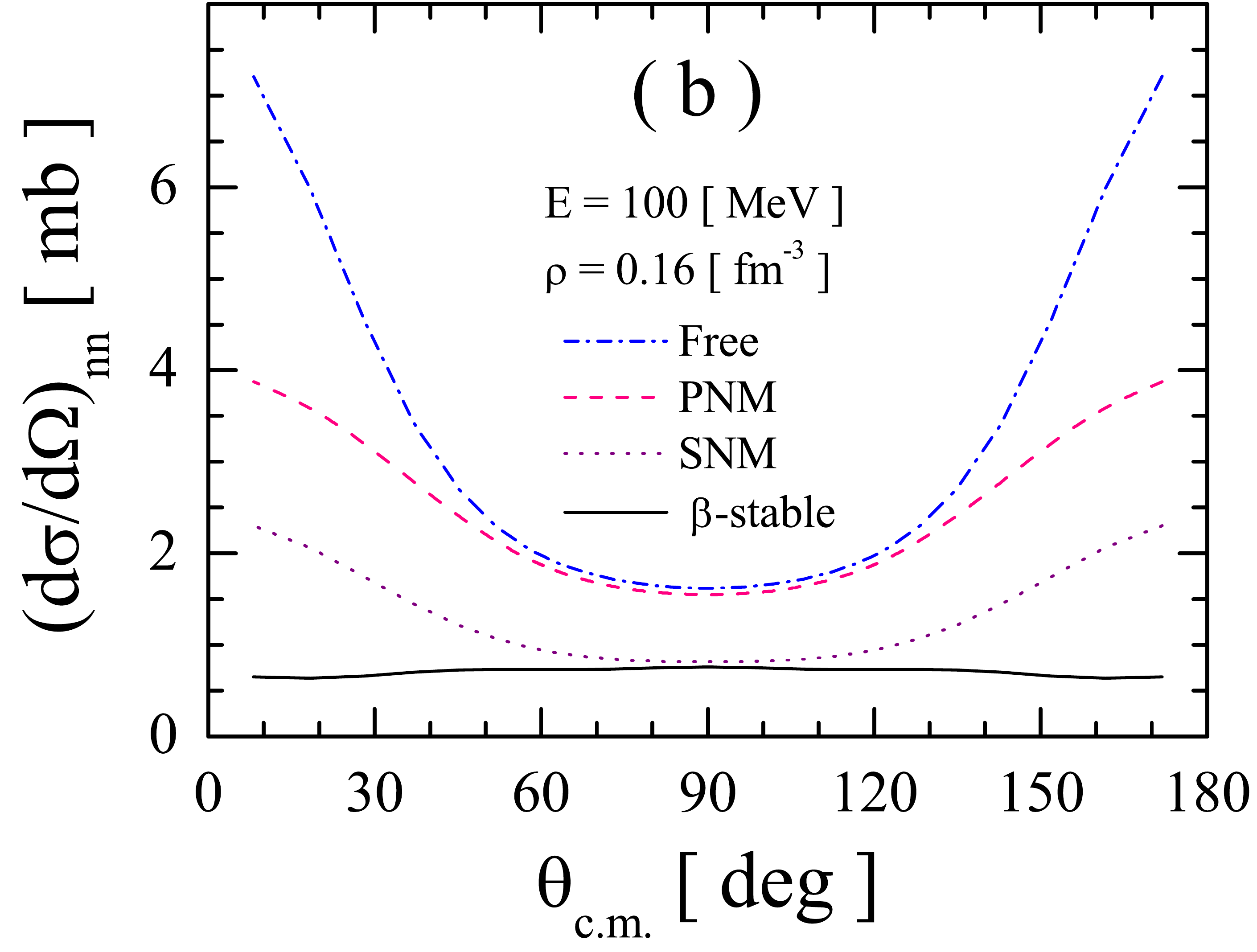}
\includegraphics[scale=0.25,clip]{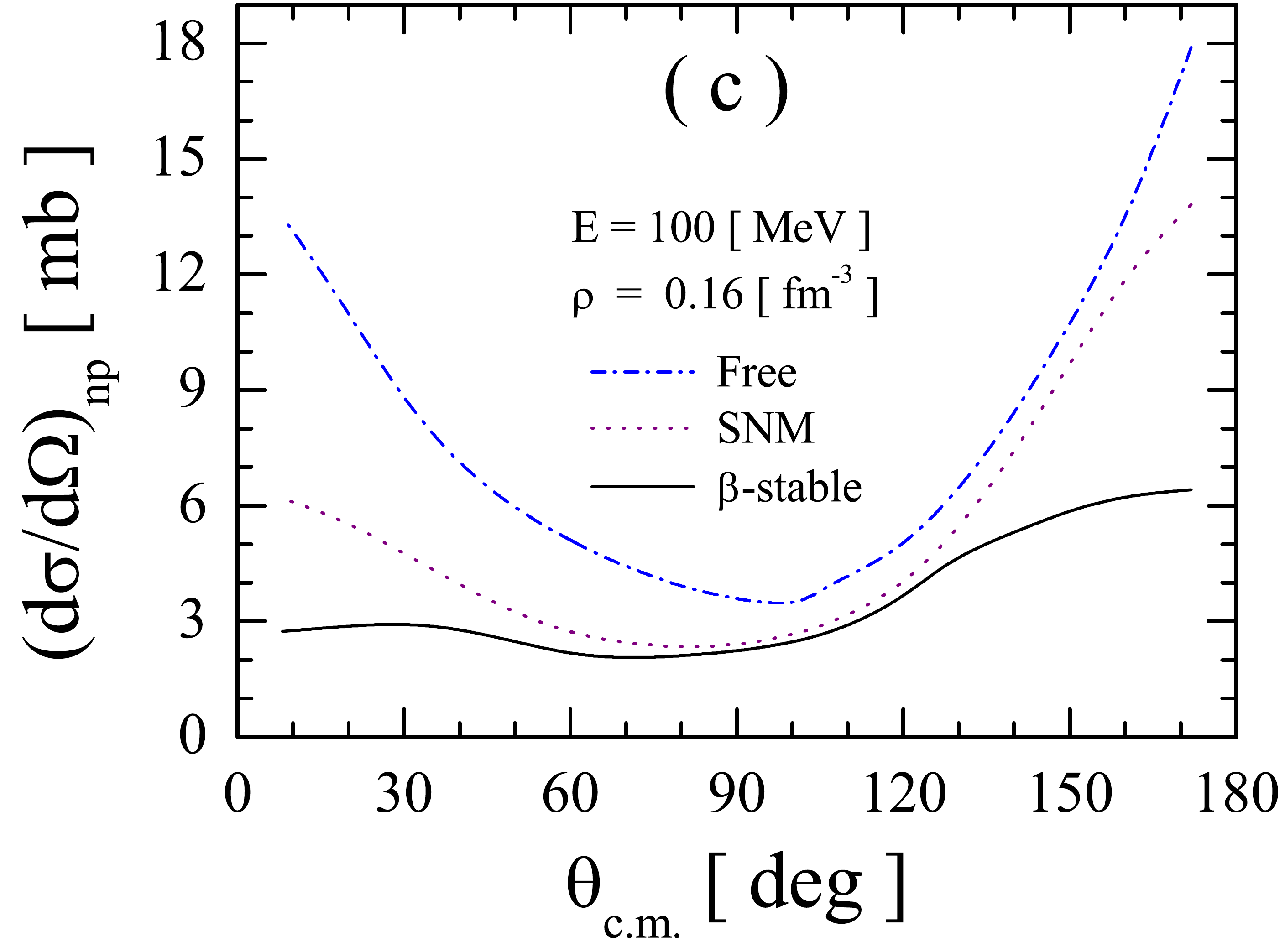}
\caption{((a) Neutron-neutron differential cross sections in pure neutron matter. (b) Neutron-neutron differential cross section in pure neutron matter, symmetric, and $\beta$-stable nuclear matter. (c) Neutron-proton differential cross section in pure neutron matter (PNM), symmetric nuclear matter (SNM), and $\beta$-stable nuclear matter. The free cross section is also plotted for comparison. Taken from ref.\,\cite{Zhang10prc}.}
\label{fig:fig3_zhang}
\end{figure}

The shear viscosity $\eta$ and the thermal conductivity $\kappa$ of nuclear matter can also be obtained from the Abrikosov-Khalatnikov approximation\,\cite{Zhang10prc,sb1,sb2}
\begin{align} \label{eta_AK}
\eta T^{2} =& \frac{1}{20} \rho v^2_{\rm{F}} W(\rho)
C(\lambda),\\
\kappa T =& \frac{1}{12} v^2_{\rm{F}} p_{\rm{F}} W(\rho) H(\mu),
\end{align}
with
\begin{align} \label{probability}
W^{-1}(\rho)=&\frac{1}{2\epsilon_{\rm{F}}}\int_0^{4\epsilon_{\rm{F}}}dE\int_0^{2\pi}
\frac{d\theta}{2\pi}\frac{1}{\sqrt{1-E/4\epsilon_{\rm{F}}}}\sigma(E,\theta).
\end{align}
In the above, $T$ is the temperature and $\rho$ is the nucleon density; $v_{\rm{F}}=p_{\rm{F}}/m^*$ is the Fermi velocity with
$m^*$ the effective mass. $C(\lambda)$ and $H(\mu)$ are correction
factors and their detailed forms can be found in refs.\,\cite{sb1,sb2}.
In terms of the $G$ matrix elements $G_{S_{z}S_{z}^{\prime}}^{S}$,  the in-medium nn (pp) and $np$ cross sections can be expressed as\,\cite{Zhang10prc,Zhang07prc}
\begin{align}
\sigma_{\rm{nn}}(E,\theta)=&
\frac{(m_n^{*})^2}{16\pi^{2}\hbar^{4}}\sum_{SS_{z}S_{z}^{\prime}}\left|G_{S_{z}S_{z}^{\prime}}^{S}
(\theta)+(-1)^{S}G_{S_{z}S_{z}^{\prime}}^{S}(\pi-\theta)\right|^{2}, \\
\sigma_{\rm{np}}(E,\theta) =& \frac{1}{16\pi^{2}\hbar^{4}}\left(\frac
{2m^*_{\rm{n}}m^*_{\rm{p}}}{m^*_{\rm{n}}+m^*_{\rm{p}}}\right)^2\sum_{SS_{z}S_{z}^{\prime}}
\left|G_{S_{z}S_{z}^{\prime}}^{S}(\theta)\right|^2.
\end{align}
In the work of Zhang et al.\,\cite{Zhang07prc,Zhang10prc}, the $G_{S_{z}S_{z}^{\prime}}^{S}$ were obtained from the Brueckner theory using the Bonn $B$ potential. Similar to the relaxation time approach discussed earlier, it is seen that the nucleon effective mass affects the transport coefficients through the Fermi velocity $v_{\rm{F}}=p_{\rm{F}}/m^*$
and the effective mass scaling of the in-medium scattering cross sections. Moreover, the $G$ matrix elements are obtained self-consistently from the nuclear interaction and thus contains the information of the nucleon effective mass as well. The resulting differential cross sections in free space, PNM, SNM and $\beta$-stable nuclear matter are displayed in Fig.\,\ref{fig:fig3_zhang}. The cross sections are more forward and backward peaked in free space or at lower densities, and more isotropic at higher densities. The nn scattering cross section becomes smaller and more isotropic with decreasing isospin asymmetry, while the $np$ scattering cross section is forward-backward asymmetric mainly because of the mixing of $S$ and $D$ waves due to the tensor force.

\begin{figure}[h!]
\centering
\includegraphics[scale=0.17,clip]{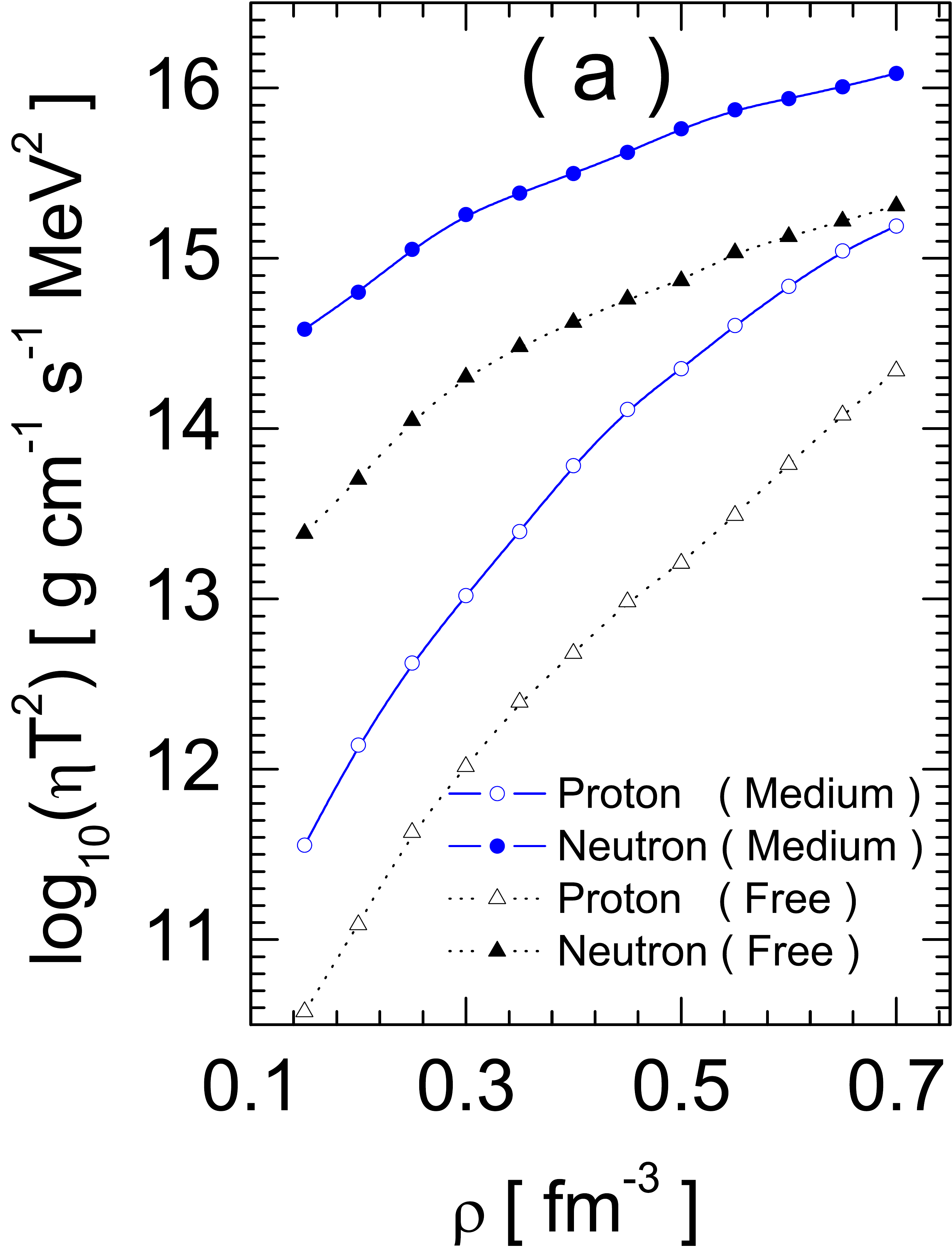}
\hspace{0.3cm}
\includegraphics[scale=0.175,clip]{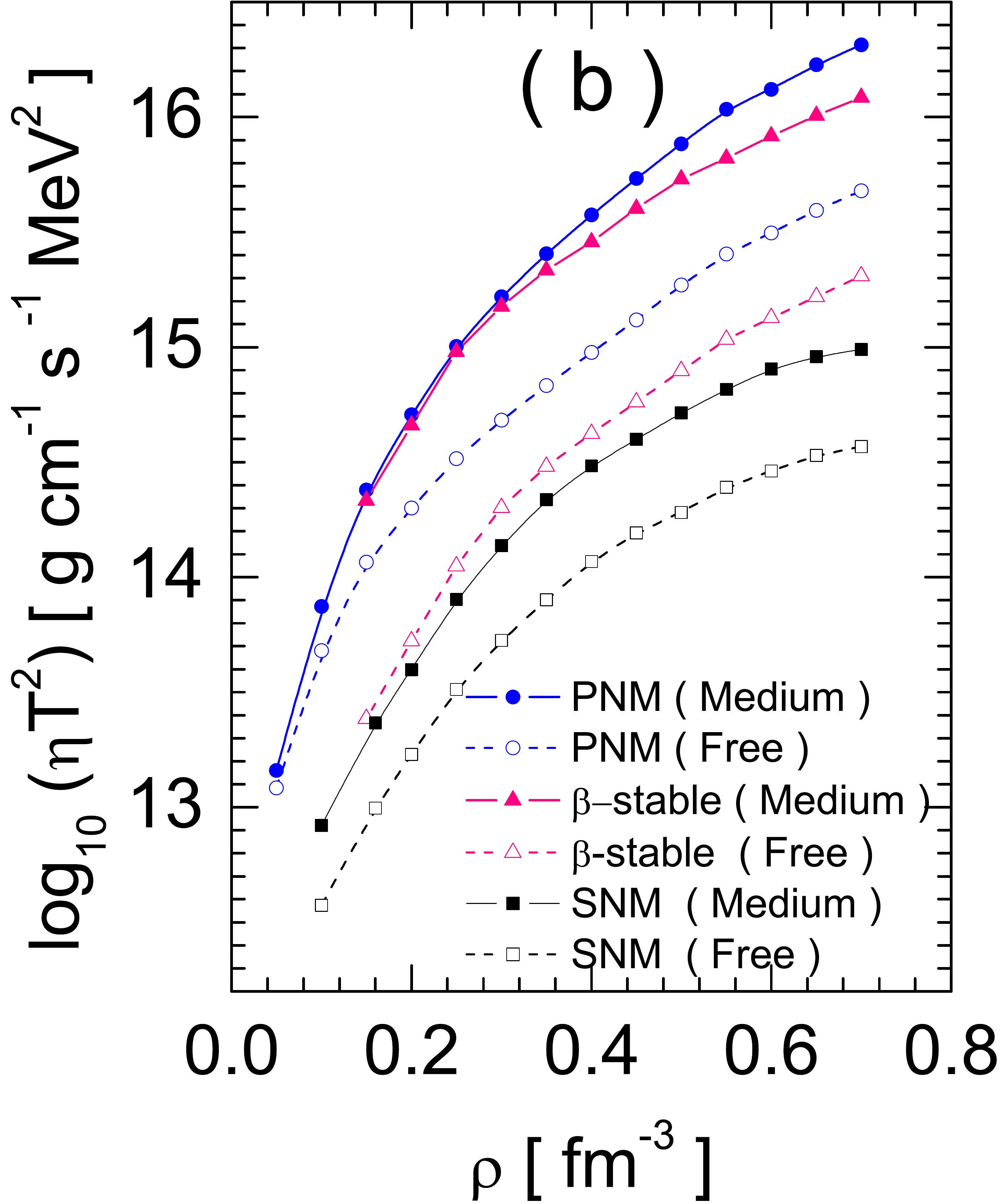}
\hspace{0.3cm}
\includegraphics[scale=0.18,clip]{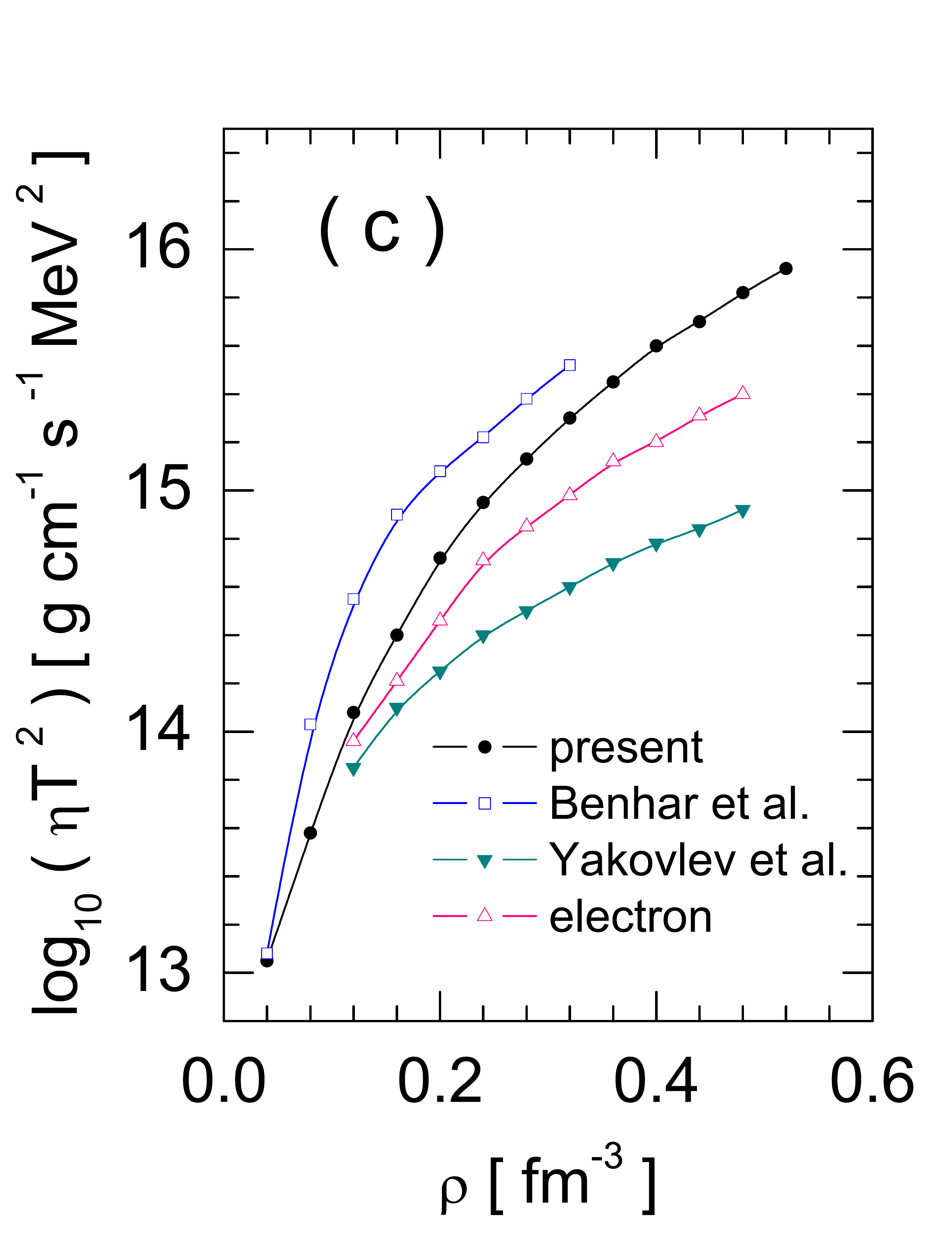}
\caption{ (a) Neutron and proton viscosity in $\beta$-stable nuclear matter. (b) Neutron viscosity in pure neutron matter (PNM), symmetric nuclear matter (SNM), and $\beta$-stable nuclear matter. (c) Comparison of the neutron viscosity from the Abrikosov-Khalatnikov approximation and the Brueckner theory with other calculations in $\beta$-stable nuclear matter. Taken from ref.\,\cite{Zhang10prc}.}
\label{fig:fig4_zhang}
\end{figure}

Fig.\,\ref{fig:fig4_zhang} displays their results of the shear viscosity from the Abrikosov-Khalatnikov approximation and the Brueckner theory. The resulting shear viscosity increases with increasing density, and the shear viscosity with in-medium cross sections is larger than that from using the free-space cross sections. Again, in neutron-rich matter such as the $\beta$-stable matter, as shown in panel (a) of Fig.\,\ref{fig:fig4_zhang}, the viscosity of neutrons is much larger than that of protons, and thus dominates the shear viscosity of the whole system. In panel (b), it is seen that the neutron viscosity becomes larger with the increasing isospin asymmetry of the system. As shown in panel (c), Benhar et al.\,\cite{benh3} predicted a larger shear viscosity. This is mainly because their model also predicted a different density dependence of the symmetry energy and a more neutron-rich $\beta$-stable nuclear matter.
Results by Yakovlev et al.\,\cite{SY} were obtained from a fixed nucleon effective mass $m^*/m=0.8$. The shear viscosity of electrons is shown to be much smaller than that of neutrons.

\begin{figure}[h!]
\centering
\includegraphics[scale=0.249,clip]{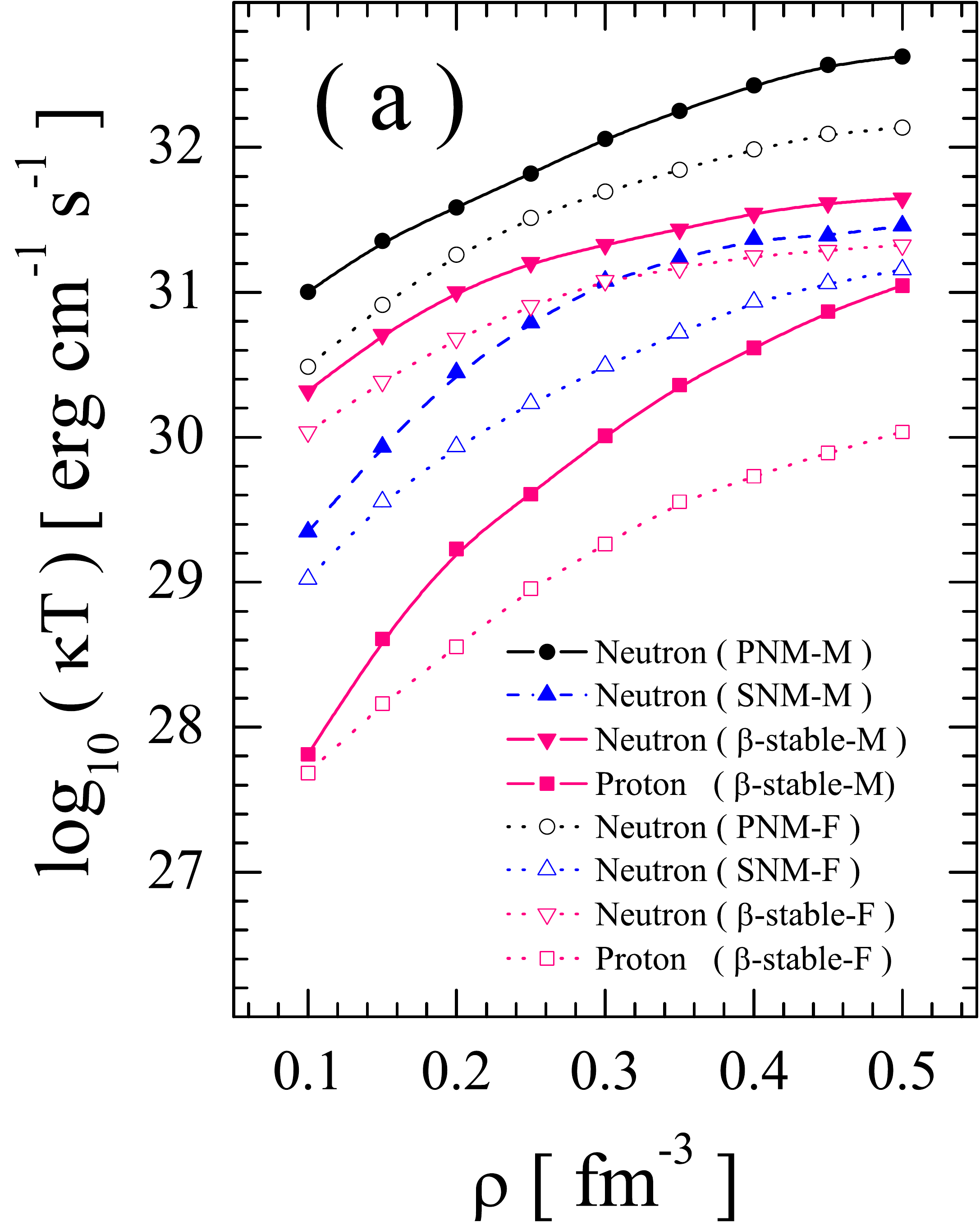}
\hspace{1.5cm}
\includegraphics[scale=0.2,clip]{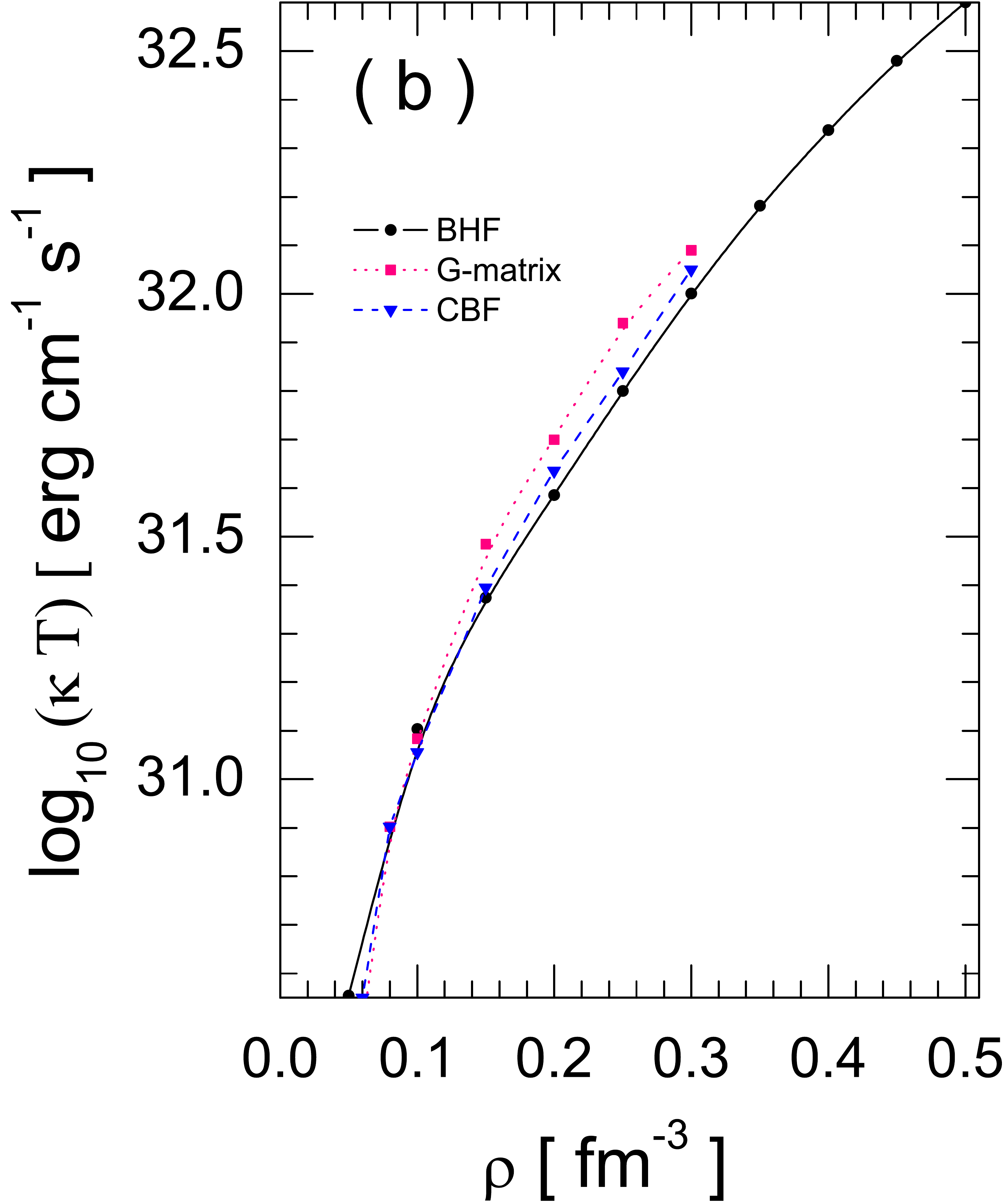}
\caption{(a) Thermal conductivity in pure neutron matter (PNM), symmetric nuclear matter (SNM), and $\beta$-stable nuclear matter. (b) Comparison of the thermal conductivity from the Abrikosov-Khalatnikov approximation and the Brueckner theory with other two methods in pure neutron matter. Taken from ref.\,\cite{Zhang10prc}.}
\label{fig:fig5_zhang}
\end{figure}

The thermal conductivity in pure neutron matter, symmetric nuclear matter, and $\beta$-stable nuclear matter is displayed in panel (a) of Fig.\,\ref{fig:fig5_zhang}. Its dependence on the density, isospin asymmetry, and the in-medium modifications of NN cross sections are qualitatively similar to the shear viscosity. As shown in panel (b), their results are similar to those obtained from other two calculations using the $G$ Matrix method and correlated-base function (CBF)\,\cite{SY,benh1}.

In summary of this section, the space-time non-locality of isovector interactions through the isovector nucleon effective mass or the neutron-proton effective mass splitting has multifaceted effects on
thermal and transport properties of neutron-rich matter. The in-medium NN cross sections  in neutron-rich matter are modified differently from those in SNM due to the neutron-proton effective mass splitting and its dependence on
the density and isospin-asymmetry of the medium. Both microscopic theories and phenomenological models have been used to study the in-medium NN cross sections. However, significant model and interaction dependences still exist.  Because the isovector nucleon effective mass affects the occupation probability of neutrons and protons differently in neutron-rich matter, essentially all thermodynamical variables can be affected by the
neutron-proton effective mass splitting. Self-consistent thermodynamical calculations have shown that the symmetry energy at finite temperature depends sensitively on whether the single-nucleon potential is momentum dependent or not and how the neutron-proton effective mass splitting varies with density and isospin asymmetry of the medium. Effects of the isoscalar effective mass on the liquid-gas phase boundary in the pressure versus isospin asymmetry plane is appreciable, while the neutron-proton effective mass splitting affects the differential isospin fractionation. However, the latter does not alter much the viscosity/entropy ratio in the gas phase. Within several different approaches, the shear viscosity $\eta$ and the thermal conductivity $\kappa$ of neutron-rich matter were found to depend on the neutron-proton effective mass splitting in non-trivial ways through the isospin dependence of both the particle flux and NN in-medium scattering cross sections in neutron-rich matter.

\section{Impacts of isovector nucleon effective mass on heavy-ion reactions at low and intermediate energies\label{HI}}
The isospin dynamics of heavy-ion reactions is determined by both the symmetry potential and the in-medium NN cross sections which depend on the isovector nucleon effective mass as we discussed in the previous section.
These two factors are both determined by the same underlying isovector interaction and closed related with each other. In particular, the isovector nuclear effective mass characterizes the momentum dependence of the symmetry potential. In fact, the symmetry energy itself depends on the isoscalar effective mass through the kinetic term as well as both the magnitude and momentum dependence of the symmetry potential as we discussed earlier in detail.

Heavy-ion reactions, especially those involving rare isotopes, provide a unique means in terrestrial laboratories to create dense neutron-rich matter, thus opportunities to study the density and momentum dependence of nuclear symmetry potential and the associated isovector nucleon effective mass at abnormal densities. Most isospin-sensitive observables make use of the fact that neutrons and protons feel the opposite symmetry (Lane) potential,
being repulsive (attractive) for neutrons (protons) at low energy/momentum but the sign may flip above certain inversion momentum as we discussed earlier. The low-intermediate energy heavy-ion reaction community has focused on constraining the magnitude and slope of the symmetry energy mostly around and below the saturation density over the last decade. Much progress has been made and indeed strong constraints have been obtained, see,
e.g., ref.\,\cite{EPJA}. However, not much about the isovector nucleon effective mass even at the saturation density has been extracted from nuclear reactions yet while both theoretical and experiments efforts are continuously being made. In this section, we review transport model predictions on reaction observables sensitive to the momentum dependence of the symmetry potential and comment on some recent efforts to constrain it by comparing with some experimental data from NSCL/MSU.

\begin{figure}[h!]
\centering
\includegraphics[width=7cm,height=7cm]{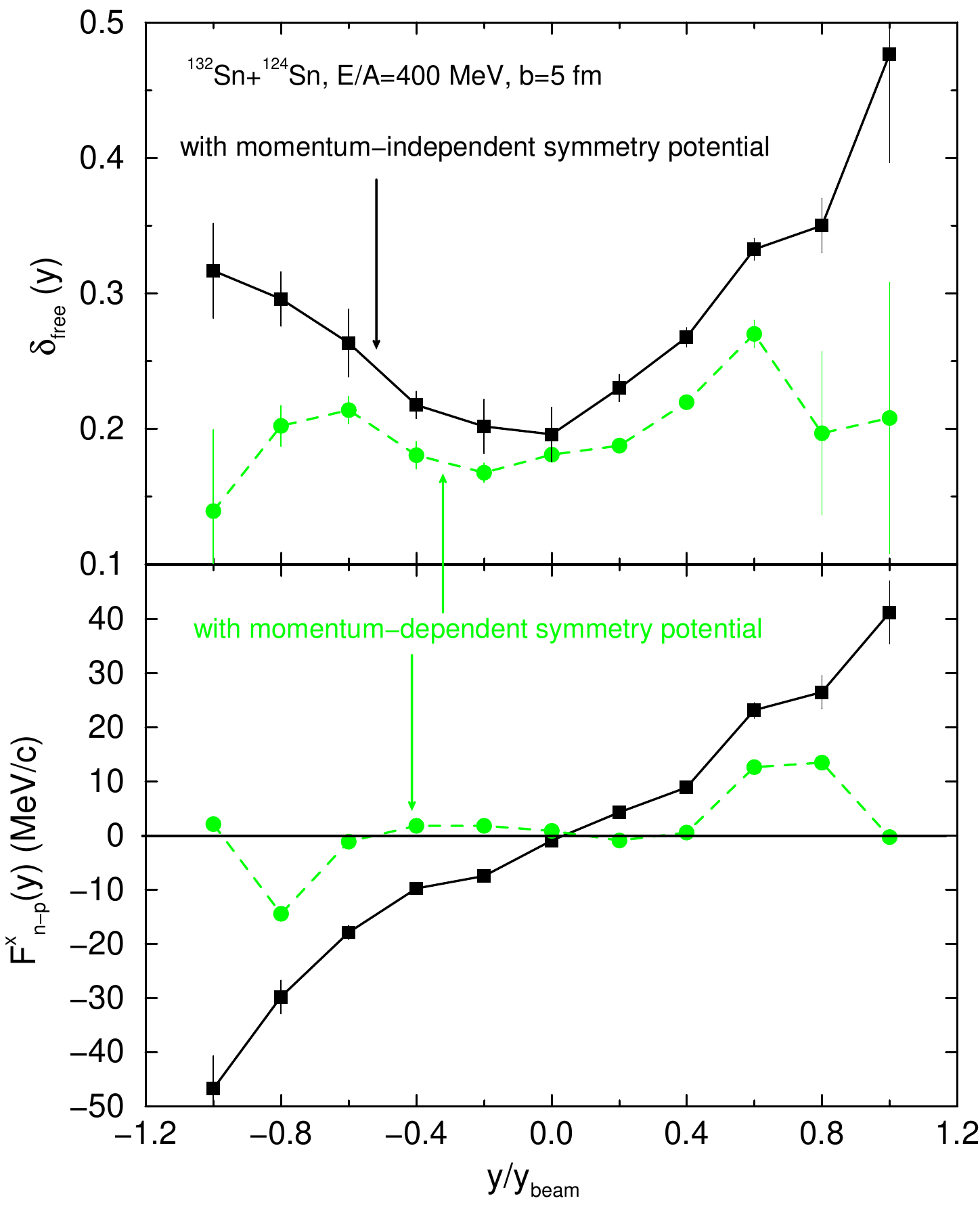}
\hspace{1cm}
\includegraphics[width=7cm,height=7cm]{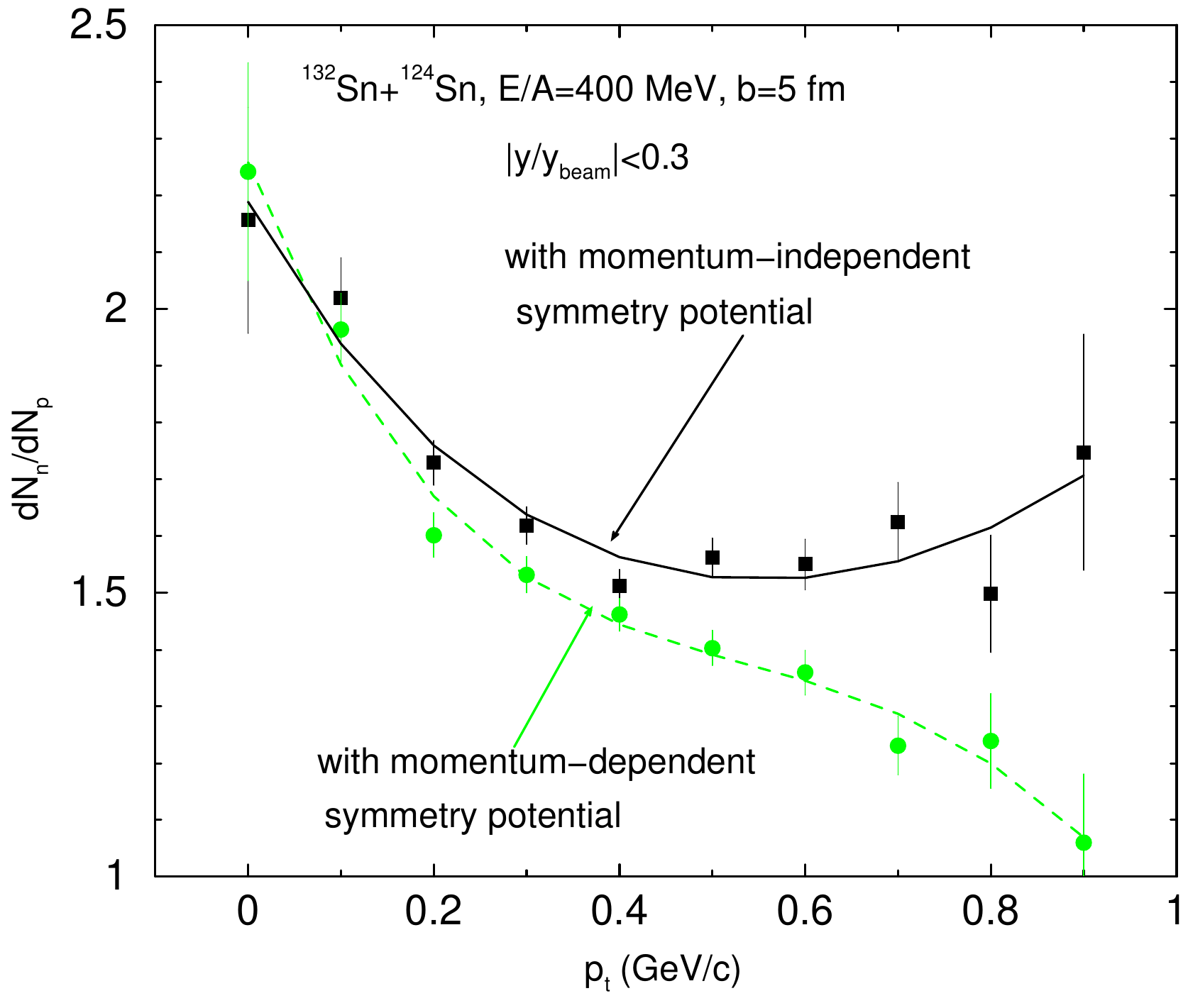}
\caption{{\protect Left: Isospin asymmetry (upper) and neutron-proton differential transverse flow
(lower) of free nucleons as a function of rapidity. Right: The ratio of free neutron to proton multiplicity as a function of transverse
momentum at midrapidity. The black solid (green dashed) lines are calculated with the
momentum-independent (-dependent) symmetry potential corresponding to $m^{*}_{n-p}=0$ ($m^{*}_{n-p}>0$). Taken from ref. \protect\cite{LiBA04}.}}
\label{dflow}
\end{figure}

\subsection{Transport model predictions on observables sensitive to the momentum dependence of the symmetry potential in heavy-ion reactions at intermediate energies}
First of all, despite of the quantitative discrepancies of transport model predictions on some observables, it is encouraging to note that several transport models have consistently predicted that the ratio of energetic nucleons
at either high rapidities or transverse momenta is sensitive to the neutron-proton effective mass splitting\,\cite{LiBA04,LiChen05,Riz05,Gio10,Feng12,Zhang14,Xie14}.
For example, shown in Fig.\,\ref{dflow} are 1) (upper left) the average isospin asymmetry $\delta_{\rm{free}}(y)$ of free nucleons and 2) (lower left) the neutron-proton differential transverse flow
\begin{equation}
F^x_{\rm{n-p}}(y)\equiv\sum_{i=1}^{N(y)}(p^x_iw_i)/N(y)
\end{equation}
where $w_i=1 (-1)$ for neutrons (protons) and $N(y)$ is the total number of free nucleons at rapidity $y$ from IBUU04 model calculations using the MDI interaction for $^{132}$Sn+$^{124}$Sn
reactions at a beam energy of 400 MeV/nucleon and an
impact parameter of 5 fm\,\cite{LiBA04}.  Shown on the right is the n/p ratio of mid-rapidity nucleons within
$|y_{\rm{cms}}/y_{\rm{beam}}|\leq 0.3$ as a function of transverse momentum $p_t$. Its overall decrease towards lower $p_t$ is due to the Coulomb force on protons.
The two calculations with and without the momentum-dependence of the symmetry potential are done with interactions constructed to give the same density dependence of nuclear symmetry energy. However, they were constructed to have either $m^{*}_{\rm{n-p}}=0$ (black) or $m^{*}_{\rm{n-p}}>0$ (green), respectively. The value of $\delta_{\rm{free}}(y)$ reflects mainly the degree of isospin fractionation between the free nucleons and the
bound ones at freeze-out of the reaction. At mid-rapidity, the $\delta_{\rm{free}}(y)$ values are close to the value expected when a complete isospin
equilibrium is established in the reaction. It is interesting to see that with $m^{*}_{\rm{n-p}}=0$, the $\delta_{\rm{free}}(y)$ is significantly higher than that with $m^{*}_{\rm{n-p}}>0$.  Moreover, the
difference tends to increase with rapidity. These features are what one expects from the different strength of the symmetry potential with or without the momentum dependence\,\cite{LiBA04}. In the case of
$m^{*}_{\rm{n-p}}=0$, the symmetry potential is a constant while it decreases with increasing momentum in the case of $m^{*}_{{\rm{n-p}}}>0$.
It is seen clearly that effects of the neutron-proton effective mass splitting are stronger on energetic nucleons at high rapidities or transverse momenta. Several other observables, such as the $^3$H-$^3$He yield ratio and their difference as a function of energy/rapidity/transverse momentum have also been studied, reaching similar conclusions, see, e.g., ref.\,\cite{Che04}.

\begin{figure}[htbp]
\centering
\includegraphics[scale=0.4]{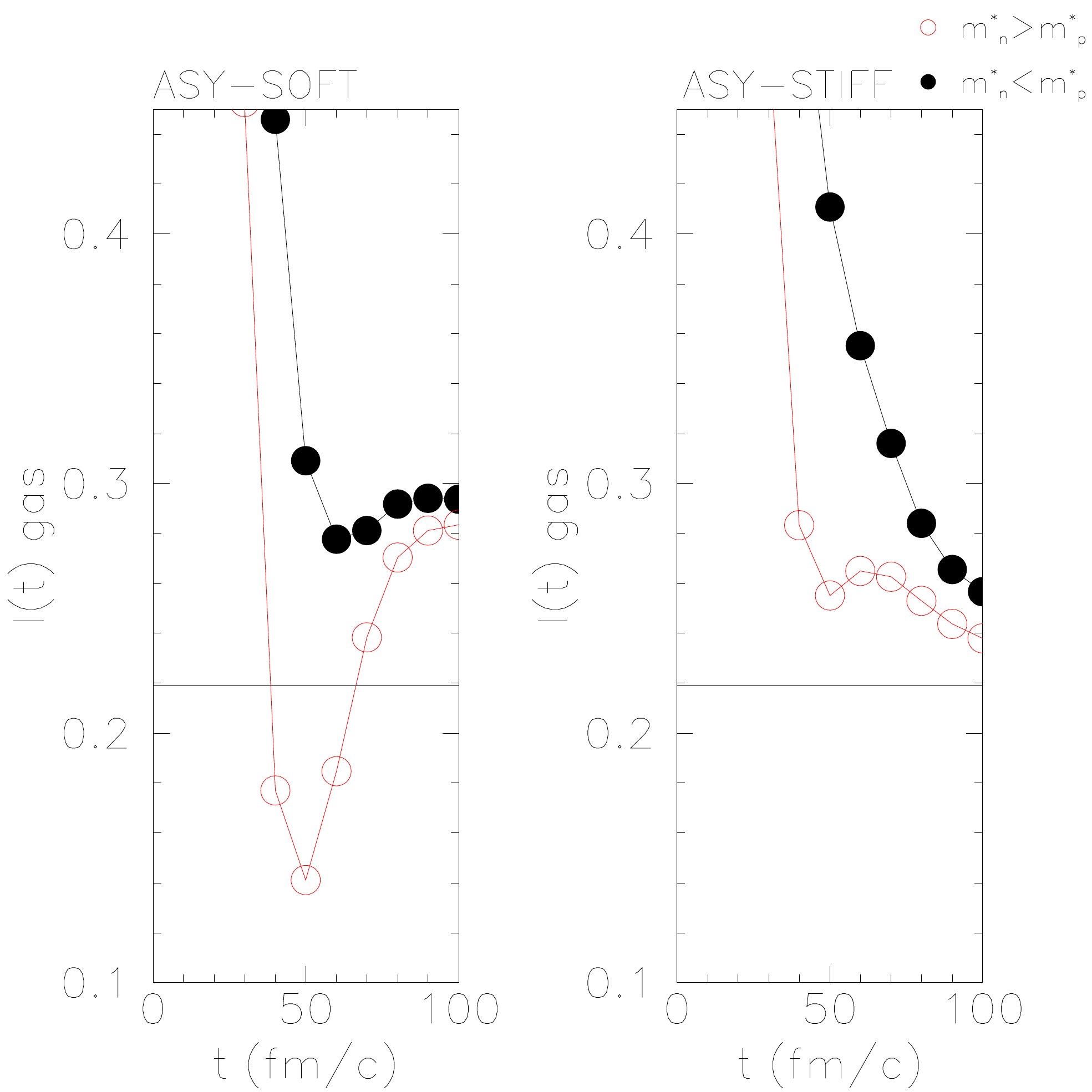}
\hspace{1cm}
\includegraphics[scale=0.7]{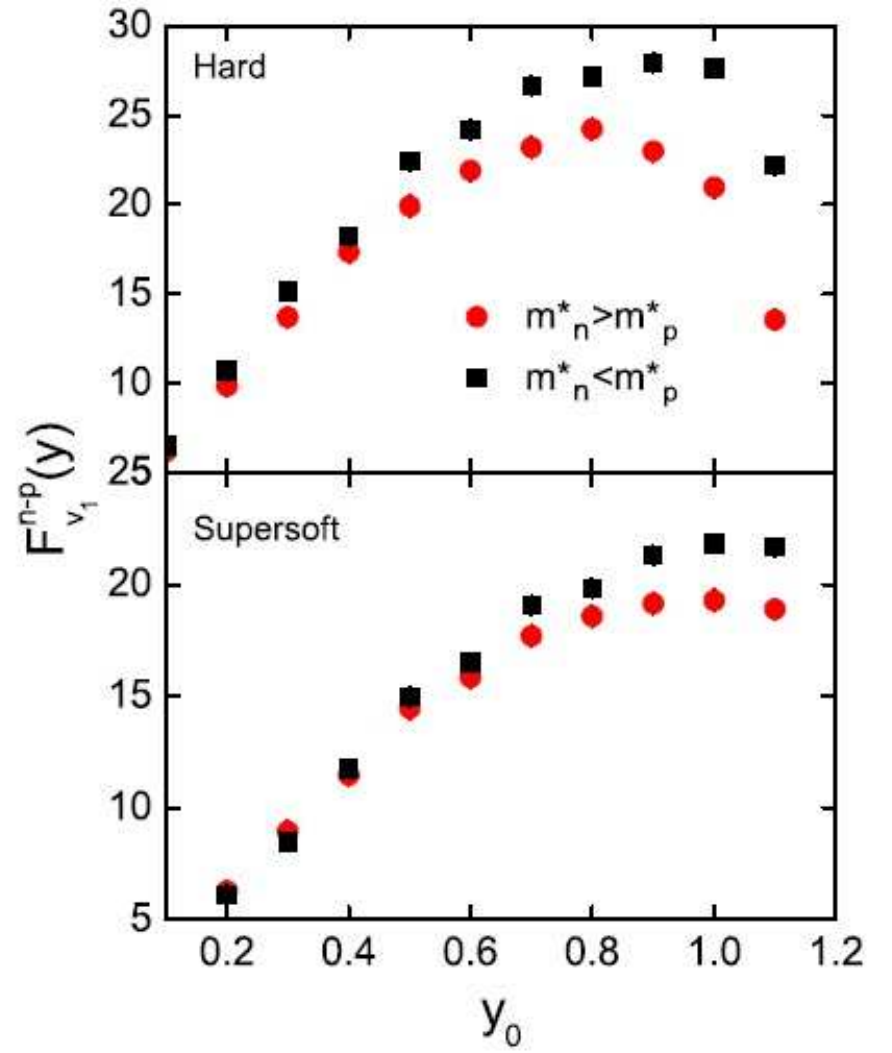}
\caption{Left: Isospin asymmetry of the gas phase as a function of time in
central $^{132}$Sn+$^{124}$Sn collisions at 50A\,MeV for two opposite choices
of mass splitting and nuclear symmetry energy. Taken from ref.\,\cite{Riz05}. Right: Rapidity dependence of neutron-proton differential directed flow for nucleons with large transverse velocities produced in mid-central $^{197}$Au+$^{197}$Au collisions at 400A\,MeV. Taken from ref.\,\cite{Xie14}.}
\label{npemitv1}
\end{figure}

The above features were further confirmed at different collision energies and from different transport models\,\cite{Riz05,Xie14}. Shown in the left part of Fig.\,\ref{npemitv1} are the time evolutions of the gas phase isospin asymmetry for different nuclear symmetry energies and neutron-proton effective mass splittings in central $^{132}$Sn+$^{124}$Sn collisions at 50A\,MeV based on simulations using a Boltzmann-Nordheim-Vlasov type transport model\,\cite{Riz05}. The gas phase is defined as nucleons with local density smaller than $\rho_0/8$. The solid horizon line shows the initial asymmetry of the system. At the early compression stage ($t<60$ fm/c), a stiffer symmetry energy leads to a higher isospin asymmetry of the gas phase, but this trend is reversed later when the low-density emission becomes important. The choice of $m_{\rm{n}}^*>m_{\rm{p}}^*$ results in the reduced emission of fast neutrons. The effect is more pronounced at the early stage ($t<60$ fm/c), when most energetic particles are emitted from the high-density phase. The right part of Fig.\,\ref{npemitv1} shows the rapidity dependence of the neutron-proton differential directed flow in mid-central $^{197}$Au+$^{197}$Au collisions at 400A\,MeV from simulations using a quantum molecular dynamics model\,\cite{Xie14}. The differential neutron-proton differential directed flow, defined as
\begin{equation}
F_{v_1}^{\rm{n-p}}(y) = \frac{N_{\rm{n}}(y)}{N(y)}v_1^{\rm{n}}(y) - \frac{N_{\rm{p}}(y)}{N(y)}v_1^{\rm{p}}(y),
\end{equation}
characterizes the difference between the neutron and proton directed flow $v_1$ weighted by their multiplicities. At 400A\,MeV when supra-saturation densities are reached in the high-density phase, a stiffer symmetry energy (upper panel) leads to a larger neutron-proton differential flow compared with a softer symmetry energy (lower panel). The case of $m_{\rm{n}}^*<m_{\rm{p}}^*$ clearly shows a larger neutron-proton differential $v_1$ compared to the $m_{\rm{n}}^*>m_{\rm{p}}^*$ case, and the difference is larger at larger reduced rapidity $y_0$.
\begin{figure}[ht]
\centering
\includegraphics[width=16cm,height=9.cm]{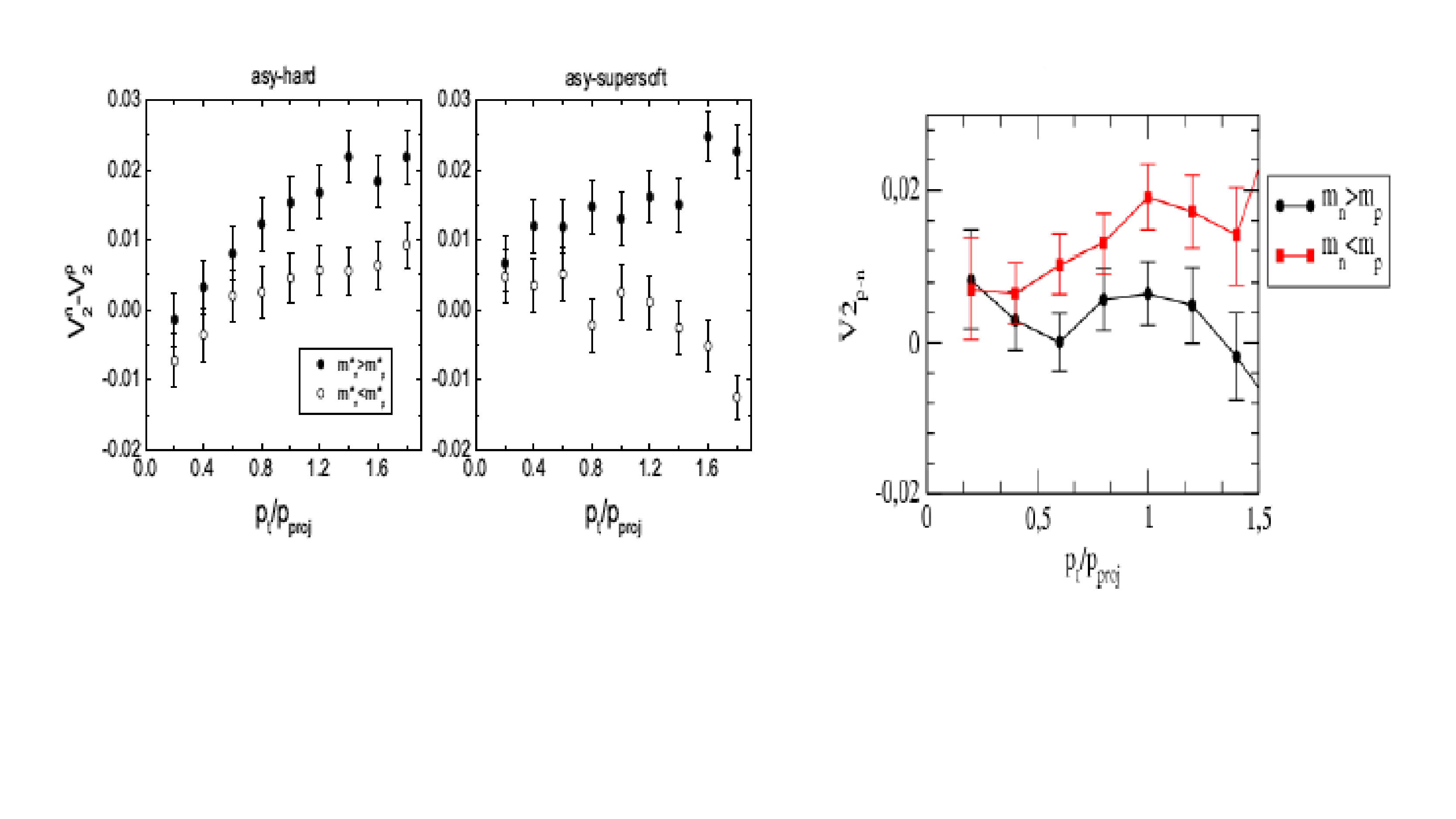}
\vspace{-2.5cm}
\caption{Left: A comparison of the difference between neutron and proton elliptic flows with the rapidity bin $|y/y_{proj}|<$0.25 for the different mass splitting in semi-central $^{197}$Au+$^{197}$Au
collisions using a quantum molecular dynamics model by the Lanzhou group \protect\cite{Feng12}. Right: Transverse momentum dependence of the difference between proton
and neutron elliptical flow flow $v_2$ at mid-rapidity ($\mid y_0 \mid < 0.3$) in semi-central Au+Au collisions at 400A\,MeV using a stochastic mean-field model with a stiff
symmetry energy by the Catania group \protect\cite{Gio10}.}
\label{v2pt}
\end{figure}

The relative elliptic flow between neutrons and protons has attracted some special attention. Shown on the left in Fig.\ \ref{v2pt} are the difference between neutron and proton elliptical flows in semi-central $^{197}$Au+$^{197}$Au collisions at 400 MeV/nucleon predicted by a quantum molecular dynamics model\,\cite{Feng12}. While on the right is the difference between the elliptical flows of protons and neutrons (note the difference in definitions) for the same reaction calculated from a stochastic mean-field model with a stiff symmetry energy\,\cite{Gio10}. Both calculations show consistently that the difference in elliptical flows of neutrons and protons are sensitive to the neutron-proton effective mass splitting. At 400A\,MeV the expansion of the participant nucleons is blocked by the spectator nucleons, so the elliptic flow is negative since free nucleons are mostly squeezed out perpendicular to the reaction plane. It is understandable that a smaller effective mass leads to larger squeeze-out effects and thus a more negative $v_2$.
\begin{figure}[htbp]
\centering
\includegraphics[scale=1.2]{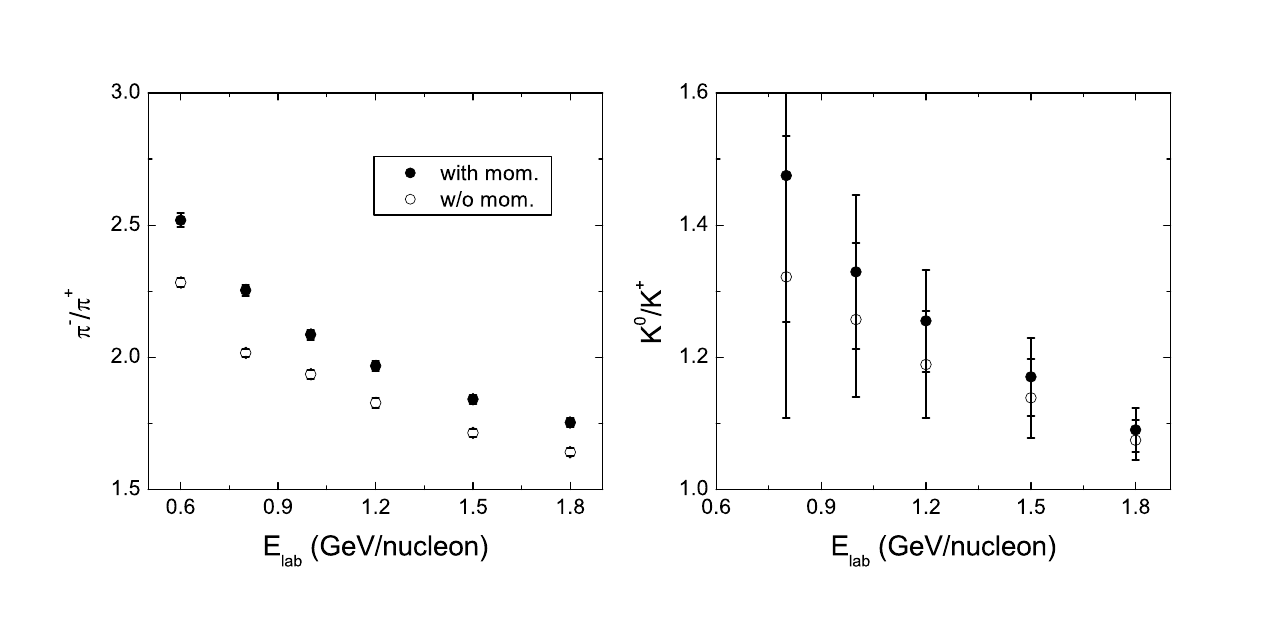}
\caption{\protect Yield ratio of $\pi^-/\pi+$ and $K^0/K+$ as a function of the beam energy in $^{197}$Au+$^{197}$Au collisions using momentum-dependent and momentum-independent symmetry potential. Taken from ref.\,\cite{Feng11prc}.}
\label{fig:Feng-pik}
\end{figure}

The momentum dependence of the mean-field potential, or the nucleon effective mass, may also affect the particle production. As shown in Fig.\,\ref{fig:Feng-pik}, the $\pi^-/\pi^+$ and $K^0/K^+$ ratio are sensitive to the momentum dependence of the symmetry potential, and the effect is larger near the threshold energy. It was argued that the neutron-proton effective mass splitting leads to more energetic neutron-neutron scatterings, resulting in the enhanced ratios of $\pi^-/\pi^+$ and $K^0/K^+$, since $\pi^-$ and $K^0$ are mostly produced by inelastic neutron-neutron scatterings in the high-density phase.

\begin{table}[htbp]
\vspace{-0.4cm}
\caption{\label{tab:table1}
Corresponding saturation properties of nuclear matter from the SLy4, SkI2, SkM*, and Gs Skyrme forces. All quantities are in MeV, except for $\rho_0$ in fm$^{-3}$ and the dimensionless effective mass ratios for nucleons, neutrons and protons. The effective mass for neutron and proton are obtained for isospin asymmetric nuclear matter with isospin asymmetry $\delta=0.2$. Taken from ref.\,\cite{Zhang14}.}
\begin{tabular}{lccccccccc}
\hline
\hline
\textrm{Para.}&
\textrm{$\rho_0$}&
\textrm{$E_0$}& \textrm{$K_0$} & \textrm{$S_0$} & \textrm{$L$} & \textrm{$K_{\rm{sym}}$} &
\textrm{$m^*/m$} &$m^*_{\rm{n}}/m$ & $m^*_{\rm{p}}/m$\\
\hline
 SLy4 & 0.160 & -15.97 & 230 &  32 & 46 & -120& 0.69& 0.68 &0.71 \\
 SkI2 & 0.158 & -15.78 & 241 &  33 & 104 & 71 & 0.68& 0.66 &0.71 \\
 SkM* & 0.160 & -15.77 & 217 &  30 & 46 & -156 & 0.79 & 0.82 &0.76 \\
 Gs   & 0.158 & -15.59 & 237 &  31 & 93 & 14 & 0.78 & 0.81 &0.76 \\
\hline
\end{tabular}
\label{Zhang}
\end{table}
It was shown that the double neutron/proton ratio is a good probe of the nuclear symmetry energy\,\cite{Li06plb,Zhang12prc}.
The interplay of the nuclear symmetry energy and the neutron-proton effective mass splitting on the ratio of free neutron/proton multiplicity has further been investigated using the Skyrme interaction based on an improved quantum molecular dynamics (ImQMD) model\,\cite{Zhang14}. In the work of Zhang et al. \,\cite{Zhang14}, four different Skyrme forces with different characteristics of the symmetry energy and neutron-proton effective mass splittings are employed. The detailed properties of these four forces at the saturation density are listed in Table.~\ref{Zhang}. The SkI2 and Gs lead to a stiffer symmetry energy compared to the SLy4 and SkM*, while the SkM* and Gs lead to $m_{\rm{n}}^*>m_{\rm{p}}^*$ and the SLy4 and SkI2 lead to $m_{\rm{n}}^*<m_{\rm{p}}^*$. As shown in the left and middle panels of Fig.\,\ref{fig:Zhang-ratio}, a softer symmetry energy and a smaller neutron effective mass than proton generally lead to a larger neutron/proton ratio. Considering the differences in $L$ and $m_{n-p}^*$ in the interactions used, it is interesting to note that the simulated results of both the single and double n/p ratio seem to indicate that effects of the neutron-proton effective mass splitting are larger than those due to the variation of the symmetry energy in the ranges explored.

\begin{figure}[h!]
\centering
\includegraphics[width=18cm,height=10cm]{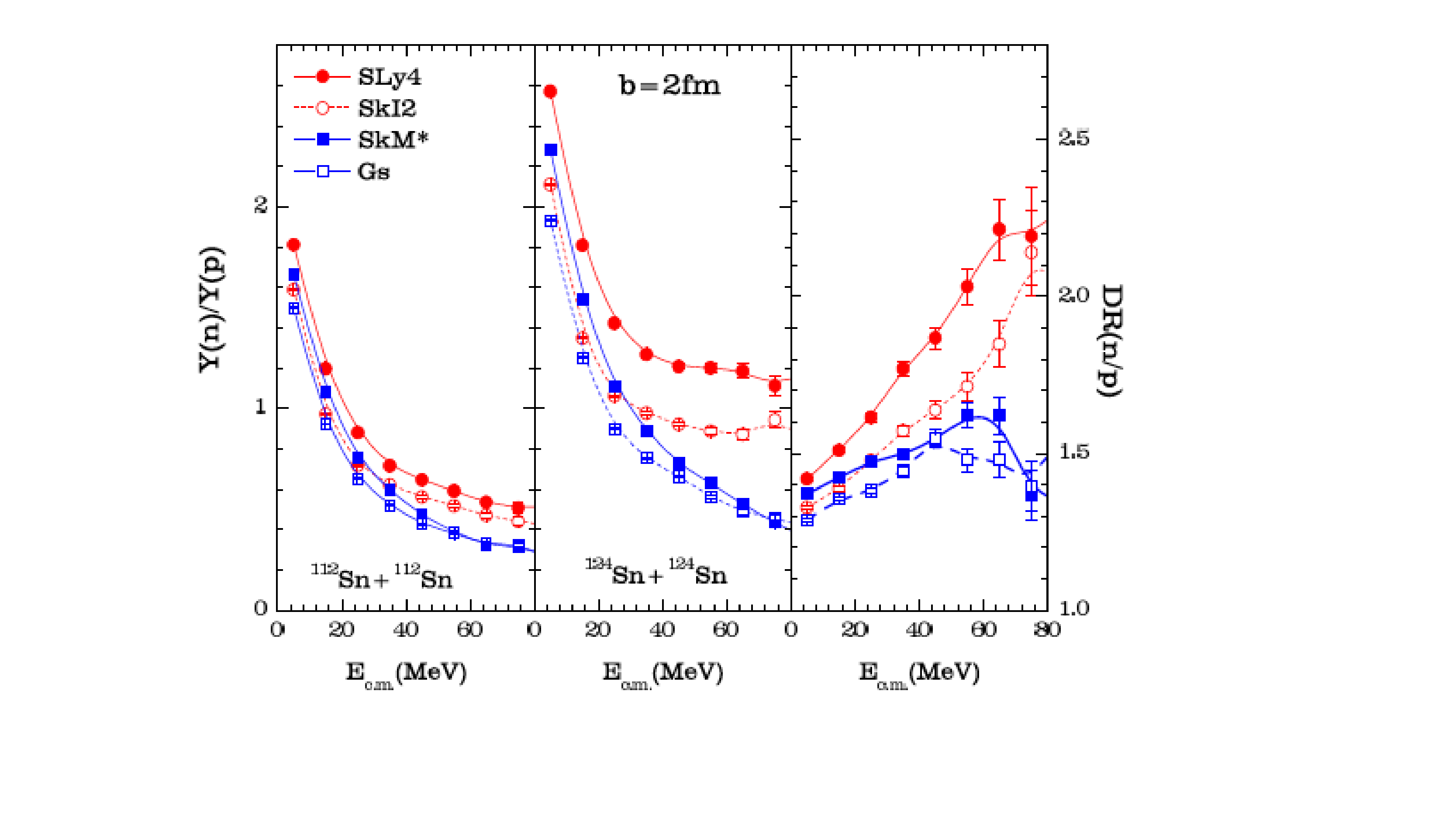}
\vspace{-1.5cm}
\caption{Left: Ratio of free neutron to proton multiplicity as a function of kinetic energy in $^{112}$Sn+$^{112}$Sn collisions at $b=2$ fm and 50A\,MeV with angular cuts $70^\circ<\theta_{\rm{c.m.}}<110^\circ$; Middle: Same as left but for $^{124}$Sn+$^{124}$Sn; Right: Double neutron/proton ratio for that from $^{124}$Sn+$^{124}$Sn to that from $^{112}$Sn+$^{112}$Sn as a function of kinetic energy. The calculated results are for SLy4 (solid circles), SkI2 (open circles), SkM* (solid squares), and Gs (open squares) Skyrme forces. Taken from ref.\,\cite{Zhang14}.}
\setlength{\belowcaptionskip}{0pt}
\label{fig:Zhang-ratio}
\end{figure}

\subsection{Recent efforts to extract the neutron-proton effective mass splitting from heavy-ion reaction data}
Given the fundamental importance of knowing accurately the momentum dependence of the symmetry potential and the corresponding neutron-proton effective mass splitting as well as the current status of transport model predictions discussed above, experiments in this field are extremely important as ultimately one has to rely on experimental data to test theories and models.
Indeed, some significant experimental efforts in this direction have been made over the last few years.  Unfortunately, to our best knowledge, no experimental observable by itself can directly tell us anything about the momentum dependence of the symmetry potential and/or the associated isovector nucleon effective mass. Comparing experimental data with model predictions (here specially transport models for heavy-ion collisions) is the only way to go forward. However, at the moment, no model can calculate completely consistently everything that has been measured. For example, most of the transport models do not describe dynamically cluster formations properly except for a few very light ones in very few codes. On the other hand, the isovector potential is very small compared to the isoscalar potential especially at high densities. Moreover, the predicted neutron-proton effective mass splitting is not only small but also proportional to the isospin asymmetry of the system. Currently, there are only few existing data from reactions not so neutron-rich that can be used to directly compare with model predictions. There are thus great challenges especially since most of the isospin-sensitive observables require simultaneous measurements of both neutrons and protons. While some interesting new physics has been learned from comparing model calculations and experimental data available, at this point no solid conclusion can be drawn about the sign of the neutron-proton effective mass splitting from these studies. As we have discussed earlier in this review, while some circumstantial evidences for a positive neutron-proton effective mass splitting  in neutron-rich matter have been accumulated from analyzing giant resonances and studying nuclear masses, etc, there is also no community consensus there even regarding the sign of neutron-proton effective mass splitting. Thus, it is not strange that the hard work of many people in the heavy-ion reaction community on this issue so far has not helped much in firmly settling it down.

\begin{figure}
\includegraphics[width=7.cm,height=8.cm]{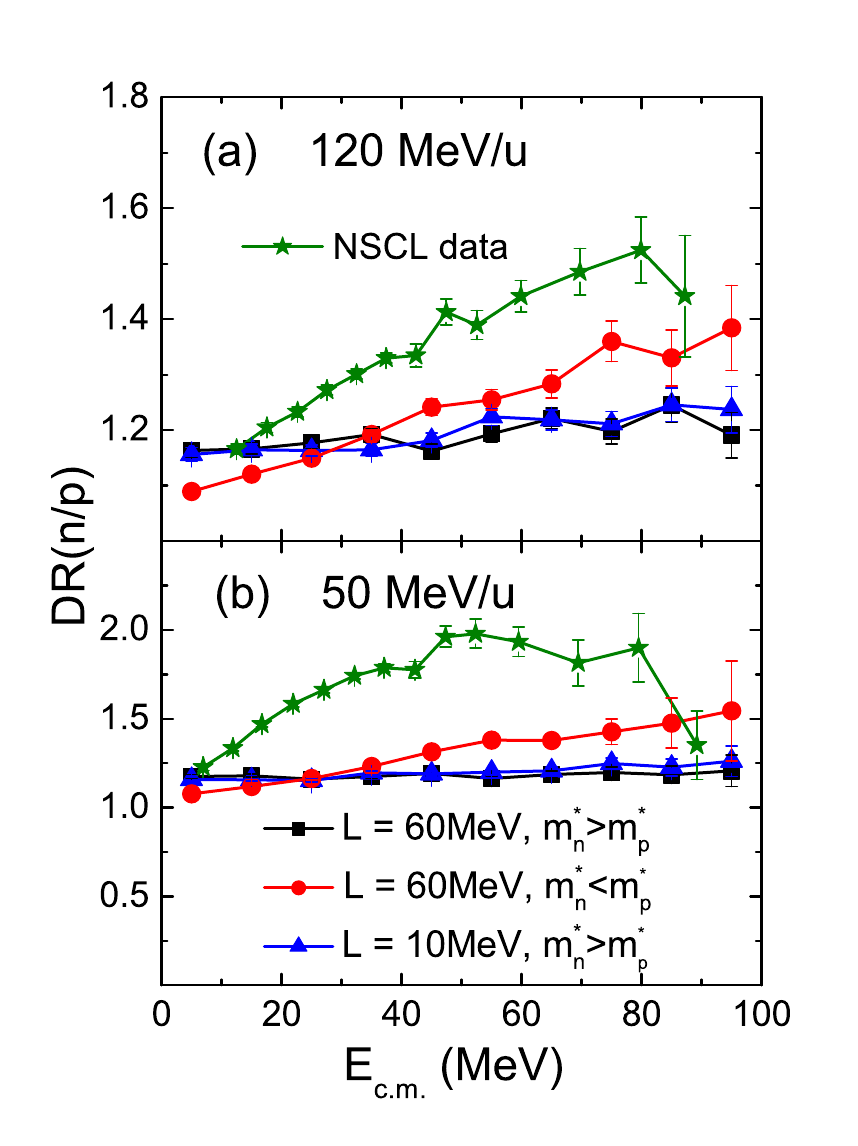}
\hspace{1.3cm}
\includegraphics[width=7cm,height=8cm]{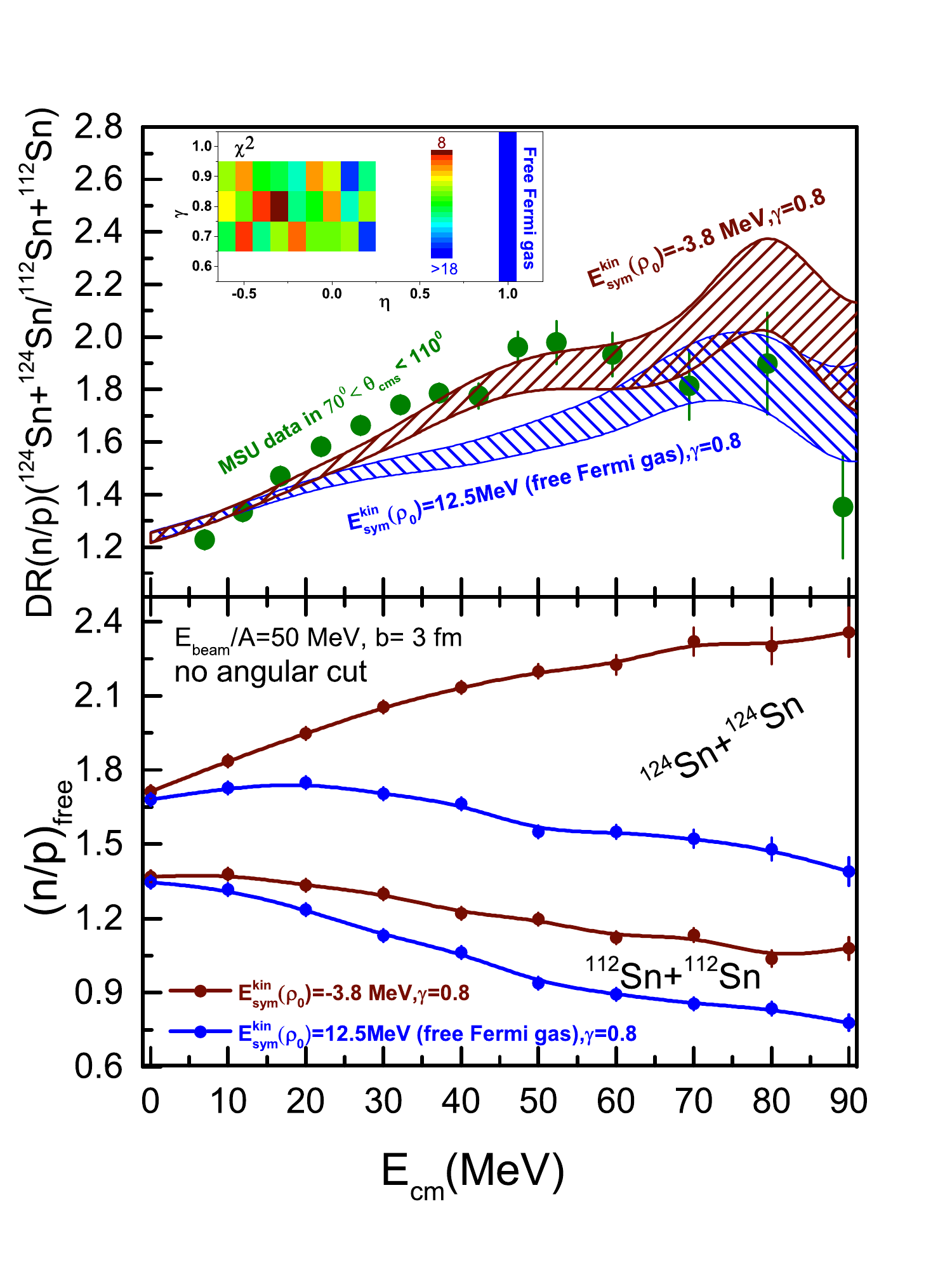}
\caption{Left: Results from IBUU11 simulation using the ImMDI interaction with different $x$ and $y$ parameters compared with the same MSU data. Taken from refs.\,\cite{Kong15prc}.
Right: The single (lower window) and double (upper window) of pre-equilibrium neutron/proton ratios in the reactions of $^{112}$Sn+$^{112}$Sn and $^{124}$Sn+$^{124}$Sn reactions at a beam energy of 50 MeV/nucleon and an impact parameter of  3 fm with different combinations of the kinetic and potential symmetry energies using the momentum-independent option of the IBUU transport model. Taken from refs. \cite{Hen14,Li15}.}
\label{fig:drnp}
\end{figure}
In the following, we make a few observations about the interesting physics extracted from a few examples of analyzing heavy-ion reactions in the efforts of constraining the neutron-proton effective mass splitting.
Earlier analyses of the free neutron/proton double ratio from central $^{124}$Sn+$^{124}$Sn and $^{112}$Sn+$^{112}$Sn collisions at 50 and 120 MeV/nucleon at the NSCL/MSU seem to indicate that protons have a slightly larger effective mass than neutrons based on comparisons with calculations using the ImQMD model\,\cite{Cou16prc}. However, the preferred Skyrme interactions predict a nucleon symmetry potential that is increasing with nucleon energy/momentum at saturation density. As we discussed extensively earlier, such a symmetry potential is in contrast to the findings from optical model analyses of nucleon-nucleus scattering data and predictions of most microscopic many-body theories. Another analysis\,\cite{Kong15prc} of the same data using the IBUU11 transport model found that indeed the assumption of $m^*_{\rm{n}}\leq m^*_{\rm{p}}$ leads to a higher neutron/proton ratio, as shown in the left part of Fig.\,\ref{fig:drnp}. Moreover, results of the IBUU11 calculations using the $E_{\rm{sym}}(\rho)$, $U_{\rm{sym,1}}(\rho,p)$ and $m^*_{\rm{n-p}}$ all within their current uncertainty ranges still under-predict significantly the NSCL/MSU data. Besides having more data with more isospin asymmetric systems, this situation clearly calls for more detailed theoretical studies with different transport models and consider possibly new mechanisms, such as the SRC effects on the symmetry potential/energy\,\cite{Hen14,Li15}. Indeed, by reducing (enhancing) the kinetic (potential) contribution to a fixed total symmetry energy consistent with the known constraints, IBUU11 calculations can increase the free neutron/proton double ratio to the level consistent with the MSU data as shown in the right window of Fig.\,\ref{fig:drnp} \cite{Hen14}. The right lower-widow shows the single neutron/proton ratios
for the reactions of $^{112}$Sn+$^{112}$Sn and $^{124}$Sn+$^{124}$Sn reactions at a beam energy of 50 MeV/nucleon and an impact parameter of  3 fm with different combinations of the kinetic and potential symmetry energies from using the momentum-independent option of the IBUU transport code. The blue lines with dots are results using the standard kinetic symmetry energy for free Fermi gas, while the brown ones are obtained by reducing the kinetic contribution to the total symmetry energy (increasing correspondingly the potential symmetry energy such that the total symmetry energy at $\rho_0$ is a constant). It is seen that the SRC-reduced kinetic symmetry energy increases
significantly the pre-equilibrium neutron/proton ratio especially for the more neutron-rich reaction system. After using the detector filter, the double neutron/proton ratios from the two reaction systems are compared in the right upper-window. It is seen that the SRC-reduced (enhanced) kinetic (potential) symmetry energy helps bring the calculations more closer to the experimental data.  We emphasize here, however, that the SRC effects have not been fully and self-consistently incorporated in the IBUU or other transport models yet. The SRC effects shown in Fig.\,\ref{fig:drnp} and those in the IBUU calculations of refs. \,\cite{Hen14,Li15} can only be used for orientation purposes.

In summary of this section, motivated by the fundamental significance of knowing the momentum dependence of the symmetry potential and the corresponding neutron-proton effective mass splitting, significant efforts have been made by the low-intermediate energy heavy-ion reaction community both theoretically and experimentally. In particular, several promising observables have been identified consistently by several groups using different transport models. However, because the symmetry potential is relatively weak compared to the isoscalar potential and most of the isospin-sensitive observables require simultaneous measurements of neutrons and charged particles, experimental measurements of these observables are challenging. Comparisons of existing data and model predictions are quite revealing and some interesting new physics has been learned. However, currently no solid conclusion
regarding the sign of the neutron-proton effective mass splitting in neutron-rich matter can be made based on these studies alone. Obviously, much more work remains to be done in this exciting field.

\section{Summary}
The multifaceted studies on nucleon effective masses in nuclear matter and their impacts on many issues in both nuclear physics and astrophysics have a very fruitful and long history. In neutron-rich matter, there are new issues related to the space-time non-locality of the isovector strong interaction. In this review, we focused on the neutron-proton effective mass splitting which is proportional to the difference of nucleon isoscalar and isovector effective masses. At the mean-field level, the neutron-proton effective mass splitting is determined by the momentum/energy dependence of the isovector (symmetry or Lane) potential which varies with density and isospin asymmetry of the medium. There are many hotly debated issues regarding the isovector nucleon effective mass or the neutron-proton effective mass splitting especially in dense neutron-rich matter. Resolutions of these issues are relevant to solving many important problems in both nuclear physics and astrophysics. In the new era of terrestrial experiments with advanced rare isotope beams as well as new astrophysical observations of neutron stars with state-of-the-art x-ray satellites and/or gravitational wave detectors,  there are great opportunities to better understand properties of the isovector strong interaction and the associated neutron-proton effective mass splitting in neutron-rich matter over a broad density range. Simultaneously, significant theoretical efforts are being devoted to predicting new physics phenomena and extracting useful information from the new experiments/observations using essentially all nuclear many-body theories and interactions available in the literature. As to the momentum/energy dependence of the symmetry potential and the corresponding neutron-proton effective mass splitting, the physics problems encountered are complicated and the model predictions are often rather diverse.

Our discussions in this review on the selected results from many interesting work done by many people are based on our limited knowledge in this very active field. Our coverages of the relevant issues are certainly incomplete and our discussions may be biased while we tried to be objective and fair. The following is a recap of the most important physics we have learned:
\begin{itemize}
\item
The momentum/energy dependence of the symmetry potential due to the space/time non-locality of isovector strong interaction is the key quantity affecting many isospin-dependent features of both nuclear structures and reactions, and our  poor knowledge about it is the fundamental cause of the diverse predictions for the isovector nucleon effective mass or the neutron-proton effective mass splitting in neutron-rich matter.

\item
Many analyses of giant resonances and masses of nuclei indicate that the nucleon isoscalar and isovector effective masses in nuclear matter at $\rho_0$  are about $m^*_{\rm{s}}/m\approx 0.8\pm 0.1$ and $m^*_{\rm{v}}/m=0.6\sim 0.93$, respectively. Their uncertainties especially for the $m^*_{\rm{v}}/m$ are still too large to determine the sign of the neutron-proton effective mass splitting in neutron-rich matter at saturation density. However, there are circumstantial evidences and credible predictions for a positive neutron-proton effective mass splitting in neutron-rich matter at saturation density based on careful analyses of different kinds of data and theories. The following table summarizes the expressions of $m_{\rm{n-p}}(\rho_0)$ either extracted from data analyses or predicted by theories well calibrated with known experimental constraints. While they all give positive values and a linear isospin dependence of $m_{\rm{n-p}}(\rho_0)$, both the magnitude and uncertainties $m_{\rm{n-p}}(\rho_0)$ are still model dependent.  
As we discussed in detail in this review, there are indeed some models/interactions predicting negative values of $m_{\rm{n-p}}(\rho_0)$. Since the corresponding symmetry potentials are then in contrast with the isovector potential from optical model analyses of huge sets of nucleon-nucleus scattering experimental data, those particular predictions for negative $m_{\rm{n-p}}(\rho_0)$ values are not summarized in the table here.

\begin{center}
\begin{table}[h!]
\centering\small
\centerline{The neutron-proton effective mass splitting $m_{\rm{n-p}}(\rho_0)$ in neutron-rich matter of isospin asymmetry $\delta$ at saturation density}\label{tabxx} \vspace*{.2cm}
\begin{tabular}{|l|l|l|}
\hline
Approach&$m_{\rm{n-p}}(\rho_0)$&Reference\\\hline
Optical model Analyses of nucleon-nucleus scattering data&$(0.41\pm0.15)\delta$&\cite{LiX15} X.H. Li {\it et al.}\\\hline
Universal nuclear energy density functional&$0.637\delta$&\cite{UNDEF} M. Kortelainen {\it et al.}\\\hline
ISGQR, IVGDR \& dipole polarizability of $^{208}$Pb using SHF+RPA&$(0.27\pm0.15)\delta$&\cite{Zha16} Z. Zhang and L.W. Chen\\\hline
ISGQR, IVGDR \& dipole polarizability of $^{208}$Pb using IBUU&$(0.216\pm0.114)\delta$&\cite{Kong17prc} K.Y. Kong {\it et al.}\\\hline
General analyses of symmetry energy using HVH theorem &$(0.27\pm0.25)\delta$&\cite{LiBA13} B.A. Li and X. Han\\\hline
Chiral effective field theory&$(0.309\pm0.227)\delta$&\cite{Hol16,Hol13} Jeremy Holt {\it et al.}\\\hline
BCPM energy functional&$0.2\delta$&\cite{Baldo16a}, M. Baldo {\it et al.}\\\hline
General analyses of energy density functional&$(0.17\pm0.24)\delta$&\cite{India17} C. Mondal {\it et al.}
\\\hline
\end{tabular}
\end{table}
\end{center}

\item
Essentially all nuclear many-body theories using basically all available two-body and three-body forces or model Lagrangians have been used to calculate the single-nucleon potential in neutron-rich nucleonic mater. The many-body theory predictions for the positive sign of $m^*_{\rm{n-p}}$ at saturation density in neutron-rich matter are rather solid. In particular, most of the calculations predict that the nucleon symmetry potential decreases with increasing nucleon energy/momentum in neutron-rich matter at $\rho_0$ qualitatively in agreement with the empirical constraint from the optical model analyses of nuclear reactions. Quantitatively, however, the predictions are still rather model and interaction dependent. Moreover, the predictions diverge quite widely at abnormal densities, especially at supra-saturation densities.

\item
The HVH theorem reveals directly and analytically the relationships among the effective masses and their own momentum dependences, symmetry potentials and symmetry energies order by order in isospin asymmetry
in both non-relativistic and relativistic frameworks. At the mean-field level, these relations are general. They can help us better understand the physics origins of the symmetry energy, neutron-proton effective
mass splitting in neutron-rich matter and their uncertainties in a model independent way. The HVH theorem also reveals the physics ingredients determining the isospin-quartic term (the fourth-order symmetry energy) in the EOS of neutron-rich matter.

\item
The extraction of nucleon E-mass from single-nucleon momentum distributions constrained by electron-nucleus and proton-nucleus scattering experiments through the application of the Migdal--Luttinger theorem is fruitful. The neutron-proton E-mass splitting is closely related to the isospin dependence of the SRC-induced HMT in single-nucleon momentum distributions. The SRC due to mainly the tensor force in the isosinglet n-p interaction channel affects the nucleon E-mass and the EOS of neutron-rich matter especially its kinetic symmetry energy. In particular, the SRC makes the symmetry energy more concave around the saturation density, leading to an isospin-dependent incompressibility of ANM in better agreement with the experimental results. Moreover, it also softens the symmetry energy at supra-saturation densities in calculations using both relativistic and non-relativistic energy density functionals. Furthermore, through the nucleon effective mass, the SRC also affects the nucleon MFP in neutron-rich matter. In particular, it enhances the nucleon MFP in SNM at saturation density by a factor of about 2. However, while neutrons are shown consistently to have a longer MFP than protons at kinetic energies below about 30 MeV, it is still unclear whether neutrons or protons have a longer MFP at higher kinetic energies in neutron-rich matter, mostly because of our poor knowledge about the isovector E-mass. It was also shown that the isospin-dependent HMT in neutron-rich matter can also affect the critical proton fraction $x^{\rm{cric}}_{\rm{p}}$ above which the direct URCA process can occur in protoneutron stars. While the $x^{\rm{cric}}_{\rm{p}}$ is about 11\% in the ${\it npe}$ matter of neutron stars without considering the HMT, the latter reduces the $x^{\rm{cric}}_{\rm{p}}$ to an average value of about 2\%.

\item
The multifaceted studies on nucleon effective masses in neutron-rich matter have multiple impacts on many issues.
For instance, the in-medium NN cross sections in neutron-rich matter are modified differently from those in SNM due to the neutron-proton effective mass splitting and its dependence on
the density and isospin asymmetry of the medium.  Because the isovector nucleon effective mass affects the occupation probability of neutrons and protons differently in neutron-rich matter, essentially all thermodynamical variables can be affected by the neutron-proton effective mass splitting. Self-consistent thermodynamical calculations have shown that the symmetry energy at finite temperatures depends sensitively on whether the single-nucleon potential is momentum dependent or not and how the neutron-proton effective mass splitting varies with density and isospin asymmetry of the medium. The neutron-proton effective mass splitting also affects the differential isospin
fractionation of liquid-gas phase transition in neutron-rich matter. Within several different approaches, the shear viscosity $\eta$ and the thermal conductivity $\kappa$ of neutron-rich matter were found to depend on the neutron-proton effective mass splitting in non-trivial ways through the isospin dependence of both the particle flux and NN in-medium scattering cross sections.

\item
Heavy-ion reactions especially those involving rare isotopes at low and intermediate energies provide the unique means to create in terrestrial laboratories dense neutron-rich matter. The dynamics and observables of these reactions
depend sensitively on the momentum dependence of the symmetry potential and the associated neutron-proton effective mass splitting through both the Vlasov term and collision integrals of transport equations.
Several observables sensitive to the neutron-proton effective mass splitting have been identified consistently by several groups using different transport models. However, for several reasons including (1) the symmetry potential is relatively weak compared to the isoscalar potential, (2) the neutron-proton effective mass splitting is small and proportional to the isospin asymmetry of the medium, and (3) most of the isospin-sensitive observables require simultaneous measurements of neutrons and charged particles accurately, experimental measurements of these observables are thus extremely challenging but certainly doable. Currently, no solid conclusion regarding even the sign of the neutron-proton effective mass splitting at saturation density has been made by comparing model predictions with available data mostly suffering from low statistics for neutrons from reactions (e.g., $^{124}$Sn+$^{124}$Sn) that are not so neutron-rich.

\end{itemize}

We have witnessed some very significant progresses made by many people working on nucleon effective masses in neutron-rich matter during the last decade. Obviously, much more work remains to be done in this multi-disciplinary area of fundamental research. There are great opportunities provided by the new facilities in both nuclear physics and astrophysics for exploring more thoroughly the nature of dense neutron-rich nucleonic matter. We are looking forward to seeing more exciting results addressing clearly all of the remaining issues regarding nucleon effective masses in neutron-rich matter in the coming decade.

\section{Acknowledgements} We would like to thank A.N. Antonov, P. Danielewicz, C.B. Das, S. Das Gupta, F.J. Fattoyev, C. Gale, W.J. Guo, O. Hen, J. W. Holt, X.T. He, W.Z. Jiang, C.M. Ko, X.H. Li, W.G. Lynch, Y.G. Ma, J.B. Natowitz, W.G. Newton, E. Piasetzky, A. W. Steiner, L.B. Weinstein, D.H. Wen, Z.G. Xiao, C. Xu, G.C. Yong and W. Zuo for helpful discussions and collaborations on some of the issues reviewed in this paper. BAL and BJC acknowledge the U.S. Department of Energy, Office of Science, under Award Number DE-SC0013702, the CUSTIPEN (China-U.S. Theory Institute for Physics with Exotic Nuclei) under the US Department of Energy Grant No. DE-SC0009971, the National Natural Science Foundation of China under Grant No. 11320101004 and the Texas Advanced Computing Center. BJC and LWC were supported in part by the Major State Basic Research Development Program (973 Program) in China under Contract Nos. 2013CB834405 and 2015CB856904, the NationalNatural Science Foundation of China under Grant Nos. 11625521, 11275125 and 11135011, the Program
for Professor of Special Appointment (Eastern Scholar) at Shanghai Institutions of Higher Learning, Key Laboratory for Particle Physics, Astrophysics and Cosmology,
Ministry of Education, China, and the Science and Technology Commission of Shanghai Municipality (11DZ2260700). JX is supported in part by the Major State Basic Research Development Program (973 Program) of China under Contract Nos. 2015CB856904 and 2014CB845401, the National Natural Science Foundation of China under Grant Nos. 11475243 and 11421505, the ``100-talent plan'' of Shanghai Institute of Applied Physics under Grant Nos. Y290061011 and Y526011011 from the Chinese Academy of Sciences, and the Shanghai Key Laboratory of Particle Physics and Cosmology under Grant No. 15DZ2272100.

\section{References}

\newcommand{\apjl}{Astrophys. J. Lett.\ }
\newcommand{\apj}{Astrophys. J. \ }
\newcommand{\prc}{Phys. Rev. C\ }
\newcommand{\prd}{Phys. Rev. D\ }
\newcommand{\mnras}{Mon. Not. R. Astron. Soc.\ }
\newcommand{\aap}{Astron. Astrophys.\ }
\newcommand{\nphysa}{Nucl. Phys. A\ }
\newcommand{\physrep}{Phys. Rep.\ }
\newcommand{\nat}{Nature\ }
\newcommand{\plb}{Phys. Lett. B\ }
\newcommand{\ijmpe}{Int. J. Mod. Phys. E\ }
\newcommand{\prl} {Phys. Rev. Lett.\ }
\newcommand{\jpg} {J. Phys. G\ }





\end{document}